\documentclass[]{report}

\usepackage{a4}
\usepackage{epsfig}
\epsfverbosetrue

\usepackage[numbers,sort&compress]{natbib}
\usepackage{amsmath}
\usepackage{amsfonts}

\newcommand{\Ent}[1]{[\mkern - 2.5 mu [#1] \mkern - 2.5 mu ]}

\def\buildrel#1\under#2{\mathrel{\mathop{\kern0pt #2}\limits_{#1}}}
\def\Res#1{{\buildrel {\scriptstyle #1} \under {\textstyle \rm Res}}\,}
\def\bbox#1{\mbox{\boldmath$#1$}}
\def\bm#1{\mbox{\boldmath$#1$}}

\def\corresponds{{\lower.2ex\hbox{=}}{\rm\kern-.75em^\triangle}}
\def\succsim{\succ\kern-.9em_\sim\kern.3em}
\def\precsim{\prec\kern-1em_\sim\kern.3em}
\def\slantfrac#1#2{\kern1em^{#1}\kern-.3em/\kern-.1em_{#2}}
\def\lfrac#1#2{{}^{#1\!}\kern-.0em/_{#2}}

\usepackage[
   a4paper,      % A4
   dvips,        % Erzeugung durch dvip
   bookmarksopen=true,
   bookmarksnumbered=true,
   pdftitle={Habilitationsschrift},
   pdfauthor={Ulrich Jentschura},
   pdfcreator={Ulrich Jentschura},
   pdfstartview={FitB},
   colorlinks,
   linkcolor=blue]{hyperref}

\begin{document}

%
%     Titel der Arbeit
%

\hrule
\vspace*{0.2cm}
\hrule
\vspace*{0.2cm}
\hrule
\vspace*{0.5cm}
\begin{center}
\sf
\color{red}
\Huge
Quantum Electrodynamic\\ Bound--State Calculations\\
and Large--Order Perturbation\\ Theory
\end{center}
\vspace*{0cm}
\begin{center}
\Large
based on a\\[2ex]
\LARGE
\color{blue}
Habilitation Thesis \\
Dresden University of Technology \\[11ex]
\color{black}
by\\
\end{center}
\vspace*{0.3cm}
\begin{center}
\color{black}
\LARGE
Ulrich Jentschura\\[3.5ex]
\color{blue}
1$^{\rm st}$ Edition June 2002\\
2$^{\rm nd}$ Edition April 2003\\{}
[with hypertext references and updates]\\
3$^{\rm rd}$ Edition April 2004\\[3ex]
\large
\color{green}
E--Mail:\\
jentschura@physik.uni-freiburg.de and ulj@nist.gov\\
\color{black}
\end{center}
\vfill
\vspace*{1cm}
\hrule
\vspace*{0.2cm}
\hrule
\vspace*{0.2cm}
\hrule

\newpage

\vspace*{7cm}

\begin{center}
\begin{minipage}{14cm}
\hrule
\vspace*{0.2cm}
\hrule
\vspace*{0.2cm}
\hrule

\vspace*{0.4cm}

\begin{center} 
This manuscript is also available -- in the form of a book --
from\\[2ex]
{\tt Shaker Verlag GmbH, Postfach 101818, 52018 Aachen,
Germany},\\
world-wide web address: {\tt http://www.shaker.de},\\
electronic-mail address: {\tt info@shaker.de}.\\[2ex]
It has been posted in the online archives with permission
of the publisher.
\end{center}

\hrule
\vspace*{0.2cm}
\hrule
\vspace*{0.2cm}
\hrule
\end{minipage}
\end{center}

\newpage

%
% Inhaltsverzeichnis
%

\tableofcontents

%
% General Overview
%

%
% Abstract 
%
\chapter*{Abstract}

This Thesis is based on quantum electrodynamic
(QED) bound-state calculations~\cite{JePa1996,JeSoMo1997,%,
JeSoIvKa1997,KaJeIvSo1998,JeSoIvKa1998,%,
KaIvJeSo1998,KaJeIvSo1998EPJ,GiEtAl1998,%,
JeMoSo1999,JeMoSo1999asil,JeMoSo1999ieee,JeMoSo2001pra,%
GoEtAl2001,SoEtAl2001,JeMoSo2001hyp,%,
JePa2002,JeSoIn2002,JeNa2002,%,
UrKrJeSo1998,JeGiVaLaWe2002,NaSaJeSo2002,LaVaJeWe2002},
as well as on investigations related to divergent series,
convergence acceleration and applications of these concepts
to physical 
problems~\cite{JeMoSoWe1999,JeBeWeSo2000,Je2000prd,JeWeSo2000,JeSo2001,
Je2001pra,JeZJ2001,JeAkMoSaSo2003,AkSaJeBeSoMo2003}.

The subjects which are discussed in this Thesis include:
the self energy of a bound electron and
the spin-dependence of QED 
corrections in bound systems (Chs.~\ref{QEDSelOne} --~\ref{QEDConclu}),
convergence acceleration techniques 
(Chs.~\ref{ConDivInt} and~\ref{ConvAccel}), and resummation methods
for divergent series with an emphasis on physical applications
(Ch.~\ref{DivIntrod}).

In {\bf Chapter~\ref{QEDSelOne},} 
we present numerical results~\cite{JeMoSo2001pra},
accurate to the level of 1~Hz for atomic hydrogen, for the
energy correction to the K and L shell states due to the
one-loop self energy. We investigate
hydrogenlike systems
with low nuclear charge number $Z=1$--$5$.
Calculations are carried out using
on-mass shell renormalization, which guarantees that
the final result is written in terms of the 
physical electron charge.
The purpose of the calculation is twofold:
first, we provide accurate  
theoretical predictions for the one-loop self energy
of K and L shell states. Second, the comparison
of the analytic and numerical approaches to the
Lamb shift calculations, which have followed separate
paths for the past few decades, is demonstrated by 
comparing the numerical values with analytic 
data obtained using the $Z\alpha$-expansion~\cite{Pa1993,JePa1996,JeSoMo1997}. 
Our calculation essentially eliminates the uncertainty due to 
the truncation of the $Z\alpha$-expansion, and 
it demonstrates the consistency 
of the numerical and analytic approaches which have 
attracted attention for more than five decades,
beginning with the seminal paper~\cite{Be1947}.
The most important numerical results are summarized 
in Table~\ref{tableFKL}.

In {\bf Chapter~\ref{QEDSelExc},} we investigate higher-order analytic
calculations of the one-photon self energy for excited atomic states.
These calculations rely on mathematical methods described in
Sec.~\ref{EpsilonMethod} which, in physical terms, lead to 
a separation of the calculation into a high- and a low-energy
part for the virtual photon. This separation does not only permit
adequate simplifications for the two energy 
regions~\cite{JePa1996,JeSoMo1997}, but  
also an adequate treatment of the infrared divergences
which plague all bound-state calculations~(see also Ch.~7 of~\cite{ItZu1980}).
The investigation represents a continuation of previous 
work on the subject~\cite{Pa1993,JePa1996,JeSoMo1997}.
The calculations are relevant for transitions to highly
excited states, which are relevant for the extraction of 
fundamental constants from the high-precision
measurements in atomic hydrogen~\cite{BeEtAl1997},
and for the estimation of QED effects in more complex atomic systems where
some of the electrons occupy highly excited states.
  
{\bf Chapter~\ref{QEDSelTwo}.} We investigate the
two-loop self energy. The calculation is based on
a generalization of the methods introduced in Sec.~\ref{EpsilonMethod}.
Historically, the two-loop self energy for bound states 
has represented a major task for theoretical evaluations.
We present an analytic 
calculation~\cite{JePa2002} of the fine-structure difference of the 
two-loop self energy for P states in atomic hydrogen in the order
$\alpha^2 \, (Z\alpha)^6 \, m_{\mathrm e}\,c^2$. This energy
difference can be parameterized by two analytic coefficients,  
known as $B_{61}$ and $B_{60}$ (see Sec.~\ref{IntroTwoLoop}). 
These coefficients describe the two-loop self energy
in the sixth order in $Z\alpha$, with an 
additional enhancement due to a large logarithm
$\ln[(Z\alpha)^{-2}]$ (in the case of the $B_{61}$-coefficient). 
The calculations are relevant for an improved theoretical
understanding of the fine-structure in hydrogenlike systems.
They are also in part relevant for current
experiments on atomic
helium~\cite{MyTa1999,MyEtAl1999,StGeHe2000,GeLoHe2001}, 
whose motivation is the determination of the fine-structure constant with 
improved accuracy. Finally, it is hoped that numerical
calculations of the two-loop effect will be carried
out in the near future for which the current analytic
evaluation will be an important consistency check.
The calculation of the analytic corrections discussed in
Sec.~\ref{QEDSelTwo} represents
a solution for the most problematic set of diagrams on the way
to advance our understanding of the fine-structure in atomic hydrogen 
to the few-Hz level.
Explicit results for the fine-structure difference of the 
$B_{61}$- and $B_{60}$-coefficients can be found in 
Eqs.~(\ref{final61}) -- (\ref{final25P}).

In {\bf Chapter~\ref{QEDSpiZer},} We investigate the spin-dependence
of the Breit hamiltonian and quantum electrodynamic effects
in general. Specifically, we consider a bound system of two
spinless particles. The calculation is motivated in part
by current experiments (DIRAC at CERN) whose aim is the experimental
study of pionium, which is the bound system of two (spinless)
oppositely charged pions. The
evaluation of the two-body relativistic
corrections of order $(Z\alpha)^4$ leads to a different
result~\cite{JeSoIn2002} than expected for a system of 
two spin-$1/2$ particles
of equal mass, e.g.~positronium. In particular, we conclude in 
Sec.~\ref{BreitEffects} that the
so-called zitterbewegung term is absent for a system of
two spinless particles
(the absence of the zitterbewegung term in 
a bound system consisting {\em of a spinless and a spin-$1/2$ 
particle} was previously pointed out in~\cite{PaKa1995}).
Final results for the relativistic correction, the
vacuum polarization, and the self energy of a system of two 
scalar particles can be found in Eqs.~(\ref{NLresult}),
(\ref{resVP}) and (\ref{reslog}). A summary of the 
QED calculations is provided in {\bf Chapter~\ref{QEDConclu}}.

In the second part of this Thesis, we discuss methods for 
{\em accelerating the convergence} of series, 
and for the {\em resummation of divergent
series}. As discussed in {\bf Chapter~\ref{ConDivInt}} and 
Sec.~\ref{BriefSurvey},
convergence acceleration is essentially based on the idea
that information hidden
in trailing digits of elements of the sequence can be used in order
to make ``educated guesses'' regarding the remainder term, which can
be used for the construction of powerful sequence transformations
(see Sec.~\ref{NonlinearST}).
In {\bf Chapter~\ref{ConvAccel}}, we discuss convergence acceleration in detail.
After a short overview of relevant mathematical methods 
(Sec.~\ref{ConvAccelConc}),
we discuss applications in applied biophysics (Sec.~\ref{sec_stat}), 
in experimental mathematics (Sec.~\ref{sec_math}),
and other applications, mainly in the evaluation of 
special functions (Sec.~\ref{sec_other}).
In particular, Secs.~\ref{sec_stat} --~\ref{sec_other} illustrate
how the combined nonlinear-condensation
transformation (CNCT) described in Sec.~\ref{CNCT}
can be used for the accelerated evaluation of
nonalternating series, with an emphasis to applications
of practical significance (Sec.~\ref{sec_stat}).

The discussion on convergence acceleration in Ch.~\ref{ConvAccel}
is complemented in {\bf Chapter~\ref{DivIntrod}} by an overview
of resummation techniques and relevant applications.
Using the Stark effect and the associated autoionization width
as a paradigmatic example, we discuss basic concepts
like the Borel resummation method and its generalizations (Sec.~\ref{sec_bp}), 
and the conformal mapping of the complex plane (Sec.~\ref{sec_co}).
We then proceed to discuss further applications of resummation 
methods like zero-dimensional model theories (Sec.~\ref{degen}), 
the QED effective action (expressed as a perturbation series in
the fine structure constant, see Sec.~\ref{DivQEDEff}), and 
the quantum-mechanical double-well problem (Sec.~\ref{DivMATIns}).
Within the context of the 
double-well problem, we perform an analytic evaluation
of higher-order corrections to the two-instanton effect
and demonstrate the consistency of numerically determined
energy levels with the instanton expansion.

We conclude with a summary of the results in 
{\bf Chapter~\ref{Conclu}},
where we also explain the interrelations and connections
between the different subjects treated in this Thesis.

%
% Part 1: QED
%
%
% QED Bound--State Calculations
%
\part{Quantum Electrodynamic Bound--State Calculations}

\typeout{}
\typeout{Part 1.}
\typeout{================================================}
\typeout{Quantum Electrodynamic Bound--State Calculations}
\typeout{================================================}
\typeout{}

\label{part1}

%
% One--Loop Self--Energy
%
\chapter{Numerical Calculation of the One--Loop Self Energy (Excited States)}

\typeout{==================================================}
\typeout{Numerical Calculation of the One--Loop Self Energy}
\typeout{==================================================}
\typeout{}

\label{QEDSelOne}
 
%
% Orientation
%
\section{Orientation}

A nonperturbative numerical evaluation of the one-photon electron self
energy for the K- and L-shell states of hydrogenlike ions with nuclear
charge numbers $Z=1$ to 5 is described. Our calculation for the $1{\rm
S}_{1/2}$ state has a numerical uncertainty of 0.8~Hz in atomic
hydrogen, and for the L-shell states ($2{\rm S}_{1/2}$, $2{\rm
P}_{1/2}$, and $2{\rm P}_{3/2}$) the numerical uncertainty is 1.0~Hz.
The method of evaluation for the ground state and for the excited
states is described in detail. The numerical results are compared to
results based on known terms in the expansion of the self energy in
powers of $Z\alpha$.

%
% Introduction
%
\section{Introduction to the Numerical Calculation of Radiative Corrections}
\label{IntroNumCalc}

The nonperturbative numerical evaluation of radiative corrections to
bound-state energy levels is interesting for two reasons. First, the
recent dramatic increase in the accuracy of experiments that measure
the transition frequencies in hydrogen and
deuterium~\cite{BeEtAl1997,UdEtAl1997,NiEtAl2000} necessitates a numerical
evaluation (nonperturbative in the binding Coulomb field) of the
radiative corrections to the spectrum of atomic systems with low
nuclear charge $Z$. Second, the numerical calculation serves as an
independent test of analytic evaluations which are based on an
expansion in the binding field with an expansion parameter $Z\alpha$.

In order to address both issues, a high-precision numerical evaluation
of the self energy of an electron in the ground state in hydrogenlike
ions has been performed~\cite{JeMoSo1999,JeMoSo2001pra,Je1999}. The approach
outlined in~\cite{JeMoSo1999} is generalized here to the L shell, and
numerical results are obtained for the ($n=2$) states $2{\rm
S}_{1/2}$, $2{\rm P}_{1/2}$ and $2{\rm P}_{3/2}$. Results are provided
for atomic hydrogen, ${\rm He}^{+}$, ${\rm Li}^{2+}$, ${\rm Be}^{3+}$,
and ${\rm B}^{4+}$.

It has been pointed out in~\cite{JeMoSo1999,Je1999} that the
nonperturbative effects (in $Z\alpha$) can be large even for
low nuclear charge and exceed the current experimental accuracy for 
atomic transitions. For example, the difference between the sum of the
analytically evaluated terms up to the order of $\alpha \,
(Z\alpha)^6$ and the final numerical result for the ground state is
roughly 27~kHz for atomic hydrogen and about 3200~kHz for ${\rm
He}^{+}$.  For the $2{\rm S}$ state the difference is $3.5$~kHz for
atomic hydrogen and 412~kHz for ${\rm He}^{+}$. The large difference
between the result obtained by an expansion in $Z\alpha$ persists even
after the inclusion of a result recently obtained in~\cite{Ka1997} for
the logarithmic term of order $\alpha\,(Z\alpha)^7\,\ln(Z\alpha)^{-2}$. 
For the ground state, the difference between the all-order numerical
result and the sum of the perturbative terms is still 13~kHz for
atomic hydrogen and 1600~kHz for ${\rm He}^{+}$. For the $2\,{\rm S}$
state, the difference amounts to $1.6$~kHz for atomic hydrogen and to
213~kHz for ${\rm He}^{+}$.

These figures should be compared to the current experimental
precision. The most accurately measured transition to date is the
1S--2S frequency in hydrogen; it has been measured with to a
relative accuracy of $1.8$ parts in $10^{14}$ or $46~{\rm
Hz}$~\cite{NiEtAl2000}. This experimental progress is due in part to
the use of frequency chains that bridge the range between optical
frequencies and the microwave cesium time standard. The accuracy of
the measurement is likely to be improved by an order of magnitude in
the near future~\cite{NiEtAl2000,HaPr2000}.  With trapped hydrogen atoms, it
should be feasible to observe the 1S--2S frequency with an
experimental linewidth that approaches the $1.3\,{\rm Hz}$ natural
width of the 2S level \cite{CeEtAl1996,KiEtAl1998}.

The apparent convergence of 
the perturbation series in $Z\alpha$ is slow. Our 
all-order numerical calculation presented here essentially
eliminates the uncertainty from unevaluated higher-order analytic
terms, and we obtain results for the self energy remainder function
$G_{\rm SE}$ with a precision of roughly $0.8 \times Z^4 \, {\rm Hz}$
for the ground state of atomic hydrogen and $1.0 \times Z^4 \, {\rm
Hz}$ for the $2{\rm S}$ state.

In the evaluation, we take advantage of resummation and convergence
acceleration techniques. The resummation techniques provide an
efficient method of 
evaluation of the Dirac-Coulomb Green function to a relative
accuracy of $10^{-24}$ over a wide parameter range~\cite{Je1999}. The
convergence acceleration techniques remove the principal numerical
difficulties associated with the singularity of the relativistic
propagators for nearly equal radial arguments~\cite{JeMoSoWe1999}.

The one-photon self energy is
about two orders of magnitude larger than the other contributions
to the Lamb shift in atomic hydrogen. Comprehensive reviews of the various
contributions to the Lamb shift in hydrogenlike atoms in the full
range of nuclear charge numbers $Z=1$--$110$ have been given
in Refs.~\cite{JoSo1985,MoPlSo1998,Mo1996,EiGrSh2001}.

%
% Method of Evaluation
%
\section{Method of Evaluation}
\label{MethodOfEvaluation}

%
% Status of Analytic Calculations
%
\subsection{Status of Analytic Calculations}
\label{StatusAnalytic}

The (real part of the) energy shift $\Delta E_{\rm SE}$ due to the
electron self energy radiative correction is usually written as
\begin{equation}
\label{ESEasF}
\Delta E_{\rm SE} = \frac{\alpha}{\pi} \, \frac{(Z \alpha)^4}{n^3} \, 
F(nl_j,Z\alpha) \, m_{\rm e} \, c^2
\end{equation}
where $F$ is a dimensionless quantity. In the following, the natural
unit system with $\hbar = c = m_{\rm e} = 1$ and $e^2 = 4\pi\alpha$ is
employed.  Note that $F(nl_j,Z\alpha)$ is a dimensionless function
which depends for a given atomic state with quantum numbers $n$, $l$
and $j$ on only one argument (the coupling $Z\alpha$).  For excited
states, the (nonvanishing) imaginary part of the self energy
is proportional to the (spontaneous) decay width of the state. We will
denote here the {\em real} part of the self energy by $\Delta E_{\rm
SE}$, exclusively.  The semi-analytic expansion of $F(nl_j,Z\alpha)$
about $Z\alpha=0$ for a general atomic state with quantum numbers $n$,
$l$ and $j$ gives rise to the expression,
\begin{eqnarray}
\label{defFLO}
F(nl_j,Z\alpha) &=& 
A_{41}(nl_j) \, \ln(Z \alpha)^{-2} \nonumber\\[2ex]
& & \;\; + A_{40}(nl_j) +
(Z \alpha) \, A_{50}(nl_j) \nonumber\\[2ex]
& & \;\; + \,
(Z \alpha)^2 \,
\left[A_{62}(nl_j) \, \ln^2(Z \alpha)^{-2} \right.
\nonumber\\[2ex]
& & \;\; \left. + A_{61}(nl_j) \,\ln(Z \alpha)^{-2} +
G_{\rm SE}(nl_j,Z\alpha) \right]\,.
\end{eqnarray}
For particular states, some of the coefficients may vanish.  Notably,
this is the case for P states, which are less singular than S states
at the origin [see Eq.~(\ref{defFLOnPj}) below].  For the $n{\rm
S}_{1/2}$ state ($l=0$, $j=1/2$), none of the terms in
Eq.~(\ref{defFLO}) vanishes, and we have,
\begin{eqnarray}
\label{defFLOnS}
F(n{\rm S}_{1/2},Z\alpha) &=&
A_{41}(n{\rm S}_{1/2}) \, \ln(Z \alpha)^{-2} \nonumber\\[2ex]
& & \;\; + A_{40}(n{\rm S}_{1/2}) +
(Z \alpha) \, A_{50}(n{\rm S}_{1/2}) \nonumber\\[2ex]
& & \;\; + \, 
(Z \alpha)^2 \, 
\left[A_{62}(n{\rm S}_{1/2}) \, \ln^2(Z \alpha)^{-2} \right. 
\nonumber\\[2ex]
& & \;\; \left. + A_{61}(n{\rm S}_{1/2}) \,\ln(Z \alpha)^{-2} + 
G_{\rm SE}(n{\rm S}_{1/2},Z\alpha) \right]\,.
\end{eqnarray}
The $A$ coefficients have two indices, the first of which denotes the
power of $Z\alpha$ [including those powers implicitly contained in
Eq.~(\ref{ESEasF})], while the second index denotes the power of the
logarithm $\ln(Z \alpha)^{-2}$.  For P states, the coefficients
$A_{41}$, $A_{50}$ and $A_{62}$ vanish, and we have
\begin{eqnarray}
\label{defFLOnPj}
F(n{\rm P}_j,Z\alpha) =
A_{40}(n{\rm P}_j) + (Z \alpha)^2 \,
\left[A_{61}(n{\rm P}_j) \, \ln(Z \alpha)^{-2} +
G_{\rm SE}(n{\rm P}_j,Z\alpha) \right]\,.
\end{eqnarray}
For S states, the self energy remainder function $G_{\rm SE}$ can be
expanded semi-analytically as
\begin{eqnarray}
\label{DefinitionOfA601S}
G_{\rm SE}(n{\rm S}_{1/2},Z\alpha) = A_{60}(n{\rm S}_{1/2})
+ (Z\alpha)\,\left[A_{71}(n{\rm S}_{1/2})\,\ln(Z \alpha)^{-2}
+ A_{70}(n{\rm S}_{1/2}) + {\rm o}(Z\alpha)\right]\,.
\end{eqnarray}
For the ``order'' symbols o and O we follow the usual 
convention~\cite{WhWa1944,Er1987}: the requirement is
${\rm O}(x)/x \to {\rm const.}$ as $x \to 0$, whereas ${\rm o}(x)$
fulfills the {\em weaker} requirement ${\rm o}(x) \to 0$ as
$x \to 0$. For example, the expression $[(Z\alpha)\,\ln(Z\alpha)]$ is
${\rm o}(Z\alpha)$ but {\em not} ${\rm O}(Z\alpha)$.
Because logarithmic terms corresponding to a nonvanishing 
$A_{83}$-coefficient must be expected in Eq.~(\ref{DefinitionOfA601S}),
the symbol 
${\rm o}(Z\alpha)$ should be used to characterize the remainder,
{\em not} ${\rm O}(Z\alpha)$. 

For P states, the semi-analytic expansion of
$G_{\rm SE}$ reads
\begin{eqnarray}
\label{A60nP}
G_{\rm SE}(n{\rm P}_j,Z\alpha) = A_{60}(n{\rm P}_j)
(Z\alpha)\,\left[A_{70}(n{\rm P}_j) + {\rm o}(Z\alpha)\right]\,.
\end{eqnarray}
The fact that $A_{71}(n{\rm P}_j)$ vanishes has been pointed out
in~\cite{Ka1997}.  We list below the analytic coefficients and the
Bethe logarithms relevant to the atomic states under
investigation. For the ground state, the coefficients $A_{41}$ and
$A_{40}$ were obtained
in~\cite{Be1947,Fe1948,Fe1949,FrWe1949,KrLa1949,Sc1949,FuMiTo1949},
the correction term $A_{50}$ was found
in~\cite{Ba1951,KaKlSc1952,BaBeFe1953}, and the higher-order binding
corrections $A_{62}$ and $A_{61}$ were evaluated
in~\cite{FrYe1958,FrYe1960,La1960,La1961a,La1961b,%,
ErYe1965a,ErYe1965b,Er1971,Sa1981,Pa1993}. The results are,
\begin{eqnarray}
\label{coeffs1S12}
A_{41}(1{\rm S}_{1/2}) & = & \frac{4}{3}\,, \nonumber\\[1ex]
A_{40}(1{\rm S}_{1/2}) & = & 
  \frac{10}{9} - \frac{4}{3} \, \ln k_0(1{\rm S})\,, 
\nonumber\\[1ex]
A_{50}(1{\rm S}_{1/2}) & = & 
4\pi\,\left[\frac{139}{128} - \frac{1}{2}\,\ln 2\right]\,,
\nonumber\\[1ex]
A_{62}(1{\rm S}_{1/2}) & = & -1\,, \nonumber\\[1ex]
A_{61}(1{\rm S}_{1/2}) & = & \frac{28}{3} \, \ln 2 - \frac{21}{20}\,.
\end{eqnarray}
The Bethe logarithm $\ln k_0(1{\rm S})$ has been evaluated 
in~\cite{KlMa1973}
and \cite{BeBrSt1950,Ha1956,ScTi1959,Li1968,Hu1969} as
\begin{equation}
\label{BetheLog1S}
\ln k_0(1{\rm S}) = 2.984~128~555~8(3).
\end{equation}
For the 2S state, we have
\begin{eqnarray}
\label{coeffs2S12}
A_{41}(2{\rm S}_{1/2}) & = & \frac{4}{3}\,, \nonumber\\[1ex]
A_{40}(2{\rm S}_{1/2}) & = &
  \frac{10}{9} - \frac{4}{3} \, \ln k_0(2{\rm S})\,,
\nonumber\\[1ex]
A_{50}(2{\rm S}_{1/2}) & = &
4\pi\,\left[\frac{139}{128} - \frac{1}{2}\,\ln 2\right]\,,
\nonumber\\[1ex]
A_{62}(2{\rm S}_{1/2}) & = & -1\,, \nonumber\\[1ex]
A_{61}(2{\rm S}_{1/2}) & = & \frac{16}{3} \, \ln 2 + \frac{67}{30}\,.
\end{eqnarray}
The Bethe logarithm $\ln k_0(2{\rm S})$ has been evaluated
(see~\cite{KlMa1973,BeBrSt1950,Ha1956,ScTi1959,Li1968,Hu1969},
the results exhibit varying accuracy) as
\begin{equation}
\label{BetheLog2S}
\ln k_0(2{\rm S}) = 2.811~769~893(3).
\end{equation}
It might be worth noting that
the value for $\ln k_0(2{\rm S})$ given in~\cite{Er1977} evidently
contains a typographical error. Our independent re-evaluation
confirms the result given in Eq.~(\ref{BetheLog2S}), which was
originally obtained in~\cite{KlMa1973} to the required precision. For
the $2{\rm P}_{1/2}$ state we have
\begin{eqnarray}
\label{coeffs2P12}
A_{40}(2{\rm P}_{1/2}) & = &
  -\frac{1}{6} - \frac{4}{3} \, \ln k_0(2{\rm P})\,,
\nonumber\\[1ex]
A_{61}(2{\rm P}_{1/2}) & = & \frac{103}{108}\,.
\end{eqnarray}
Note that a general analytic result for the logarithmic correction
$A_{61}$ as a function of the bound state quantum numbers $n$, $l$ and
$j$ can be inferred from Eq.~(4.4a) of~\cite{ErYe1965a,ErYe1965b} upon
subtraction of the vacuum polarization contribution implicitly
contained in the quoted equation. The Bethe logarithm for the 2P
states reads~\cite{KlMa1973,DrSw1990}
\begin{equation}
\label{BetheLog2P}
\ln k_0(2{\rm P}) = -0.030~016~708~9(3)\,.
\end{equation}
Because the Bethe logarithm is an inherently nonrelativistic
quantity, it is spin-independent and therefore independent
of the total angular momentum $j$ for a given orbital angular momentum
$l$. For the $2{\rm
P}_{3/2}$ state the analytic coefficients are
\begin{eqnarray}
\label{coeffs2P32}
A_{40}(2{\rm P}_{3/2}) & = &
  \frac{1}{12} - \frac{4}{3} \, \ln k_0(2{\rm P})\,,
\nonumber\\[1ex]
A_{61}(2{\rm P}_{3/2}) & = & \frac{29}{90}\,.
\end{eqnarray}
We now consider the limit of the function $G_{\rm SE}(Z\alpha)$ as
$Z\alpha \to 0$.  The higher-order terms in the potential expansion
(see Fig.~\ref{ExactVExpansion} below) and relativistic corrections to
the wavefunction both generate terms of higher order in $Z\alpha$
which are manifest in Eq.~(\ref{defFLO}) in the form 
of the nonvanishing function $G_{\rm SE}(Z\alpha)$
which summarizes the effects of the relativistic corrections
to the bound electron wave function
and of higher-order terms in the potential expansion. For very soft virtual
photons, the potential expansion fails and generates an infrared
divergence which is cut off by the atomic momentum scale,
$Z\alpha$. This cut-off for the {\em infrared} divergence is one of
the mechanisms which lead to the logarithmic terms in
Eq.~(\ref{defFLO}). Some of the nonlogarithmic terms in relative order
$(Z\alpha)^2$ in Eq.~(\ref{defFLO}) are generated by the relativistic
corrections to the wave function. The function $G_{\rm SE}$ does not
vanish, but approaches a constant in the limit $Z\alpha\to 0$. This
constant can be determined by analytic or semi-analytic calculations;
it is referred to as the $A_{60}$ coefficient, i.e.
\begin{equation}
A_{60}(nl_j) = G_{\rm SE}(nl_j,0)\,.
\end{equation}
The evaluation of the coefficient $A_{60}(1{\rm S}_{1/2})$ has drawn
a lot of attention for a long time~\cite{ErYe1965a,ErYe1965b,%,
Er1971,Sa1981,Pa1993}.
For the 2S state, there is currently only one accurate analytic result
available,
\begin{equation}
\label{DiffA602S12}
A_{60}(2{\rm S}_{1/2}) = -31.840\,47(1) \qquad
  \mbox{(see Ref.~\cite{Pa1993}).}
\end{equation}
For the $2{\rm P}_{1/2}$ state, the analytically obtained result is
\begin{equation}
\label{A602P12}
A_{60}(2{\rm P}_{1/2}) = -0.998\,91(1) \qquad
  \mbox{(see Ref.~\cite{JePa1996}),} 
\end{equation}
and for the $2{\rm P}_{3/2}$ state, we have 
\begin{equation}
\label{A602P32}
A_{60}(2{\rm P}_{3/2}) = -0.503\,37(1) \qquad
  \mbox{(see Ref.~\cite{JePa1996}),}  
\end{equation}
The analytic evaluations essentially rely on an expansion of the
relativistic Dirac-Coulomb propagator in powers of the binding field,
i.e.~in powers of Coulomb interactions of the electron with the
nucleus.  In numerical evaluations, the binding field is treated
nonperturbatively, and no expansion is performed.

%
% Formulation of the Numerical Problem
%
\subsection{Formulation of the Numerical Problem}
\label{FormulationNumerical}

Numerical cancellations are severe for small nuclear charges.  In
order to understand the origin of the numerical cancellations it is
necessary to consider the renormalization of the self energy.  The
renormalization procedure postulates that the self energy is
essentially the effect on the bound electron due to the self
interaction with its own radiation field, minus the same effect on a
free electron which is absorbed in the mass of the electron and
therefore not observable. 
The self energy of the bound electron is the residual
effect obtained after the subtraction of two large quantities. Terms
associated with renormalization counterterms are of order $1$ in the
$Z\alpha$-expansion, whereas the residual effect is of order
$(Z\alpha)^4$ [see Eq.~(\ref{ESEasF})]. This corresponds to a loss of
roughly $9$ significant digits at $Z=1$. Consequently, even the
precise evaluation of the one-photon self energy in a Coulomb field
presented in \cite{Mo1992} extends only down to $Z = 5$. Among the
self energy corrections in one-loop and higher-loop order, numerical
cancellations in absolute terms are most severe for the {\em one}-loop
problem because of the large size of the effect of the one-loop
self energy correction on the spectrum.

%
% Figure: Integration Contour
%
\begin{figure}[htb]
\begin{center}
\begin{minipage}{14.0cm}
\centerline{\mbox{\epsfysize=9.0cm\epsffile{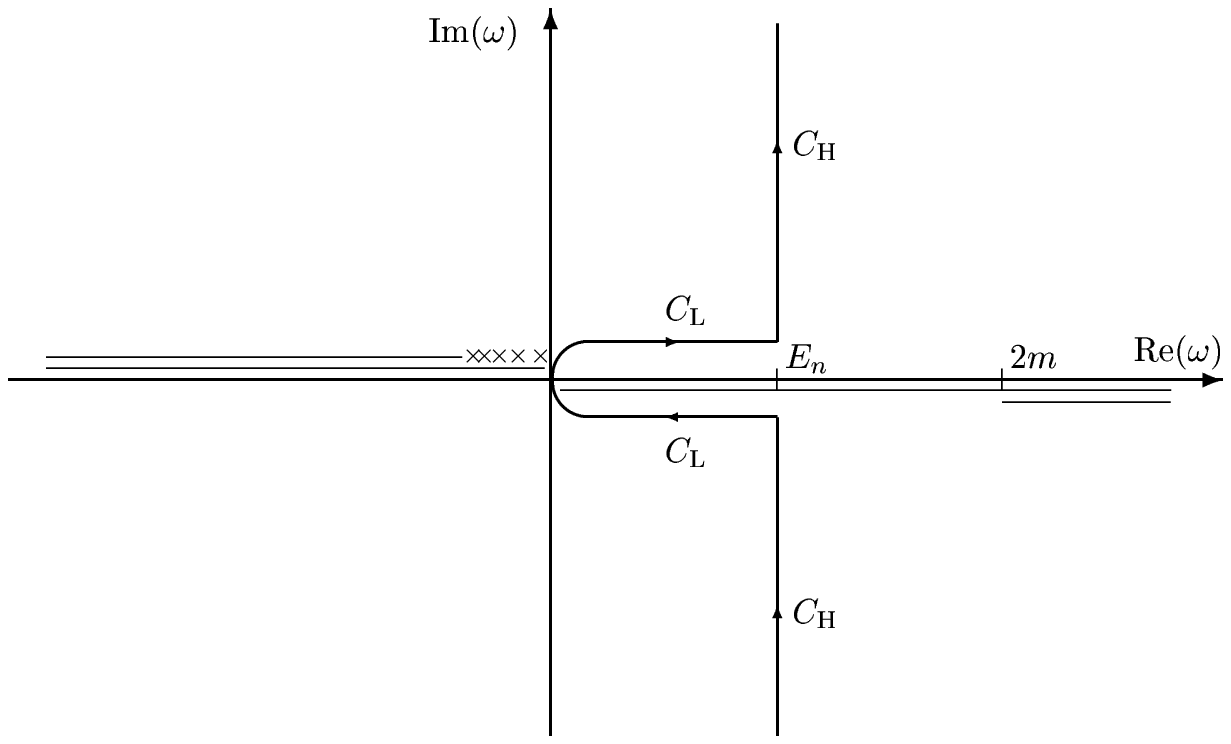}}}
\caption{\label{IntegrationContour}
Integration contour ${\cal C}$ for the integration over the energy
$\omega = E_n - z$ of the virtual photon. The contour ${\cal C}$
consists of the low-energy contour $C_{\rm L}$ and the high-energy
contour $C_{\rm H}$. Lines shown displaced
directly below and above the real axis
denote branch cuts from the photon and electron propagator. Crosses
denote poles originating from the discrete spectrum of the electron
propagator.  The contour used in this work corresponds to the one used
in~{\protect\cite{Mo1974a}}.}
\end{minipage}
\end{center}
\end{figure}

For our high-precision numerical evaluation, we start from the
regularized and renormalized expression for the one-loop self energy
of a bound electron,
\begin{eqnarray}
\label{deltaESEM}
\Delta E_{\rm SE} &=& \lim_{\Lambda\to\infty}
\left\{ {\rm i}\, e^2 \, {\rm Re} \,
\int_{C_{\rm F}} \! \frac{d\omega}{2 \pi} 
\int \! \! \frac{d^3 \bbox{k}}{(2 \pi)^3} \,
D_{\mu\nu}(k^2,\Lambda) \right.  \nonumber\\[1ex]
& & \left. \times \left< \bar{\psi} \left| \gamma^{\mu} \,
  \frac{1}{\not{\! p} - \not{\! k} - 1 - \gamma^0 V }
    \gamma^{\nu} \, \right| \psi \right>  - \Delta m \right\} 
\nonumber\\[2ex]
&=& \lim_{\Lambda\to\infty}
\bigg\{ {\rm -i}\, e^2 \, {\rm Re} \,
\int_{\cal C} \! \frac{d\omega}{2 \pi}
\int \! \! \frac{d^3 \bbox{k}}{(2 \pi)^3} \,
D_{\mu\nu}(k^2,\Lambda)  \nonumber\\[1ex]
& & \times \left< \psi \left| \, \alpha^{\mu} \,\,
{\rm e}^{{\rm i}\bbox{\scriptstyle k}\cdot \bbox{\scriptstyle x}} \,\,
G(E_n - \omega) \, \alpha^{\nu} \,\, 
{\rm e}^{-{\rm i}\bbox{\scriptstyle k}\cdot \bbox{\scriptstyle x}} \,\,
\right| \psi \right>  - \Delta m \bigg\}\,,
\end{eqnarray}
where $G$ denotes the Dirac-Coulomb propagator,
\begin{equation}
\label{DefinitionOfG}
G(z) \; = \; \frac{1}{\bbox{\alpha}\cdot \bbox{p}
  + \beta + V - z }\,,
\end{equation}
and $\Delta m$ is the $\Lambda$-dependent 
(cutoff-dependent) one-loop mass-counter term,
\begin{equation}
\label{MassCounterTerm}
\Delta m = \frac{\alpha}{\pi} \,
\left(\frac{3}{4}\,\ln \Lambda^2 + \frac{3}{8} \right) \,
\langle \beta \rangle\,.
\end{equation}
The photon propagator $D_{\mu\nu}(k^2,\Lambda)$ in
Eq.~(\ref{deltaESEM}) in Feynman gauge reads
\begin{equation}
\label{PhotonPropagator}
D_{\mu\nu}(k^2,\Lambda) = 
  - \left(\frac{g_{\mu\nu}}{k^2 + {\rm i}\,\epsilon} - 
          \frac{g_{\mu\nu}}{k^2 - \Lambda^2 + {\rm i}\,\epsilon}\right)\,.
\end{equation}
The contour $C_{\rm F}$ in Eq.~(\ref{deltaESEM}) is the Feynman
contour, whereas the contour $\cal C$ is depicted in
Fig.~\ref{IntegrationContour}.  The contour $\cal C$ is employed for
the $\omega$-integration in the current evaluation [see the last line
of Eq.~(\ref{deltaESEM})]. The energy variable $z$ in
Eq.~(\ref{DefinitionOfG}) therefore assumes the value
\begin{equation} 
\label{DefOfZ}
z = E_n - \omega\,,
\end{equation}
where $E_n$ is the Dirac energy of the atomic state, and $\omega$
denotes the complex-valued energy of the virtual photon. It is
understood that the limit $\Lambda~\to~\infty$ is taken {\em after}
all integrals in Eq.~(\ref{deltaESEM}) are evaluated.

%
% Figure: Separation of low-energy part
%
\begin{figure}[thb!]
\begin{center}
\begin{minipage}{15.0cm}
\centerline{\mbox{\epsfysize=7.0cm\epsffile{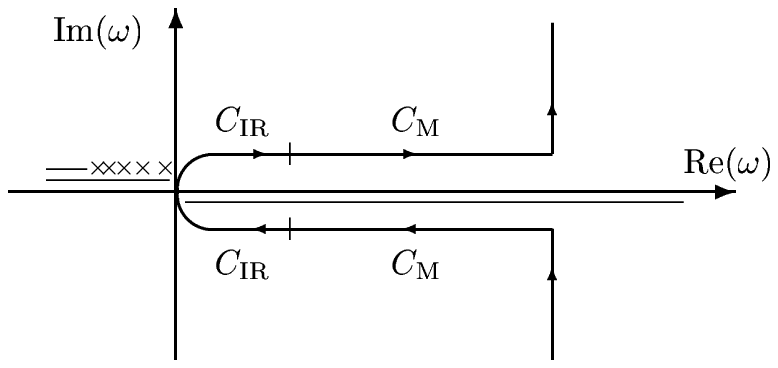}}}
\caption{\label{SeparationLow} 
Separation of the low-energy contour $C_{\rm L}$ into the infrared
part $C_{\rm IR}$ and the middle-energy part $C_{\rm M}$.  As in
Fig.~\ref{IntegrationContour}, the lines directly above and below the
real axis denote branch cuts from the photon and electron
propagator. Strictly speaking, the figure is valid only for the ground
state. For excited states, some of the crosses, which denote poles
originating from the discrete spectrum of the electron propagator, are
positioned to the right of the line ${\rm Re} \, \omega = 0$.  These
poles are subtracted in the numerical evaluation.}
\end{minipage}
\end{center}
\end{figure}

The integration contour for the complex-valued energy of the virtual
photon $\omega$ in this calculation is the contour $\cal C$ employed
in \cite{Mo1974a,Mo1974b,Mo1982,Mo1992} and depicted in
Fig.~\ref{IntegrationContour}.  The integrations along the low-energy
contour $C_{\rm L}$ and the high-energy contour $C_{\rm H}$ in
Fig.~\ref{IntegrationContour} give rise to the low- and the
high-energy contributions $\Delta E_{\rm L}$ and $\Delta E_{\rm H}$ to
the self energy, respectively. Here, we employ a further separation of
the low-energy integration contour $C_{\rm L}$ into an infrared
contour $C_{\rm IR}$ and a middle-energy contour $C_{\rm M}$ shown in
Fig.~\ref{SeparationLow}.  This separation gives rise to a separation
of the low-energy part $\Delta E_{\rm L}$ into the infrared part
$\Delta E_{\rm IR}$ and the middle-energy part $\Delta E_{\rm M}$,
\begin{equation}
\Delta E_{\rm L}=\Delta E_{\rm IR}+\Delta E_{\rm M}\,.
\end{equation}
For the low-$Z$ systems discussed here, all complications which arise
for excited states due to the decay into the ground state are
relevant only for the infrared part.  Except for the further
separation into the infrared and the middle-energy part, the same
basic formulation of the self energy problem as in~\cite{Mo1974a} is
used.  This leads to the following separation,
\[
\begin{array}{ll}
\omega \in (0,\lfrac{1}{10}\,E_n) \pm {\rm i}\,\epsilon & :  
\mbox{infrared part $\Delta E_{\rm IR}$,} \\[1ex]
\omega \in (\lfrac{1}{10}\,E_n, E_n) \pm {\rm i}\,\epsilon & : 
\mbox{middle-energy part $\Delta E_{\rm M}$,} \\[1ex]
\omega \in E_n + {\rm i}\,(-\infty,+\infty) & :  
\mbox{high-energy part $\Delta E_{\rm H}$.}
\end{array}
\]
Integration along these contours gives rise to the infrared, the
middle-energy, and the high-energy contributions to the energy
shift. For all of these contributions, lower-order terms are
subtracted in order to obtain the contribution to the self energy
of order $(Z\alpha)^4$. We obtain for the infrared part,
\begin{eqnarray}
\label{DefinitionOfFIR}
\Delta E_{\rm IR} &=& \frac{\alpha}{\pi} \,
\left[
\frac{21}{200} \langle \beta \rangle +
\frac{43}{600} \langle V \rangle +
\frac{(Z\alpha)^4}{n^3} \, F_{\rm IR}(nl_j,Z\alpha) \right]\,,
\end{eqnarray}
where $F_{\rm IR}(nl_j,Z\alpha)$ is a dimensionless function of order
one. The middle-energy part is recovered as
\begin{eqnarray}
\label{DefinitionOfFM}
\Delta E_{\rm M} &=& \frac{\alpha}{\pi} \,
\left[
\frac{279}{200} \langle \beta \rangle +
\frac{219}{200} \langle V \rangle +
\frac{(Z\alpha)^4}{n^3} \, F_{\rm M}(nl_j,Z\alpha)
\right]\,,
\end{eqnarray}
and the high-energy part reads~\cite{Mo1974a,Mo1974b}
\begin{eqnarray}
\label{DefinitionOfFH}
\Delta E_{\rm H} &=& \Delta m + \frac{\alpha}{\pi} \,
\left[ - \frac{3}{2} \langle \beta \rangle
- \frac{7}{6} \langle V \rangle 
+ \frac{(Z \alpha)^4}{n^3} \, F_{\rm H}(nl_j,Z\alpha)
\right]\,.
\end{eqnarray}
The infrared part is discussed in Sec.~\ref{InfraRed}. The
middle-energy part is divided into a middle-energy subtraction term
$F_{\rm MA}$ and a middle-energy remainder $F_{\rm MB}$. The
subtraction term $F_{\rm MA}$ is discussed in
Sec.~\ref{MiddleEnergySubtraction}, the remainder term $F_{\rm MB}$ is
treated in Sec.~\ref{MiddleEnergyRemainder}. We recover the
middle-energy term as the sum
\begin{equation}
\label{DecompositionOfFM}
F_{\rm M}(nl_j,Z\alpha) \; = \;
F_{\rm MA}(nl_j,Z\alpha) +
F_{\rm MB}(nl_j,Z\alpha)\,.
\end{equation}
A similar separation is employed for the
high-energy part. The high-energy part is divided into a subtraction
term $F_{\rm HA}$, which is evaluated in
Sec.~\ref{HighEnergySubtraction}, and the high-energy remainder
$F_{\rm HB}$, which is discussed in
Sec.~\ref{HighEnergyRemainder}. The sum of the subtraction term and
the remainder is
\begin{equation}
\label{DecompositionOfFH}
F_{\rm H}(nl_j,Z\alpha) \; = \;
F_{\rm HA}(nl_j,Z\alpha) +
F_{\rm HB}(nl_j,Z\alpha)\,.
\end{equation}
The total energy shift is given as
\begin{eqnarray}
\label{net}
\Delta E_{\rm SE} & = &
\Delta E_{\rm IR} + \Delta E_{\rm M} + E_{\rm H} - \Delta m 
\nonumber\\[2ex]
& = & \frac{\alpha}{\pi} \frac{(Z\alpha)^4}{n^3} \,
\left[F_{\rm IR}(nl_j,Z\alpha) +
F_{\rm M}(nl_j,Z\alpha) + 
F_{\rm H}(nl_j,Z\alpha)\right]\,.
\end{eqnarray}
The scaled self energy function $F$ defined in Eq.~(\ref{ESEasF}) is 
therefore obtained as
\begin{eqnarray}
F(nl_j,Z\alpha) &=& F_{\rm IR}(nl_j,Z\alpha)
+ F_{\rm M}(nl_j,Z\alpha) + F_{\rm H}(nl_j,Z\alpha)\,.
\end{eqnarray}
In analogy to the approach described in~\cite{Mo1974a,Mo1982,Mo1992},
we define the low-energy part as the sum of the infrared part and the
middle-energy part,
\begin{eqnarray}
\label{DefinitionOfFL}
\Delta E_{\rm L} & = & \Delta E_{\rm IR} + \Delta E_{\rm M} 
\nonumber\\[1ex]
& = & \frac{\alpha}{\pi}
\left[\frac{3}{2} \langle \beta \rangle +
\frac{7}{6} \langle V \rangle +
\frac{(Z \alpha)^4}{n^3} \, F_{\rm L}(nl_j,Z\alpha)\right]\,,
\end{eqnarray}
where
\begin{equation}
\label{FLasSum}
F_{\rm L}(nl_j,Z\alpha) = 
F_{\rm IR}(nl_j,Z\alpha) +
F_{\rm M}(nl_j,Z\alpha)\,.
\end{equation}
The limits for the functions $F_{\rm L}(nl_j,Z\alpha)$ and 
$F_{\rm H}(nl_j,Z\alpha)$ as $Z\alpha \to 0$ were obtained
in~\cite{Mo1974b,Mo1973phd,Je1999}.

%
% Treatment of the divergent terms
%
\subsection{Treatment of the divergent terms}
\label{RegularizationPrescription}

\begin{figure}[htb]
\begin{center}
\begin{minipage}{14.0cm}
\centerline{\mbox{\epsfysize=7.0cm\epsffile{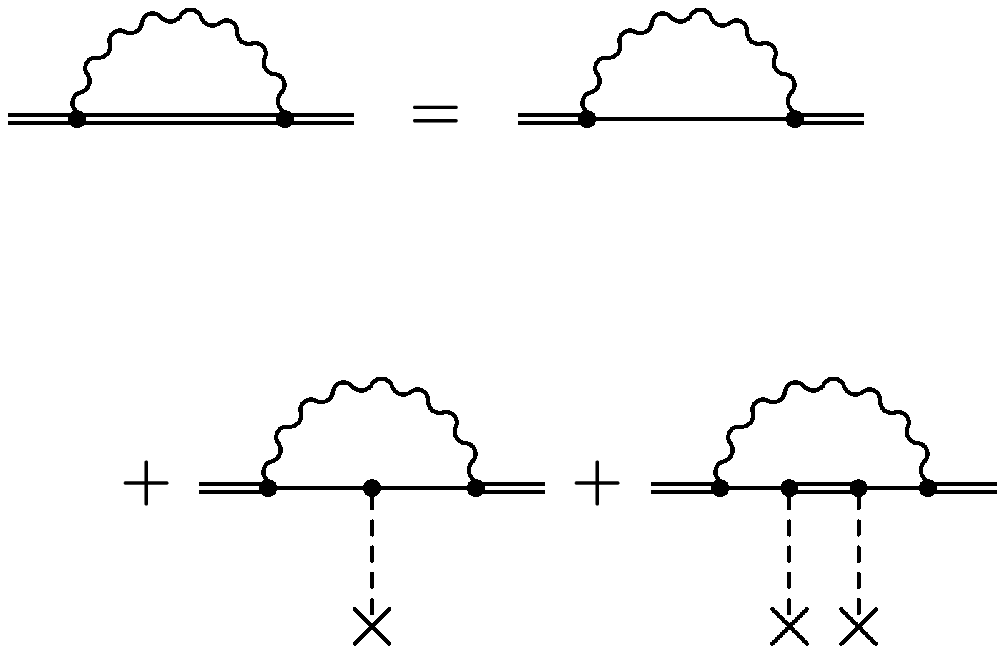}}}
\caption{\label{ExactVExpansion}
The exact expansion of the bound electron propagator in powers of the
binding field leads to a zero-potential, a one-potential and a
many-potential term.  The dashed lines denote Coulomb photons, the
crosses denote the interaction with the (external) binding field.}
\end{minipage}
\end{center}
\end{figure}

The free electron propagator,
\begin{equation}
\label{DefinitionOfF}
F = \frac{1}{\bbox{\alpha} \cdot \bbox{p}
  + \beta - z }\,,
\end{equation}
and the full electron propagator $G$ defined in
Eq.~(\ref{DefinitionOfG}), fulfill the following identity which is of
particular importance for the validity of the method used in the
numerical evaluation of the all-order binding correction to the Lamb
shift,
\begin{equation}
\label{ExactExpansion}
G = F - F\,V\,F + F\,V\,G\,V\,F\,.
\end{equation}
This identity leads naturally to a separation of the one-photon
self energy into a zero-vertex, a single-vertex and a many-vertex term.
This is represented diagrammatically in Fig.~\ref{ExactVExpansion}.

%
% Figure: ExactVExpansion
%

All ultraviolet divergences which occur in the one-photon problem
(mass counter term and vertex divergence) are generated by the
zero-vertex and the single-vertex terms. The many-vertex term is
ultraviolet safe.  Of crucial importance is the observation that one
may additionally simplify the problem by replacing the one-potential
term with an approximate expression in which the potential is
``commuted to the outside''. The approximate expression generates all
divergences and all terms of lower order than $\alpha\,(Z\alpha)^4$
present in the one-vertex term. Unlike the raw one-potential term, it
is amenable to significant further simplification and can be reduced
to {\em one}-dimensional numerical integrals which can be evaluated
easily (a straightforward formulation of the self energy problem
requires a {\em three}-dimensional numerical integration).
Without this significant improvement, an all-order calculation would
be much more difficult at low nuclear charge, because the lower-order
terms would introduce significant further numerical cancellations.

Furthermore, the special approximate resolvent can
be used effectively for an efficient subtraction
scheme in the middle-energy part of the calculation.  In the infrared
part, such a subtraction is not used because it would introduce
infrared divergences.

We now turn to the construction of the special approximate resolvent,
which will be referred to as $G_{\rm A}$ and will be used in this
calculation to isolate the ultraviolet divergences in the high-energy
part (and to provide subtraction terms in the middle-energy
part). It is based on an approximation to the first two terms on the
right-hand side of Eq.~(\ref{ExactExpansion}).  The so-called
one-potential term $FVF$ in Eq.~(\ref{ExactExpansion}) is
approximated by an expression in which the potential terms $V$ are
commuted to the outside,
\begin{equation}
- FVF \approx -\frac{1}{2}\,\left\{V,F^2\right\}\,.
\end{equation}
Furthermore, the following identity is used,
\begin{eqnarray}
F^2 &=& 
  \left( \frac{1}{\bbox{\alpha}\cdot\bbox{p} + \beta - z} \right)^2 
\nonumber\\[1ex]
&=& \frac{1}{\bbox{p}^2 + 1 - z^2} +
\frac{2\,z\,(\beta + z)}{\left(\bbox{p}^2 + 1 - z^2\right)^2} 
\nonumber\\[1ex]
& & \quad\quad\quad\quad\quad\quad\quad\quad\quad
  + \frac{2\,z\,(\bbox{\alpha}\cdot\bbox{p})}
  {\left(\bbox{p}^2 + 1 - z^2\right)^2}\,.
\end{eqnarray}
In $2\times2$ spinor space, this expression may be divided into a
diagonal and a non-diagonal part. The diagonal part is
\begin{equation}
{\rm diag}(F^2) = \frac{1}{\bbox{p}^2 + 1 - z^2} +
\frac{2\,z\,(\beta + z)}{\left(\bbox{p}^2 + 1 - z^2\right)^2}\,.
\end{equation}
The off-diagonal part is given by
\[
F^2 - {\rm diag}(F^2) = \frac{2\,z\,(\bbox{\alpha}\cdot\bbox{p})}
  {\left(\bbox{p}^2 + 1 - z^2\right)^2}\,.
\]
We define the resolvent $G_{\rm A}$ as
\begin{equation}
\label{DefinitionOfGA}
G_{\rm A} = F - \frac{1}{2}\,\left\{V,{\rm diag}\left(F^2\right)\right\}\,.
\end{equation}
All divergences which occur in the self energy are generated by the
simplified propagator $G_{\rm A}$.  We define the propagator $G_{\rm
B}$ as the difference of $G$ and $G_{\rm A}$,
\begin{eqnarray}
\label{DefinitionOfGB}
G_{\rm B} &=& G - G_{\rm A} \nonumber\\[1ex]
&=& \frac{1}{2}\,\left\{V,{\rm diag}(F^2)\right\} - F\,V\,F 
    + F\,V\,G\,\,V\,F\,.
\end{eqnarray}
$G_{\rm B}$ does not generate any divergences and leads to the
middle-energy remainder discussed in Sec.~\ref{MiddleEnergyRemainder}
and the high-energy remainder (Sec.~\ref{HighEnergyRemainder}).

%
% The Low-Energy Part
%
\section{The Low-Energy Part}
\label{LowEnergyPart}

%
% The Infrared Part
%
\subsection{The Infrared Part}
\label{InfraRed}

The infrared part is given by
\begin{eqnarray}
\label{deltaEIR}
\lefteqn{\Delta E_{\rm IR} = 
{\rm -i}\, e^2 \, {\rm Re} \,
\int_{C_{\rm IR}} \! \frac{{\rm d}\omega}{2 \pi}
\int \! \! \frac{{\rm d}^3 \bbox{k}}{(2 \pi)^3} \,
D_{\mu\nu}(k^2)} \nonumber\\[2ex]
& & \times \left< \psi \left| \, \alpha^{\mu} \,
{\rm e}^{{\rm i}\bbox{\scriptstyle k}\cdot \bbox{\scriptstyle x}} \,\,
G(E_n - \omega) \, \alpha^{\nu} \,\,
{\rm e}^{-{\rm i}\bbox{\scriptstyle k}\cdot \bbox{\scriptstyle x}} \,\,
\right| \psi \right> \,,
\end{eqnarray}
where relevant definitions of the symbols can be found in
Eqs.~(\ref{deltaESEM}--\ref{PhotonPropagator}), the contour $C_{\rm
IR}$ is as shown in Fig.~\ref{SeparationLow}, and the unregularized
version of the photon propagator
\begin{equation}
\label{PhotonPropagatorUnreg}
D_{\mu\nu}(k^2) = 
- \frac{\displaystyle g_{\mu\nu}}{\displaystyle k^2 + {\rm i}\,\epsilon} 
\end{equation}
may be used. The infrared part comprises the following integration
region for the virtual photon (contour $C_{\rm IR}$
in Fig.~\ref{SeparationLow}),
\begin{equation}
\left.
\begin{array}{l}
\omega \in \left(0, \lfrac{1}{10}\,E_n\right) \pm {\rm i}\,\epsilon \\[1ex]  
z \in \left(\lfrac{9}{10}\,E_n, E_n\right) \pm {\rm i}\,\epsilon
\end{array}
\right\}
\mbox{infrared part $\Delta E_{\rm IR}$}\,.
\end{equation}
Following Secs.~2 and 3 of~\cite{Mo1974a}, we write $\Delta E_{\rm
IR}$ as a three-dimensional integral [see, e.g., Eqs.~(3.4), (3.11)
and (3.14) {\em ibid.}]
\begin{eqnarray}
\label{defEIR}
\Delta E_{\rm IR} = \frac{\alpha}{\pi} \, \frac{E_n}{10} 
- \frac{\alpha}{\pi} \, ({\rm P. V.}) \, 
\int_{\lfrac{9}{10}\,E_n}^{E_n} {\rm d}z 
\int_0^\infty {\rm d} x_1 \, x_1^2 
\int_0^\infty {\rm d} x_2 \, x_2^2 \,\, {\cal M}_{\rm IR}(x_2,x_1,z)\,,
\end{eqnarray}
where
\begin{eqnarray}
\label{defMIR}
{\cal M}_{\rm IR}(x_2,x_1,z) =
\sum_\kappa \sum_{i,j=1}^{2} 
f_{\bar \imath}(x_2) \,
G^{ij}_{\kappa}(x_2,x_1,z) \, 
f_{\bar \jmath}(x_1) \, 
A^{ij}_{\kappa}(x_2,x_1)\,.
\end{eqnarray}
Here, the quantum number $\kappa$ is the Dirac angular quantum number
of the intermediate state, 
\begin{equation}
\label{DefOfkappa}
\kappa = 2 \, (l-j) \, (j+1/2)\,,
\end{equation}
where $l$ is the orbital angular momentum quantum number and $j$ is
the total angular momentum of the bound electron. The functions
$f_i(x_2)$ ($i=1,2$) are the radial wave functions defined in
Eq.~(A.4) in~\cite{Mo1974a} for an arbitrary bound state (and in
Eq.~(A.8) in~\cite{Mo1974a} for the 1S state). We define 
${\bar \imath} = 3 - i$. The functions $G^{ij}_{\kappa}(x_2,x_1,z)$ ($i,j=1,2$)
are the radial Green functions, which result from a decomposition of
the electron Green function defined in Eq.~(\ref{DefinitionOfG}) into
partial waves.  The explicit formulas are given in Eq.~(A.16)
in~\cite{Mo1974a}, and we do not discuss them in any further detail,
here.

The photon angular functions $A^{ij}_{\kappa}$ ($i,j=1,2$) are defined
in Eq.~(3.15) of Ref.~\cite{Mo1974a} for an arbitrary bound state. In
Eq.~(3.17) in~\cite{Mo1974a}, specific formulas are given for the 1S
state.  In Eqs.~(2.2), (2.3) and (2.4) of~\cite{Mo1982}, the
special cases of ${\rm S}_{1/2}$, ${\rm P}_{1/2}$ and ${\rm P}_{3/2}$
states are considered. Further relevant formulas for excited states
can be found in~\cite{MoKi1992}. The photon angular functions depend
on the energy argument $z$, but this dependence is usually
suppressed. The summation over $\kappa$ in Eq.~(\ref{defMIR}) extends
over all negative and all positive integers, excluding zero.  We
observe that the integral is symmetric under the interchange of the
radial coordinates $x_2$ and $x_1$, so that
\begin{eqnarray}
\label{IntegralEIR}
\Delta E_{\rm IR} = \frac{\alpha}{\pi} \, \frac{E_n}{10} - 
\frac{2\,\alpha}{\pi} \, ({\rm P. V.}) 
\int_{\lfrac{9}{10}\,E_n}^{E_n} {\rm d}z 
\int_0^\infty {\rm d} x_1 \, x_1^2 
\int_0^{x_1} {\rm d} x_2 \, x_2^2 \,\, {\cal M}_{\rm IR}(x_2,x_1,z)\,.
\end{eqnarray}
The following variable substitution,
\begin{equation}
\label{DefinitionOfR}
r = x_2/x_1\,,\;\;\; y = a\,x_1\,,
\end{equation}
is made, so that $r \in (0,1)$ and $y \in (0,\infty)$.  The scaling
variable $a$ is defined as
\begin{equation}
\label{DefinitionOfA}
a = 2 \,\sqrt{1-E_n^2}\,.
\end{equation}
The Jacobian is
\begin{equation}
\left|\frac{\partial(x_2,x_1)}{\partial(r,y)}\right| =
\left| \begin{array}{cc}
\frac{\textstyle \partial x_2}{\textstyle \partial r} & 
\frac{\textstyle \partial x_1}{\textstyle \partial r} \\[2ex]
\frac{\textstyle \partial x_2}{\textstyle \partial y} & 
\frac{\textstyle \partial x_1}{\textstyle \partial y} 
\end{array} \right| = \frac{y}{a^2}\,.
\end{equation}
The function $S_{\rm IR}$ is given by,
\begin{eqnarray}
\label{defSIR}
S_{\rm IR}(r,y,z) &=& - \frac{2 \, r^2 \, y^5}{a^6} \, 
{\cal M}_{\rm IR}\left(\frac{r\,y}{a},\frac{y}{a},z\right) \nonumber\\[2ex]
&=& - \frac{2 \, r^2 \, y^5}{a^6} \, 
\sum_{|\kappa|=1}^{\infty} \sum_{\kappa = \pm |\kappa|} 
\sum_{i,j=1}^{2} f_{\bar \imath}\left(\frac{r\,y}{a}\right) \nonumber\\[1ex]
& & \, \times \,\,
G^{ij}_{\kappa}\left(\frac{r\,y}{a},\frac{y}{a},z\right) \, 
f_{\bar \jmath}\left(\frac{y}{a}\right) \, 
A^{ij}_{\kappa}\left(\frac{r\,y}{a},\frac{y}{a}\right) \nonumber\\[2ex]
&=& - \frac{2 \, r^2 \, y^5}{a^6} \,  
\sum_{|\kappa|=1}^{\infty} T_{{\rm IR},|\kappa|}(r,y,z)\,,
\end{eqnarray}
where in the last line we define implicitly the terms $T_{{\rm
IR},|\kappa|}$ for $|\kappa| = 1,\dots,\infty$ as
\begin{eqnarray}
\label{DefinitionOfTIRkappa}
& & T_{{\rm IR},|\kappa|}(r,y,z) =
\sum_{\kappa = \pm |\kappa|} 
\sum_{i,j=1}^{2} \nonumber\\[1ex]
& & \quad f_{\bar \imath}\left(\frac{r\,y}{a}\right) \,
G^{ij}_{\kappa}\left(\frac{r\,y}{a},\frac{y}{a},z\right) \, 
f_{\bar \jmath}\left(\frac{y}{a}\right) \, 
A^{ij}_{\kappa}\left(\frac{r\,y}{a},\frac{y}{a}\right)\,.
\end{eqnarray}
Using the definition~(\ref{defSIR}), we obtain for $\Delta E_{\rm IR}$,
\begin{eqnarray}
\label{EvaluationOfEIR}
\Delta E_{\rm IR} &=& \frac{\alpha}{\pi}\,\frac{E_n}{10} + 
\frac{\alpha}{\pi} \, ({\rm P.V.})
\int_{\lfrac{9}{10}\,E_n}^{E_n} {\rm d}z 
\nonumber\\[1ex]
& & \int_0^1 {\rm d}r \, \int_0^\infty {\rm d}y \, 
S_{\rm IR}(r,y,z)\,.
\end{eqnarray}
The specification of the principal value (P.V.) is necessary for the
excited states of the L shell, because of the poles along the
integration contour which correspond to the spontaneous decay into the
ground state.  Here we are exclusively concerned with the real part of
the energy shift, as specified in
Eq.~(\ref{deltaEIR}), which is equivalent to the specification of the
principal value in~(\ref{EvaluationOfEIR}). 
Evaluation of the integral over $z$ is facilitated by the
subtraction of those terms which generate the singularities along the
integration contour (for higher excited states, there can be numerous
bound state poles, as pointed out in~\cite{MoKi1992,JeSoMo1997}). For
the 2S and $2{\rm P}_{1/2}$ states, only the pole contribution from
the ground state must be subtracted. For the $2{\rm P}_{3/2}$ state,
pole contributions originating from the 1S, the 2S and the $2{\rm
P}_{1/2}$ states must be taken into account. The numerical evaluation
of the subtracted integrand proceeds along ideas outlined
in~\cite{Mo1982,MoKi1992} and is not discussed here in any further
detail.

The scaling parameter $a$ for the integration over $y$ is chosen to
simplify the exponential dependence of the function $S$ defined in
Eq.~(\ref{defSIR}). The main exponential dependence is given by the
relativistic radial wave functions (upper and lower components).  Both
components [$f_1(x)$ and $f_2(x)$] vary approximately as (neglecting
relatively slowly varying factors)
\[
\exp\left(-a\,x/2\right) 
\;\;\;\;\;\;\;\; \mbox{(for large $x$)}\,.
\]
The scaling variable $a$, expanded in powers of $Z\alpha$, is
\begin{eqnarray}
\label{ExpansionOfA}
a &=& 2\,\sqrt{1-E_n^2} \nonumber\\[2ex]
&=& 2\,\sqrt{1 - \left(1 - \frac{(Z\alpha)^2}{2\,n^2} + 
{\rm O}\left[(Z\alpha)^4\right]\right)^2} \nonumber\\[2ex]
&=& 2\,\frac{Z\alpha}{n} + {\rm O}\!\left[(Z\alpha)^3\right]\,.
\end{eqnarray}
Therefore, $a$ is just twice the {\em inverse} of the Bohr radius
$n/(Z\alpha)$ in the nonrelativistic limit. The product
\[
f_{\bar \imath}\left(\frac{r y}{a}\right) \times
f_{\bar \jmath}\left(\frac{y}{a}\right)
\;\;\;\;\;\;\;\;\; \mbox{for arbitrary ${\bar \imath},{\bar \jmath} \in \{1,2\}$}
\]
[which occurs in Eq.~(\ref{defSIR})] depends on the radial arguments
approximately as
\[
e^{-y} \times \exp\left[\lfrac{1}{2}\,(1-r)\,y\right] 
\;\;\;\;\;\;\;\;\; \mbox{(for large $y$)}\,.
\]
Note that the main dependence as given by the term $\exp(-y)$ is
exactly the weight factor of the Gau\ss{}-Laguerre integration
quadrature formula. The deviation from the exact $\exp(-y)$--type
behavior becomes smaller as $r \to 1$. This is favorable because the
region near $r = 1$ gives a large contribution to the integral
in~(\ref{EvaluationOfEIR}).

%
% Table FIRKL
%
\begin{table}[thb!]
\begin{center}
\begin{minipage}{16cm}
\begin{center}
\begin{tabular}{lr@{.}lr@{.}lr@{.}lr@{.}l}
\hline
\hline
\multicolumn{1}{c}{$Z$} &
\multicolumn{2}{c}
  {\rule[-3mm]{0mm}{8mm}$F_{\rm IR}(1{\rm S}_{1/2},Z\alpha)$} &
\multicolumn{2}{c}
  {\rule[-3mm]{0mm}{8mm}$F_{\rm IR}(2{\rm S}_{1/2},Z\alpha)$} &
\multicolumn{2}{c}
  {\rule[-3mm]{0mm}{8mm}$F_{\rm IR}(2{\rm P}_{1/2},Z\alpha)$} &
\multicolumn{2}{c}
  {\rule[-3mm]{0mm}{8mm}$F_{\rm IR}(2{\rm P}_{3/2},Z\alpha)$}
\\
\hline
$1$ & 
 $7$ & $236~623~736~8(1)$ & 
 $7$ & $479~764~180(1)$ &
 $0$ & $085~327~852(1)$ &
 $0$ & $082~736~497(1)$ \\
$2$  &
 $5$ & $539~002~119~1(1)$ & 
 $5$ & $782~025~637(1)$ &
 $0$ & $086~073~669(1)$ &
 $0$ & $083~279~461(1)$ \\
$3$  &
 $4$ & $598~155~821~8(1)$ &
 $4$ & $840~923~962(1)$ &
 $0$ & $087~162~510(1)$ &
 $0$ & $084~091~830(1)$ \\
$4$  &
 $3$ & $963~124~140~6(1)$ &
 $4$ & $205~501~798(1)$ &
 $0$ & $088~543~188(1)$ &
 $0$ & $085~140~788(1)$ \\
$5$  &
 $3$ & $493~253~319~4(1)$ & 
 $3$ & $735~114~958(1)$ &
 $0$ & $090~180~835(1)$ &
 $0$ & $086~403~178(1)$ \\
\hline
\hline
\end{tabular}
\caption{\label{tableFIRKL} Infrared part for the K and
L shell states,
$F_{\rm IR}(1{\rm S}_{1/2},Z\alpha)$,
$F_{\rm IR}(2{\rm S}_{1/2},Z\alpha)$,
$F_{\rm IR}(2{\rm P}_{1/2},Z\alpha)$,
and $F_{\rm IR}(2{\rm P}_{3/2},Z\alpha)$,
evaluated for low-$Z$
hydrogen(-like) ions. The calculations are performed with the
numerical value of $\alpha^{-1} = 137.036$
for the fine structure constant.}
\end{center}
\end{minipage}
\end{center}
\end{table}

The sum over $|\kappa|$ in Eq.~(\ref{defSIR}) is carried out locally,
i.e.~for each set of arguments $r,y,z$. The sum over $|\kappa|$ is
absolutely convergent. For $|\kappa| \to \infty$, the convergence of
the sum is governed by the asymptotic behavior of the Bessel functions
which occur in the photon functions $A^{ij}_{\kappa}$ ($i,j=1,2$) [see
Eqs.~(3.15) and (3.16) in~\cite{Mo1974a}]. The photon functions
contain products of two Bessel functions of the form ${\cal
J}_l(\rho_{2/1})$ where ${\cal J}_l$ stands for either $j_l$ or
$j'_l$, the index $l$ is in the range
$l~\in~\{|\kappa|-1,|\kappa|,|\kappa| + 1\}$. The argument is either
$\rho_2 = (E_n - z)\,x_2$ or $\rho_1 = (E_n - z)\,x_1$. The asymptotic
behavior of the two relevant Bessel functions for large $l$ (and
therefore large $|\kappa|$) is
\begin{eqnarray}
\label{BesselAsymptotics1}
j'_l(x) & = & 
\frac{l}{x}\,\frac{x^l}{(2 l + 1)!!} \,
\left[1 + {\rm O}\left( \frac{1}{l} \right) \right]
\quad \mbox{and}\\
\label{BesselAsymptotics2}
j_l(x) & = & 
\frac{x^l}{(2 l + 1)!!} \,
\left[1 + {\rm O}\left( \frac{1}{l} \right) \right]\,.
\end{eqnarray}
This implies that when $\min\{\rho_2,\rho_1\} = \rho_2 < l$, the
function ${\cal J}_l(\rho_2)$ vanishes with increasing $l$
approximately as $({\rm e}\,\rho_2/2 l)^l$. This rapidly converging asymptotic
behavior sets in as soon as $l \approx |\kappa| > \rho_2 =
r\,\omega\,y/a$ [see Eqs.~(\ref{DefOfZ}) 
and~(\ref{DefinitionOfTIRkappa})].
Due to the rapid convergence for $|\kappa| > \rho_2$, the maximum
angular momentum quantum number $|\kappa|$ in the numerical
calculation of the infrared part is less than $3~000$. Note that
because $z~\in~{\bf (}\lfrac{9}{10}\,E_n,E_n{\bf )}$ in the infrared
part, $\omega < \lfrac{1}{10}\,E_n$.

The integration scheme is based on a crude estimate of the dependence
of the integrand $S_{\rm IR}(r,y,z)$ defined in~Eq.~(\ref{defSIR}) on
the integration variables $r$, $y$ and $z$. The main contribution to
the integral is given by the region where the arguments of the
Whittaker functions as they occur in the Green function~[see
Eq.~(A.16)~in~\cite{Mo1974a}] are much larger than the Dirac angular
momentum,
\[
2\,c\,\frac{y}{a} \gg |\kappa|
\]
(see also p.~56 of~\cite{Mo1974b}).  We assume the asymptotic form of
the Green function given in Eq.~(A.3) in~\cite{Mo1974b} 
applies and attribute a factor
\[
\exp[-(1-r)\,c\,y/a]
\]
to the radial Green functions $G^{ij}_{\kappa}$ as they occur in
Eq.~(\ref{defSIR}). Note that relatively slowly varying factors are
replaced by unity. The products of the radial wave functions $f_{\bar
\imath}$ and $f_{\bar \jmath}$, according to the discussion following
Eq.~(\ref{ExpansionOfA}), behave as
\[
e^{-y} \, \exp\left[\lfrac{1}{2}\,(1-r)\,y\right]
\]
for large $y$. The photon functions $A^{ij}_{\kappa}$ in
Eq.~(\ref{defSIR}) give rise to an approximate factor
\begin{equation}
\frac{\sin[(1-r)\,(E_n-z)\,y/a]}{(1-r)}\,.
\end{equation}
Therefore [see also Eq.~(2.12) in~\cite{Mo1974b}], we base our
choice of the integration routine on the approximation
\begin{eqnarray}
\label{approxsir}
e^{-y} \, \exp\left[-\left(\frac{c}{a} - 
  \frac{1}{2}\right)\,(1-r)\,y\right] 
\nonumber\\[2ex]
\quad\quad
  \times \frac{\sin\left[(1-r)\,\left\{(E_n-z)\,y/a\right\}\right]}{(1-r)}
\end{eqnarray}
for $S_{\rm IR}$. The three-dimensional integral in~(\ref{EvaluationOfEIR}) 
is evaluated by successive Gaussian
quadrature.  Details of the integration procedure can be found
in~\cite{Je1999}.

In order to check the numerical stability of the results, the
calculations are repeated with three different values of the fine
structure constant $\alpha$,
\begin{equation}
\label{DefinitionOfAlphas}
\begin{array}{rcll}
\alpha_< & = & 1/137.036~000~5 \,, & \\
\alpha_0 & = & 1/137.036~000~0 & \mbox{and}\\
\alpha_> & = & 1/137.035~999~5 \,. &
\end{array}
\end{equation}
These values of the fine-structure constant
are close to the 1998 CODATA recommended value of
$\alpha^{-1} = 137.035~999~76(50)$~\cite{MoTa2000}. The calculation
was parallelized using the Message Passing Interface (MPI) and carried
out on a cluster of Silicon Graphics workstations and on an IBM 9276
SP/2 multiprocessor system. The results for the
infrared part, $F_{\rm IR}$ defined in Eq.~(\ref{DefinitionOfFIR}), are
given in~Table~\ref{tableFIRKL} for a value of $\alpha^{-1} =
\alpha_0^{-1} = 137.036$. This value of $\alpha$ will be used
exclusively in the numerical evaluations presented here. For numerical
results obtained by employing the values of $\alpha_<$ and $\alpha_>$
[see Eq.~(\ref{DefinitionOfAlphas})] we refer to~\cite{Je1999}.

%
% The Middle-Energy Subtraction Term
%
\subsection{The Middle-Energy Subtraction Term}
\label{MiddleEnergySubtraction}

The middle-energy part is given by
\begin{eqnarray}
\label{deltaEM}
\lefteqn{\Delta E_{\rm M} = 
{\rm -i}\, e^2 \, 
\int_{C_{\rm M}} \! \frac{{\rm d}\omega}{2 \pi}
\int \! \! \frac{{\rm d}^3 \bbox{k}}{(2 \pi)^3} \,
D_{\mu\nu}(k^2)} \nonumber\\[2ex]
& & \times \left< \psi \left| \, \alpha^{\mu} \,\,
{\rm e}^{{\rm i}\bbox{\scriptstyle k}\cdot \bbox{\scriptstyle x}} \,\,
G(E_n - \omega) \, \alpha^{\nu} \,\, 
{\rm e}^{-{\rm i}\bbox{\scriptstyle k}\cdot \bbox{\scriptstyle x}} \,\,
\right| \psi \right> \,,
\end{eqnarray}
where relevant definitions of the symbols can be found in
Eqs.~(\ref{deltaESEM})--(\ref{PhotonPropagator})
and in Eq.~(\ref{PhotonPropagatorUnreg}).
The middle-energy part comprises the following integration region for the
virtual photon (contour $C_{\rm M}$ in Fig.~\ref{SeparationLow}),
\begin{equation}
\left.
\begin{array}{l}
\omega \in \left(\lfrac{1}{10}\,E_n,E_n\right) \pm {\rm i}\,\epsilon \\[1ex]  
z \in \left(0,\lfrac{9}{10}\,E_n\right) \pm {\rm i}\,\epsilon
\end{array}
\right\}
\mbox{middle-energy part $\Delta E_{\rm M}$}\,.
\end{equation}
The numerical evaluation of the middle-energy part is simplified
considerably by the decomposition of the relativistic Dirac-Coulomb
Green function $G$ as
\begin{equation}
\label{DecompositionOfG}
G \;\; = \;\; G_{\rm A} \; + \; G_{\rm B}\,,
\end{equation}
where $G_{\rm A}$ is defined in (\ref{DefinitionOfGA}) and represents
the sum of an approximation to the so-called zero- and one-potential
terms generated by the expansion of the Dirac-Coulomb Green function
$G$ in powers of the binding field $V$.  We define the middle-energy
subtraction term $F_{\rm MA}$ as the expression obtained upon
substitution of the propagator $G_{\rm A}$ for $G$ in
Eq.~(\ref{deltaEM}).  The propagator $G_{\rm B}$ is simply calculated
as the difference of $G$ and $G_{\rm A}$ [see
Eq.~(\ref{DefinitionOfGB})]. A substitution of the propagator $G_{\rm
B}$ for $G$ in Eq.~(\ref{deltaEM}) leads to the middle-energy
remainder $F_{\rm MB}$ which is discussed in
Sec.~\ref{MiddleEnergyRemainder}. We provide here the explicit
expressions
\begin{eqnarray}
\label{deltaEMA}
\lefteqn{\Delta E_{\rm MA} =
{\rm -i}\, e^2 \,
\int_{C_{\rm M}} \! \frac{{\rm d}\omega}{2 \pi}
\int \! \! \frac{{\rm d}^3 \bbox{k}}{(2 \pi)^3} \,
D_{\mu\nu}(k^2)} \nonumber\\[2ex]
& & \times \left< \psi \left| \, \alpha^{\mu} \,
{\rm e}^{{\rm i}\bbox{\scriptstyle k}\cdot \bbox{\scriptstyle x}} \,\,
G_{\rm A}(E_n - \omega) \, \alpha^{\nu} \, 
{\rm e}^{-{\rm i}\bbox{\scriptstyle k}\cdot \bbox{\scriptstyle x}} \,\,
\right| \psi \right> 
\end{eqnarray}
and
\begin{eqnarray}
\label{deltaEMB}
\lefteqn{\Delta E_{\rm MB} =
{\rm -i}\, e^2 \,
\int_{C_{\rm M}} \! \frac{{\rm d}\omega}{2 \pi}
\int \! \! \frac{{\rm d}^3 \bbox{k}}{(2 \pi)^3} \,
D_{\mu\nu}(k^2)} \nonumber\\[2ex]
& & \times \left< \psi \left| \, \alpha^{\mu} \,
{\rm e}^{{\rm i}\bbox{\scriptstyle k}\cdot \bbox{\scriptstyle x}} \,\,
G_{\rm B}(E_n - \omega) \, \alpha^{\nu} \,
{\rm e}^{-{\rm i}\bbox{\scriptstyle k}\cdot \bbox{\scriptstyle x}} \,\,
\right| \psi \right> \,.
\end{eqnarray}
Note that the decomposition of the Dirac-Coulomb Green function as
in~(\ref{DecompositionOfG}) is not applicable in the infrared part,
because of numerical problems for ultra-soft photons (infrared
divergences). Rewriting (\ref{deltaEMA}) appropriately
into a three-dimensional integral~\cite{Mo1974a,Mo1974b,Je1999},
we have
\begin{eqnarray}
\label{IntegralEMA}
\lefteqn{\Delta E_{\rm MA} = \frac{\alpha}{\pi} \, \frac{9}{10} \, E_n - 
\frac{2\,\alpha}{\pi} \, 
\int_{0}^{\lfrac{9}{10}\,E_n} \!\!\!\! {\rm d}z }
\nonumber\\[2ex] 
& & \int_0^\infty \! {\rm d} x_1 \, x_1^2 \,
\int_0^{x_1} \! {\rm d} x_2 \, x_2^2 \,\, {\cal M}_{\rm MA}(x_2,x_1,z)\,.
\end{eqnarray}
The function ${\cal M}_{\rm MA}(x_2,x_1,z)$ is defined in analogy to
the function ${\cal M}_{\rm IR}(x_2,x_1,z)$ defined in
Eq.~(\ref{defMIR}) for the infrared part. Also, we define a function
${\cal S}_{\rm MA}(x_2,x_1,z)$ in analogy to the function ${\cal
S}_{\rm IR}(x_2,x_1,z)$ given in Eq.~(\ref{defSIR}) for the infrared
part, which will be used in Eq.~(\ref{DefinitionOfEMA}) below.
We have,
\begin{eqnarray}
\label{defSMA}
S_{\rm MA}(r,y,z) &=& - \frac{2 \, r^2 \, y^5}{a^6} \, 
{\cal M}_{\rm MA}\left(\frac{r\,y}{a},\frac{y}{a},z\right) \nonumber\\[2ex]
&=& - \frac{2 \, r^2 \, y^5}{a^6} \, 
\sum_{|\kappa|=1}^{\infty} \sum_{\kappa = \pm |\kappa|} 
\sum_{i,j=1}^{2} f_{\bar \imath}\left(\frac{r\,y}{a}\right) 
\nonumber\\[1ex]
& & \, \times \,\,
G^{ij}_{{\rm A},\kappa}\left(\frac{r\,y}{a},\frac{y}{a},z\right) \, 
f_{\bar \jmath}\left(\frac{y}{a}\right) \, 
A^{ij}_{\kappa}\left(\frac{r\,y}{a},\frac{y}{a}\right) \nonumber\\[2ex]
&=& - \frac{2 \, r^2 \, y^5}{a^6} \,
\sum_{|\kappa|=1}^{\infty} T_{{\rm MA},|\kappa|}(r,y,z)\,.
\end{eqnarray}
The expansion of the propagator $G_{\rm A}$ into partial waves is
given in Eqs.~(5.4) and (A.20) in \cite{Mo1974a} and in Eqs.~(D.37) and (D.42)
in~\cite{Je1999}. This expansion leads to the component
functions $G^{ij}_{{\rm A},\kappa}$. The terms $T_{{\rm MA},|\kappa|}$ in
the last line of Eq.~(\ref{defSMA}) read
\begin{eqnarray}
\label{DefinitionOfTMAkappa}
T_{{\rm MA},|\kappa|}(r,y,z) = \sum_{\kappa = \pm |\kappa|} 
\sum_{i,j=1}^{2} \, f_{\bar \imath}\left(\frac{r\,y}{a}\right) \,
G^{ij}_{{\rm A},\kappa}\left(\frac{r\,y}{a},\frac{y}{a},z\right) \, 
f_{\bar \jmath}\left(\frac{y}{a}\right) \, 
A^{ij}_{\kappa}\left(\frac{r\,y}{a},\frac{y}{a}\right)\,.
\end{eqnarray}
With these definitions, the middle-energy subtraction term $\Delta
E_{\rm MA}$ can be written as
\begin{eqnarray}
\label{DefinitionOfEMA}
\Delta E_{\rm MA} = \frac{\alpha}{\pi}\,\frac{9}{10}\,E_n 
+ \frac{\alpha}{\pi} \, \int_0^{\lfrac{9}{10}\,E_n} \!\!\!\! {\rm d}z
\int_0^\infty \! {\rm d}y \, 
\int_0^1 \! {\rm d}r \,\, 
S_{\rm MA}(r,y,z)\,.
\end{eqnarray}
The subtracted lower-order terms yield,
\begin{eqnarray}
\label{DefinitionOfFMA}
\Delta E_{\rm MA} = \frac{\alpha}{\pi} \,
\left[
\frac{279}{200} \langle \beta \rangle +
\frac{219}{200} \langle V \rangle +
\frac{(Z\alpha)^4}{n^3}\,F_{\rm MA}(nl_j,Z\alpha)
\right]\,.
\end{eqnarray}
The three-dimensional integral (\ref{DefinitionOfEMA}) is evaluated 
by successive Gaussian quadrature. Details of the 
integration procedure can be found in~\cite{Je1999}.
The numerical results are summarized in the
Table~\ref{tableFMKL}.

%
% tableFMKL
%
\begin{table}[thb!]
\begin{center}
\begin{minipage}{15cm}
\begin{center}
{\small
\begin{tabular}{lr@{.}lr@{.}lr@{.}lr@{.}l}
\hline
\hline
\multicolumn{1}{c}{$Z$} &
\multicolumn{2}{c}
  {\rule[-3mm]{0mm}{8mm}$F_{\rm MA}(1{\rm S}_{1/2},Z\alpha)$} &
\multicolumn{2}{c}
  {\rule[-3mm]{0mm}{8mm}$F_{\rm MA}(2{\rm S}_{1/2},Z\alpha)$} &
\multicolumn{2}{c}
  {\rule[-3mm]{0mm}{8mm}$F_{\rm MA}(2{\rm P}_{1/2},Z\alpha)$} &
\multicolumn{2}{c}
  {\rule[-3mm]{0mm}{8mm}$F_{\rm MA}(2{\rm P}_{3/2},Z\alpha)$}
\\
\hline
$1$  &
 $2$ & $699~379~904~5(1)$ &
 $2$ & $720~878~318(1)$ &
 $0$ & $083~207~314(1)$ &
 $0$ & $701~705~240(1)$ \\
$2$  &
 $2$ & $659~561~381~1(1)$ &
 $2$ & $681~820~660(1)$ &
 $0$ & $084~208~832(1)$ &
 $0$ & $701~850~024(1)$ \\
$3$  &
 $2$ & $623~779~453~0(1)$ &
 $2$ & $647~262~568(1)$ &
 $0$ & $085~831~658(1)$ &
 $0$ & $702~091~147(1)$ \\
$4$  &
 $2$ & $591~151~010~1(1)$ &
 $2$ & $616~290~432(1)$ &
 $0$ & $088~040~763(1)$ &
 $0$ & $702~426~850(1)$ \\
$5$  & 
 $2$ & $561~096~522~1(1)$ &
 $2$ & $588~297~638(1)$ &
 $0$ & $090~803~408(1)$ &
 $0$ & $702~854~461(1)$ \\
\hline
\hline
\multicolumn{1}{c}{$Z$} &
\multicolumn{2}{c}
  {\rule[-3mm]{0mm}{8mm}$F_{\rm MB}(1{\rm S}_{1/2},Z\alpha)$} &
\multicolumn{2}{c}
  {\rule[-3mm]{0mm}{8mm}$F_{\rm MB}(2{\rm S}_{1/2},Z\alpha)$} &
\multicolumn{2}{c}
  {\rule[-3mm]{0mm}{8mm}$F_{\rm MB}(2{\rm P}_{1/2},Z\alpha)$} &
\multicolumn{2}{c}
  {\rule[-3mm]{0mm}{8mm}$F_{\rm MB}(2{\rm P}_{3/2},Z\alpha)$}
\\
\hline
$1$  &
 $ 1$ & $685~993~923~2(1)$ & 
 $ 1$ & $784~756~705(2)$ &
 $ 0$ & $771~787~771(2)$ &
 $-0$ & $094~272~681(2)$ \\
$2$  &
 $ 1$ & $626~842~294~5(1)$ & 
 $ 1$ & $725~583~798(2)$ &
 $ 0$ & $770~778~394(2)$ &
 $-0$ & $094~612~071(2)$ \\
$3$  &
 $ 1$ & $571~406~090~7(1)$ & 
 $ 1$ & $670~086~996(2)$ &
 $ 0$ & $769~153~314(2)$ &
 $-0$ & $095~165~248(2)$ \\
$4$  &
 $ 1$ & $519~082~768~6(1)$ & 
 $ 1$ & $617~650~004(2)$ &
 $ 0$ & $766~954~435(2)$ &
 $-0$ & $095~922~506(2)$ \\
$5$  &
 $ 1$ & $469~482~409~0(1)$ & 
 $ 1$ & $567~873~140(2)$ &
 $ 0$ & $764~220~149(2)$ &
 $-0$ & $096~874~556(2)$ \\
\hline
\hline
\multicolumn{1}{c}{$Z$} &
\multicolumn{2}{c}
  {\rule[-3mm]{0mm}{8mm}$F_{\rm M}(1{\rm S}_{1/2},Z\alpha)$} &
\multicolumn{2}{c}
  {\rule[-3mm]{0mm}{8mm}$F_{\rm M}(2{\rm S}_{1/2},Z\alpha)$} &
\multicolumn{2}{c}
  {\rule[-3mm]{0mm}{8mm}$F_{\rm M}(2{\rm P}_{1/2},Z\alpha)$} &
\multicolumn{2}{c}
  {\rule[-3mm]{0mm}{8mm}$F_{\rm M}(2{\rm P}_{3/2},Z\alpha)$}
\\
\hline
$1$  &
 $4$ & $385~373~827~7(1)$ &
 $4$ & $505~635~023(2)$ &
 $0$ & $854~995~085(2)$ &
 $0$ & $607~432~559(2)$ \\
$2$  &
 $4$ & $286~403~675~7(1)$ &
 $4$ & $407~404~458(2)$ &
 $0$ & $854~987~226(2)$ &
 $0$ & $607~237~953(2)$ \\
$3$  &
 $4$ & $195~185~543~6(1)$ &
 $4$ & $317~349~564(2)$ &
 $0$ & $854~984~972(2)$ &
 $0$ & $606~925~899(2)$ \\
$4$  &
 $4$ & $110~233~778~8(1)$ &
 $4$ & $233~940~436(2)$ &
 $0$ & $854~995~198(2)$ &
 $0$ & $606~504~344(2)$ \\
$5$  &
 $4$ & $030~578~931~1(1)$ &
 $4$ & $156~170~778(2)$ &
 $0$ & $855~023~557(2)$ &
 $0$ & $605~979~905(2)$ \\
\hline
\hline
\end{tabular}}
\caption{\label{tableFMKL} Numerical results for the
middle-energy subtraction term $F_{\rm MA}$,
the middle-energy remainder term $F_{\rm MB}$
and the middle-energy term $F_{\rm M}$.
The middle-energy term $F_{\rm M}$ is given as the sum
$F_{\rm M}(nl_j,Z\alpha) =
F_{\rm MA}(nl_j,Z\alpha) + F_{\rm MB}(nl_j,Z\alpha)$
[see also Eqs.(\ref{DefinitionOfFM}),
(\ref{DefinitionOfFMA}) and (\ref{DefinitionOfFMB})].}
\end{center}
\end{minipage}
\end{center}
\end{table}

%
% The Middle-Energy Remainder
%
\subsection{The Middle-Energy Remainder}
\label{MiddleEnergyRemainder}

The remainder term in the middle-energy part involves the propagator
$G_{\rm B}$ defined in Eq.~(\ref{DefinitionOfGB}), $G_{\rm B} = G -
G_{\rm A}$, where $G$ is defined in (\ref{DefinitionOfG}) and $G_{\rm
A}$ is given in (\ref{DefinitionOfGA}).  In analogy to the
middle-energy subtraction term, the middle-energy remainder can be
rewritten as a three-dimensional integral,
\begin{eqnarray}
\label{EvaluationOfEMB}
\Delta E_{\rm MB} &=& \frac{\alpha}{\pi} \, 
  \int_0^{\lfrac{9}{10}\,E_n} {\rm d}z 
\int_0^1 {\rm d}r \, \int_0^\infty {\rm d}y \,
S_{\rm MB}(r,y,z)\,,
\end{eqnarray}
where
\begin{eqnarray}
\label{defSMB}
S_{\rm MB}(r,y,z) &=& - \frac{2 \, r^2 \, y^5}{a^6} \, 
\sum_{|\kappa|=1}^{\infty} \sum_{\kappa = \pm |\kappa|} 
\sum_{i,j=1}^{2} \, f_{\bar \imath}\left(\frac{r\,y}{a}\right) \,
G^{ij}_{{\rm B},\kappa}\left(\frac{r\,y}{a},\frac{y}{a},z\right) \, 
f_{\bar \jmath}\left(\frac{y}{a}\right) \, 
A^{ij}_{\kappa}\left(\frac{r\,y}{a},\frac{y}{a}\right)
\nonumber\\[2ex]
&=& - \frac{2 \, r^2 \, y^5}{a^6} \,
\sum_{|\kappa|=1}^{\infty} T_{{\rm MB},|\kappa|}(r,y,z)\,,
\end{eqnarray}
where we implicitly define the terms $T_{{\rm MB},|\kappa|}(r,y,z)$
in analogy with the infrared part Eq.~(\ref{defSIR}). 
The functions $G^{ij}_{{\rm B},\kappa}$ are obtained as the difference
of the expansion of the full propagator $G$ and the simplified
propagator $G_{\rm A}$ into angular momenta,
\begin{equation}
G^{ij}_{{\rm B},\kappa} = G^{ij}_{\kappa} -
  G^{ij}_{{\rm A},\kappa}
\end{equation}
where the $G^{ij}_{\kappa}$ are listed in Eq.~(A.16) in~\cite{Mo1974a}
and in Eq.~(D.43) in~\cite{Je1999}, and the $G^{ij}_{{\rm A},\kappa}$
have already been defined in Eqs.~(5.4) and (A.20) in \cite{Mo1974a} and in
Eqs.~(D.37) and (D.42) in~\cite{Je1999}.  There are no lower-order
terms to subtract, and therefore
\begin{equation}
\label{DefinitionOfFMB}
\Delta E_{\rm MB} = \frac{\alpha}{\pi} \,
\frac{(Z\alpha)^4}{n^3}\,F_{\rm MB}(nl_j,Z\alpha)\,.
\end{equation}
The three-dimensional integral (\ref{EvaluationOfEMB}) is evaluated by
successive Gaussian quadrature. Details of the integration procedure
are provided in~\cite{Je1999}. Numerical results for the middle-energy
remainder $F_{\rm MB}$, are summarized in Table~\ref{tableFMKL} for
the K- and L-shell states.

%
% tableFLKL
%
\begin{table}[hb]
\begin{center}
\begin{minipage}{16cm}
\begin{center}
\begin{tabular}{lr@{.}lr@{.}lr@{.}lr@{.}l}
\hline
\hline
\multicolumn{1}{c}{$Z$} &
\multicolumn{2}{c}
  {\rule[-3mm]{0mm}{8mm}$F_{\rm L}(1{\rm S}_{1/2},Z\alpha)$} &
\multicolumn{2}{c}
  {\rule[-3mm]{0mm}{8mm}$F_{\rm L}(2{\rm S}_{1/2},Z\alpha)$} &
\multicolumn{2}{c}
  {\rule[-3mm]{0mm}{8mm}$F_{\rm L}(2{\rm P}_{1/2},Z\alpha)$} &
\multicolumn{2}{c}
  {\rule[-3mm]{0mm}{8mm}$F_{\rm L}(2{\rm P}_{3/2},Z\alpha)$}
\\
\hline
$1$  &
 $11$ & $621~997~564~5(1)$ &
 $11$ & $985~399~203(2)$ &
 $ 0$ & $940~322~937(2)$ &
 $ 0$ & $690~169~056(2)$ \\
$2$  &
  $9$ & $825~405~794~7(1)$ &
 $10$ & $189~430~095(2)$ &
 $ 0$ & $941~060~895(2)$ &
 $ 0$ & $690~517~414(2)$ \\
$3$  &
  $8$ & $793~341~365~4(1)$ &
 $ 9$ & $158~273~526(2)$ &
 $ 0$ & $942~147~482(2)$ &
 $ 0$ & $691~017~729(2)$ \\
$4$  &
  $8$ & $073~357~919~4(1)$ &
 $ 8$ & $439~442~234(2)$ &
 $ 0$ & $943~538~386(2)$ &
 $ 0$ & $691~645~132(2)$ \\
$5$  &
  $7$ & $523~832~250~6(1)$ &
 $ 7$ & $891~285~736(2)$ &
 $ 0$ & $945~204~392(2)$ &
 $ 0$ & $692~383~083(2)$ \\
\hline
\hline
\end{tabular}
\caption{\label{tableFLKL} Low-energy
part $F_{\rm L}$ for the K- and L-shell states
$F_{\rm L}(1{\rm S}_{1/2},Z\alpha)$,
$F_{\rm L}(2{\rm S}_{1/2},Z\alpha)$,
$F_{\rm L}(2{\rm P}_{1/2},Z\alpha)$,
and $F_{\rm L}(2{\rm P}_{3/2},Z\alpha)$,
evaluated for low-$Z$
hydrogenlike ions.}
\end{center}
\end{minipage}
\end{center}
\end{table}

For the middle-energy part, the separation into a subtraction and a
remainder term has considerable computational advantages which become
obvious upon inspection of Eqs.~(\ref{DefinitionOfFMA})
and~(\ref{DefinitionOfFMB}).  The subtraction involves a propagator
whose angular components can be evaluated by 
recursion~\cite{Mo1974b,Je1999}, which is computationally
time-consuming. Because the subtraction term involves lower-order components
[see Eq.~(\ref{DefinitionOfFM})], it has to be evaluated to high
precision numerically (in a typical case, a relative accuracy of
$10^{-19}$ is required).  This high precision requires in turn a large
number of integration points for the Gaussian quadratures, which is
possible only if the numerical evaluation of the integrand is not
computationally time-consuming.  For the remainder term, no lower-order
terms have to be subtracted, and the relative accuracy required of the
integrals is in the range of $10^{-11}\dots10^{-9}$.  A numerical
evaluation to this smaller level of precision is feasible although the
calculation of the Green function $G_{\rm B}$ is computationally more
expensive than that of $G_{\rm A}$~\cite{Mo1974a,Mo1974b,Je1999}.  The
separation of the high-energy part into a subtraction term and a
remainder term, which is discussed in Sec.~\ref{HighEnergyPart}, is
motivated by analogous considerations as for the middle-energy part.
In the high-energy part, this separation is even more important than
in the middle-energy part, because of the occurrence of infinite terms
which need to be subtracted analytically before a numerical evaluation
can proceed [see Eq.~(\ref{DeltaEHArecovered}) below].

We now summarize the results for the middle-energy part.
The middle-energy part is the sum of the middle-energy subtraction
term $F_{\rm MA}$ and the middle-energy remainder $F_{\rm MB}$ [see
also Eq.~(\ref{DecompositionOfFM})]. Numerical results are summarized
in Table~\ref{tableFMKL} for the K- and L-shell states. The low-energy
part $F_{\rm L}$ is defined as the sum of the infrared contribution
$F_{\rm IR}$ and the middle-energy contribution $F_{\rm M}$ [see
Eq.~(\ref{FLasSum})]. The results for $F_{\rm L}$ are provided in the
Table~\ref{tableFLKL} for the K- and L-shell states.  The 
limits for the low-energy part as a function of the bound state
quantum numbers can be found in Eq.~(7.80) of~\cite{Je1999},
\begin{eqnarray}
F_{\rm L}(nl_j,Z\alpha) &=&
  \frac{4}{3}\,\delta_{l,0}\,\ln(Z\alpha)^{-2} - \frac{4}{3}\,\ln k_0(n,l)
+ \left(\ln2 - \frac{11}{10}\right) \, \frac{1}{n} \nonumber\\
& & + \left(2\,\ln2 - \frac{16}{15}\right) \, \frac{1}{2\,l + 1}
+ \left(\frac{3}{2}\,\ln2 - \frac{7}{4}\right)\,
\frac{1}{\kappa\,(2\,l+1)}
\nonumber \\
& & + \left(-\frac{3}{2}\,\ln2 + \frac{9}{4}\right)\,\frac{1}{|\kappa|}
+ \left(\frac{4}{3}\,\ln2 - \frac{1}{3}\right) \delta_{l,0}
\nonumber \\
& & + \left(\ln2 - \frac{5}{6}\right) \frac{n-2\,l-1}{n\,(2\,l+1)} +
{\rm O}(Z\alpha)\,.
\end{eqnarray}
The leadings asymptotics for the
states under investigation are,
\begin{eqnarray}
\label{AsympFL}
F_{\rm L}(1{\rm S}_{1/2},Z \alpha) &=& 
 (4/3) \, \ln (Z\alpha)^{-2} - 1.554~642 + {\rm O}(Z\alpha) \,,
\nonumber\\[2ex]
F_{\rm L}(2{\rm S}_{1/2},Z \alpha) &=&  
 (4/3) \, \ln (Z\alpha)^{-2} - 1.191~497 + {\rm O}(Z\alpha) \,,
\nonumber\\[2ex] 
F_{\rm L}(2{\rm P}_{1/2},Z \alpha) &=&
 0.940~023 + {\rm O}(Z\alpha) \,,
\nonumber\\[2ex] 
F_{\rm L}(2{\rm P}_{3/2},Z \alpha) &=&
 0.690~023 + {\rm O}(Z\alpha) \,.
\end{eqnarray}
These asymptotics are consistent with the numerical data in
Table~\ref{tableFLKL}. For S states, the
low-energy contribution $F_{\rm L}$ diverges logarithmically as
$Z\alpha\to0$, whereas for P states, $F_{\rm L}$ approaches a constant
as $Z\alpha\to0$. The leading logarithm is a consequence of an
infrared divergence cut off by the atomic momentum scale. It is a
nonrelativistic effect which is generated by the nonvanishing
probability density of S waves at the origin in the 
nonrelativistic limit. The presence of the logarithmic behavior
for S states [nonvanishing $A_{41}$-coefficient, 
see Eqs.~(\ref{defFLO}) and (\ref{defFLOnS})] and its
absence for P states
is reproduced consistently by the data in~Table~\ref{tableFLKL}.

%
% The High-Energy Part
%
\section{The High-Energy Part}
\label{HighEnergyPart}

%
% The High-Energy Subtraction Term
%
\subsection{The High-Energy Subtraction Term}
\label{HighEnergySubtraction}

The high-energy part is given by
\begin{eqnarray}
\label{deltaEH}
\Delta E_{\rm H} &=& 
- \lim_{\Lambda\to\infty}
{\rm i}\, e^2 \,
\int_{C_{\rm H}} \! \frac{d\omega}{2 \pi}
\int \! \! \frac{d^3 \bbox{k}}{(2 \pi)^3} \,
D_{\mu\nu}(k^2,\Lambda) \nonumber\\[2ex]
& & \times \left< \psi \left| \, \alpha^{\mu} \,
{\rm e}^{{\rm i}\bbox{\scriptstyle k}\cdot \bbox{\scriptstyle x}} \,\,
G(E_n - \omega) \, \alpha^{\nu} \, 
{\rm e}^{-{\rm i}\bbox{\scriptstyle k}\cdot \bbox{\scriptstyle x}} \,\,
\right| \psi \right> \,,
\end{eqnarray}
where relevant definitions of the symbols can be found in
Eqs.~(\ref{deltaESEM})--(\ref{PhotonPropagator}), and
the contour $C_{\rm H}$ is as shown in Fig.~\ref{IntegrationContour}.
The high-energy part comprises the following integration region for
the virtual photon,
\begin{equation}
\label{SpecEnergyHigh}
\left.
\begin{array}{l}
\omega \in 
\left(E_n-{\rm i}\,\infty,E_n+{\rm i}\,\infty\right) \\[1ex]  
z \in \left(-{\rm i}\,\infty,{\rm i}\,\infty\right) 
\end{array}
\right\}
\mbox{high-energy part $\Delta E_{\rm H}$}\,.
\end{equation}
The separation of the high-energy part into a subtraction term and a
remainder is accomplished as in the middle-energy part [see
Eq.~(\ref{DecompositionOfG})] by writing the full Dirac-Coulomb Green
function $G$ [Eq.~(\ref{DefinitionOfG})] as $G = G_{\rm A} + G_{\rm
B}$.  We define the high-energy subtraction term $F_{\rm HA}$ as the
expression obtained upon substitution of the propagator $G_{\rm A}$
for $G$ in Eq.~(\ref{deltaEH}), and a substitution of the propagator
$G_{\rm B}$ for $G$ in Eq.~(\ref{deltaEH}) leads to the high-energy
remainder $F_{\rm HB}$ which is discussed in
Sec.~\ref{HighEnergyRemainder}. The subtraction term (including all
divergent contributions) is generated by $G_{\rm A}$, the high-energy
remainder term corresponds to $G_{\rm B}$.  We have
\begin{eqnarray}
\label{deltaEHA}
\Delta E_{\rm HA} &=& - \lim_{\Lambda\to\infty}
{\rm i}\, e^2 \,
\int_{C_{\rm H}} \! \frac{d\omega}{2 \pi}
\int \! \! \frac{d^3 \bbox{k}}{(2 \pi)^3} \,
D_{\mu\nu}(k^2,\Lambda) \nonumber\\[2ex]
& &  \times \left< \psi \left| \, \alpha^{\mu} \,
{\rm e}^{{\rm i}\bbox{\scriptstyle k}\cdot \bbox{\scriptstyle x}} \,\,
G_{\rm A}(E_n - \omega) \, \alpha^{\nu} \, 
{\rm e}^{-{\rm i}\bbox{\scriptstyle k}\cdot \bbox{\scriptstyle x}} \,\,
\right| \psi \right> 
\end{eqnarray}
and
\begin{eqnarray}
\label{deltaEHB}
\Delta E_{\rm HB} &=& 
{\rm -i}\, e^2 \,
\int_{C_{\rm H}} \! \frac{d\omega}{2 \pi}
\int \! \! \frac{d^3 \bbox{k}}{(2 \pi)^3} \,
D_{\mu\nu}(k^2) \nonumber\\[1ex]
& & \times \left< \psi \left| \, \alpha^{\mu} \,
{\rm e}^{{\rm i}\bbox{\scriptstyle k}\cdot \bbox{\scriptstyle x}} \,\,
G_{\rm B}(E_n - \omega) \, \alpha^{\nu} \, 
{\rm e}^{-{\rm i}\bbox{\scriptstyle k}\cdot \bbox{\scriptstyle x}} \,\,
\right| \psi \right> \,.
\end{eqnarray}
The contribution $\Delta E_{\rm HA}$ corresponding to $G_{\rm A}$ can
be separated further into a term $\Delta E^{(1)}_{\rm HA}$, which
contains all divergent contributions, and a term $\Delta E^{(2)}_{\rm
HA}$, which comprises contributions of lower order than $(Z\alpha)^4$,
but is convergent as $\Lambda \to \infty$.  This separation is
described in detail in~\cite{Mo1974a,Mo1973phd}.  We have
\begin{equation}
\Delta E_{\rm HA} = \Delta E^{(1)}_{\rm HA} +
\Delta E^{(2)}_{\rm HA}\,.
\end{equation}
We obtain for $\Delta E^{(1)}_{\rm HA}$, which contains a 
logarithmic divergence as $\Lambda \to \infty$,
\begin{equation}
\Delta E^{(1)}_{\rm HA} = \frac{\alpha}{\pi} \,
\left[ \left(\frac{3}{4}\,\ln \Lambda^2 - \frac{9}{8} \right) \,
\langle \beta \rangle +
\left(\frac{1}{2}\,\ln2 - \frac{17}{12} \right) \, \langle V \rangle 
+ \frac{(Z\alpha)^4}{n^3} \, 
F^{(1)}_{\rm HA}(nl_j,Z\alpha) \right]\,.
\end{equation}
For the contribution $F^{(1)}_{\rm HA}$, an explicit analytic result
is obtained in Eq.~(4.15) in~\cite{Mo1974a}. This contribution is
therefore not discussed in any further detail, here. The contribution
$\Delta E^{(2)}_{\rm HA}$ contains lower-order terms,
\begin{eqnarray}
\Delta E^{(2)}_{\rm HA} &=& \frac{\alpha}{\pi} \,
\left[\left(-\frac{1}{2}\,\ln2 + \frac{1}{4} \right) \, \langle V \rangle
+ \frac{(Z\alpha)^4}{n^3} \,
F^{(2)}_{\rm HA}(nl_j,Z\alpha) \right]\,.
\end{eqnarray}
Altogether we have
\begin{eqnarray}
\label{DeltaEHArecovered}
\Delta E_{\rm HA} &=&
\Delta E^{(1)}_{\rm HA} + \Delta E^{(2)}_{\rm HA}
\nonumber\\[2ex]
&=& \frac{\alpha}{\pi} \,
\left[ \left(\frac{3}{4}\,\ln \Lambda^2 - \frac{9}{8} \right) \,
\langle \beta \rangle -
\frac{7}{6} \, \langle V \rangle + 
\frac{(Z\alpha)^4}{n^3} \,
F_{\rm HA}(nl_j,Z \alpha) \right]\,.
\end{eqnarray}
The scaled function $F_{\rm HA}(nl_j,Z \alpha)$ is given as
\begin{equation}
F_{\rm HA}(nl_j,Z \alpha) =
F^{(1)}_{\rm HA}(nl_j,Z \alpha) +
F^{(2)}_{\rm HA}(nl_j,Z \alpha)\,.
\end{equation}
The term $\Delta
E^{(2)}_{\rm HA}$ falls naturally into a sum of
four contributions~\cite{Mo1974a},
\begin{equation}
\Delta E^{(2)}_{\rm HA} = T_1 + T_2 + T_3 + T_4
\end{equation}
where
\begin{eqnarray}
T_1 &=& -\frac{1}{10} \langle V \rangle + \frac{(Z\alpha)^4}{n^3} \, 
h_1(nl_j,Z \alpha)\,, \nonumber\\
T_2 &=& \left(\frac{7}{20} - \frac{1}{2} \, \ln2 \right)
\, \langle V \rangle + 
\frac{(Z\alpha)^4}{n^3} \, h_2(nl_j,Z \alpha)\,, \nonumber\\
T_3 &=& \frac{(Z\alpha)^4}{n^3} \, h_3(nl_j,Z \alpha)\,, \nonumber\\
T_4 &=& \frac{(Z\alpha)^4}{n^3} \, h_4(nl_j,Z \alpha)\,.
\end{eqnarray}
The functions $h_i$ ($i=1,2,3,4$) are defined in Eqs.~(4.18), 
(4.19) and (4.21) in~\cite{Mo1974a} (see also Eq.~(3.6) in~\cite{Mo1982}).
The evaluation of the high-energy subtraction term proceeds as outlined
in~\cite{Mo1974a,Mo1974b,Mo1982}, albeit with an increased accuracy
and improved calculational methods 
in intermediate steps of the calculation in order to overcome the
severe numerical cancellations in the low-$Z$ region.
We recover $F^{(2)}_{\rm HA}$ as the sum
\begin{eqnarray}
& & F^{(2)}_{\rm HA}(nl_j,Z \alpha) = 
h_1(nl_j,Z\alpha) + h_2(nl_j,Z\alpha) +
h_3(nl_j,Z\alpha) + h_4(nl_j,Z\alpha)\,.
\end{eqnarray}
The scaled function $F_{\rm HA}(nl_j,Z \alpha)$ [see also
Eqs.~(\ref{DefinitionOfFH}) and (\ref{DecompositionOfFH})] is obtained as
\begin{equation}
F_{\rm HA}(nl_j,Z \alpha) = 
F^{(1)}_{\rm HA}(nl_j,Z \alpha) + 
F^{(2)}_{\rm HA}(nl_j,Z \alpha)\,.
\end{equation}
The limits of the contributions $F^{(1)}_{\rm HA}(nl_j,Z \alpha)$
and $F^{(2)}_{\rm HA}(nl_j,Z\alpha)$ as $(Z\alpha)\to 0$ have been
investigated in~\cite{Mo1974a,Mo1973phd,Mo1982}.  For the contribution
$F^{(1)}_{\rm HA}(nl_j,0)$, the result can be found in Eq.~(3.5)
in~\cite{Mo1982}.  The limits of the functions
$h_i(nl_j,Z\alpha)$ ($i=1,2,3,4$) as $Z \alpha \to 0$ are given as a
function of the atomic state quantum numbers in Eq.~(3.8)
in~\cite{Mo1982}.  For the scaled high-energy subtraction term $F_{\rm
HA}$, the limits read (see Eq.~(3.9) in~\cite{Mo1982})
\begin{eqnarray}
\label{limFHAgen}
F_{\rm HA}(nl_j,Z\alpha) &=& 
\left(\frac{11}{10} - \ln2\right) \, \frac{1}{n} +
\left(\frac{16}{15} - 2 \, \ln2\right) \, \frac{1}{2\,l + 1}
\nonumber\\
& & + \left(\frac{1}{2}\,\ln2 - \frac{1}{4}\right) \,
\frac{1}{\kappa\,(2\,l+1)}
+ \left(\frac{3}{2}\,\ln2 - \frac{9}{4}\right)\,\frac{1}{|\kappa|}
+ {\rm O}(Z\alpha)\,.
\end{eqnarray}
Therefore, the explicit forms of the 
limits for the states under investigation are,
\begin{eqnarray}
\label{limFHA}
F_{\rm HA}(1{\rm S}_{1/2},Z \alpha) &=& 
-1.219~627 + {\rm O}(Z\alpha)\,, \nonumber\\[2ex]
F_{\rm HA}(2{\rm S}_{1/2},Z \alpha) &=&
-1.423~054 + {\rm O}(Z\alpha)\,, \nonumber\\[2ex]
F_{\rm HA}(2{\rm P}_{1/2},Z \alpha) &=&
-1.081~204 + {\rm O}(Z\alpha)\,, \nonumber\\[2ex]
F_{\rm HA}(2{\rm P}_{3/2},Z \alpha) &=&
-0.524~351 + {\rm O}(Z\alpha)\,. 
\end{eqnarray}
Numerical results for $F_{\rm HA}$, which are presented in
Table~\ref{tableFHKL}, exhibit consistency with the limits in
Eq.~(\ref{limFHA}).

%
% tableFHKL
%
\begin{table}[htb!]
\begin{center}
\begin{minipage}{16cm}
\begin{center}
\begin{tabular}{lr@{.}lr@{.}lr@{.}lr@{.}l}
\hline
\hline
$Z$ & 
\multicolumn{2}{c}{\rule[-3mm]{0mm}{8mm} 
  $F_{\rm HA}(1{\rm S}_{1/2},Z\alpha)$} &
\multicolumn{2}{c}{\rule[-3mm]{0mm}{8mm} 
  $F_{\rm HA}(2{\rm S}_{1/2},Z\alpha)$} &
\multicolumn{2}{c}{\rule[-3mm]{0mm}{8mm} 
  $F_{\rm HA}(2{\rm P}_{1/2},Z\alpha)$} &
\multicolumn{2}{c}{\rule[-3mm]{0mm}{8mm} 
  $F_{\rm HA}(2{\rm P}_{3/2},Z\alpha)$} \\
\hline
$1$ &
 $-1$ & $216~846~660~6(1)$ & 
 $-1$ & $420~293~291(1)$ &
 $-1$ & $081~265~954(1)$ &
 $-0$ & $524~359~802(1)$ \\
$2$ &
 $-1$ & $214~322~536~9(1)$ & 
 $-1$ & $417~829~864(1)$ &
 $-1$ & $081~451~269(1)$ &
 $-0$ & $524~385~053(1)$ \\
$3$ &
 $-1$ & $212~026~714~1(1)$ & 
 $-1$ & $415~635~310(1)$ &
 $-1$ & $081~760~224(1)$ &
 $-0$ & $524~427~051(1)$ \\
$4$ &
 $-1$ & $209~942~847~4(1)$ &
 $-1$ & $413~693~422(1)$ &
 $-1$ & $082~192~995(1)$ &
 $-0$ & $524~485~727(1)$ \\
$5$ &
 $-1$ & $208~059~033~6(1)$ &
 $-1$ & $411~992~480(1)$ &
 $-1$ & $082~749~845(1)$ &
 $-0$ & $524~561~017(1)$ \\
\hline
\hline
$Z$ & 
\multicolumn{2}{c}{\rule[-3mm]{0mm}{8mm} 
  $F_{\rm HB}(1{\rm S}_{1/2},Z\alpha)$} &
\multicolumn{2}{c}{\rule[-3mm]{0mm}{8mm} 
  $F_{\rm HB}(2{\rm S}_{1/2},Z\alpha)$} &
\multicolumn{2}{c}{\rule[-3mm]{0mm}{8mm} 
  $F_{\rm HB}(2{\rm P}_{1/2},Z\alpha)$} &
\multicolumn{2}{c}{\rule[-3mm]{0mm}{8mm} 
  $F_{\rm HB}(2{\rm P}_{3/2},Z\alpha)$} \\
\hline
$1$ &
 $-0$ & $088~357~254(1)$ &
 $-0$ & $018~280~727(5)$ &
 $ 0$ & $014~546~64(1)$ &
 $-0$ & $042~310~69(1)$ \\
$2$ &
 $-0$ & $082~758~206(1)$ &
 $-0$ & $012~729~99(1)$ &
 $ 0$ & $014~574~21(1)$ &
 $-0$ & $042~296~81(1)$ \\
$3$ &
 $-0$ & $076~811~229(1)$ &
 $-0$ & $006~861~02(1)$ &
 $ 0$ & $014~620~51(1)$ &
 $-0$ & $042~273~58(1)$ \\
$4$ &
 $-0$ & $070~590~991(1)$ &
 $-0$ & $000~746~40(1)$ &
 $ 0$ & $014~685~82(1)$ &
 $-0$ & $042~240~92(1)$ \\
$5$ &
 $-0$ & $064~146~139(1)$ &
 $ 0$ & $005~567~16(1)$ &
 $ 0$ & $014~770~52(1)$ &
 $-0$ & $042~198~76(1)$ \\
\hline
\hline
$Z$ & 
\multicolumn{2}{c}{\rule[-3mm]{0mm}{8mm} 
  $F_{\rm H}(1{\rm S}_{1/2},Z\alpha)$} &
\multicolumn{2}{c}{\rule[-3mm]{0mm}{8mm} 
  $F_{\rm H}(2{\rm S}_{1/2},Z\alpha)$} &
\multicolumn{2}{c}{\rule[-3mm]{0mm}{8mm} 
  $F_{\rm H}(2{\rm P}_{1/2},Z\alpha)$} &
\multicolumn{2}{c}{\rule[-3mm]{0mm}{8mm} 
  $F_{\rm H}(2{\rm P}_{3/2},Z\alpha)$} \\
\hline
$1$ &
 $-1$ & $305~203~915(1)$ &
 $-1$ & $438~574~018(5)$ &
 $-1$ & $066~719~31(1)$ &
 $-0$ & $566~670~50(1)$ \\
$2$ &
 $-1$ & $297~080~743(1)$ &
 $-1$ & $430~559~85(1)$ &
 $-1$ & $066~877~06(1)$ &
 $-0$ & $566~681~86(1)$ \\
$3$ &
 $-1$ & $288~837~943(1)$ &
 $-1$ & $422~496~33(1)$ &
 $-1$ & $067~139~72(1)$ &
 $-0$ & $566~700~63(1)$ \\
$4$ &
 $-1$ & $280~533~839(1)$ &
 $-1$ & $414~439~82(1)$ &
 $-1$ & $067~507~18(1)$ &
 $-0$ & $566~726~65(1)$ \\
$5$ &
 $-1$ & $272~205~173(1)$ &
 $-1$ & $406~425~32(1)$ &
 $-1$ & $067~979~33(1)$ &
 $-0$ & $566~759~78(1)$ \\
\hline
\hline
\end{tabular}
\caption{\label{tableFHKL} Numerical results for the
high-energy subtraction term $F_{\rm HA}$
and the high-energy remainder term $F_{\rm HB}$.
The high-energy term $F_{\rm H}$ is given as the sum
$F_{\rm H}(nl_j,Z\alpha) =
F_{\rm HA}(nl_j,Z\alpha) + F_{\rm HB}(nl_j,Z\alpha)$.}
\end{center}
\end{minipage}
\end{center}
\end{table}

%
% The High-Energy Remainder
%
\subsection{The High-Energy Remainder}
\label{HighEnergyRemainder}

The remainder term in the high-energy part involves the
propagator $G_{\rm B}$ defined in Eq.~(\ref{DefinitionOfGB}),
$G_{\rm B} = G - G_{\rm A}$,
where $G$ is defined in (\ref{DefinitionOfG}) and $G_{\rm A}$ is given
in (\ref{DefinitionOfGA}). The evaluation proceeds in complete
analogy to the calculation of the middle-energy remainder 
term (Sec.~\ref{MiddleEnergyRemainder}). 
The only difference lies in the different integration region
for the photon energy, which is given -- for the high-energy
part -- in Eq.~(\ref{SpecEnergyHigh}).
The photon energy integration variable $z$
is conveniently expressed as 
\begin{equation}
\label{SubstitutionHB}
z \to {\rm i}\,u \;\;\;\;\;\;\; \mbox{where} \;\;\;\;\;\;\;
u = \frac{1}{2}\,\left(\frac{1}{t} - t\right)\,.
\end{equation}
The method of integration is described in~\cite{Je1999,JeMoSo2001pra},
and we do not discuss any further details, here. We focus instead
on the convergence acceleration technique used in the 
evaluation.

In analogy with the middle-energy remainder,
we may write the integrand $S_{\rm HB}$ which is defined
in complete analogy to (\ref{defSMB}) as a sum over angular
momenta (``partial waves'')
\begin{equation}
\label{defSasymp}
S_{\rm HB} \propto \sum_{|\kappa|=1}^{\infty} T_{{\rm HB},|\kappa|}\,,
\end{equation}
where the $T_{{\rm HB},|\kappa|}$ are
defined in analogy to Eq.~(\ref{defSMB}).
Here, $|\kappa|$ represents the modulus
of the Dirac angular momentum quantum number of the virtual
intermediate state. 
The asymptotic behaviour of the $T_{{\rm HB},|\kappa|}$
for large $|\kappa|$ is [see Eq.~(4.7) in
\cite{Mo1974b}]
\begin{equation}
\label{defTasymp}
T_{{\rm HB},|\kappa|} = \frac{r^{2\,|\kappa|}}{|\kappa|} \,
 \left[{\rm const.} + {\rm O}\!\left(\frac{1}{|\kappa|}\right)\right]\,,
\end{equation}
where ``${\rm const.}$'' is independent of $|\kappa|$. 
The series in Eq.~(\ref{defSasymp}) is slowly
convergent for $r$ close to one, and the region near $r=1$ is known to
be problematic in numerical evaluations. 

It is found that the convergence of the series (\ref{defTasymp})
series near $r=1$ can be
accelerated very efficiently using the combined nonlinear-condensation
transformation (see~\cite{JeMoSoWe1999} and Sec.~\ref{CNCT})
applied to the
series $\sum_{k=0}^{\infty} t_k$ where $t_k = T_{{\rm HB},k+1}$
[see Eqs.~(\ref{defSasymp}) and~(\ref{defTasymp})]. 
The combined transformation (combination of
the condensation transformation and the Weniger transformation) was
found to be applicable to a wide range of slowly convergent monotone
series (series whose terms have the same sign), and many examples for
its application were given in 
Ref.~\cite{JeMoSoWe1999,AkSaJeBeSoMo2003}. For the
numerical treatment of radiative corrections in low-$Z$ systems, the
transformation has the advantage of removing the principal 
numerical difficulties
associated with the slow convergence of angular momentum
decompositions of the propagators near their singularity for equal
radial arguments.

All that remains to be discussed in the current section
is the low-$Z$ limit of the energy shift
\begin{equation}
\label{DefinitionOfFHB}
\Delta E_{\rm HB} = \frac{\alpha}{\pi} \,
\frac{(Z\alpha)^4}{n^3}\,F_{\rm HB}(nl_j,Z\alpha)\,.
\end{equation}
For the high-energy remainder $F_{\rm HB}$, the limits as
$Z\alpha\to0$ read [see Eq.~(4.15) in \cite{Mo1982}]
\begin{eqnarray}
\label{limFHBgen}
F_{\rm HB}(nl_j,Z\alpha) &=& 
\frac{1}{2\,l+1} \,
\left[\left(\frac{17}{18} - \frac{4}{3}\,\ln2\right) \delta_{l,0} 
+ \left(\frac{3}{2} - 2\,\ln2\right) \frac{1}{\kappa} \right.
\nonumber\\
&&
\quad \left. +
\left(\frac{5}{6} - \ln2\right) \frac{n-2\,l-1}{n}\right] +
{\rm O}(Z\alpha)\,.
\end{eqnarray}
For the atomic states under investigation, this leads to
\begin{eqnarray}
\label{limFHB}
F_{\rm HB}(1{\rm S}_{1/2},Z\alpha) &=& 
-0.093~457 + {\rm O}(Z\alpha) \,,\nonumber\\[2ex]
F_{\rm HB}(2{\rm S}_{1/2},Z\alpha) &=&
-0.023~364 + {\rm O}(Z\alpha) \,,\nonumber\\[2ex]
F_{\rm HB}(2{\rm P}_{1/2},Z\alpha) &=&
0.014~538 + {\rm O}(Z\alpha) \,,\nonumber\\[2ex]
F_{\rm HB}(2{\rm P}_{3/2},Z\alpha) &=&
-0.042~315 + {\rm O}(Z\alpha) \,.
\end{eqnarray}
%

%
% Results for the High-Energy Part
%
\subsection{Results for the High-Energy Part}

The limit of the function $F_{\rm H}$ as $Z\alpha\to0$
can be derived easily from the Eqs.~(\ref{limFHAgen}),
(\ref{limFHBgen}) as a function of the bound state quantum
numbers. It reads 
\begin{eqnarray}
F_{\rm H}(nl_j,Z\alpha) &=& 
\left(\frac{11}{10} - \ln2\right) \, \frac{1}{n}
+ \left(\frac{16}{15} - 2\,\ln2\right) \, \frac{1}{2\,l + 1} 
\nonumber\\[1ex]
& & + \left(-\frac{3}{2}\,\ln2 + \frac{5}{4}\right)\, 
\frac{1}{\kappa\,(2\,l+1)} +
\left(\frac{3}{2}\,\ln2 - \frac{9}{4}\right)\,\frac{1}{|\kappa|} 
\nonumber \\[1ex]
& & + 
\left(\frac{17}{18} - \frac{4}{3}\,\ln2\right) \delta_{l,0} +
\left(\frac{5}{6} - \ln2\right) \frac{n-2\,l-1}{n\,(2\,l+1)} +
{\rm O}(Z\alpha)\,.
\end{eqnarray}
For the atomic states investigated here, 
this expression yields the numerical values, 
\begin{eqnarray}
\label{limFH}
F_{\rm H}(1{\rm S}_{1/2},Z\alpha) &=&
- 1.313~085 + {\rm O}(Z\alpha)\,, \nonumber\\[2ex]
F_{\rm H}(2{\rm S}_{1/2},Z\alpha) &=&
- 1.446~418 + {\rm O}(Z\alpha)\,, \nonumber\\[2ex]
F_{\rm H}(2{\rm P}_{1/2},Z\alpha) &=&
- 1.066~667 + {\rm O}(Z\alpha)\,, \nonumber\\[2ex]
F_{\rm H}(2{\rm P}_{3/2},Z\alpha) &=&
- 0.566~667 + {\rm O}(Z\alpha)\,.
\end{eqnarray}
Numerical results for the high-energy part
\begin{equation}
F_{\rm H}(nl_j,Z\alpha) = 
F_{\rm HA}(nl_j,Z\alpha) + 
F_{\rm HB}(nl_j,Z\alpha)
\end{equation}
are summarized in Table~\ref{tableFHKL}. Note the apparent consistency
of the numerical results in Table~\ref{tableFHKL} with their analytically
obtained low-$Z$ limits in Eq.~(\ref{limFH}).

%
% Comparison to Analytic Calculations
%
\section{Comparison to Analytic Calculations}
\label{ComparisonAnalytic}

The numerical results for the scaled function
$F(nl_j,Z\alpha)$ describing the self energy 
defined in Eq.~(\ref{ESEasF}) are given in
Table~\ref{tableFKL}, together with the results for the nonperturbative
function $G_{\rm SE}(nl_j,Z\alpha)$,
which is implicitly defined in Eq.~(\ref{defFLO}). 
Results are provided for K and L shell states. The numerical
results for the remainder $G_{\rm SE}$ are obtained by
subtracting the analytic
lower-order terms listed in Eq.~(\ref{defFLO}) from the
complete numerical result for the scaled function 
$F(nl_j,Z\alpha)$. No additional fitting is performed. 

Analytic and numerical results at low
$Z$ can be compared by considering the remainder
function $G_{\rm SE}$. Note that an inconsistency in any
of the analytically obtained lower-order terms would be likely
to manifest itself in a grossly inconsistent dependence of
$G_{\rm SE}(nl_j,Z\alpha)$ on its argument $Z\alpha$;
this is not observed.
For S states, the following analytic
model for $G_{\rm SE}$ is commonly assumed,
which is motivated in part
by a renormalization-group analysis~\cite{MaSt2000} and is
constructed
in analogy with the pattern of the analytic coefficients $A_{ij}$
in Eq.~(\ref{defFLO}) and~(\ref{defFLOnS}),
\begin{eqnarray}
G_{\rm SE}(n{\rm S}_{1/2},Z\alpha) &=& 
A_{60}(n{\rm S}_{1/2}) 
+ (Z\alpha) \, \left[A_{71}(n{\rm S}_{1/2}) \, \ln(Z\alpha)^{-2} 
+ A_{70}(n{\rm S}_{1/2}) \right] 
\nonumber\\[1ex]
& & + (Z\alpha)^2 \, \left[
A_{83}(n{\rm S}_{1/2}) \, \ln^3(Z\alpha)^{-2} +
A_{82}(n{\rm S}_{1/2}) \, \ln^2(Z\alpha)^{-2} \right.
\nonumber\\[1ex]
& & \left. + A_{81}(n{\rm S}_{1/2}) \, \ln(Z\alpha)^{-2} 
+ A_{80}(n{\rm S}_{1/2}) \right]\,.
\end{eqnarray}
The (probably nonvanishing) $A_{83}$ coefficient, which 
introduces a triple logarithmic singularity at $Z\alpha=0$, 
hinders an accurate comparison of numerical and 
analytic data for $G_{\rm SE}$. 
A somewhat less singular behavior is expected of the 
difference
\begin{equation}
\label{DefDeltaGSE}
\Delta G_{\rm SE}(Z\alpha) = 
G_{\rm SE}(2{\rm S}_{1/2},Z\alpha) -
G_{\rm SE}(1{\rm S}_{1/2},Z\alpha)\,,
\end{equation}
because the leading logarithmic coefficients in any given order
of $Z\alpha$ are generally assumed to be equal for all
S states, which would mean in particular
\begin{equation}
A_{71}(1{\rm S}_{1/2}) = A_{71}(2{\rm S}_{1/2})
\end{equation}
and
\begin{equation}
A_{83}(1{\rm S}_{1/2}) = A_{83}(2{\rm S}_{1/2})\,.
\end{equation}
Now we define $\Delta A_{kl}$ as the difference of the values of the 
analytic coefficients for the two lowest S states,
\begin{equation}
\label{DefDeltaA}
\Delta A_{kl} = 
A_{kl}(2{\rm S}_{1/2}) -
A_{kl}(1{\rm S}_{1/2})\,.
\end{equation}
The function $\Delta G_{\rm SE}$ defined in
Eq.~(\ref{DefDeltaGSE}) can be assumed to have the following
semi-analytic expansion around $Z \alpha = 0$,
\begin{eqnarray}
\label{ExpansionDeltaGSE}
\Delta G_{\rm SE}(Z\alpha) &=& 
\Delta A_{60} + (Z\alpha) \, \Delta A_{70} + (Z\alpha)^2 \,
\left[\Delta A_{82} \, \ln^2(Z\alpha)^{-2} \right.
\nonumber\\[1ex]
& & \qquad \left. 
+ \Delta A_{81} \, \ln(Z\alpha)^{-2} 
+ \Delta A_{80} 
+ {\rm o}(Z\alpha) \right]\,.
\end{eqnarray}
In order to detect possible inconsistencies in the 
numerical and analytic data for $G_{\rm SE}$,  
we difference the data for $\Delta G_{\rm SE}$,
i.e., we consider the following finite difference approximation
to the derivative of the function $\Delta G_{\rm SE}$,
\begin{equation}
\label{DefgZ}
g(Z) = \Delta G_{\rm SE}\bbox{(} (Z+1)\,\alpha\bbox{)} -
\Delta G_{\rm SE}\bbox{(}Z\alpha\bbox{)}\,.
\end{equation}
It is perhaps not {\em a priori} obvious why this 
combination leads to a sensible comparison of numerical
and analytic data. This will be explained in the sequel.

We denote the analytic
and numerical limits of $\Delta G_{\rm SE}(Z\alpha)$ as
$Z\alpha \to 0$ as $\Delta A^{\rm ana}_{60}$ and $\Delta A^{\rm num}_{60}$,
respectively. Of course, we hope that these will turn
out to be equal. However, we temporarily leave open the possibility of 
an inconsistency between numerical and analytic data by keeping
$\Delta A^{\rm num}_{60}$ and $\Delta A^{\rm ana}_{60}$
as distinct variables.
In order to illustrate how a discrepancy
could be detected by investigating the function
$g(Z)$, we consider special cases
of the function $\Delta G_{\rm SE}(Z\alpha)$ and $g(Z)$.
For $Z=0$, the exact result can be inferred
exclusively and uniquely using the analytic approach of Ref.~\cite{Pa1993},
and we have
\begin{equation}
\Delta G_{\rm SE}(0) = \Delta A^{\rm ana}_{60}\,,
\end{equation}
whereas for $Z=1$, which is determined by numerical data,
\begin{equation}
\Delta G_{\rm SE}(\alpha) \approx \Delta A^{\rm num}_{60} +
  \alpha\, \Delta A_{70}\,,
\end{equation}
with a possibly different limit $\Delta A^{\rm num}_{60}$
as $Z \to 0$, and for $Z=2$,
\begin{equation}
\Delta G_{\rm SE}(2\alpha) \approx \Delta A^{\rm num}_{60} +
  2\,\alpha\,\Delta A_{70}\,,
\end{equation}
where we ignore higher-order analytic terms. Hence, for $Z=0$, we have 
\begin{eqnarray}
\label{kink}
g(0) = \Delta G_{\rm SE}(\alpha) - \Delta G_{\rm SE}(0)
= \Delta A^{\rm num}_{60} - \Delta A^{\rm ana}_{60} +
  \alpha\,\left[ \Delta A_{70} + {\rm o}(Z\alpha) \right]\,.
\end{eqnarray}
A possible inconsistency (nonvanishing 
$\Delta A^{\rm num}_{60} - \Delta A^{\rm ana}_{60}$) would 
influence the value of $g(0)$.
For $Z=1$, the value of $g$ is determined solely by numerical data,
\begin{equation}
g(1) = \Delta G_{\rm SE}(2\alpha) - \Delta G_{\rm SE}(\alpha) 
= \alpha\,\left[ \Delta A_{70} + {\rm o}(Z\alpha) \right]\,,
\end{equation}
and for $Z=2$, we have
\begin{equation}
\label{g2}
g(2) = \Delta G_{\rm SE}(3\alpha) - \Delta G_{\rm SE}(2\alpha)         
= \alpha\,\left[ \Delta A_{70} + {\rm o}(Z\alpha) \right]\,.
\end{equation}
Of course, analogous relations hold for $Z > 2$.
This means that a meaningful comparison of the analytic and numerical
approaches can be made by comparing the value $g(0)$, which is
influenced by the term $\Delta A^{\rm num}_{60} - \Delta A^{\rm ana}_{60}$,
to the other values $g(1),\dots,g(4)$, which are {\em independent}
of the difference $\Delta A^{\rm num}_{60} - \Delta A^{\rm ana}_{60}$.
We recall that the numerical data
from Table~\ref{tableFKL} lead to the evaluation of the five values
$g(0)$, $g(1)$, $g(2)$, $g(3)$ and $g(4)$. 
A plot of the function $g(Z)$ serves two purposes: (i) the values
$g(1),\dots,g(4)$ should exhibit apparent convergence to
some limiting value $\alpha\,\Delta A_{70}$ as $Z \to 0$, and this can
be verified by inspection of the plot, and (ii)
a discrepancy between the analytic and numerical approaches --
as explained above --
would result in a nonvanishing value for 
$\Delta A^{\rm num}_{60} - \Delta A^{\rm ana}_{60}$, and it would
introduce an inconsistency between the trend in the values of
$g(1),\dots,g(4)$, and the value of $g(0)$ [see Eq.~(\ref{kink})].

%
% tableFKL
%
\begin{table}[thb!]
\begin{center}
\begin{minipage}{16cm}
\begin{center}
\begin{tabular}{lr@{.}lr@{.}lr@{.}lr@{.}l}
\hline
\hline
$Z$ & 
\multicolumn{2}{c}{\rule[-3mm]{0mm}{8mm} 
  $F(1{\rm S}_{1/2},Z\alpha)$} &
\multicolumn{2}{c}{\rule[-3mm]{0mm}{8mm} 
  $F(2{\rm S}_{1/2},Z\alpha)$} &
\multicolumn{2}{c}{\rule[-3mm]{0mm}{8mm} 
  $F(2{\rm P}_{1/2},Z\alpha)$} &
\multicolumn{2}{c}{\rule[-3mm]{0mm}{8mm} 
  $F(2{\rm P}_{3/2},Z\alpha)$} \\
\hline
$1$ &
 $10$ & $316~793~650(1)$ &
 $10$ & $546~825~185(5)$ &
 $-0$ & $126~396~37(1)$ &
 $ 0$ & $123~498~56(1)$ \\
$2$ &
  $8$ & $528~325~052(1)$ &
  $8$ & $758~870~25(1)$ &
 $-0$ & $125~816~16(1)$ &
 $ 0$ & $123~835~55(1)$ \\
$3$ &
  $7$ & $504~503~422(1)$ &
  $7$ & $735~777~20(1)$ &
 $-0$ & $124~992~24(1)$ &
 $ 0$ & $124~317~10(1)$ \\
$4$ &
  $6$ & $792~824~081(1)$ &
  $7$ & $025~002~41(1)$ &
 $-0$ & $123~968~79(1)$ &
 $ 0$ & $124~918~48(1)$ \\
$5$ &
  $6$ & $251~627~078(1)$ &
  $6$ & $484~860~42(1)$ &
 $-0$ & $122~774~94(1)$ &
 $ 0$ & $125~623~30(1)$ \\
\hline
\hline
$Z$ & 
\multicolumn{2}{c}{\rule[-3mm]{0mm}{8mm} 
  $G_{\rm SE}(1{\rm S}_{1/2},Z\alpha)$} &
\multicolumn{2}{c}{\rule[-3mm]{0mm}{8mm} 
  $G_{\rm SE}(2{\rm S}_{1/2},Z\alpha)$} &
\multicolumn{2}{c}{\rule[-3mm]{0mm}{8mm} 
  $G_{\rm SE}(2{\rm P}_{1/2},Z\alpha)$} &
\multicolumn{2}{c}{\rule[-3mm]{0mm}{8mm} 
  $G_{\rm SE}(2{\rm P}_{3/2},Z\alpha)$} \\
\hline
$1$ &
 $-30$ & $290~24(2)$ &
 $-31$ & $185~15(9)$ &
 $ -0$ & $973~5(2)$ &
 $ -0$ & $486~5(2)$ \\
$2$ &
 $-29$ & $770~967(5)$ &
 $-30$ & $644~66(5)$ &
 $ -0$ & $949~40(5)$ &
 $ -0$ & $470~94(5)$ \\
$3$ &
 $-29$ & $299~170(2)$ &
 $-30$ & $151~93(2)$ &
 $ -0$ & $926~37(2)$ &
 $ -0$ & $456~65(2)$ \\
$4$ &
 $-28$ & $859~222(1)$ &
 $-29$ & $691~27(1)$ &
 $ -0$ & $904~12(1)$ &
 $ -0$ & $443~13(1)$ \\
$5$ &
 $-28$ & $443~372~3(8)$ &
 $-29$ & $255~033(8)$ &
 $ -0$ & $882~478(8)$ &
 $ -0$ & $430~244(8)$ \\
\hline
\hline
\end{tabular}
\caption{\label{tableFKL} Numerical results for
the scaled function $F$ [defined in Eq.~(\ref{ESEasF})]
and the remainder function $G_{\rm SE}$
implicitly defined in Eq.~(\ref{defFLO}).}
\end{center}
\end{minipage}
\end{center}
\end{table}

%
% Plot of 2S1S
%
\begin{figure}[htb!]
\begin{center}
\begin{minipage}{12.0cm}
\centerline{\mbox{\epsfysize=7.0cm\epsffile{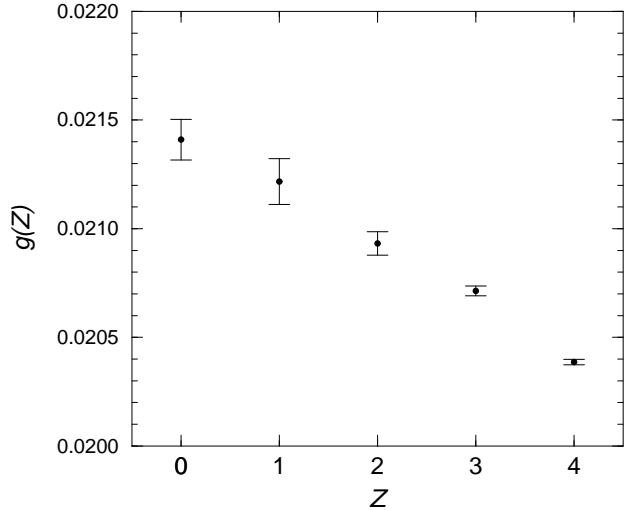}}}
\caption{\label{Plot2S1S} Plot of the function $g(Z)$ defined in
Eq.~(\protect\ref{DefgZ}) in the
region of low nuclear charge.
For the evaluation of the 
data point at $Z = 0$, a value of $A_{60}(1{\rm S}_{1/2}) =
-30.92415(1)$ is employed~\protect\cite{Pa1993,JeMoSo1999,Pa1998priv}.}
\end{minipage}
\end{center}
\end{figure}

Among the different evaluations of $A_{60}$ for the ground state, the
result in~\cite{Pa1993} has the smallest quoted uncertainty.  In
Fig.~\ref{Plot2S1S} we display a plot of $g(Z)$ for low nuclear charge
$Z$. A value of $A_{60}(1{\rm S}_{1/2}) = 
A^{\rm ana}_{60}(1{\rm S}_{1/2}) = 
-30.92415(1)$~\cite{Pa1993,JeMoSo1999,Pa1998priv}
is used in Fig.~\ref{Plot2S1S}.
The results indicate very good agreement between the numerical and
analytic approaches to the Lamb shift in the low-$Z$ region up to the
level of a few Hz in frequency units for the low-lying
atomic states (where $n$ is the principal quantum number).
The error bars represent the numerical uncertainty of the 
values in Table~\ref{tableFKL}, which correspond to an 
uncertainty on the level of $1.0 \times Z^4 \, {\rm Hz}$ in
frequency units.

Analytic work on the correction $A_{60}$ has
extended over three decades~\cite{ErYe1965a,ErYe1965b,Er1971,Sa1981,Pa1993}.  
The complication arises that although the calculations are in general
analytic, some remaining one-dimensional integrations could not be
evaluated analytically because of the nature of the 
integrals [see e.g.~Eq.~(6.96) in~\cite{Pa1993}]. Therefore a
step-by-step comparison of the analytic calculations is difficult.  An
additional difficulty is the isolation of those analytic terms which
contribute in a given order in $Z\alpha$, i.e.~the isolation of only
those terms which contribute to $A_{60}$. The apparent consistency 
of the numerical and analytic data in Fig.~\ref{Plot2S1S} represents
an independent consistency check on the rather
involved analytic calculations.
 
%
% Plot of 2P1/2
%
\begin{figure}[htb!]
\begin{center}
\begin{minipage}{12.0cm}
\centerline{\mbox{\epsfysize=7.0cm\epsffile{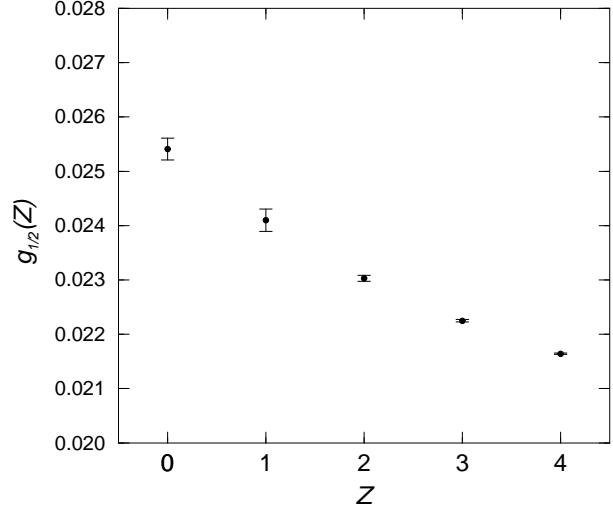}}}
\caption{\label{Plot2P12} Comparison of numerical data
and analytically evaluated higher-order binding corrections
for the $2{\rm P}_{1/2}$ state.
We plot the function $g_{1/2}(Z)$ defined
in Eq.~(\ref{Defgj}) in the region of low $Z$. The numerical data
appear to be consistent with the analytic result of
$A_{60}(2{\rm P}_{1/2}) = -0.99891(1)$
obtained in~\protect\cite{JePa1996}.}

\end{minipage}
\end{center}
\end{figure}

Our numerical results are not inconsistent with the
analytic result \cite{Ka1997} for a higher-order logarithm,
\begin{equation}
A_{71} = \pi\,\left(\frac{139}{64}-\ln2\right) = 4.65,
\end{equation}
although they do not necessarily confirm it. As
in~\cite{JeMoSo1999}, we obtain as an estimate $A_{71} = 5.5(1.0)$
(from the fit to the numerical data for both S states). Logarithmic
terms corresponding to the probably nonvanishing $A_{83}$ coefficient
should be taken into account for a consistent fit of the
corrections to $G_{\rm SE}$. These highly singular terms are difficult
to handle with a numerical fitting procedure. The terms 
$A_{83}$, $A_{82}$ and $A_{81}$ furnish
three more free parameters for the numerical fit, where only five data
points are available (in addition to the quantities $A_{60}$, $A_{71}$
and $A_{70}$, which may also be regarded as free parameters for the
fitting procedure).  The determination of $A_{60}$ by a fit from the
numerical data is much more stable than the determination of
the logarithmic correction $A_{71}$. We briefly note that our
all-order evaluation essentially eliminates the uncertainty due to the
unknown higher-order analytic terms. It is interesting to note that
the same numerical 
methods are employed for both the S and P states in our all-order
(in $Z\alpha$) calculation, whereas the analytic treatment
of S and P states differs~\cite{Pa1993,JePa1996}.

The comparison of numerical and analytic result is much 
less problematic for P states, because the function
$G_{\rm SE}$ is less singular [see Eqs.~(\ref{defFLOnPj})
and~(\ref{A60nP})].
For the 2P states, 
we observe that the function $G_{\rm SE}(2{\rm P}_j,Z\alpha)$ 
has the same semi-analytic expansion about $Z\alpha=0$ as the 
function $\Delta G_{\rm SE}(Z\alpha)$ defined for S states in
Eq.~(\ref{DefDeltaGSE}). We have
\begin{eqnarray}
G_{\rm SE}(2{\rm P}_j,Z\alpha) &=& 
A_{60}(2{\rm P}_j) + (Z\alpha) \, A_{70}(2{\rm P}_j)
+ (Z\alpha)^2 \, \left[ A_{82}(2{\rm P}_j) \, \ln^2(Z\alpha)^{-2} \right.
\nonumber\\[1ex]
& & \qquad \left. 
+ A_{81}(2{\rm P}_j) \, \ln(Z\alpha)^{-2} 
+ A_{80}(2{\rm P}_j) 
+ {\rm o}(Z\alpha) \right]\,.
\end{eqnarray}
Hence, we plot the function
\begin{equation}
\label{Defgj}
g_j(Z) =
G_{\rm SE}\bbox{(}2{\rm P}_j,(Z+1)\,\alpha\bbox{)} -
G_{\rm SE}\bbox{(}2{\rm P}_j,Z\alpha\bbox{)}
\end{equation}
for $j=1/2$ in the region of low $Z$, with the notion that
an inconsistent analytic result for $A_{60}(2{\rm P}_j)$ would lead to
irregularity at $Z=0$, in analogy with the
S states. The numerical data
shown in Fig.~\ref{Plot2P12} appears to be
consistent with the analytic results of
\begin{equation}
A_{60}(2{\rm P}_{1/2}) = -0.998\,91(1) \,.
\end{equation}
obtained in~\cite{JePa1996}. For the $2{\rm P}_{3/2}$ state,
see Fig.~6 of Ref.~\cite{JeMoSo2001pra}.
In this context it may be interesting to
note that analytic results obtained in~\cite{JePa1996,JeSoMo1997}
for the higher-order binding corrections to 2P, 3P and 4P states have
recently been confirmed indirectly~\cite{Pa1999}.

%
% Analytic Self--Energy Calculations for Excited States
%
\chapter{Analytic Self--Energy Calculations for Excited States}
\label{QEDSelExc}

\typeout{========================================}
\typeout{Analytic Calculations for Excited States}
\typeout{========================================}
\typeout{}

%
% Orientation
%
\section{Orientation}
\label{ORISelExc}

Analytic calculations of the one-loop self energy in bound systems 
have a long history, starting from Bethe's seminal paper~\cite{Be1947},
and have meanwhile extended over more  than five decades.
The calculations are based on an expansion of the effect into powers of the 
parameter $Z\alpha$, where $Z$ is the nuclear charge and $\alpha$
is the fine-structure constant
(see e.g.~\cite{Pa1993,JePa1996,JeSoMo1997,Je1999}). 
The expansion is semi-analytic, i.e.
it involves powers of $Z\alpha$ and $\ln[(Z\alpha)^{-2}]$.
Terms which are of higher than leading order in $Z\alpha$ are commonly
referred to as the binding corrections (see also Ch.~\ref{QEDSelTwo}).
A certain nonlogarithmic binding correction, known as the 
$A_{60}$-coefficient, has been of particular interest. 
For relevant formulas and definitions regarding $A_{60}$,
we refer to Ch.~\ref{QEDSelOne},
especially to Eqs.~(\ref{ESEasF}) -- (\ref{A60nP}).

In this chapter, we present results of an analytic evaluation
of higher-order binding corrections to the one-loop self energy
for excited bound states of hydrogenlike systems. We focus
on states with principal quantum numbers $n = 2 \dots 5$ and
Dirac angular quantum numbers $\kappa = 1,-2,2,\dots,4,-5$.
The current calculation represents 
a continuation of previous investigations on
the subject~\cite{JePa1996,JeSoMo1997}.

The improved results for $A_{60}$ coefficients rely essentially
on a more compact code for the analytic calculations, written
in the computer-algebra package {\em Mathematica}~\cite{Wo1988}, 
which enables the corrections to be evaluated 
semi-automatically. Intermediate expressions with some 
10,000 terms are encountered, and the complexity of the 
calculations sharply increases with the principal quantum number 
$n$, and, as far as the complexity of
the angular momentum algebra is concerned,
with the orbital angular momentum $l = |\kappa + 1/2| - 1/2$
of the bound electron. 
Calculations were performed,
in part, on IBM RISC/6000 workstations, both at the 
National Institute of Standards and Technology and at the 
Laboratoire Kastler--Brossel, and profited from the 
availability of 2 GigaBytes of Random Access Memory.

Of crucial importance was the development of convergence acceleration
methods which were used extensively for the evaluation of remaining
one-dimensional integrals which could not be done analytically.
These integrals are analogous to expressions encountered in
previous work~\cite{JePa1996,JeSoMo1997}.
The numerically evaluated contributions
involve slowly convergent hypergeometric series,
and -- in more extreme cases -- infinite series 
over partial derivatives of hypergeometric functions, and
generalizations of Lerch's $\Phi$ transcendent~\cite{Ol1974,Ba1953vol1}. 
We have found
the Combined Nonlinear--Condensation Transformation~\cite{JeMoSoWe1999}
to be a versatile method for the efficient numerical evaluation
of the slowly convergent series encountered in the evaluation of the 
$A_{60}$-coefficients. 

Results are given below for the $A_{60}$-coefficients with an absolute
precision of $10^{-9}$. As explained in detail 
in~\cite{Pa1993,JePa1996,JeSoMo1997}, the calculation
of the one-loop self energy falls 
naturally into a high- and a low-energy part 
($F_H$ and $F_L$, respectively). In turn, the low-energy
part can be separated naturally into the nonrelativistic dipole and
the nonrelativistic quadrupole part, and into relativistic corrections
to the current, to the hamiltonian, to the wavefunction, and 
to the energy of the bound state. For some states, severe numerical
cancellations are observed between these (low-energy)
contributions to the low-energy part of the self energy.
In the final step of the calculation, we observe additional
numerical cancellations between the 
high- and the low-energy part for the finite contributions
to $A_{60}$, beyond the necessary cancellation
of the divergent contributions which depend
on the scale-separation parameter $\epsilon$;
this parameter and the mathematical method which is
employed in the expansion of the self energy
are described and explained
in Sec.~\ref{EpsilonMethod}. For some of the investigated atomic states,
the absolute magnitude of the
$A_{60}$-coefficients is as small as $10^{-3}$, 
although the largest single contribution
is always on the order of $10^{-2}$. This illustrates the
cancellations between the finite contributions to $A_{60}$.

%
% epsilon method
%
\section{The epsilon-Method}
\label{EpsilonMethod}

We discuss here, by way of example,
the $\epsilon$ method employed
in the analytic calculation of the self energy in bound systems.
This method is very suitable~\cite{Pa1993,JePa1996,JeSoMo1997}
for the separation of the two different
energy scales for virtual photons: the nonrelativistic domain, in which the 
virtual photon assumes values of the order of the atomic binding energy,
and the relativistic domain, in which the virtual photon assumes values
of the order of the electron rest mass. Different approximation schemes and
different asymptotic expansions are
adequate for the two different domains. Without these approximations and
expansions, the analytic evaluation of either the high- or the low-energy
part would not be feasible. Therefore, the $\epsilon$ method is also
used for the treatment of the highly excited atomic states 
which are the subject of (this part) of the thesis.
At the same time, the model example discussed here 
is meant to illustrate the usefulness of the ``$\epsilon$ method''
in a more general context.

We will consider here a model problem with one ``virtual photon''.
The separation into high-and low-energy photons necessitates the temporary
introduction of a parameter $\epsilon$; the dependence on $\epsilon$ cancels  
when the high- and the low-energy parts are added together. We have,
\begin{eqnarray}
\mbox{nonrelativistic domain} & \ll \epsilon \ll & 
\mbox{electron rest mass} \\
(Z \alpha)^2 \, m_{\mathrm{e}} & \ll \epsilon \ll & m_{\mathrm{e}}\,,
\end{eqnarray}
where $\alpha$ is the fine structure constant, 
and $Z$ is the nuclear charge. The high-energy part
is associated with photon energies $\omega > \epsilon$, and the 
low-energy part is associated with photon energies $\omega < \epsilon$.

In order to illustrate the procedure, we discuss a simple,
one-dimensional example: the evaluation of 
\begin{equation}
I(\beta) = 
\int_0^1 \sqrt{\frac{\omega^2 + \beta^2}{1 - \omega^2}} \, 
{\mathrm{d}}\omega\,.
\end{equation}
where the integration variable $\omega$ might be interpreted as the 
``energy'' of a ``virtual photon''. The integral $I$ can be expressed in
terms of special functions,
\begin{equation}
I(\beta) = \beta\,E\left(-\frac{1}{\beta^2}\right) = 
\beta \, \frac{\pi}{2} \,
{}_2 F_1 \left(-\frac{1}{2}, \frac{1}{2}; 1;  -\frac{1}{\beta^2}\right)\,,
\end{equation}
where $E$ is the complete elliptic integral of the second kind, and 
${}_2 F_1$ denotes a hypergeometric function. An alternative integral
representation reads $I(\beta) = \int_0^{\pi/2} \sqrt{\beta^2 + 
\sin^2(\omega)} \, {\mathrm{d}}\omega$.

The purpose of the calculation is
to derive a semi-analytic expansion of $I(\beta)$ in powers of 
$\beta$ and $\ln \beta$. The fine structure constant $\alpha$ takes the 
r\^{o}le 
of the expansion parameter $\beta$ in actual 
calculations of the self energy. 
We discuss first the ``high-energy part'' of the calculation. 
It is given by the expression
\begin{eqnarray}
\label{IH}
I_{\mathrm{H}}(\beta) &=& 
\int_\epsilon^1 \sqrt{\frac{\omega^2 + \beta^2}{1 - \omega^2}} \,
{\mathrm{d}}\omega\,.
\end{eqnarray}
For $\omega > \epsilon$, we may expand
\begin{equation}
\label{expansion1}
\sqrt{\omega^2 + \beta^2} = \omega + \frac{\beta^2}{2\,\omega} +
  \frac{\beta^4}{8\,\omega^3} + \mathcal{O}(\beta^6)\,,
\end{equation}
but this expansion is not applicable in higher
orders to the domain $0 < \omega < \epsilon$ because of the appearance
of inverse powers of $\omega$ (analogous to an
``infrared divergence'' in QED).

The separation parameter $\epsilon$ acts an infrared regulator.
After expanding in $\beta$ [see Eq.~(\ref{expansion1})], the 
resulting integrals in each order of $\beta$ can be evaluated
analytically. Subsequently, we expand every term in
the $\beta$-expansion in powers of $\epsilon$ up to
the order $\epsilon^0$, i.e.~we keep only the divergent and constant
terms in $\epsilon$. The result is
\begin{eqnarray}
\label{IHresult}
I_{\mathrm{H}}(\beta,\epsilon) &=& 1 + \beta^2 \, \left\{ \frac{1}{2} \,
\ln \left(\frac{2}{\epsilon}\right) +
{\mathcal{O}}(\epsilon) \right\} \nonumber\\[1ex]
& & + \beta^4 \, \left\{ - \frac{1}{16\, \epsilon^2} -
\frac{1}{16} \, \ln\left( \frac{2}{\epsilon} \right) +
\frac{1}{32} + {\mathcal{O}}(\epsilon) \right\} 
\nonumber\\[1ex]
& & + \beta^6 \, \left\{ \frac{1}{64\, \epsilon^4} +
\frac{1}{64 \, \epsilon^2} +
\frac{3}{128} \, \ln\left( \frac{2}{\epsilon} \right) - \frac{7}{512} +
{\mathcal{O}}(\epsilon) \right\} \nonumber\\[1ex]
& & + {\mathcal{O}}(\beta^8) \,.
\end{eqnarray}
Here, the ``${\mathcal{O}}$''-symbol identifies a contribution for which
${\mathcal{O}}(x)/x \to {\mathrm{const.}}$ as 
$x \to 0$, whereas the ``${\mathrm{o}}$''-symbol identifies
the weaker requirement ${\mathrm{o}}(x) \to 0$ as 
$x \to 0$; this is consistent with the 
standard notation (see e.g.~\cite{Er1987}).

The contribution $I_{\mathrm{H}}(\beta)$ corresponds to the 
``high-energy part'' in an analytic calculation
of the self energy, where
the propagator of the bound electron is explicitly expanded in
powers of the fine structure constant $\alpha$.
Now we turn to the ``low-energy part''. The expression for the 
low-energy part ($0 < \omega < \epsilon$) reads
\begin{eqnarray}
\label{IL}
I_{\mathrm{L}}(\beta) &=& 
\int_0^\epsilon \sqrt{\frac{\omega^2 + \beta^2}{1 - \omega^2}} \,
{\mathrm{d}}\omega\,.
\end{eqnarray}
The expansion (\ref{expansion1}) is not applicable in this energy
domain; we therefore have to keep the numerator of the integrand
$\sqrt{\omega^2 + \beta^2}$ in unexpanded form. However, we can expand
the denominator $\sqrt{1 - \omega^2}$ of the integrand in powers
of $\omega$; because $0 < \omega <\epsilon$ (with $\epsilon$ small),
this expansion in $\omega$ is in fact an expansion in $\beta$ --
although the situation is somewhat problematic in the sense that
every term in the $\omega$-expansion gives rise to terms of arbitrarily
high order in the $\beta$-expansion [see also Eq.~(\ref{problematic})
below]. 

The term $\sqrt{\omega^2 + \beta^2}$ is analogous to the 
Schr\"{o}dinger--Coulomb propagator in the 
evaluation of the self energy 
-- it has to be kept in an unexpanded form --, whereas the expansion

\begin{equation}
\label{expansion2}
\frac{1}{\sqrt{1 - \omega^2}} = 1 + \frac{\omega^2}{2} + 
\frac{3}{8}\,\omega^4 + {\mathcal{O}}(\omega^6)
\end{equation}
corresponds to the expansion into the $(Z\alpha)$-expansion
in the low-energy part.

Every term in the expansion (\ref{expansion2}) gives rise to arbitrarily
high-order corrections in $\beta$, but it starts with the power
$\omega^n \to \beta^{n+2}$. For example, we have for the leading
term of order $\omega^0 = 1$ from Eq.~(\ref{expansion2}),
\begin{eqnarray}
\label{problematic}
\lefteqn{\int_0^\epsilon \sqrt{\omega^2 + \beta^2} \, {\mathrm{d}}\omega =
\beta^2 \, \left\{ \frac{1}{2}\,
\ln\left(\frac{2}{\beta} \, \epsilon \right) + 
\frac{1}{4} + {\mathcal{O}}(\epsilon) \right\}} \nonumber\\[1ex]
& & + \beta^4 \, \left\{ \frac{1}{16 \, \epsilon^2} +
{\mathcal{O}}(\epsilon) \right\} +
\beta^6 \, \left\{ - \frac{1}{64 \, \epsilon^4} +
{\mathcal{O}}(\epsilon) \right\} + {\mathcal{O}}(\beta^8)\,.
\end{eqnarray}
Note that the terms generated in the orders $\beta^4$ and $\beta^6$
are needed to cancel divergent contributions in respective
orders of $\beta$ from the high-energy part given in
Eq.~(\ref{expansion1}). 
The term of order $\omega^2$ from (\ref{expansion2}) results in
\begin{eqnarray}
\frac{1}{2}\, \int_0^\epsilon \omega^2 \, \sqrt{\omega^2 + \beta^2} \, 
{\mathrm{d}}\omega &=&
\beta^4 \, \left\{ -\frac{1}{16} \, 
\ln\left(\frac{2}{\beta} \, \epsilon \right) + 
\frac{1}{64} + {\mathcal{O}}(\epsilon) \right\} \nonumber\\[1ex]
& & + \beta^6 \, \left\{ - \frac{1}{64 \, \epsilon^2} + 
{\mathcal{O}}(\epsilon) \right\} +
{\mathcal{O}}(\beta^8)\,.
\end{eqnarray}
Altogether, we obtain for the low-energy part,
\begin{eqnarray}
\label{ILresult}
I_{\mathrm{L}}(\beta,\epsilon) &=& \beta^2 \, 
\left\{ \frac{1}{2} \, \ln \left(\frac{2}{\beta} \, \epsilon\right) +
\frac{1}{4} + {\mathcal{O}}(\epsilon) \right\} \nonumber\\[1ex]
& & + \beta^4 \, \left\{ \frac{1}{16\, \epsilon^2} -
\frac{1}{16} \, \ln\left( \frac{2}{\beta} \, \epsilon \right) +
\frac{1}{64} + {\mathcal{O}}(\epsilon) \right\} 
\nonumber\\[1ex]
& & + \beta^6 \, \left\{ - \frac{1}{64\, \epsilon^4} -
\frac{1}{64 \, \epsilon^2} +
\frac{3}{128} \, \ln\left( \frac{2}{\beta} \, \epsilon \right) -
\frac{5}{512} +
{\mathcal{O}}(\epsilon) \right\} \nonumber\\[1ex]
& & + {\mathcal{O}}(\beta^8 \, \ln \beta) \,.
\end{eqnarray}
When the high-energy part (\ref{IHresult}) and the low-energy part 
(\ref{ILresult}) are added, the dependence on $\epsilon$ cancels,
and we have
\begin{eqnarray}
I(\beta) &=& I_{\mathrm{H}}(\beta, \epsilon) + 
I_{\mathrm{L}}(\beta, \epsilon) \nonumber\\[2ex]
&=& 1 + \beta^2 \,
\left\{ \frac{1}{2}\,
\ln\left(\frac{4}{\beta}\right) + \frac{1}{4} \right\} +
\beta^4 \, \left\{ -\frac{1}{16}\, 
\ln\left(\frac{4}{\beta}\right) + \frac{3}{64} \right\}
\nonumber\\[2ex]
& & + \beta^6 \, \left\{ \frac{3}{128}\, 
\ln\left(\frac{4}{\beta}\right) - \frac{3}{128} \right\}
+ {\mathcal{O}}(\beta^8 \, \ln \beta)\,.
\end{eqnarray}
In order to illustrate the analogy with the self energy calculation
presented here, we would like to point out that
the dependence on $\epsilon$ cancels out in the final result
which is the sum of the high-energy part $E_{\mathrm H}$
given in Eq.~(\ref{FH2P12}) [for the two-loop case
see Eq.~(\ref{EH})] and the low-energy part
$E_{\mathrm L}$ in Eq.~(\ref{FL2P12}) 
[for the two-loop case see Eq.~(\ref{EL})].

%
% Results for High-- and Low--Energy Contributions
%
\section{Results for High-- and Low--Energy Contributions}
\label{ResHighLow}

%
% $P_{1/2}$ states ($\kappa = 1$)
%
\subsection{P\_1/2 states (kappa = 1)}
\label{resultsP12}

Here and in the following subsections,
we present results which have been obtained for the high-
and low-energy energy parts of the
self energy or all states with principal
quantum number $n = 2 \dots 5$ and orbital angular
quantum number $l > 0$.
These results, combined with numerical calculations for
low-, intermediate and high nuclear charge, allow to reliably estimate
quantum electrodynamic (QED) radiative corrections for any atomic state with
principal quantum number $n \leq 5$.
For the definition of the scaled self energy function $F$
and of the analytic $A$-coefficients,
see Eqs.~(\ref{ESEasF}) -- (\ref{A60nP}).
The results are given in terms of the expansion parameter
$Z\alpha$ and the scale-separation parameter $\epsilon$,
whose r\^{o}le in the calculation
is explained in Sec.~\ref{EpsilonMethod}.

For ${\rm P}_{1/2}$ states, we obtain for the 
following results for the high- and low-energy parts.
The $2{\rm P}_{1/2}$ high-energy part is:
\begin{eqnarray}
\label{FH2P12}
& & F_H (2{\rm P}_{1/2}) = -\frac{1}{6} + (Z\alpha)^2 \, 
  \left[\frac{4177}{21600} - \frac{2}{9\,\epsilon} -
    \frac{103}{180} \, \ln(2\epsilon)\right]\,,
\end{eqnarray}
$2{\rm P}_{1/2}$ low-energy part:
\begin{equation}
\label{FL2P12}
F_L (2{\rm P}_{1/2}) = -\frac{4}{3} \, \ln k_0 (2{\rm P}) +
\left(Z \alpha \right)^2 \left[ -0.795~649~812(1) +
\frac{2}{9 \, \epsilon} +
\frac{103}{180} \, 
\ln\left(\frac{\epsilon}{(Z \alpha)^2}\right)\right]\,,
\end{equation}
$3{\rm P}_{1/2}$ high-energy part:
\begin{equation}
F_H (3{\rm P}_{1/2}) = -\frac{1}{6} + (Z\alpha)^2 \, 
  \left[\frac{6191}{24300} - \frac{20}{81\,\epsilon} -
    \frac{268}{405} \, \ln(2\epsilon)\right]\,,
\end{equation}
$3{\rm P}_{1/2}$ low-energy part:
\begin{equation}
F_L (3{\rm P}_{1/2}) = - \frac{4}{3} \ln k_0(3{\rm P}) +
(Z \alpha)^2 \, \left[ -0.944~288~447(1) + \frac{20}{81 \, \epsilon}
+ \frac{268}{405} \, 
\ln\left(\frac{\epsilon}{(Z \alpha)^2}\right)\right]\,,
\end{equation}
$4{\rm P}_{1/2}$ high-energy part:
\begin{equation}
F_H (4{\rm P}_{1/2}) = -\frac{1}{6} + (Z\alpha)^2 \,
  \left[\frac{24409}{86400} - \frac{23}{90\,\epsilon} -
    \frac{499}{720} \, \ln(2\epsilon)\right]\,,
\end{equation}
$4{\rm P}_{1/2}$ low-energy part:
\begin{equation}
F_L(4{\rm P}_{1/2}) = - \frac{4}{3} \ln k_0(4{\rm P}) +
  (Z \alpha)^2 \, \left[ -0.997~810~211(1) + 
    \frac{23}{90 \, \epsilon} \, + \frac{499}{720} \, 
      \ln\left(\frac{\epsilon}{(Z \alpha)^2}\right)\right]\,,
\end{equation}
$5{\rm P}_{1/2}$ high-energy part:
\begin{equation}
F_H (5{\rm P}_{1/2}) = -\frac{1}{6} + (Z\alpha)^2 \,
  \left[\frac{20129}{67500} - \frac{292}{1125\,\epsilon} -
    \frac{796}{1125} \, \ln(2\epsilon)\right]\,,
\end{equation}
$5{\rm P}_{1/2}$ low-energy part:
\begin{equation}
F_L(5{\rm P}_{1/2}) = - \frac{4}{3} \ln k_0(5{\rm P}) +
(Z \alpha)^2 \, \left[ -1.023~991~781(1) + 
\frac{292}{1125 \, \epsilon} \,
+ \frac{796}{1125} \, 
\ln\left(\frac{\epsilon}{(Z \alpha)^2}\right)\right]\,,
\end{equation}
Here and in the following, $\ln k_0$ is the Bethe logarithm for which
very accurate values are 
available~\cite{BeBrSt1950,Ha1956,ScTi1959,Li1968,Hu1969,KlMa1973,DrSw1990}.

%
% $P_{3/2}$ states ($\kappa = -2$)
%
\subsection{P\_3/2 states (kappa = -2)}
\label{resultsP32}

The obtained results read as follows.
The $2{\rm P}_{3/2}$ high-energy part reads:
\begin{equation}
F_H (2{\rm P}_{3/2}) = \frac{1}{12} + (Z\alpha)^2 \,
  \left[\frac{6577}{21600} - \frac{2}{9\,\epsilon} -
    \frac{29}{90} \, \ln(2\epsilon)\right]\,,
\end{equation}
$2{\rm P}_{3/2}$ low-energy part:
\begin{equation}
F_L(2{\rm P}_{3/2}) = -\frac{4}{3} \, \ln k_0 (2{\rm P}) +
\left(Z \alpha \right)^2 \left[ -0.584~516~780(1) + 
\frac{2}{9 \, \epsilon} +
\frac{29}{90} \,
\ln\left(\frac{\epsilon}{(Z \alpha)^2}\right)\right]\,,
\end{equation}
$3{\rm P}_{3/2}$ high-energy part:
\begin{equation}
F_H (3{\rm P}_{3/2}) = \frac{1}{12} + (Z\alpha)^2 \,
  \left[\frac{67903}{194400} - \frac{20}{81\,\epsilon} -
    \frac{148}{405} \, \ln(2\epsilon)\right]\,,
\end{equation}
$3{\rm P}_{3/2}$ low-energy part:
\begin{equation}
F_L(3{\rm P}_{3/2}) = - \frac{4}{3} \ln k_0(3{\rm P}) +
(Z \alpha)^2 \, \left[ -0.693~566~427(1) +
\frac{20}{81 \, \epsilon} +
\frac{148}{405} \, 
\ln\left(\frac{\epsilon}{(Z \alpha)^2}\right)\right]\,,
\end{equation}
$4{\rm P}_{3/2}$ high-energy part:
\begin{equation}
F_H (4{\rm P}_{3/2}) = \frac{1}{12} + (Z\alpha)^2 \,
  \left[\frac{31399}{86400} - \frac{23}{90\,\epsilon} -
    \frac{137}{360} \, \ln(2\epsilon)\right]\,,
\end{equation}
$4{\rm P}_{3/2}$ low-energy part:
\begin{equation}
F_L(4{\rm P}_{3/2}) = - \frac{4}{3} \ln k_0(4{\rm P}) +
(Z \alpha)^2 \, \left[ -0.730~579~137(1) +
\frac{23}{90 \, \epsilon} +
\frac{137}{360} \, 
\ln\left(\frac{\epsilon}{(Z \alpha)^2}\right)\right]\,,
\end{equation}
$5{\rm P}_{3/2}$ high-energy part:
\begin{equation}
F_H (5{\rm P}_{3/2}) = \frac{1}{12} + (Z\alpha)^2 \,
  \left[\frac{199387}{540000} - \frac{292}{1125\,\epsilon} -
    \frac{436}{1125} \, \ln(2\epsilon)\right]\,,
\end{equation}
$5{\rm P}_{3/2}$ low-energy part:
\begin{equation}
F_L(5{\rm P}_{3/2}) = - \frac{4}{3} \ln k_0(5{\rm P}) +
(Z \alpha)^2 \, \left[ -0.747~615~653(1) + 
\frac{292}{1125 \, \epsilon} + 
\frac{436}{1125} \, 
\ln\left(\frac{\epsilon}{(Z \alpha)^2}\right)\right]\,.
\end{equation}

%
% $D_{3/2}$ states ($\kappa = 2$)
%
\subsection{D\_3/2 states (kappa = 2)}
\label{resultsD32}

The results obtained for states with $\kappa = 2$ read as follows.
$3{\rm D}_{3/2}$ high-energy part:
\begin{equation}
F_H (3{\rm D}_{3/2}) = -\frac{1}{20} + (Z\alpha)^2 \,
  \left[-\frac{1721}{194400} - \frac{4}{405\,\epsilon} -
    \frac{4}{405} \, \ln(2\epsilon)\right]\,,
\end{equation}
$3{\rm D}_{3/2}$ low-energy part:
\begin{equation}
F_L (3{\rm D}_{3/2}) =
-\frac{4}{3} \ln k_0(3{\rm D}) + 
  (Z\alpha)^2 \,
  \left[ 0.021~250~354(1) + \frac{4}{405\,\epsilon} +
   \frac{4}{405} \, 
   \ln\left( \frac{\epsilon}{(Z\alpha)^2} \right) \right]\,,
\end{equation}
$4{\rm D}_{3/2}$ high-energy part:
\begin{equation}
F_H (4{\rm D}_{3/2}) = -\frac{1}{20} + (Z\alpha)^2 \,
  \left[-\frac{829}{86400} - \frac{1}{90\,\epsilon} -
    \frac{1}{90} \, \ln(2\epsilon)\right]\,,
\end{equation}
$4{\rm D}_{3/2}$ low-energy part:
\begin{equation}
F_L (4{\rm D}_{3/2}) =
-\frac{4}{3} \ln k_0(4\mathrm{D}) +
  (Z\alpha)^2 \,
  \left[ 0.022~882~528(1) + \frac{1}{90\,\epsilon} +
    \frac{1}{90} \, \ln\left( \frac{\epsilon}{(Z\alpha)^2} \right) \,
     \right]\,,
\end{equation}
$5{\rm D}_{3/2}$ high-energy part:
\begin{equation}
F_H (5{\rm D}_{3/2}) = -\frac{1}{20} + (Z\alpha)^2 \,
  \left[-\frac{35947}{3780000} - \frac{92}{7875\,\epsilon} -
    \frac{92}{7875} \, \ln(2\epsilon)\right]\,,
\end{equation}
$5{\rm D}_{3/2}$ low-energy part:
\begin{equation}
F_L (5{\rm D}_{3/2}) =
-\frac{4}{3} \ln k_0(5{\rm D}) +
  (Z\alpha)^2 \,
  \left[ 0.023~759~683(1) + \frac{92}{7875\,\epsilon} + 
   \frac{92}{7875} \, 
\ln\left( \frac{\epsilon}{(Z\alpha)^2} \right)\right]\,.
\end{equation}

%
% $D_{5/2}$ states ($\kappa = -3$)
%
\subsection{D\_5/2 states (kappa = -3)}
\label{resultsD52}

For ${\rm D}_{5/2}$ states with Dirac quantum number $\kappa = -3$,
the results read as follows. For the $3{\rm D}_{5/2}$ high-energy part,
we have:
\begin{equation}
F_H (3{\rm D}_{5/2}) = \frac{1}{30} + (Z\alpha)^2 \,
  \left[\frac{371}{24300} - \frac{4}{405\,\epsilon} -
    \frac{4}{405} \, \ln(2\epsilon)\right]\,,
\end{equation}
$3{\rm D}_{5/2}$ low-energy part:
\begin{equation}
F_L (3{\rm D}_{5/2}) = -\frac{4}{3} \ln k_0(3{\rm D}) +
  (Z\alpha)^2 \,
  \left[ 0.019~188~397(1) + \frac{4}{405\,\epsilon} +
   \frac{4}{405} \, 
\ln\left( \frac{\epsilon}{(Z\alpha)^2} \right)\right]\,,
\end{equation}
$4{\rm D}_{5/2}$ high-energy part:
\begin{equation}
F_H (4{\rm D}_{5/2}) = \frac{1}{30} + (Z\alpha)^2 \,
  \left[\frac{53}{2880} - \frac{1}{90\,\epsilon} -
    \frac{1}{90} \, \ln(2\epsilon)\right]\,,
\end{equation}
$4{\rm D}_{5/2}$ low-energy part:
\begin{equation}
F_L (4{\rm D}_{5/2}) =
-\frac{4}{3} \ln k_0(4{\rm D}) +
  (Z\alpha)^2 \,
  \left[ 0.020~710~720(1) + \frac{1}{90\,\epsilon} +
    \frac{1}{90} \, 
\ln\left( \frac{\epsilon}{(Z\alpha)^2} \right)\right]\,,
\end{equation}
$5{\rm D}_{5/2}$ high-energy part:
\begin{equation}
F_H (5D_{5/2}) = \frac{1}{30} + (Z\alpha)^2 \,
  \left[\frac{3097}{157500} - \frac{92}{7875\,\epsilon} -
    \frac{92}{7875} \, \ln(2\epsilon)\right]\,,
\end{equation}
$5{\rm D}_{5/2}$ low-energy part:
\begin{equation}
F_L (5D_{5/2}) =
-\frac{4}{3} \ln k_0(5\mathrm{D}) +
  (Z\alpha)^2 \,
  \left[ 0.021~511~798(1) + \frac{92}{7875\,\epsilon} + 
    \frac{92}{7875} \, \ln\left( \frac{\epsilon}{(Z\alpha)^2} \right)
      \right]\,.
\end{equation}
 
%
% $F_{5/2}$ states ($\kappa = 3$)
%
\subsection{F\_5/2 states (kappa = 3)}
\label{resultsF52}

Now we turn our attention to $F_{5/2}$ states. We have for
the $4{\rm F}_{5/2}$ high-energy part:
\begin{equation}
F_H (4{\rm F}_{5/2}) = -\frac{1}{42} + (Z\alpha)^2 \,
  \left[-\frac{493}{235200} - \frac{1}{630\,\epsilon} -
    \frac{1}{630} \, \ln(2\epsilon)\right]\,,
\end{equation}
$4{\rm F}_{5/2}$ low-energy part:
\begin{equation}
F_L (4{\rm F}_{5/2}) = -\frac{4}{3} \ln k_0(4\mathrm{F}) +
  (Z\alpha)^2 \,
  \left[  0.005~523~310(1) + \frac{1}{630\,\epsilon} +
    \frac{1}{630} \, \ln\left( \frac{\epsilon}{(Z\alpha)^2} \right) 
    \right]\,,
\end{equation}
$5{\rm F}_{5/2}$ high-energy part:
\begin{equation}
F_H (5{\rm F}_{5/2}) =  -\frac{1}{42} + (Z\alpha)^2 \,
  \left[-\frac{2657}{1102500} - \frac{2}{1125\,\epsilon} -
    \frac{2}{1125} \, \ln(2\epsilon)\right]\,,
\end{equation}
$5{\rm F}_{5/2}$ low-energy part:
\begin{equation}
F_L (5{\rm F}_{5/2}) = -\frac{4}{3} \ln k_0(5\mathrm{F}) +
  (Z\alpha)^2 \,
  \left[ 0.006~045~398(1) + \frac{2}{1125\,\epsilon} +
    \frac{2}{1125} \, \ln\left( \frac{\epsilon}{(Z\alpha)^2} \right) 
    \right]\,,
\end{equation}

%
% $F_{7/2}$ states ($\kappa = -4$)
%
\subsection{F\_7/2 states (kappa = -4)}
\label{resultsF72}

States with $\kappa = -4$ yield the following results.
We obtain for the $4{\rm F}_{7/2}$ state (high-energy part): 
\begin{equation}
F_H (4{\rm F}_{7/2}) = \frac{1}{56} + (Z\alpha)^2 \,
  \left[\frac{25357}{8467200} - \frac{1}{630\,\epsilon} -
    \frac{1}{630} \, \ln(2\epsilon)\right]\,,
\end{equation}
$4{\rm F}_{7/2}$ low-energy part:
\begin{equation}
F_L (4\mathrm{F}_{7/2}) =
-\frac{4}{3} \ln k_0(4\mathrm{F}) +
  (Z\alpha)^2 \,
  \left[ 0.005~180~461(1) + \frac{1}{630\,\epsilon} +
    \frac{1}{630} \, \ln\left( \frac{\epsilon}{(Z\alpha)^2} \right) 
     \right]\,,
\end{equation}
$5{\rm F}_{7/2}$ high-energy part:
\begin{equation}
F_H (5{\rm F}_{7/2}) = \frac{1}{56} + (Z\alpha)^2 \,
  \left[\frac{774121}{211680000} - \frac{2}{1125\,\epsilon} -
    \frac{2}{1125} \, \ln(2\epsilon)\right]\,,
\end{equation}
$5{\rm F}_{7/2}$ low-energy part:
\begin{equation}
F_L (5{\rm F}_{7/2}) =
-\frac{4}{3} \ln k_0(5\mathrm{F}) +
  \frac{2\,(Z\alpha)^2}{1125\,\epsilon} +
  (Z\alpha)^2 \,
  \left[ 0.005~662~248(1) + 
    \frac{2}{1125} \, \ln\left( \frac{\epsilon}{(Z\alpha)^2} \right) 
     \right]\,.
\end{equation}

%
% $G_{7/2}$ states ($\kappa = 4$)
%
\subsection{G\_7/2 states (kappa = 4)}
\label{resultsG72}

${\rm G}_{7/2}$ and 
${\rm G}_{9/2}$ states involve the most problematic 
angular momentum algebra of
all atomic states considered here. We obtain,
\begin{equation}
\label{FH5G72}
F_H (5\mathrm{G}_{7/2}) = -\frac{1}{72} + (Z\alpha)^2 \,
  \left[-\frac{4397}{6048000} - \frac{2}{4725\,\epsilon} -
    \frac{2}{4725} \, \ln(2\epsilon)\right]\,,
\end{equation}
and for the low-energy part,
\begin{equation}
\label{FL5G72}
F_L (5\mathrm{G}_{7/2}) =
-\frac{4}{3} \ln k_0(5\mathrm{G}) +
  (Z\alpha)^2 \,
  \left[ 0.001~834~827(1) + \frac{2}{4725\,\epsilon} +
    \frac{2}{4725} \, \ln\left( \frac{\epsilon}{(Z\alpha)^2} \right) 
     \right]\,.
\end{equation}

%
% $G_{9/2}$ states ($\kappa = -5$)
%
\subsection{G\_9/2 states (kappa = -5)}
\label{resultsG92}

We conclude by investigating $G_{9/2}$ states, where the
high-energy part reads ($n = 5$)
\begin{equation}
F_H (5\mathrm{G}_{9/2}) = \frac{1}{90} + (Z\alpha)^2 \,
  \left[\frac{269}{283500} - \frac{2}{4725\,\epsilon} -
    \frac{2}{4725} \, \ln(2\epsilon)\right]\,,
\end{equation}
and the low-energy part reads,
\begin{equation}
\label{FL5G92}
F_L (5\mathrm{G}_{9/2}) =
-\frac{4}{3} \ln k_0(5\mathrm{G}) +
  (Z\alpha)^2 \,
  \left[ 0.001~757~471(1) + \frac{2}{4725\,\epsilon} +
    \frac{2}{4725} \, \ln\left( \frac{\epsilon}{(Z\alpha)^2} \right) 
     \right]\,.
\end{equation}

%
% Summary of Results
%
\section{Summary of Results}
\label{SummaryOfResults}

By adding the high- and low-energy parts given in 
Secs.~\ref{resultsP12} --~\ref{resultsG92}, we obtain the following results
for the self energy.  
For all atomic states,
the dependence on the scale-separating parameter $\epsilon$ cancels
when the low- and the high-energy parts are added, in agreement
with the principal idea of the $\epsilon$-method described in 
Sec.~\ref{EpsilonMethod}. The coefficients $A_{40}$, $A_{61}$ and $A_{60}$
can then be read off according to the definitions
contained in Eqs.~(\ref{ESEasF}) -- (\ref{A60nP}).

\begin{table}[htb]
\begin{center}
\begin{minipage}{14cm}
\begin{center}
\begin{tabular}{cccccc}
\hline
\hline
\multicolumn{6}{c}{\rule[-3mm]{0mm}{8mm} $P_{1/2}$ state ($\kappa = 1$)} \\
\multicolumn{1}{c}{\rule[-3mm]{0mm}{8mm} $n$} &
\multicolumn{1}{c}{\rule[-3mm]{0mm}{8mm} $A_{40}$} &
\multicolumn{1}{c}{\rule[-3mm]{0mm}{8mm} $A_{40}$ (num)} &
\multicolumn{1}{c}{\rule[-3mm]{0mm}{8mm} $A_{61}$} &
\multicolumn{1}{c}{\rule[-3mm]{0mm}{8mm} $A_{61}$ (num)} &
\multicolumn{1}{c}{\rule[-3mm]{0mm}{8mm} $A_{60}$ (num)} \\
\hline
\rule[-3mm]{0mm}{8mm}
2 & $-\frac{1}{6} - \frac{4}{3} \, \ln k_0(2P)$
& -0.126\,644
& $\frac{103}{108}$
& 0.572\,222
& -0.998~904~402(1)\\
\rule[-3mm]{0mm}{8mm}
3 & $-\frac{1}{6} - \frac{4}{3} \, \ln k_0(3P)$
& -0.115\,746
& $\frac{268}{405}$
& 0.661\,728
& -1.148~189~956(1)\\
\rule[-3mm]{0mm}{8mm}
4 & $-\frac{1}{6} - \frac{4}{3} \, \ln k_0(4P)$
& -0.110\,727
& $\frac{499}{720}$
& 0.693\,056
& -1.195~688~142(1)\\
\rule[-3mm]{0mm}{8mm}
5 & $-\frac{1}{6} - \frac{4}{3} \, \ln k_0(5P)$
& -0.107\,954
& $\frac{796}{1125}$
& 0.707\,556
& -1.216~224~512(1) \\
\hline
\hline
\end{tabular}
\caption{\label{tableP12} Results for ${\rm P}_{1/2}$ states.
The coefficients $A_{40}$ and $A_{61}$ are rational numbers;
in the columns entitled ``(num)'', we give the numerical values
to six decimals. The $A_{60}$-coefficients in the last column
constitute the main result of the investigation presented in 
the current chapter.
As explained in Sec.~\ref{ORISelExc}, these coefficients involve
numerical integrals which cannot be done analytically. }
\end{center}
\end{minipage}
\end{center}
\end{table}

\begin{table}[htb]
\begin{center}
\begin{minipage}{14cm}
\begin{center}
\begin{tabular}{cccccc}
\hline
\hline
\multicolumn{6}{c}{\rule[-3mm]{0mm}{8mm} $P_{3/2}$ state ($\kappa = -2$)} \\
\multicolumn{1}{c}{\rule[-3mm]{0mm}{8mm} $n$} &
\multicolumn{1}{c}{\rule[-3mm]{0mm}{8mm} $A_{40}$} &
\multicolumn{1}{c}{\rule[-3mm]{0mm}{8mm} $A_{40}$ (num)} &
\multicolumn{1}{c}{\rule[-3mm]{0mm}{8mm} $A_{61}$} &
\multicolumn{1}{c}{\rule[-3mm]{0mm}{8mm} $A_{61}$ (num)} &
\multicolumn{1}{c}{\rule[-3mm]{0mm}{8mm} $A_{60}$ (num)} \\
\hline
\rule[-3mm]{0mm}{8mm}
2 & $\frac{1}{12} - \frac{4}{3} \, \ln k_0(2P)$
& 0.123\,355
& $\frac{29}{90}$
& 0.322\,222
& -0.503~373~465(1) \\
\rule[-3mm]{0mm}{8mm}
3 & $\frac{1}{12} - \frac{4}{3} \, \ln k_0(3P)$
& 0.134\,254
& $\frac{148}{405}$
& $0.365\,432$
& -0.597~569~388(1) \\
\rule[-3mm]{0mm}{8mm}
4 & $\frac{1}{12} - \frac{4}{3} \, \ln k_0(4P)$
& 0.139\,273
& $\frac{137}{360}$
& 0.380\,556
& -0.630~945~796(1) \\
\rule[-3mm]{0mm}{8mm}
5 & $\frac{1}{12} - \frac{4}{3} \, \ln k_0(5P)$
& 0.142\,046
& $\frac{426}{1125}$
& 0.387\,556
& -0.647~013~509(1) \\
\hline
\hline
\end{tabular}
\caption{\label{tableP32}
Analytic results for ${\rm P}_{3/2}$ states.}
\end{center}
\end{minipage}
\end{center}
\end{table}

\begin{table}[htb]
\begin{center}
\begin{minipage}{14cm}
\begin{center}
\begin{tabular}{ccccccc}
\hline
\hline
\multicolumn{6}{c}{\rule[-3mm]{0mm}{8mm} $D_{3/2}$ state ($\kappa = 2$)} \\
\multicolumn{1}{c}{\rule[-3mm]{0mm}{8mm} $n$} &
\multicolumn{1}{c}{\rule[-3mm]{0mm}{8mm} $A_{40}$} &
\multicolumn{1}{c}{\rule[-3mm]{0mm}{8mm} $A_{40}$ (num)} &
\multicolumn{1}{c}{\rule[-3mm]{0mm}{8mm} $A_{61}$} &
\multicolumn{1}{c}{\rule[-3mm]{0mm}{8mm} $A_{61}$ (num)} &
\multicolumn{1}{c}{\rule[-3mm]{0mm}{8mm} $A_{60}$ (num)} \\
\hline
\rule[-3mm]{0mm}{8mm}
3 & $- \frac{1}{20} - \frac{4}{3} \, \ln k_0(3D)$
& -0.043\,024
& $\frac{4}{405}$
& 0.009\,877
& 0.005~551~575(1) \\
\rule[-3mm]{0mm}{8mm}
4 & $- \frac{1}{20} - \frac{4}{3} \, \ln k_0(4D)$
& -0.041\,012
& $\frac{1}{90}$
& 0.011\,111
& 0.005~587~985(1) \\
\rule[-3mm]{0mm}{8mm}
5 & $- \frac{1}{20} - \frac{4}{3} \, \ln k_0(5D)$
& -0.039\,866
& $\frac{92}{7875}$
& 0.011\,682
& 0.006~152~175(1) \\
\hline
\hline
\end{tabular}
\caption{\label{tableD32}
Analytic results for ${\rm D}_{3/2}$ states.}
\end{center}
\end{minipage}
\end{center}
\end{table}

\begin{table}[htb]
\begin{center}
\begin{minipage}{14cm}
\begin{center}
\begin{tabular}{cccccc}
\hline
\hline
\multicolumn{6}{c}{\rule[-3mm]{0mm}{8mm} $D_{5/2}$ state ($\kappa = -3$)} \\
\multicolumn{1}{c}{\rule[-3mm]{0mm}{8mm} $n$} &
\multicolumn{1}{c}{\rule[-3mm]{0mm}{8mm} $A_{40}$} &
\multicolumn{1}{c}{\rule[-3mm]{0mm}{8mm} $A_{40}$ (num)} &
\multicolumn{1}{c}{\rule[-3mm]{0mm}{8mm} $A_{61}$} &
\multicolumn{1}{c}{\rule[-3mm]{0mm}{8mm} $A_{61}$ (num)} &
\multicolumn{1}{c}{\rule[-3mm]{0mm}{8mm} $A_{60}$ (num)} \\
\hline
\rule[-3mm]{0mm}{8mm}
3 & $\frac{1}{30} - \frac{4}{3} \, \ln k_0(3D)$
& 0.040\,310
& $\frac{4}{405}$
& 0.009\,877
& $0.027~609~989(1)$ \\
\rule[-3mm]{0mm}{8mm}
4 & $\frac{1}{30} - \frac{4}{3} \, \ln k_0(4D)$
& 0.042\,321
& $\frac{1}{90}$
& 0.011\,111
& $0.031~411~862(1)$ \\
\rule[-3mm]{0mm}{8mm}
5 & $\frac{1}{30} - \frac{4}{3} \, \ln k_0(5D)$
& 0.043\,468
& $\frac{92}{7875}$
& 0.011\,682
& $0.033~077~571(1)$ \\
\hline
\hline
\end{tabular}
\caption{\label{tableD52}
Analytic results for ${\rm D}_{5/2}$ states.}
\end{center}
\end{minipage}
\end{center}
\end{table}

\begin{table}[htb]
\begin{center}
\begin{minipage}{14cm}
\begin{center}
\begin{tabular}{ccccccc}
\hline
\hline
\multicolumn{6}{c}{\rule[-3mm]{0mm}{8mm} $F_{5/2}$ state ($\kappa = 3$)} \\
\multicolumn{1}{c}{\rule[-3mm]{0mm}{8mm} $n$} &
\multicolumn{1}{c}{\rule[-3mm]{0mm}{8mm} $A_{40}$} &
\multicolumn{1}{c}{\rule[-3mm]{0mm}{8mm} $A_{40}$ (num)} &
\multicolumn{1}{c}{\rule[-3mm]{0mm}{8mm} $A_{61}$} &
\multicolumn{1}{c}{\rule[-3mm]{0mm}{8mm} $A_{61}$ (num)} &
\multicolumn{1}{c}{\rule[-3mm]{0mm}{8mm} $A_{60}$ (num)} \\
\hline
\rule[-3mm]{0mm}{8mm}
4 & $-\frac{1}{42} - \frac{4}{3} \, \ln k_0(4F)$
& -0.021\,498
& $\frac{1}{630}$
& 0.001\,587             
& $0.002~326~988(1)$ \\
\rule[-3mm]{0mm}{8mm}
5 & $-\frac{1}{42} - \frac{4}{3} \, \ln k_0(5F)$
& -0.020\,873
& $\frac{2}{1125}$
& 0.001\,778
& $0.002~403~159(1)$ \\
\hline
\hline
\end{tabular}
\caption{\label{tableF52}
Analytic results for ${\rm F}_{5/2}$ states.}
\end{center}
\end{minipage}
\end{center}
\end{table}

\begin{table}[htb]
\begin{center}
\begin{minipage}{14cm}
\begin{center}
\begin{tabular}{cccccc}
\hline
\hline
\multicolumn{6}{c}{\rule[-3mm]{0mm}{8mm} $F_{7/2}$ state ($\kappa = -4$)} \\
\multicolumn{1}{c}{\rule[-3mm]{0mm}{8mm} $n$} &
\multicolumn{1}{c}{\rule[-3mm]{0mm}{8mm} $A_{40}$} &
\multicolumn{1}{c}{\rule[-3mm]{0mm}{8mm} $A_{40}$ (num)} &
\multicolumn{1}{c}{\rule[-3mm]{0mm}{8mm} $A_{61}$} &
\multicolumn{1}{c}{\rule[-3mm]{0mm}{8mm} $A_{61}$ (num)} &
\multicolumn{1}{c}{\rule[-3mm]{0mm}{8mm} $A_{60}$ (num)} \\
\hline
\rule[-3mm]{0mm}{8mm}
4 & $\frac{1}{56} - \frac{4}{3} \, \ln k_0(4F)$
& 0.020\,169
& $\frac{1}{630}$
& 0.001\,587
& $0.007~074~960(1)$ \\
\rule[-3mm]{0mm}{8mm}
5 & $\frac{1}{56} - \frac{4}{3} \, \ln k_0(5F)$
& 0.020\,793
& $\frac{2}{1125}$
& 0.001\,778
& $ 0.008~087~021(1)$ \\
\hline
\hline
\end{tabular}
\caption{\label{tableF72}
Analytic results for ${\rm F}_{7/2}$ states.}
\end{center}
\end{minipage}
\end{center}
\end{table}

\begin{table}[htb]
\begin{center}
\begin{minipage}{14cm}
\begin{center}
\begin{tabular}{ccccccc}
\hline
\hline
\multicolumn{6}{c}{\rule[-3mm]{0mm}{8mm} $G_{7/2}$ state ($\kappa = 4$)} \\
\multicolumn{1}{c}{\rule[-3mm]{0mm}{8mm} $n$} &
\multicolumn{1}{c}{\rule[-3mm]{0mm}{8mm} $A_{40}$} &
\multicolumn{1}{c}{\rule[-3mm]{0mm}{8mm} $A_{40}$ (num)} &
\multicolumn{1}{c}{\rule[-3mm]{0mm}{8mm} $A_{61}$} &
\multicolumn{1}{c}{\rule[-3mm]{0mm}{8mm} $A_{61}$ (num)} &
\multicolumn{1}{c}{\rule[-3mm]{0mm}{8mm} $A_{60}$ (num)} \\
\hline
\rule[-3mm]{0mm}{8mm}
5 & $- \frac{1}{72} - \frac{4}{3} \, \ln k_0(5G)$
& -0.012\,860
& $\frac{2}{4725}$
& 0.000\,423
& $0.000~814~414(1)$ \\
\hline
\hline
\multicolumn{6}{c}{\rule[-3mm]{0mm}{8mm} $G_{9/2}$ state ($\kappa = -5$)} \\
\multicolumn{1}{c}{\rule[-3mm]{0mm}{8mm} $n$} &
\multicolumn{1}{c}{\rule[-3mm]{0mm}{8mm} $A_{40}$} &
\multicolumn{1}{c}{\rule[-3mm]{0mm}{8mm} $A_{40}$ (num)} &
\multicolumn{1}{c}{\rule[-3mm]{0mm}{8mm} $A_{61}$} &
\multicolumn{1}{c}{\rule[-3mm]{0mm}{8mm} $A_{61}$ (num)} &
\multicolumn{1}{c}{\rule[-3mm]{0mm}{8mm} $A_{60}$ (num)} \\
\hline
\rule[-3mm]{0mm}{8mm}
5 & $\frac{1}{90} - \frac{4}{3} \, \ln k_0(5G)$
& 0.012\,141
& $\frac{2}{4725}$
& 0.000\,423
& $0.002~412~929(1)$ \\
\hline
\hline
\end{tabular}
\caption{\label{tableG}
Analytic results for ${\rm G}_{7/2}$ and ${\rm G}_{9/2}$ states.}
\end{center}
\end{minipage}
\end{center}
\end{table}

%
% Typical Cancellations
%
\section{Typical Cancellations}
\label{TypicalCancellations}

As already emphasized in Sec.~\ref{ORISelExc},
severe numerical cancellations are encountered in the
calculation of the $A_{60}$-coefficients.
These cancellations will be illustrated here in 
Tables~\ref{cancel1} and~\ref{cancel2} for the
$5{\rm G}_{7/2}$ state.

\begin{table}[htb]
\begin{center}
\begin{minipage}{14cm}
\begin{center}
\begin{tabular}{cr@{.}l}
\hline
\hline
\multicolumn{3}{c}{Contributions to the Low-Energy Part ($5{\rm G}_{7/2}$)}\\
\hline
$A_{60}$-contribution due to $F_{\rm nq}$          &
   0.002~875~830~9(5) \\
$A_{60}$-contribution due to $F_{\rm \delta y}$    &
  -0.001~083~109~4(5) \\
$A_{60}$-contribution due to $F_{\rm \delta H}$    &
  -0.008~917~782~1(5) \\
$A_{60}$-contribution due to $F_{\rm \delta E}$    &
   0.004~920~556~0(5) \\
$A_{60}$-contribution due to $F_{\rm \delta \phi}$ &
   0.004~039~332~1(5) \\
$A_{60}$ [from low-energy, cf. Eq.~(\ref{FL5G72})]    & 
  $ 0$ & $001~834~827(1)$ \\
\hline
\hline
\end{tabular}
\caption{\label{cancel1}
As explained in Refs.~\cite{JePa1996,JeSoMo1997},
the low-energy part of the self energy
can be separated naturally into the nonrelativistic dipole term
(which does not contribute to $A_{60}$), and into the following
contributions: the nonrelativistic quadrupole part $F_{\rm{nq}}$, the
relativistic corrections to the current $F_{\rm{\delta y}}$, 
and relativistic corrections to the hamiltonian $F_{\rm{\delta H}}$, 
to the bound-state energy $F_{\rm{\delta E}}$ and
to the wavefunction $F_{\rm{\delta \phi}}$. Observe that the 
total contribution to $A_{60}$ to the low-energy part which read
$0.001~834~827(1)$ is almost an order-of-magnitude smaller than
the biggest individual contribution (from $F_{\rm{\delta H}}$),
due to mutual cancellations.}
\end{center}
\end{minipage}
\end{center}
\end{table}

\begin{table}[htb]
\begin{center}
\begin{minipage}{14cm}
\begin{center}
\begin{tabular}{cr@{.}l}
\hline
\hline
$A_{60}(F_H)$  & $-0$ & $001~020~413$ \\
$A_{60}(F_L)$  & $ 0$ & $001~834~827(1)$ \\
$A_{60}$       & $ 0$ & $000~814~415(1)$ \\
\hline
\hline
\end{tabular}
\caption{\label{cancel2}
An additional numerical cancellation occurs when 
the finite contributions to $A_{60}$ originating
from the low-energy part [$A_{60}(F_L)$,
see Eq.~(\ref{FL5G72})] and the
high-energy part [$A_{60}(F_H)$,
see Eq.~(\ref{FH5G72})] are added. The result
for the total $A_{60}$ is also contained in Table~\ref{tableG}.}
\end{center}
\end{minipage}
\end{center}
\end{table}

%
% Observations
%
\section{Observations}
\label{Observations}

It is hoped that the analytic calculations described in the 
previous sections will be supplemented in the near future
by a numerical treatment of the problem, which follows
the lines indicated in Ch.~\ref{QEDSelOne}.
Work performed recently on relevant angular 
functions~\cite{LBInMo2001} is expected to become useful in that
context.
In order to facilitate the comparisons and consistency checks
which will be of relevance for the numerical calculations,
we report here on a number of 
observations which should facilitate the comparison of numerical 
and analytic approaches to the self energy problem. They are
relevant for all atomic states under investigation here.
 
(i) Our analytic calculations for the high-energy part establish
the structure of nonvanishing $A_{40}$, $A_{61}$ and $A_{60}$-coefficients
for the states under investigation.
 
(ii) For a given $\kappa$ quantum number, the coefficients increase in
absolute magnitude with increasing principal quantum number $n$. 
This is the case for Bethe logarithms, for $A_{61}$ coefficients and 
also for $A_{60}$ coefficients. This behavior is consistent
with the phenomenon that relativistic corrections to the 
self energy become more important as $n$ increases.
 
(iii) For a given principal quantum number $n$, the coefficients decrease 
in absolute magnitude with increasing $\kappa$. As a function of the nuclear
charge number $Z$, the plots of $F(Z\alpha)$ against $Z$ should become
``more flat'' as $\kappa$ increases for a given $n$~\cite{InMo1998hyp}.
 
(iv) The $A_{61}$ coefficients, which represent the dominant relativistic
correction to $F(Z\alpha)$, all have the same (positive) sign for states with 
$\kappa$ quantum numbers $\kappa = 2,-3,3,-4,4,-5$. This ``same-sign''
pattern is also true of the 
$A_{60}$ coefficients. Because $A_{61}$ is the {\em dominant}
correction in the order $(Z\alpha)^6$ due to the enhancement by the
large logarithm, this observation explains why, in general,
the qualitative dependence (on the nuclear charge number $Z$)
of the self energy correction
for the highly excited states is the same for the states with
$\kappa = 2,-3,3,-4,4,-5$ (see~\cite{InMo1998hyp}).

(v) $A_{61}$-coefficients are different for $P_{1/2}$ and $P_{3/2}$ states,
but the spin dependence vanishes for higher orbital angular momentum $L$.
Consequently, $A_{61}$-coefficients are the same e.g.~for $D_{3/2}$ and 
$D_{5/2}$ states, or for $F_{5/2}$ and
$F_{7/2}$ states etc. This can be easily checked against numerical
results.

Preliminary results indicate satisfactory consistency between
the analytic and numerical approaches to the self energy problem
for highly excited states~\cite{LBInPriv2002}.

%
% The Two--Loop Self--Energy
%
\chapter{The Two--Loop Self--Energy}
\label{QEDSelTwo}

\typeout{==========================}
\typeout{The Two--Loop Self--Energy}
\typeout{==========================}
\typeout{}

%
% Orientation
%
\section{Orientation}

In the current Chapter, we investigate
two-loop higher-order binding corrections to the fine structure,
which contribute to the spin-dependent part of the Lamb shift.
Our calculation focuses on the so-called ``two-loop self energy''
involving two virtual closed photon loops. For bound states,
this correction has proven to be notoriously difficult to evaluate.
The calculation of the binding corrections
to the bound-state two-loop self energy
is simplified by a separate treatment
of hard and soft virtual photons. The two photon-energy
scales are matched at the
end of the calculation. We present
results for the fine-structure difference of the two-loop self energy
through the order of $\alpha^8$.

%
% Introduction
%
\section{Introduction to the Two--Loop Self--Energy}
\label{IntroTwoLoop}

\begin{figure}[htb]
\begin{center}
\begin{minipage}{10cm}
\centerline{\mbox{\epsfysize=7.7cm\epsffile{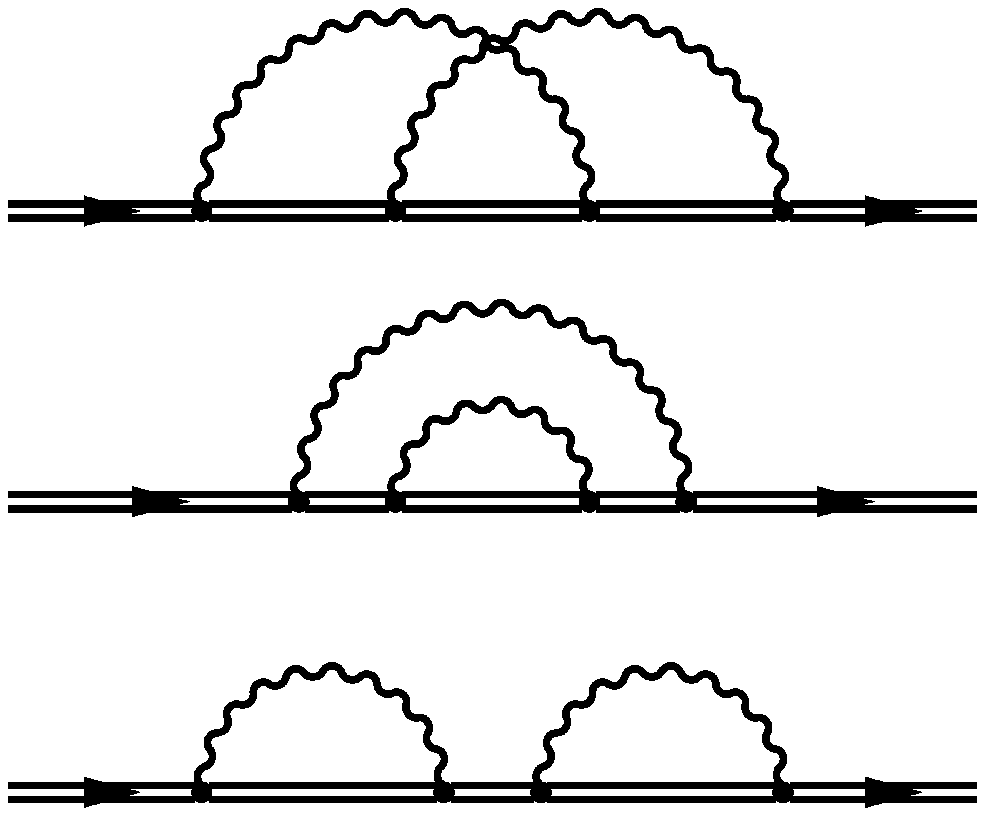}}}
\caption{\label{fig1} Feynman diagrams representing the
two-photon electron self energy.
The double line denotes the bound electron propagator.
The arrow of time is from left to right.}
\end{minipage}
\end{center}
\end{figure}

As already discussed in Secs.~\ref{StatusAnalytic}
and~\ref{ORISelExc} --~\ref{ResHighLow},
radiative corrections can be described 
-- for atomic systems with low nuclear charge number -- by a nonanalytic
expansion in powers of the three parameters (i) $\alpha$ (the fine-structure
constant), (ii) the product $Z\alpha$ ($Z$ is the nuclear charge
number), and (iii) the logarithm $\ln[(Z\alpha)^{-2}]$.
The expansion in powers of $\alpha$, which is the perturbation
theory parameter in quantum electrodynamics (QED), corresponds to the 
number of loops in the diagrams. The bound-state effects are taken
into account by the expansions in the two latter parameters (see also
Ch.~\ref{QEDSelExc}).
Higher-order terms in the
expansions in powers of $Z\alpha$ and $\ln[(Z\alpha)^{-2}]$
are referred to as the ``binding corrections''.
One of the historically most problematic sets of Feynman diagrams
in the treatment of the Lamb shift for atomic systems 
has been the radiative correction due to two closed
virtual-photon loops shown in Fig.~\ref{fig1}.

%
% Fig. 1
%

Let us recall at this point that even the evaluation
of higher-order binding corrections to the one-loop self energy
-- see also Ch.~\ref{QEDSelExc} --,
which {\em a priori} should represent a less involved calculational
challenge, has represented a problem for analytic evaluations
for over three 
decades~\cite{ErYe1965ab,Er1971,Sa1981,Pa1993,JePa1996,JeSoMo1997,JeMoSo1999}.
The energy shifts of the bound states due to the radiative 
corrections are conveniently expressed by
expansion coefficients corresponding
to the powers of $Z\alpha$ and $\ln[(Z\alpha)^{-2}]$;
the naming convention is that the power of $Z\alpha$ and 
the power of the logarithm are indicated as indices to the 
analytic coefficients~[see also Eq.~(\ref{DefESE})] below.
Because the expansion in both the one-loop and the two-loop
case starts with the fourth power of $Z\alpha$, the non-vanishing
coefficients carry indices $A_{kl}$ and $B_{kl}$ for the 
one- and two-loop cases, respectively 
(with $k\geq 4$ -- see~\cite{SaYe1990,MoPlSo1998} for comprehensive 
reviews).

Logarithmic corrections with $l \geq 1$ can sometimes be inferred
separately in a much simplified approach, e.g.~by considering infrared
divergent contributions to electron form factors. By contrast,
the higher-order non-logarithmic
coefficients represent a considerable calculational challenge.
Realistically, i.e.~with the help of current computer
algebra systems~\cite{Wo1988}, one can hope to 
evaluate non-logarithmic coefficients of sixth order
in $Z\alpha$. Complete results for the one-loop higher-order
correction $A_{60}$ for S and P states have only been
available recently~\cite{Pa1993,JePa1996,JeSoMo1997}.
Calculational difficulties have by now precluded
a successful evaluation of the corresponding coefficient
$B_{60}$ for the two-loop effect. Ground-work for the
evaluation of $B_{60}$ was laid in~\cite{Pa2001}.
Here, we are concerned with the evaluation of the fine-structure
differences of the logarithmic and non-logarithmic coefficients
$B_{6L}$ (where $L = 0,1,2$), i.e.~with the 
$n{\mathrm P}_{3/2}$--$n{\mathrm P}_{1/2}$ difference of these
coefficients.

Using natural Gaussian units 
($\hbar = c = \epsilon_0 = 1$), as it is customary for the current type of
calculation, we write the two-photon self energy in the 
$Z\alpha$-expansion for P states in terms of $B$-coefficients as
\begin{equation}
\label{DefESE}
\Delta E_{\mathrm{SE}} = \left(\frac{\alpha}{\pi}\right)^2 \,
(Z\alpha)^4 \, \frac{m}{n^3} \, \bigg[ B_{40} 
+ (Z\alpha)^2 \,
\left[ B_{62} \, \ln^2(Z\alpha)^{-2} 
+ B_{61} \, \ln(Z\alpha)^{-2} + B_{60} \right] + {\mathcal R} \bigg]\,,
\end{equation}
where the remainder ${\mathcal R}$ is of order 
${\mathcal O}(Z\alpha)^3$.
Relevant Feynman diagrams are shown in Fig.~\ref{fig1}.

Here, $m$ denotes the electron mass (we write
Eq.~(\ref{DefESE}) in the non-recoil limit, i.e.~for an infinite
nuclear mass). The double logarithmic $B_{62}$-coefficient 
is spin-independent, so that
we have $\Delta_{\mathrm{fs}} B_{62} = 0$.
Here, we evaluate the fine-structure
differences
\begin{eqnarray}
\Delta_{\mathrm{fs}} B_{61} &=& 
  B_{61}(n {\mathrm P}_{3/2}) - B_{61}(n {\mathrm P}_{1/2})\,,
\nonumber\\[1ex]
\Delta_{\mathrm{fs}} B_{60} &=& 
  B_{60}(n {\mathrm P}_{3/2}) - B_{60}(n {\mathrm P}_{1/2})\,. 
\end{eqnarray}
We follow the convention that
$\Delta_{\mathrm{fs}} X \equiv X(n {\mathrm P}_{3/2}) - 
X(n {\mathrm P}_{1/2})$ denotes the ``fine-structure part'' of a
given quantity $X$.
For $\Delta_{\mathrm{fs}} B_{61}$ and $\Delta_{\mathrm{fs}} B_{60}$,
we provide complete results.
It is perhaps worth noting that the two-loop self energy for bound states 
has represented a considerable challenge for theoretical evaluations.
Our investigation represents a continuation of previous 
work on the two-loop problem 
(see e.g.~\cite{ApBr1970,Pa1994prl,EiKaSh1995,Pa2001}).
It is probably a triviality to express that
technical difficulties in the calculation and its description
in the following Sections cannot be avoided. 

For the description of the self energy radiative effects -- mediated by hard
virtual photons --, we use the modified Dirac hamiltonian
\begin{equation}
\label{HDm}
H_{\mathrm D}^{(m)} = \bbox{\alpha} \cdot 
  \left[\bbox{p} -{\mathrm e} \, F_1(\Delta) \, \bbox{A}\right]
     + \beta\,m + {\mathrm e} \, F_1(\Delta) \, \phi +
       F_2(\Delta) \, \frac{e}{2\,m} \, \left({\mathrm i}\, 
         \bbox{\gamma} \cdot
         \bbox{E} - \beta \, \bbox{\sigma} \cdot \bbox{B} \right)\,,
\end{equation}
which approximately describes an electron subject to an
external scalar potential $\phi \equiv \phi(\bbox{r})$
and an external vector potential $\bbox{A} \equiv \bbox{A}(\bbox{r})$. 
This modified hamiltonian is still local in coordinate space. 
The Dirac matrices in (\ref{HDm}) are to be understood in the
standard (Dirac) representation~\cite{ItZu1980} (in the sequel, we will also
use the non-covariant notation $\beta \equiv \gamma^0$ and 
$\alpha^i \equiv \gamma^0 \gamma^i$). 

The argument $\Delta$ of the electron form factors $F_1$ and $F_2$
in Eq.~(\ref{HDm}) 
is to be interpreted as a Laplacian operator acting on all
quantities to the right (but not on the wave function of the 
bound electron in evaluating $H_{\mathrm D}^{(m)} | \psi \rangle$). 
In momentum space, the action of the hamiltonian
$H_{\mathrm D}^{(m)}$ is described by the convolution 
$\left[H_{\mathrm D}^{(m)} \psi\right](\bbox{p}')
= \int {\mathrm d}^3 p/(2\pi)^3 \, 
H_{\mathrm D}^{(m)}(\bbox{p}'-\bbox{p}) \, \psi(\bbox{p})$.
The form factors -- in momentum space -- assume arguments according to the 
replacement $\Delta \to - \bbox{q}^2 \equiv - (\bbox{p}'-\bbox{p})^2$.
In Eq.~(\ref{HDm}), radiative corrections are taken into account in the 
sense of an effective theory via the inclusion of the
on-shell form factors $F_1$ and $F_2$. Although the bound electron
is not an on-shell particle, the modified hamiltonian
(\ref{HDm}) can still approximately account for significant
radiative systems with low nuclear charge number $Z$. 
Of course, the hamiltonian (\ref{HDm})
cannot offer a complete description of the bound electron.
Recoil effects cannot be described by a one-particle 
equation {\em in principle}, and vacuum-polarization
effects are not contained in Eq.~(\ref{HDm}). However, the effective description
of self energy radiative corrections mediated by hard virtual photons given by 
Eq.~(\ref{HDm}) will turn out to be useful in the context
of the current investigation of the two-loop self energy.
 
Both of the form factors $F_1$ and $F_2$ entering in
Eq.~(\ref{HDm}) are infrared divergent, but this divergence
is cut off in a natural way at the atomic 
binding energy scale $(Z\alpha)^2\,m$.
The fact that on-shell form factors can describe radiative corrections
to the fine structure --
mediated by high-energy virtual photons -- has been demonstrated explicitly
in~\cite{Pa1999}. The modified Dirac hamiltonian (\ref{HDm}) and 
the associated modified Dirac equation have been introduced 
-- in the one-loop approximation -- in Ch.~7 of~\cite{ItZu1980} 
[see for example Eqs.~(7-77) and (7-103) {\em ibid.}].
The low-energy part of the calculation is carried out using 
nonrelativistic approximations in the spirit of the simplified
treatment introduced in the previous one- and two-loop
calculations~\cite{Pa1993,JePa1996,JeSoMo1997,Pa1998,Pa2001}.
This approach was inspired, in part, by various attempts to 
formulate simplified low-energy (``nonrelativistic'') approximations
to quantum electrodynamics (``NRQED''), see 
e.g.~\cite{BB1984,CaLe1986}. Both the
high-energy and the low-energy contributions are matched at the 
separation scale $\epsilon$ whose r\^{o}le in the calculation
is illustrated by the mathematical model example discussed
in Sec.~\ref{EpsilonMethod}.

In a two-loop calculation, either of the two virtual
photons may have a high or low energy as compared to the 
separation scale $\epsilon$.
{\em A priori}, this necessitates~\cite{Pa2001} a separation 
of the calculation into three different contributions: (i) both
photon energies large, (ii) one photon with a large and one with a 
small energy, and (iii) both photons with small energies. 
For the particular problem at hand (the fine-structure
differences of $B_{61}$ and $B_{60}$), we are in the fortunate position
that effects caused by hard virtual photons (i) are described by the
modified Dirac hamiltonian (\ref{HDm}), whereas the low-energy
part discussed in Sec.~\ref{lep} below comprises both  
remaining contributions (ii) and (iii). 

We will continue by analyzing the two-loop form factors which enter into
Eq.~(\ref{HDm}); these will be treated in Sec.~\ref{twoloopff}. The calculation
will be split into two parts: the high-energy part discussed in Sec.~\ref{hep}
and the low-energy part, which is treated along ideas 
introduced in~\cite{CaLe1986} in Sec.~\ref{lep}. 

%
% Two-loop form factors
%
\section{Two-loop Form Factors}
\label{twoloopff}

In order to analyze the modified Dirac hamiltonian (\ref{HDm}) 
through two-loop order, we first have to investigate certain 
expansion coefficients of the electronic 
$F_1$ and $F_2$ form factors which are
thoroughly discussed in the seminal papers~\cite{Re1972a,Re1972b}. 
For the momentum transfer $q^2$ which is the argument of the 
two functions $F_1 \equiv F_1(q^2)$ and $F_2 \equiv F_2(q^2)$,
we use the convention $q^2 = q_\mu q^\mu = (q^0)^2 - \bbox{q}^2$.
The variable $t$ in~\cite{Re1972a,Re1972b} is given 
as $t = q^2$. When we evaluate radiative corrections to the 
binding Coulomb field which is mediated by space-like virtual
photons, we have $q^2 = - \bbox{q}^2$ because for $q^0 = 0$.
We use the conventions (see Eq.~(1.2) in~\cite{Re1972a}):
\begin{equation}
F_1(t) = 1 + \sum_{n=1}^{\infty} \left(\frac{\alpha}{\pi}\right)^n
  F_1^{(2n)}(t) \,,\quad
F_2(t) = \sum_{n=1}^{\infty} \left(\frac{\alpha}{\pi}\right)^n
  F_2^{(2n)}(t)\,.
\end{equation}
One and two-loop effects are denoted by upper 
indices 2 and 4, respectively. This notation is motivated
by the observation that two-loop effects are of forth 
order in the quantum electrodynamic interaction Lagrangian
$- e\, \bar{\psi}\, \gamma^\mu A_\mu \, \psi$ 
(in the Furry picture, which is used for the description of bound states,
the Coulomb interaction is taken out of the interaction Lagrangian). 

\begin{figure}[htb]
\begin{center}
\begin{minipage}{12cm}
\centerline{\mbox{\epsfysize=6.7cm\epsffile{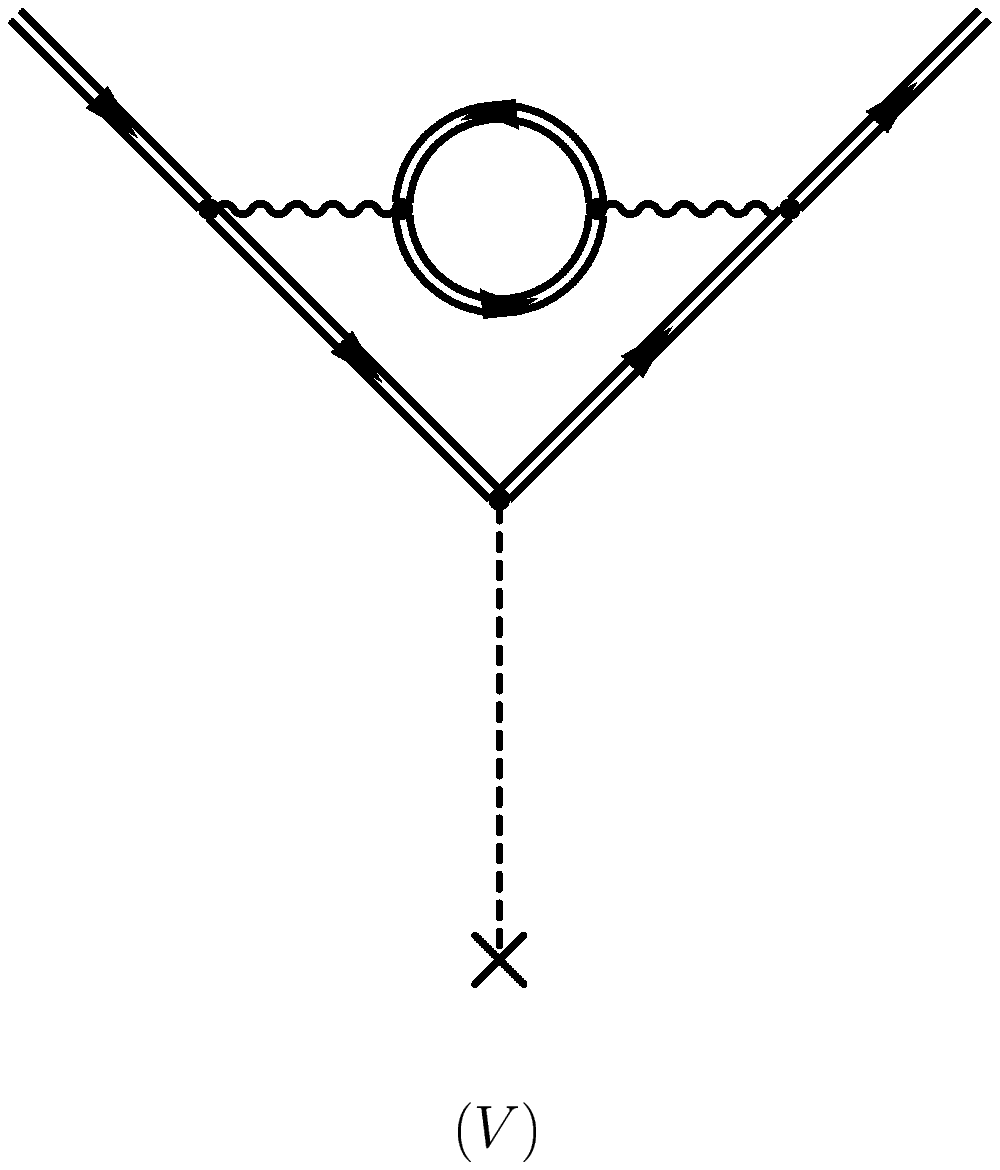}}}
\caption{\label{fig2} Combined self energy vacuum-polarization diagram
(denoted ``$V$'' in the text).}
\end{minipage}
\end{center}
\end{figure}

There are two different points of view regarding the choice of
diagrams to be included in the two-loop form factors, depending on 
whether the self energy vacuum polarization diagram~Fig.~\ref{fig2}
is included in the calculation or not. We will discuss both cases
and give results with and without the diagram shown in~Fig.~\ref{fig2}
taken into account.  

%
% Fig. 2
%

First, we discuss results obtained for $F_1$ {\em including} the
combined self energy vacuum polarization diagram. In this case, the
known results for the slope $F_1'(0)$  
and for $F_2(0)$, through two-loop order, read as follows. From
Eq.~(1.11) of~\cite{Re1972a}, we have:
\begin{eqnarray}
m^2 F_1'(0) &=& \frac{\alpha}{\pi} \, 
\left[- \frac{1}{3} \, \ln\left(\frac{\lambda}{m}\right) -  
  \frac{1}{8}\right] \nonumber\\[1ex] 
& & + \left(\frac{\alpha}{\pi}\right)^2 \, 
  \left[ - \frac{4819}{5184} - \frac{49}{72} \, \zeta(2) +
    3\,\zeta(2)\,\ln 2 - \frac{3}{4} \, \zeta(3) \right]\,,
\end{eqnarray}
where the forth-order coefficient has the numerical value
\begin{equation}
\label{resultF1prime}
m^2 F_1'^{(4)}(0) = 0.469\,941\,487\,460\,.
\end{equation}
According to Eq.~(1.7) in~\cite{Re1972a}, the value of $F_2(0)$,
through two-loop order, reads
\begin{equation}
F_2(0) = \frac{1}{2}\,\frac{\alpha}{\pi} 
+ \left(\frac{\alpha}{\pi}\right)^2 \, 
  \left[ \frac{197}{144} + \frac{1}{2} \, \zeta(2) 
   - 3 \, \zeta(2) \, \ln 2 + \frac{3}{4}\,\zeta(3) \right]\,,
\end{equation}
where the two-loop coefficient has the numerical value
\begin{equation}
\label{resultF2}
F_2^{(4)}(0) = -0.328\,478\,965\,579\,.
\end{equation}
We now turn to the discussion of the slope $F_2'^{(4)}(0)$.
In view of Eq.~(1.20) of~\cite{Re1972a} (see also~\cite{YeFrSu1961}),
we have (up to two-loop order)
\begin{equation}
\label{yefrsu}
F_2(t) = \frac{\alpha}{\pi} \, {\mathcal F}_2^{(2)}(t) +
  \left(\frac{\alpha}{\pi}\right)^2 \,
     \left[ \ln \frac{\lambda}{m} \, B(t) \, {\mathcal F}_2^{(2)}(t)
        + {\mathcal F}_2^{(4)}(t) \right]\,,
\end{equation}
where the coefficients ${\mathcal F}$ are by definition infrared safe and
\begin{equation}
\label{specificf2}
{\mathcal F}_2^{(2)}(0) = \frac{1}{2} \,,\quad
B(t) = - \frac{t}{3\,m^2} - \frac{t^2}{20\,m^2} +
  {\mathcal O}(t^3)\,.
\end{equation}
Equations (\ref{yefrsu}) and (\ref{specificf2})
uniquely determine the infrared divergent contribution to $F_2'^{(4)}(0)$.
An analytic expressions for ${\mathcal F}_2^{(4)}(t)$, $t$ spacelike, 
has recently been obtained~\cite{MaRe2001,BoMaRe2003,MaRe2003,BoMaRe2004} 
in terms of harmonic polylogarithms~\cite{ReVa2000,GeRe2001}. As a byproduct, 
an analytic expression for the slope ${\mathcal F}_2'^{(4)}(0)$
was found. The result reads
\begin{eqnarray}
\label{resultF2prime}
m^2 \, F_2'^{(4)}(0) &=& 
  - \frac{1}{6} \, \ln\left(\frac{\lambda}{m}\right)
    + m^2 \, {\mathcal F}_2'^{(4)}(0)\,, \nonumber\\[2ex]
m^2 \, {\mathcal F}_2'^{(4)}(0) &=& 
\frac{1751}{2160} 
+ \frac{13}{20} \, \zeta(2)
- \frac{23}{10} \, \zeta(2) \, \ln 2
+ \frac{23}{40} \, \zeta(3)\,.
\end{eqnarray}
A numerical result for ${\mathcal F}_2'^{(4)}(0)$, complementing
the above analytic expression,
can easily be derived in combining Eq.~(1.20), Eq.~(1.30),
and Eq.~(3.2) in~\cite{Re1972a}, as will be explained in 
the sequel. The dispersion relation (1.30) in~\cite{Re1972a} reads,
\begin{equation}
\label{dispersion}
{\mathrm{Re}} \, F_2(t) = - \frac{4 m^2}{t - 4 m^2} \, F_2(0) +
\frac{1}{\pi} \, \frac{t}{t - 4 m^2} \, P \int\limits_{4 m^2}^{\infty}
\frac{{\mathrm d} t'}{t' - t} \, \frac{t' - 4 m^2}{t'} \,
{\mathrm{Im}} F_2(t')\,,
\end{equation}
where $P$ denotes the Cauchy principal value.
Equation~(\ref{dispersion}) applies also if we single out the 
two-loop effect and differentiate at zero momentum transfer, 
and we obtain for the slope $F_2'^{(4)}(0)$ the relation
\begin{equation}
\label{F2prime1}
m^2 \, F_2'^{(4)}(0) = \frac{1}{4} \, F_2^{(4)}(0) +
\frac{1}{4 \, \pi} \, P \int\limits_{4 m^2}^{\infty}
{\mathrm d} t' \, \frac{4 m^2 - t'}{t'^2} \,
{\mathrm{Im}} F_2'^{(4)}(t') = 
\frac{1}{4} \, F_2^{(4)}(0) + {\mathcal T}\,,
\end{equation}
where $F_2^{(4)}(0)$ is given in Eq.~(\ref{resultF2}).
The second term on the right-hand side, denoted by ${\mathcal T}$,
can be evaluated using the result for ${\mathrm{Im}} \, F_2^{(4)}(x)$
presented in Eq.~(3.2) in~\cite{Re1972a}; it reads
\begin{equation}
\label{F2prime2}
{\mathcal T} = - \int\limits_0^1 {\mathrm d} x \frac{(1-x)^3}{x\,(1+x)}\,
  {\mathrm{Im}} \, F_2^{(4)}(x) = 
  - \frac{1}{6} \, \ln\left(\frac{\lambda}{m}\right) +
     0.030\,740\,507\,833(1) \,.
\end{equation}
Here, the last error is due to numerical integration,
and use is made of the natural variable~\cite{Re1972a}
\begin{equation}
x = \frac{1 - \sqrt{1 - 4 m^2/t}}{1 + \sqrt{1 - 4 m^2/t}}\,.
\end{equation}
In combining the result of Eq.~(\ref{resultF2}) with 
Eqs.~(\ref{F2prime1}) and (\ref{F2prime2}), the result
$m^2 \, {\mathcal F}_2'^{(4)}(0) = - 0.051\,379\,233\,561(1)$
is obtained which is in agreement with (\ref{resultF2prime}).

Now we will provide results for the form factors
obtained {\em excluding} the self energy vacuum-polarization
graph $V$ shown in Fig.~\ref{fig2}. These results refer to the 
pure two-photon self energy diagrams shown in Fig.~\ref{fig1}.
The two-loop self energy diagrams independently form a gauge-invariant set.
They represent a historically problematic correction, and are the main
subject of our investigation.
The combined self energy vacuum-polarization diagram,
according to Eqs. (1.9) and (1.10) in~\cite{Re1972b} --
taking into account the subtracted dispersion relation (1.30)
of~\cite{Re1972a} -- leads to the following corrections:
\begin{eqnarray}
F_1'^{(4),V}(0) &=& 
  -\frac{1099}{1296} + \frac{77}{144} \, \zeta(2) =
  0.031\,588\,972\,474\,,
\nonumber\\[3ex]
F_2^{(4),V}(0) &=& 
  \frac{119}{36} - 2 \, \zeta(2) =
  0.015\,687\,421\,859\,,
\nonumber\\[3ex]
F_2'^{(4),V}(0) &=& 
  \frac{311}{216} - \frac{7}{8} \, \zeta(2) =
  0.000\,497\,506\,323\,.
\end{eqnarray}
For the pure self energy graphs, which we would
like to denote by the symbol $S$, we therefore obtain the 
following results,
\begin{eqnarray}
\label{seF1p0}
m^2 \, F_1'^{(4),S}(0) &=& 
  - \frac{47}{576} - \frac{175}{144}\,\zeta(2)
  + 3\,\zeta(2) \, \ln 2 - \frac{3}{4} \, \zeta(3) =
  0.438\,352\,514\,986\,,\\[3ex]
\label{seF20}
F_2^{(4),S}(0) &=& 
  - \frac{31}{16} + \frac{5}{2}\,\zeta(2)
  - 3\,\zeta(2) \, \ln 2 + \frac{3}{4} \, \zeta(3) =
  -0.344\,166\,387\,438\,,\\[3ex]
\label{seF2p0}
m^2 \, F_2'^{(4),S}(0)  &=&
  - \frac{1}{6} \, \ln\left(\frac{\lambda}{m}\right)
    - \frac{151}{240} + \frac{61}{40}\,\zeta(2)
    - \frac{23}{10} \, \zeta(2) \, \ln 2 
    + \frac{23}{40} \, \zeta(3)
\nonumber\\[1ex]
&=&
  - \frac{1}{6} \, \ln\left(\frac{\lambda}{m}\right)
    - 0.051\,876\,739\,885 
\equiv
  - \frac{1}{6} \, \ln\left(\frac{\lambda}{m}\right)
    + {\mathcal F}_2'^{(4),S}(0)\,.
\end{eqnarray}
where the latter equality defines ${\mathcal F}_2'^{(4),S}(0)$ in 
analogy with Eqs.~(\ref{yefrsu}) and (\ref{resultF2prime}).

%
% High-energy part
%
\section{High--Energy Part}
\label{hep}

Based on the modified Dirac hamiltonian (\ref{HDm}),
corrections to the energy of the bound Dirac particle
can be inferred.
We will refer to the energy corrections 
attributable to the $F_1$ and $F_2$ form factors
as $E_1$ and $E_2$, respectively. For $E_1$, we have
\begin{equation}
E_1 = \left< \left[ F_1(-\bbox{q}^2) - 1 \right] \,
  e\,\phi \right>_{\mathrm{fs}} \,,
\end{equation}
where the index fs refers to the fine-structure terms, i.e.~to the 
result obtained by subtracting the value of the matrix element for
a $n{\mathrm P}_{3/2}$ state from the value of the same matrix 
element evaluated on a $n{\mathrm P}_{1/2}$ state.  
A matrix element $\left< A \right>_{\mathrm{fs}}$ of a
given operator $A$ is evaluated as
\[
\left< A \right>_{\mathrm{fs}} \equiv
\left< \psi_{{\mathrm{nP}_{3/2}}} \bigg| \, A \, \bigg| 
\psi_{{\mathrm{nP}_{3/2}}} \right> -
\left< \psi_{{\mathrm{nP}_{1/2}}} \bigg| \, A \, \bigg|
\psi_{{\mathrm{nP}_{1/2}}} \right>\,.
\]
Note that in evaluating these matrix elements, 
$\psi^{+}$ (the hermitian conjugate of the 
Dirac wave function $\psi$) should be used (not the 
Dirac adjoint $\bar{\psi} = \psi^{+} \gamma^0$). 
The Dirac wave functions $\psi$ are expanded in powers of $(Z\alpha)$ up to the
order relevant for the current investigation of two-loop effects.
This expansion avoids potential
problems associated with the logarithmic divergence of the Dirac
wave function at the origin. 

For $E_1$, up to the order of $(Z \alpha)^6$, we have 
\begin{equation}
E_1 = 4 \pi Z \alpha \, F_1'^{(4)}(0) \, 
  \left< \delta^{(3)}(\bbox{r}) \right>_{\mathrm{fs}}\,.
\end{equation}
For P states, the nonrelativistic 
(Schr\"{o}dinger) wave function -- the leading term in the 
$Z\alpha$-expansion of the Dirac wave function -- 
vanishes at $\bbox{r} = 0$, but the first relativistic
correction gives a finite contribution, resulting in
\begin{equation}
\langle \delta^{(3)}(\bbox{r}) \rangle_{\mathrm{fs}} = 
  - \frac{n^2 - 1}{4\,n^5} \, (Z\alpha)^5 \, m^3\,.
\end{equation}
This leads -- again up to the order of $(Z \alpha)^6$ --
to the following result for $E_1$,
\begin{equation}
\label{E1}
E_1 = \left(\frac{\alpha}{\pi}\right)^2 \,
  \frac{(Z\alpha)^6}{n^3} \, \left[ - F_1'^{(4)}(0) \,
     \frac{n^2 - 1}{n^2} \right] \, m^3\,.
\end{equation}
Observe that the derivative of the $F_1$ form factor has a physical
dimension of $1/m^2$ in natural units, giving the correct
physical dimension for $E_1$. 
The correction due to $F_2$ in (\ref{HDm}) reads,
\begin{equation}
\label{E2}
E_2 = \langle F_2(-\bbox{q}^2) \,
  \frac{e}{2 m} \, \mathrm{i}\, \bbox{\gamma} \cdot \bbox{E} 
     \rangle_{\mathrm{fs}}\,.
\end{equation}
A particle in an external binding Coulomb
field feels an electric field 
$\bbox{E} = {\mathrm i}\,(Z e) \, \bbox{q}/\bbox{q}^2$ 
-- in momentum space -- or 
$\bbox{E} = - (Z e) \, \bbox{r}/(4 \pi r^3)$
in coordinate space. 
Vacuum polarization corrections to 
$\bbox{E} = - (Z e) \, \bbox{r}/(4 \pi r^3)$
lead to higher-order effects. The correction
$E_2$ splits up in a natural way into two
contributions $E_{2a}$ and $E_{2b}$
which are associated with $F_2(0)$ and the slope $F'_2(0)$,
respectively. $E_{2a}$ reads
\begin{equation}
E_{2a} = \frac{Z \alpha}{2 m} \, F_2^{(4)}(0) \, 
\left< - \mathrm{i} \, 
\frac{\bbox{\gamma} \cdot \bbox{r}}{r^3} \right>_{\mathrm{fs}}\,.
\end{equation}
The evaluation of the matrix element leads to
\begin{equation}
\left< - \mathrm{i} \, 
\frac{\bbox{\gamma} \cdot \bbox{r}}{r^3} \right>_{\mathrm{fs}}
  = \left\{ \frac{(Z\alpha)^3}{n^3} + 
    \left[ \frac{487}{360} + \frac{5}{4 n} - \frac{23}{10 n^2}\right]
       \frac{(Z\alpha)^5}{n^3} \right\} \, m^2\,.
\end{equation}
For the purpose of the current investigation,
the $(Z\alpha)^6$-component of $E_{2a}$ is selected only:
\begin{equation}
\label{E2a}
E_{2a} = \left(\frac{\alpha}{\pi}\right)^2 \,
  \frac{(Z\alpha)^6}{n^3} \, \left[ F_2^{(4)}(0) \,
     \left( \frac{487}{720} + \frac{5}{8 n} - 
        \frac{23}{20 n^2} \right) \right]\,m\,.
\end{equation}
The matrix element $E_{2b}$ can be expressed as
\begin{equation}
E_{2b} = \frac{4 \pi Z \alpha}{2 m} \, F_2'^{(4)}(0) \, 
\langle \bbox{\gamma} \cdot \bbox{q} \rangle_{\mathrm{fs}}\,.
\end{equation}
A transformation into coordinate space leads to
\begin{equation}
\langle \bbox{\gamma} \cdot \bbox{q} \rangle_{\mathrm{fs}} = 
{\mathrm i} \, \left[ \frac{\partial}{\partial \bbox{x}} 
\left(\psi^{+}(\bbox{x}) \bbox{\gamma} \psi(\bbox{x}) \right) 
  \right]_{x=0,{\mathrm{fs}}}
  =  - \frac{n^2 - 1}{\pi\,n^5} \, (Z\alpha)^5\,m^4\,.
\end{equation}
As a function of the principal quantum number $n$,
the result for $E_{2b}$ reads:
\begin{equation}
\label{E2bVersion1}
E_{2b} = \left(\frac{\alpha}{\pi}\right)^2 \,
  \frac{(Z\alpha)^6}{n^3} \, \left[ - 2 \, F_2'^{(4)}(0) \,
     \frac{n^2 - 1}{n^2} \right]\,m^3\,.
\end{equation}                
This result involves the infrared divergent slope
of the $F_2$ form factor [see Eqs.~(\ref{resultF2prime}) and (\ref{seF2p0})].
We are thus faced with 
the problem of matching the infrared divergence of the slope of 
the $F_2$ form factor, expressed in terms of the fictitious photon mass
$\lambda$, with the usual (energy matching parameter) $\epsilon$
introduced originally in~\cite{Pa1993}.
This can be done in two ways: (i) by matching the infrared divergence 
of the rate of soft bremsstrahlung, calculated with a fictitious photon
mass $\lambda$, to a result of the same calculation, carried out with
an explicit infrared cut-off $\epsilon$ for the photon energy.
This way of calculation is described on pp.~361--362 of~\cite{ItZu1980}.
It leads to the result
\begin{equation}
\label{matching1}
\ln \frac{\lambda}{2 \, \epsilon} = - \frac{5}{6}\,.
\end{equation}
The matching procedure (ii) consists in a comparison of the result
of the application of the formalism considered above, and its
application to the high-energy part of the ground state Lamb shift,
which is in leading order given by the infrared divergence of the 
$F_1$ form factor, and the result obtained by direct calculation
of this high-energy part in a non-covariant formalism with an 
explicit energy cut-off $\epsilon$, as it has been carried out 
in~\cite{Pa1993}. This second matching procedure leads to the following
result -- in agreement with (\ref{matching1}) --,
\begin{equation}
\label{matching2}
\ln \frac{m}{\lambda} - \frac{3}{8} = \ln \frac{m}{2 \epsilon} + 
  \frac{11}{24}\,.
\end{equation}
So, we are led to the replacement
\begin{equation}
\label{replacement}
- \ln \frac{\lambda}{m} \to 
  \ln \frac{m}{2 \epsilon} + \frac{5}{6}  
\end{equation}
A comparison with the results in Eqs.~(\ref{resultF2prime}), (\ref{seF2p0}),
and~(\ref{E2b}) reveals that the
logarithmic divergence for the fine-structure difference is given by
a term
\begin{equation}
\label{logterm}
  - \frac{n^2 - 1}{3 n^2} \, \ln \frac{m}{2 \epsilon} \,,
\end{equation}
so that we may anticipate at this stage the result for 
$\Delta_{\mathrm{fs}} B_{\mathrm 61}$,
\begin{equation}
\label{DeltaB61}
\Delta_{\mathrm{fs}} B_{\mathrm 61} = 
  - \frac{n^2 - 1}{3 n^2} \,.
\end{equation}
Based on (\ref{E2bVersion1}) and (\ref{replacement}),
we can express $E_{2b}$ in terms of $\epsilon$ and 
${\mathcal F}_2'^{(4)}(0)$,
\begin{equation}
\label{E2b}
E_{2b} = \left(\frac{\alpha}{\pi}\right)^2 \,
  \frac{(Z\alpha)^6}{n^3} \, \left[ 
     - \frac{1}{3}\,\frac{n^2-1}{n^2}\,\ln\frac{m}{2\epsilon} - 
          \left(\frac{5}{18} + 2\,{\mathcal F}_2'^{(4)}(0)\,m^2 \right) \, 
    \frac{n^2-1}{n^2} \right]\,m\,.
\end{equation}

There is a third correction due to the effect of {\em two} one-loop 
corrections on the electron vertices. Because we are only interested
in the fine structure, we isolate the terms which are proportional
to the spin-orbit coupling, and we obtain
\begin{equation}
\label{defE3}
E_3 = \left< \left[2 F_2(0)\right] \, H_{\mathrm{fs}} \,
  \left( \frac{1}{E - H} \right)' \,
\left[2 F_2(0)\right] \, H_{\mathrm{fs}} \right>_{\mathrm{fs}}\,,
\end{equation}
where
\begin{equation}
\label{Hfs}
H_{\mathrm{fs}} = \frac{Z\alpha}{4 m^2 r^3} \, 
  \bbox{\sigma} \cdot \bbox{L}\,,
\end{equation}
and $1/(E-H)'$ is the nonrelativistic, spin-independent
reduced Schr\"{o}dinger--Coulomb Green function~\cite{SwDr1991a,SwDr1991b}.
The only spin-dependence in (\ref{defE3})
occurs in the coupling $\bbox{\sigma} \cdot
\bbox{L}$, and it can be taken into account by an overall factor,
\begin{equation}
\label{spindependence}
\left< (\bbox{\sigma} \cdot \bbox{L})^2 \right>_{\mathrm{fs}} = -3\,.
\end{equation}
We are therefore led to consider the ``spin-independent
version'' of the matrix element which occurs in Eq.~(\ref{defE3}) 
and obtain the following result,
\begin{equation}
\label{matE3}
\left< \frac{Z\alpha}{4 m^2 r^3} \left(\frac{1}{E-H}\right)'
\frac{Z\alpha}{4 m^2 r^3} \right>_{\mathrm{nP}} =
  \left( -\frac{227}{8640} - \frac{1}{96 n} 
     + \frac{1}{80 n^2} \right) \, \frac{(Z \alpha)^6 \, m}{n^3}\,.
\end{equation}
The spin-dependence can be easily restored by considering 
Eq.~(\ref{spindependence}). The index ``nP'' in Eq.~(\ref{matE3}) means 
that the matrix element is evaluated with the nonrelativistic,
spin-independent (Schr\"{o}dinger) wave function. Alternatively,
on may evaluate with either the ${\mathrm{nP}}_{1/2}$ or the
${\mathrm{nP}}_{3/2}$ Dirac wave function and expand up to the leading order
in $(Z\alpha)$.

The evaluation of (\ref{matE3}) can proceed e.g.~by solving the differential 
equation which defines the correction to the wave function induced
by $H_{\mathrm{fs}}$, and subsequent direct evaluation of the 
resulting standard integrals using computer algebra~\cite{Wo1988}. 
The final result for $E_3$ reads,
\begin{equation}
\label{E3}
E_3 = \left( \frac{\alpha}{\pi}\right)^2 \, \frac{(Z\alpha)^6}{n^3} \,
  \left[ \frac{227}{2880} + \frac{1}{32 n} - \frac{3}{80 n^2} \right]\,m\,.
\end{equation}
This concludes the discussion of the high-energy part. 
The final result for the high-energy part is
\begin{equation}
\label{EH}
E_{\mathrm H} = E_1 + E_{2a} + E_{2b} + E_3\,,
\end{equation}
where $E_1$, $E_{2a}$, $E_{2b}$, $E_3$ are given in 
Eqs.~(\ref{E1}), (\ref{E2a}), (\ref{E2b}), (\ref{E3}), respectively.

%
% The low-energy part
%
\section{Low--Energy Part}
\label{lep}

The low-energy part consists essentially of two contributions.
Both effects, denoted here $E_4$ and $E_5$,
can be obtained by a suitable variation 
of the low-energy part of the {\em one-loop} self energy, by considering the 
spin-dependent effects introduced by a 
{\em further} one-loop electron anomalous 
magnetic moment interaction.
The first of the two terms, $E_4$, is caused by spin-dependent
higher-order effects in the one-loop self energy, which receive
additional corrections due to the anomalous magnetic moment
of the electron.
The second term, $E_5$, is due to an anomalous magnetic moment correction
to the electron transition current, which can also be seen
as a correction to the radiation field of the electron
due to its anomalous magnetic moment.

The leading-order low-energy part (see~\cite{Pa1993}) reads
\begin{equation}
\label{NREL}
E_{\mathrm L} = -\frac{2 \alpha}{3 \pi m} \,
\int_0^\epsilon {\mathrm d}\omega \, \omega \,
  \left< \phi \left| \bbox{p} \, \frac{1}{H - (E - \omega)} \,
     \bbox{p} \right| \phi \right>\,.
\end{equation}
In order to isolate the fine-structure effects,
we should now consider corrections to the wave function, to the current,
to the hamiltonian and to the energy of the bound state due to the 
spin-dependent relativistic (spin-orbit) hamiltonian 
\begin{equation}
\label{calH}
{\mathcal H} =
F_2(0) \, \frac{e}{2 m} \, {\mathrm i} \, \bbox{\gamma} \cdot \bbox{E} =
\frac{\alpha (Z \alpha)}{4 \pi m} \, 
\frac{-{\mathrm i} \, \bbox{\gamma} \cdot \bbox{r}}{r^3}\,.
\end{equation}
The above hamiltonian ${\mathcal H}$
is the last term in the modified Dirac hamiltonian [right-hand side
of Eq.~(\ref{HDm})],
approximated for a particle bound in a Coulomb field with the
$F_2$ form factor evaluated at zero momentum. The electric
field $\bbox{E}$ in (\ref{calH}) corresponds to the binding
Coulomb interaction. The hamiltonian (\ref{calH}) 
describes the modification of the spin-orbit interaction due to 
the anomalous magnetic moment of the electron.

The nonrelativistic limit of ${\mathcal H}$ is the spin-orbit
coupling $H_{\mathrm{fs}}$ given in Eq.~(\ref{Hfs}),
multiplied by a factor $2 F_2^{(2)}(0) = \alpha/\pi$
(the additional factor 2 finds an explanation
in~\cite{BeSa1957}).
The resulting hamiltonian
\begin{equation}
\label{Heff}
H_{\mathrm{eff}} = \frac{\alpha}{\pi}\, H_{\mathrm{fs}} =
\frac{\alpha}{\pi}\, \frac{Z\alpha}{4 m^2 r^3} \,
  \bbox{\sigma} \cdot \bbox{L}
\end{equation}
takes into account magnetic vertex corrections in the 
framework of an effective theory. Denoting the variation of  
the expression (\ref{NREL}) mediated by $H_{\mathrm{eff}}$ with the symbol
$\delta_{\mathrm{eff}}$ -- in the spirit of
the notation introduced in~\cite{Pa2001} --, we obtain the contribution
\begin{equation}
\label{defE4a}
E_{4a} = \delta_{\mathrm{eff}} 
\left\{ - \frac{2 \alpha}{3 \pi m} \,
\int_0^\epsilon {\mathrm d}\omega \, \omega \,
  \left< \phi \left| \, \bbox{p} \, \frac{1}{H - (E - \omega)} \,
     \bbox{p} \, \right| \phi \right> \right\}\,.
\end{equation}
Following the notation introduced in~\cite{JePa1996,JeSoMo1997},
the contribution $E_{4a}$ is the sum of the fine-structure effects
created by the wave-function-correction $F_{\delta\phi}$, 
the first relativistic correction to the energy $F_{\delta E}$,
and the correction due to the relativistic hamiltonian $F_{\delta H}$,
each multiplied by a factor $\alpha/\pi$. The final result for $E_4$ is
\begin{equation}
\label{E4aspec}
E_{4a} = \left(\frac{\alpha}{\pi}\right)^2 \, (Z\alpha)^4 \,
\frac{m}{n^3} \, \left(\Delta_{\mathrm{fs}} F_{\delta\phi} + 
\Delta_{\mathrm{fs}} F_{\delta E} +
\Delta_{\mathrm{fs}} F_{\delta H} \right)\,.
\end{equation}
There is a further correction to the nonrelativistic effective coupling
to the radiation field due to the ``anomalous 
spin-orbit hamiltonian'' (\ref{calH}).
The correction, in the nonrelativistic limit, can be derived by considering
a Foldy--Wouthuysen transformation which by definition diagonalizes  
the hamiltonian~(\ref{calH}) in spinor space and also 
leads to corrections to the current according to
\begin{equation}
\label{FWcorrection}
\alpha^i \to U \, \alpha^i U^{-1} \,, \quad
U = \exp\left(-{\mathrm i} \frac{\beta {\mathcal H}}{2 m} \right) \,.
\end{equation}
Here, $\beta$ and $\alpha^i$ are standard Dirac matrices~\cite{ItZu1980},
$i$ is a spatial index, and ${\mathcal H}$ is given in (\ref{calH}).
The calculation is carried
out along ideas introduced in~\cite{JePa1996} and leads to the result
\begin{equation}
\delta \bbox{j}_{4b} = \frac{\alpha}{\pi} \, \frac{Z\alpha}{2 m r^3} \,\,
\bbox{\sigma} \times \bbox{r}\,,
\end{equation}
as a relativistic correction to the electron current
which is simply $\alpha^i$ in the relativistic
formalism and  $p^i/m$ in the leading nonrelativistic 
approximation.
Again, following the notation introduced in~\cite{JePa1996,JeSoMo1997},
the resulting additional contribution is
\begin{equation}
\label{E4bspec}
E_{4b} = \left(\frac{\alpha}{\pi}\right)^2 \, (Z\alpha)^4 \,
\frac{m}{n^3} \, \Delta_{\mathrm{fs}} F_{\delta y}\,.
\end{equation}
The sum of (\ref{E4aspec}) and (\ref{E4bspec}) is just the 
$(Z\alpha)^6$--component of the 
fine-structure difference of the one-loop self energy 
from~\cite{JePa1996,JeSoMo1997}, multiplied
by an additional factor $\alpha/\pi$. It can also be written as
\begin{equation}
\label{E4}
E_4 = E_{4a} + E_{4b} = \left( \frac{\alpha}{\pi} \right)^2 \,
\frac{(Z\alpha)^6 \, m}{n^3} \, \left[ - \frac{n^2 - 1}{3 n^2} \, 
  \ln \frac{2 \epsilon}{(Z\alpha)^2\,m} 
     + \frac{n^2 - 1}{n^2} \, \Delta_{\mathrm{fs}}{{\ell}}_4(n) \right]\,,
\end{equation}
where $\Delta_{\mathrm{fs}}{{\ell}}_4(n)$ could be interpreted as a 
relativistic generalization of a
Bethe logarithm, which is $n$-dependent. However, a significant
numerical fraction of the $n$-dependence can be eliminated 
if the factor $(n^2 - 1)/n^2$ is taken out of the 
final result. The evaluation of $\Delta_{\mathrm{fs}}{{\ell}}_4(n)$ 
has recently been performed 
in~\cite{JeEtAl2003} with improved numerical methods 
(see e.g.~\cite{JeMoSoWe1999}), and the following results have been obtained:
\begin{eqnarray}
\Delta_{\mathrm{fs}}{\ell}_4(2) &=& 0.512~559~768(1)\,, \nonumber\\
\Delta_{\mathrm{fs}}{{\ell}}_4(3) &=& 0.511~978~815(1)\,, \nonumber\\
\Delta_{\mathrm{fs}}{{\ell}}_4(4) &=& 0.516~095~539(1)\,, \nonumber\\
\Delta_{\mathrm{fs}}{{\ell}}_4(5) &=& 0.519~976~941(1)\,, 
\end{eqnarray}
where the uncertainty is due to numerical integration.

There is, as stated above, a further
correction due to the explicit modification of the 
transition current due to the anomalous magnetic moment;
it can be obtained through the replacement
\begin{equation}
\alpha^i \to \alpha^i + F_2(0) \, 
\frac{{\mathrm i}\,\beta\,\sigma^{i\nu}}{2 m} \, q_\nu
\end{equation}
and must be considered in addition to the correction (\ref{FWcorrection}).
A careful consideration of the nonrelativistic limit of this 
correction to the current, including retardation effects, leads to the
result
\begin{equation}
\delta \bbox{j}_{5} = \frac{\alpha}{2 \pi} \, \frac{Z\alpha}{2 m r^3} \,\,
\bbox{\sigma} \times \bbox{r}\,.
\end{equation}
Consequently, we find that the correction is
effectively $F_2(0)$ times the retardation corrections to the
transition current $F_{\delta y}$ found in~\cite{JePa1996,JeSoMo1997}.
We obtain
\begin{equation}
E_5 = \left(\frac{\alpha}{\pi}\right)^2 \, (Z\alpha)^4 \,
\frac{m}{n^3} \, \frac{\Delta_{\mathrm{fs}} F_{\delta y}}{2}\,.
\end{equation}
In analogy with $E_4$, this correction can favorably be rewritten as
\begin{equation}
\label{E5}
E_5 = \left( \frac{\alpha}{\pi} \right)^2 \,
\frac{(Z\alpha)^6 \, m}{n^3} \, \left[ 
\frac{n^2 - 1}{n^2} \, \Delta_{\mathrm{fs}}{{\ell}}_5(n) \right]\,,
\end{equation}
On the basis of~\cite{JePa1996,JeSoMo1997,JeEtAl2003}, we obtain
\begin{eqnarray}
\Delta_{\mathrm{fs}}{{\ell}}_5(2) &=& -0.173~344~868(1)\,, \nonumber\\
\Delta_{\mathrm{fs}}{{\ell}}_5(3) &=& -0.164~776~514(1)\,, \nonumber\\
\Delta_{\mathrm{fs}}{{\ell}}_5(4) &=& -0.162~263~216(1)\,, \nonumber\\
\Delta_{\mathrm{fs}}{{\ell}}_5(5) &=& -0.161~165~602(1)\,. 
\end{eqnarray}
The final result for the low-energy part is
\begin{equation}
\label{EL}
E_{\mathrm L} = E_4 + E_5\,,
\end{equation}
with $E_4$ and $E_5$ being given in Eqs.~(\ref{E4}) and (\ref{E5}),
respectively. 

We can now understand why it was possible to join the two contributions
with ``mixed'' and ``low-and-low'' energy virtual photons (ii) and (iii),
which were discussed in Sec.~\ref{IntroTwoLoop}, into a joint 
``low-energy part''. The reason is simple:
The effective hamiltonian (\ref{Heff}) has no
infrared divergence, because it involves the low-energy limit
of the magnetic form factor $F_2$, which is infrared safe in
one-loop order according to Eq.~(\ref{yefrsu}). Because the main
contribution to the quantity $F_2(0)$ is caused by hard virtual photons,
it is also justified to say that the contribution of ``low-and-low''
energy virtual photons vanishes at the order of interest for the 
current calculation (fine-structure difference). In higher-loop order,
the further infrared divergence acquired by $F_2$ would lead to an infrared
divergence in the effective hamiltonian constructed in analogy with 
Eq.~(\ref{Heff}); this infrared divergence would have 
to be attributed to a ``mixed'' contribution. 

%
% Results for the Two--Loop Corrections
%
\section{Results for the Two--Loop Corrections}
\label{resconcl}

We have obtained analytic results for higher-order correction to the two-loop 
self energy of P states in hydrogen-like systems.  
In our calculation,
we have analyzed the electron form factors through two-loop
order in Sec.~\ref{twoloopff}, and we have split the calculation into 
a high-energy part with two hard virtual photons discussed in
Sec.~\ref{hep}, and a low-energy part with at least one soft virtual
photon analyzed in Sec.~\ref{lep}.
The final result for the contribution to the 
fine-structure energy difference 
is obtained by adding the high-energy contributions 
$E_1$ -- $E_3$ given in
Eqs.~(\ref{E1}), (\ref{E2a}), (\ref{E2b}), (\ref{E3}), 
and the low-energy effects 
$E_4$ and $E_5$ from Eqs.~(\ref{E4}) and (\ref{E5}).
The dependence on $\epsilon$ cancels out in the final result
which is the sum of the high-energy part $E_{\mathrm H}$
given in Eq.~(\ref{EH}) and the low-energy part 
$E_{\mathrm L}$ defined in Eq.~(\ref{EL}).
This is also evident when considering explicitly
the Eqs.~(\ref{E2b}) and (\ref{E4}). The final results for the analytic
coefficients of order $\alpha^2 (Z\alpha)^6$ read 
\begin{equation}
\label{final61}
\Delta_{\mathrm{fs}} B_{\mathrm 61} =
  - \frac{n^2 - 1}{3 n^2} \,.
\end{equation}
[see also Eq.~(\ref{DeltaB61})] and
\begin{eqnarray}
\label{final}
\lefteqn{\Delta_{\mathrm{fs}} B_{\mathrm 60} =
   \left( \frac{227}{2880} + \frac{1}{32 n} - \frac{3}{80 n^2} \right) +
   F_2^{(4),S}(0) \, \left( \frac{487}{720} + \frac{5}{8 n} -
        \frac{23}{20 n^2} \right)}
\nonumber\\[1ex]
& &  + \frac{n^2 - 1}{n^2} \,
       \left[- \left( F_1'^{(4),S}(0) + 
            2\, {\mathcal F}_2'^{(4),S}(0) \right) \, m^2 
          - \frac{5}{18} + \Delta_{\mathrm{fs}}{{\ell}}_4(n) +
            \Delta_{\mathrm{fs}}{{\ell}}_5(n) \right]\,,
\end{eqnarray}
where explicit numerical results for $F_1'^{(4),S}(0)$,
$F_2^{(4),S}(0)$ and ${\mathcal F}_2'^{(4),S}(0)$ can be found in 
Eqs.~(\ref{seF1p0}), (\ref{seF20}) and (\ref{seF2p0}),
respectively. This result refers to the pure self energy diagrams
in Fig.~\ref{fig1}. The result reads numerically for the principal
quantum numbers $n=2$--5,
\begin{eqnarray}
\Delta_{\mathrm{fs}} B_{\mathrm 60}(2) &=& -0.361~196~470(1)\,,\\
\Delta_{\mathrm{fs}} B_{\mathrm 60}(3) &=& -0.411~156~068(1)\,,\\
\Delta_{\mathrm{fs}} B_{\mathrm 60}(4) &=& -0.419~926~624(1)\,,\\
\Delta_{\mathrm{fs}} B_{\mathrm 60}(5) &=& -0.419~832~876(1)\,.
\label{final25P}
\end{eqnarray}
If it is desired to add in the combined self energy
vacuum-polarization diagram from Fig.~\ref{fig2}, then the 
form-factor results from Eqs.~(\ref{resultF1prime}), (\ref{resultF2}) and
(\ref{resultF2prime}) instead of
the pure self energy results given in Eqs.~(\ref{seF1p0}), (\ref{seF20}) 
and (\ref{seF2p0}) have to be used in evaluating (\ref{final}).
When including the combined self energy
vacuum-polarization diagram from Fig.~\ref{fig2},
there is no further low-energy contribution, so that
the alternative set of numerical values for the form factors
from Eqs.~(\ref{resultF1prime}), (\ref{resultF2}) and
(\ref{resultF2prime}) fully takes into account the additional 
effect of the diagram in Fig.~\ref{fig1} on the fine-structure 
in the order of $\alpha^2\,(Z\alpha)^6$.

%
% Spinless Particles and Bound Systems within QED
%
\chapter{Spinless Particles in Bound--State Quantum Electrodynamics}
\label{QEDSpiZer}

\typeout{==========================================================}
\typeout{Spinless Particles in Bound--State Quantum Electrodynamics}
\typeout{==========================================================}
\typeout{}

%
% Orientation
%
\section{Orientation}
We describe in this Chapter a simplified derivation of the 
relativistic corrections of order $\alpha^4$ 
for a bound system consisting of two 
spinless particles. We devote special attention to pionium,
the bound system of two oppositely charged pions. The leading quantum 
electrodynamic (QED) correction to the 
energy levels is of the order of $\alpha^3$ and due to electronic
vacuum polarization. We analyze further corrections due to
the self energy of the pions, and due to recoil effects,
and we give a complete result for the 
scalar-QED leading logarithmic corrections 
which are due to virtual loops involving only the scalar constituent
particles (the pions); these corrections are of order 
$\alpha^5 \,\ln \alpha$ for S states.

%
% Introduction
%
\section{Introduction to Spinless QED and Pionium}
\label{IntroSpiZer}

Exotic bound systems like pionium~\cite{AfEtAl1993,AfEtAl1994} 
(the bound system
of two oppositely charged pions) offer interesting possibilities
for studies of fundamental properties
of quantum mechanical bound states: the interplay
between strong-interaction corrections and quantum electrodynamic
corrections is of prime interest,
and the small length scales characteristic of the 
heavy particles make it possible to explore effects of the virtual 
excitations of the quantum fields in previously unexplored
kinematical regimes~\cite{KaJeIvSo1998,JeSoIvKa1997,KaIvJeSo1998}.
We do not wish to hide the fact that any potential high-precision
experiments in this area are faced with various 
experimental difficulties.
Our calculations address QED corrections to the spectrum 
of bound systems whose constituent particles are spinless;
relativistic corrections to the decay lifetime of pionium
have recently been discussed in~\cite{HeHeTrBa2001}
in the context of the DIRAC experiment at CERN.

Here, we report on results regarding the spectrum of a bound 
system consisting of two spinless particles.
We apply the simplified calculational scheme employed in
\cite{PaKa1995} for the relativistic and recoil corrections
to a bound system of two ``non-Dirac'' particles to the case
of two interacting spinless particles (see Sec.~\ref{BreitEffects}). 
We then recall known results
on leading-order vacuum polarization corrections in Sec.~\ref{VacuumEffects}
and clarify the relative order-of-magnitude of the one- and two-loop
electronic vacuum polarization, the relativistic and recoil corrections
and the self energies in pionium (also in Sec.~\ref{VacuumEffects}). 
We then provide
an estimate for the self energy in Sec.~\ref{SelfEnergyEffects},
and we analyze the leading recoil correction of order
$\alpha^5$ (the Salpeter correction) which leads us to complete
results for the scalar-QED logarithmic corrections of order
$\alpha^5 \ln\alpha$.

%
% Breit hamiltonian
%
\section{Breit hamiltonian for Spinless Particles}
\label{BreitEffects}

We start from the Lagrangian for a 
charged spinless field coupled to 
the electromagnetic field [see equations~(6-50) -- 
(6-51b) of~\cite{ItZu1980}],
\begin{equation}
\label{Lagrangian}
{\mathcal L}(x) = 
\left[\left( \partial_\mu - {\mathrm i} e A_\mu \right) 
\phi^{*}(x) \right]\,
\left( \partial^\mu + {\mathrm i} e A^\mu \right) \phi(x) -
m^2 \phi^{*}(x) \, \phi(x) - 
\frac{1}{4} \, F_{\mu\nu}(x) \, F^{\mu\nu}(x)\,,
\end{equation}
where the field strength tensor $F_{\mu\nu}$ reads
$F_{\mu\nu}(x) = \partial_\mu A_\nu(x) - \partial_\nu A_\mu(x)$.
We use natural Gaussian units with $\hbar = c = \epsilon_0 = 1$.
The transition current for a free spinless particle 
($A^\mu = 0$) can be inferred from (\ref{Lagrangian}); it reads
in momentum space 
\begin{equation}
\label{currentjmu}
j^\mu(p',p) = \phi^{*}(p') \, (p'^\mu + p^\mu) \, \phi(p)\,.
\end{equation}
This current now has to be expressed in terms of nonrelativistic
wave functions. Specifically, the $j^0$-component has to reproduce
the normalization of the nonrelativistic (Schr\"{o}dinger)
wave function. By contrast, according to Eq.~(\ref{currentjmu})
the zero-component of the current reads $2 m\,\phi^{*} \phi$ 
in the nonrelativistic limit $p'^0 \to m\,, p^0 \to m$. 
The nonrelativistic wave functions
are normalized according to
\begin{equation}
\int {\mathrm d}^3 x \, \phi^{*}_{\mathrm S}(\bbox{x}) 
\phi_{\mathrm S}(\bbox{x}) = 1\,.
\end{equation}
It is therefore evident that we cannot simply associate the relativistic
wave function $\phi$ with $\phi_{\mathrm S}$; rather, we should 
define according to Eqs.~(13) -- (14) of~\cite{PaKa1995}
\begin{equation}
\phi(\bbox{p}) = 
\frac{\phi_{\mathrm S}(p^0,\bbox{p})}{\sqrt{2 p^0}}\,,
\end{equation}
where $p^0 = \sqrt{\bbox{p}^2 + m^2} \approx m$
is the energy of the free nonrelativistic
particle (in deriving low-energy
effective interactions, one always expands about {\em free}-particle
amplitudes; all interactions are treated as perturbations;
note the analogy to nonrelativistic QED -- NRQED -- 
for spinor particles~\cite{CaLe1986}).
The Klein--Gordon current, in the 
presence of external fields, reads in contrast to (\ref{currentjmu})
\begin{equation}
\label{currentjmuAmu}
j^\mu(p',p) = \phi^{*}(p') \, (p'^\mu + p^\mu - 2 \, e A^\mu) \, \phi(p)\,.
\end{equation}
The zero-component of this current can be interpreted
as a charge density, which is not necessarily positive definite.
Questions related to the normalization of the 
Klein--Gordon wave functions in this case are discussed in detail 
in~\cite{KlRa1975,BaLa1975,KlRa1975reply,FlSo1984}.

In terms of the Schr\"{o}dinger wave function,
the current is given as
\begin{eqnarray}
\label{j0}
j^0(\bbox{p}', \bbox{p}) &=&
\phi^{*}_{\mathrm S}(\bbox{p'}) \, 
\phi_{\mathrm S}(\bbox{p})\,, \\[2ex]
\label{ji}
j^i(\bbox{p}', \bbox{p}) &=& 
\phi^{*}_{\mathrm S}(\bbox{p'}) \, 
\frac{p^i + p'^i}{2\,m} \, 
\phi_{\mathrm S}(\bbox{p})\,,
\end{eqnarray}
where $m$ is the mass of the particle.
The atomic momenta $p^i$ and $p'^i$ in Eq.~(\ref{ji}) are
of order $Z\alpha$. As shown below, interactions
involving the spatial components $j^i$ of the
transition current give rise to relativistic
contributions of order $(Z\alpha)^4$ to the spectrum.
This is exactly the order of magnitude that is the subject of the
current investigation.
Therefore, although Eq.~(\ref{ji}) is only valid up to 
corrections of relative order $(Z\alpha)^2$, these 
can be neglected because the further corrections contribute
to the energy levels at the order of $(Z\alpha)^6$.
Specifically, we can expect corrections proportional to 
$(p^i \, \bbox{p}^2)$ to the current $j^i$ when a systematic expansion of the 
nonrelativistic current is performed; these terms are analogous 
to those obtained for relativistic corrections to the 
current of spinor particles which can be obtained via a
Foldy--Wouthuysen transformation~\cite{JePa1996,LaPa2001}.

In the following, the index S on the wave function will be 
dropped, and the nonrelativistic amplitudes describing the two interacting
particles (with electric charges $e_1$ and $e_2$) 
will be denoted as $\phi_1$ and $\phi_2$, respectively.
Following~\cite{PaKa1995},
the Breit hamiltonian $U(\bbox{p}_1, \bbox{p}_2, \bbox{q})$
in momentum space 
is related to the invariant scattering amplitude $M$
and to the photon propagator $D_{\mu\nu}(q)$ in the following
way [see also equation (83,8) in~\cite{BeLiPi1991}]:
\begin{eqnarray}
M &=& e_1 \, e_2 \, j_1^\mu(\bbox{p}'_1, \bbox{p}_1) \, 
D_{\mu\nu}(\bbox{q}) \, 
j_2^\nu(\bbox{p}'_2, \bbox{p}_2) \nonumber\\[2ex]
&=& - \phi_1^{*}(\bbox{p}'_1) \, \phi_2^{*}(\bbox{p}'_2) \,
\left[ \frac{e_1\,e_2}{\bbox{q}^2} + U(\bbox{p}_1, \bbox{p}_2, \bbox{q})
\right]\,
\phi_1(\bbox{p}_1) \, \phi_2(\bbox{p}_2)
\end{eqnarray}
where $\bbox{q} = \bbox{p}'_2 - \bbox{p}_2 = - (\bbox{p}_1 - \bbox{p}'_1)$.
We employ a Coulomb-gauge photon propagator,
\begin{equation}
D_{00}(\bbox{q}) = - \frac{1}{\bbox{q}^2} \,, \quad
D_{ij}(\bbox{q}) = - \frac{1}{\bbox{q}^2 - \bbox{\omega}^2}\,
\left[ \delta^{ij} - \frac{q^i \, q^j}{\bbox{q}^2} \right]
\approx - \frac{1}{\bbox{q}^2}\,
\left[ \delta^{ij} - \frac{q^i \, q^j}{\bbox{q}^2} \right]\,,
\end{equation}
where we can neglect the energy of the virtual photon in
$D_{ij}$ for the 
derivation of next-to-leading order relativistic corrections.
The invariant scattering amplitude $M$ then reads
\begin{eqnarray}
\frac{M}{e_1\,e_2} &=& 
- \phi^{*}_1(\bbox{p}'_1) \, \phi^{*}_2(\bbox{p}'_2) \, \frac{1}{\bbox{q}^2}\,
\phi_1(\bbox{p}_1) \, \phi_2(\bbox{p}_2) \nonumber\\[1ex]
& & \;\;\; + \phi^{*}_1(\bbox{p}'_1) \, \phi^{*}_2(\bbox{p}'_2) \,
\left[ \frac{p^i_1 + p'^i_1}{2\, m_1} \,
\frac{p^j_2 + p'^j_2}{2\, m_2} \, \frac{1}{\bbox{q}^2} \,
\left[ \delta^{ij} - \frac{q^i \, q^j}{\bbox{q}^2} \right]\,
\right] \, \phi_1(\bbox{p}_1) \, \phi_2(\bbox{p}_2) \,.
\end{eqnarray}
We therefore identify
\begin{eqnarray}
U(\bbox{p}_1, \bbox{p}_2, \bbox{q}) &=& 
- \frac{e_1\,e_2}{4 m_1 m_2} \, \frac{ (2 p_1^i - q^i) \,
(2 p_2^j + q^j)}{\bbox{q}^2} \,
\left[ \delta^{ij} - \frac{q^i \, q^j}{\bbox{q}^2} \right]
\nonumber\\[2ex]
&=& - \frac{e_1\,e_2}{4 m_1 m_2} \, \left\{ 
\frac{(2 \bbox{p}_1 - \bbox{q}) \cdot (2 \bbox{p}_2 + \bbox{q})}{\bbox{q}^2} 
\right. \nonumber\\[1ex]
& & \;\;\; \left. - \frac{(2 \bbox{p}_1 \cdot \bbox{q} - \bbox{q}^2) \, 
(2 \bbox{p}_2 \cdot \bbox{q} + \bbox{q}^2)}{\bbox{q}^4} \right\} \,.
\nonumber\\[2ex]
&=&
- \frac{e_1\,e_2}{m_1 \, m_2} \, 
\left[ \frac{\bbox{p}_1 \cdot \bbox{p}_2}{\bbox{q}^2} 
- \frac{(\bbox{p}_1 \cdot \bbox{q}) \,
(\bbox{p}_2 \cdot \bbox{q})}{\bbox{q}^4} \right]\,.
\end{eqnarray}
We now transform to the center-of-mass frame in which 
$\bbox{p}_1 = -\bbox{p}_2 = \bbox{p}$, so that the expression
for $U(\bbox{p}_1, \bbox{p}_2, \bbox{q})$ becomes even simpler,
\begin{eqnarray}
\label{penultimate}
U(\bbox{p}, -\bbox{p}, \bbox{q}) =
\frac{e_1\,e_2}{m_1 \, m_2} \, 
\left[ \frac{\bbox{p}^2}{\bbox{q}^2} 
- \frac{(\bbox{p} \cdot \bbox{q})^2}{\bbox{q}^4} \right]\,.
\end{eqnarray}
The formula (83,13) of~\cite{BeLiPi1991}
can now be employed in evaluating the Fourier transform,
\begin{equation}
\int \frac{{\mathrm d}^3 q}{(2 \pi)^3} \,
\exp\left({\mathrm i} \, \bbox{q} \cdot \bbox{r} \right) \,
\frac{4 \pi (\bbox{a} \cdot \bbox{q}) \, (\bbox{b} \cdot \bbox{q})}{\bbox{q}^4}
= \frac{1}{2 r}\,\left[ \bbox{a} \cdot \bbox{b} -
\frac{(\bbox{a} \cdot \bbox{r}) (\bbox{b} \cdot \bbox{r})}{r^2} \right]\,.
\end{equation}
The Breit hamiltonian, which we would like to 
denote by $H_{\mathrm B}$, is obtained by adding to the
Fourier transform of (\ref{penultimate})
the relativistic correction to the kinetic energy.
Denoting with $\bbox{\hat{p}} = - {\rm i} \,
{\partial}/{\partial \bbox{x}}$ the momentum operator in the
coordinate-space representation, we obtain 
\begin{eqnarray}
\label{hbreit}
H_{\mathrm B}(\bbox{r}, \bbox{\hat{p}}) =
- \frac{\bbox{\hat{p}}^4}{8\,m_1^3} - \frac{\bbox{\hat{p}}^4}{8\,m_2^3} 
+ \frac{e_1 \, e_2}{8 \pi r} \, \frac{\bbox{\hat{p}}^2}{m_1 \, m_2} 
+ \frac{e_1 \, e_2}{8 \pi r^3} \, \frac{\bbox{r} \cdot
(\bbox{r} \cdot \bbox{\hat{p}}) \,\, \bbox{\hat{p}}}{m_1 m_2}\,.
\end{eqnarray}
In the order of $(Z\alpha)^4$,
there is no contribution due to virtual annihilation 
for spinless particles; corrections of this type would enter only for 
positronium and dimuonium~\cite{KaIvJeSo1998} because they 
are caused by the spin-dependent part of the transition
current [see Eqs.~(83,20) and (82,22) of~\cite{BeLiPi1991}],
which is absent for spinless particles. For S states, virtual
annihilation is altogether prohibited by angular momentum conservation.

The matrix elements of the 
Breit hamiltonian (\ref{hbreit}) for spinless particles can be evaluated
on nonrelativistic bound states via computational techniques
outlined in Sec.~A3 of Ch.~1 of~\cite{BeSa1957}.
For $m_1 = m_2 = m$ and $e_1\,e_2 = - 4 \pi Z \alpha$,
we obtain 
\begin{equation}
\label{NLresult}
E_{\mathrm{B}} = - \frac{(Z\alpha)^2 \, m}{4 n^2} 
- \frac{(Z\alpha)^4 \, m}{2 \, n^3} \,
\left[ \frac{1}{2 l + 1} - \frac{1}{4} \, \delta_{l0} - 
\frac{11}{32\,n} \right]
\end{equation}
as the Breit energy for 
the energy levels of the bound system of two 
spinless particles, including relativistic 
corrections of order $(Z \alpha)^4$.
Here, we keep $Z$ as a parameter which denotes the nuclear charge number
in a bound system. Of course,
for two particles each carrying an elementary charge,
$Z$ has to be set to unity. The fine-structure constant is 
denoted by $\alpha$. The result (\ref{NLresult}) agrees
with previous calculations~\cite{BaGl1955,BrItZJ1969,Na1972,Ow1994,%
HaOw1994,PaKa1995}, notably with Eq.~(38) of~\cite{Na1972}. 

It is instructive to compare the result (\ref{NLresult}) with the 
known result for a single-particle system of mass $m/2$ satisfying
the Klein--Gordon equation, bound to a nucleus with 
charge $Z e$. According to Eq.~(2-86) of~\cite{ItZu1980}, 
we obtain the ``Klein--Gordon energy'' (KG) 
\begin{equation}
\label{KGresult}
E_{\mathrm{KG}} = - \frac{(Z\alpha)^2 \, m}{4 n^2}
- \frac{(Z\alpha)^4 \, m}{2 \, n^3} \,
\left[ \frac{1}{2 l + 1} - \frac{3}{8\,n} \right]\,.
\end{equation}
The two results (\ref{NLresult}) and
(\ref{KGresult}) are manifestly different in the order of
$(Z\alpha)^4$.

From (\ref{hbreit}) we conclude that the zitterbewegung term is absent
for spinless particles. However, this statement is in need of further
explanation because a considerable variety of physical
interpretations exists in the literature
with regard to the zitterbewegung term. 
We briefly expand:
The Dirac $\alpha$-matrices fulfill $\alpha = {\mathrm i} [H_{\mathrm D},x]$ 
($H_{\rm D}$ is the Dirac hamiltonian)
as the relativistic generalization of the velocity operator. 
By contrast, in the 
nonrelativistic formalism, we have the analogous 
relation $p/m  = {\mathrm i} [H_{\mathrm S} ,x]$ where $H_{\mathrm S}$
is the Schr\"{o}dinger hamiltonian.
Since the $\bbox{\alpha}$--matrices have eigenvalues $\pm 1$, the 
magnitude of the velocity of the electron -- at face value --
is equal to the velocity
of light at any given instant. On p.~106 of~\cite{He1950}, it is argued that
``the explanation for this fact is that the electron carries out a fast 
irregular motion (``zitterbewegung'') -- which is responsible for the 
spin -- whereas the mean velocity is given by the momentum
$\bbox{p}/m$''.  Note that the introduction of the Dirac matrix 
formalism is necessitated by the need to describe the internal degrees
of freedom of the particle -- the spin.
On p.~71 of~\cite{ItZu1980}, it is shown that
the zitterbewegung term can be traced back to the positional
fluctuations $\langle \delta \bbox{r}^2 \rangle \sim 1/m^2$ of the electron,
and a connection is drawn to the Darwin term which results naturally 
in the context of the Foldy--Wouthuysen transformed Dirac hamiltonian.
On pp.~117--118 of~\cite{Sa1967Adv} and 
p.~62 of~\cite{ItZu1980}, it is argued that the momentum 
$\bbox{p}$ of a Dirac wave packet can be associated in a natural
way with the group velocity, but that in addition to the group
velocity term, there exist highly oscillatory terms which represent
the zitterbewegung.  
Similarly, on pp.~139--140 of~\cite{Sa1967Adv}, it is shown that
the zitterbewegung term can also be interpreted as arising from
the interaction of the atomic electron with virtual electron-positron
pairs created in the Coulomb 
field of the nucleus. This virtual electron-positron
pair-creation is subject to the uncertainty principle and can
occur only for time intervals of the order of
$\Delta t \sim \hbar/(2 m c^2)$ (where we temporarily restore
the factor $\hbar$). At the time the original atomic electron
fills up the vacated negative-energy state (the bound-electron wave-function
has negative-energy components), the escalated electron (which forms
part of the virtual pair) is at most 
a distance $c \Delta t \sim 1/m$ away from the original electron.
This distance is precisely of the order of magnitude of the 
fluctuations of the electron coordinate and consistent
with the discussion on p.~71 of~\cite{ItZu1980}.
All these interpretations elucidate different
aspects of the same problem.

In the context of the Breit hamiltonian,
we would like to adhere to the definition
that the zitterbewegung term is the term of order $(Z\alpha)^4$ in 
the Breit hamiltonian generated by a contribution which is manifestly
proportional of $\delta(\bbox{r})$ in coordinate space 
(or a constant in momentum space). Such a term is absent in 
the result (\ref{hbreit}).  
For spin-$1/2$ particles, such a term is generated by the multiplication
of the photon propagator (proportional to $1/\bbox{q}^2$) with 
the zero-component of the transition current  which is given
for a spin-$1/2$ particle as [see equation (4) of~\cite{PaKa1995}]
\begin{equation}
\bar{u}' \gamma^0 \bar{u} =
w^{*} \, \left( 1 - \frac{\bbox{q}^2}{8\,m^2} +
\frac{{\mathrm i}\, \bbox{\sigma} \cdot \bbox{p}' \times \bbox{p}}
  {4\,m^2} \right) \, w\,.
\end{equation}
Here, $u$ is the bispinor amplitude for the bound particle,
and $w$ is the bound-state Schr\"{o}dinger wave function related by
\begin{equation}
u = \left( \begin{array}{c} 
\left(1 - 
\frac{\displaystyle \bbox{p}^2}{\displaystyle 8\,m^2}\right) w \\[3ex]
\frac{\displaystyle \bbox{\sigma}\cdot\bbox{p}}{\displaystyle 2 m} w 
\end{array} \right)
\end{equation}
according to equation (3) of~\cite{PaKa1995}.
One might wonder why a term proportional to $\delta_{l0}$,
apparently generated by a $\delta$-function in coordinate space,      
prevails in the Breit energy (\ref{NLresult}).
This term arises naturally when evaluating a matrix element
of the structure 
$\langle \phi_{\mathrm S} |
(\bbox{r} \cdot (\bbox{r} \cdot \bbox{\hat{p}}) \,
\bbox{\hat{p}})/r^3 | \phi_{\mathrm S} \rangle$
(last term of equation~(\ref{hbreit}))
on the nonrelativistic wave function $\phi_{\mathrm S}$ and
should {\em not} be associated with the zitterbewegung.

Vacuum polarization corrections and the self energy, as well
as corrections due to the strong interaction, are not included in 
(\ref{NLresult}). These corrections will be discussed in the two 
following sections.

%
% Vacuum Polarization
%
\section{Vacuum Polarization Effects}
\label{VacuumEffects}

As pointed out by various
authors (e.g.~\cite{Pu1957,%,
BoRi1982,EiSo2000plb,EiSo2000prd,LaBu1998,JeSoIvKa1998,KaJeIvSo1998EPJ}), 
the electronic vacuum polarization enters already at the order of $\alpha^3$ 
[more precisely, $\alpha\,(Z\alpha)^2$]
in bound systems with spinless particles, because the spinless 
particles are much heavier than the electron, which means that 
the Bohr radius of the bound system is roughly
of the same order of magnitude
as the Compton wavelength of the electron.
The Compton wavelength of the electron, however, is the fundamental 
length scale at which the charge of any bound particle is screened
by the electronic vacuum polarization. 

The vacuum polarization (VP) correction to energy levels 
has been evaluated~\cite{Pu1957,EiSo2000plb,EiSo2000prd,%,
JeSoIvKa1998,KaIvJeSo1998,LaBu1998}
with nonrelativistic wave functions. We recall
that the leading-order VP correction (due to the Uehling potential)
can be expressed as
\begin{equation}
\label{defVP}
\Delta E = \langle \psi | V_U | \psi \rangle =
\frac{\alpha}{\pi}\,C_E\,E_\psi\,,
\end{equation}
where
\begin{equation}
\label{Schroedinger}
E_\psi = -\frac{(Z\,\alpha)^2\,m}{4\,n^2}
\end{equation}
is the Schr\"odinger binding energy for a two-body system with two 
particles each of mass $m$ [first term on the right-hand side
of (\ref{NLresult})]. For the $C_E$ coefficients, we recall
the following known results~\cite{EiSo2000plb,EiSo2000prd,KaIvJeSo1998},
\begin{equation}
\label{resVP}
C_E({\mathrm{1S}}) = 0.22 \,, \quad
C_E({\mathrm{2S}}) = 0.10\,.
\end{equation}
Two-loop vacuum polarization effects enter at a
relative order $\alpha^2$ in pionium and are therefore
of the same order of magnitude as the relativistic
corrections mediated by the Breit interaction (discussed in
Sec.~\ref{BreitEffects}). The self energy correction which is 
discussed in the following section is even smaller, but of considerable
theoretical interest. 

%
% Self Energy Effects
%
\section{Effects due to Scalar QED}
\label{SelfEnergyEffects}

As shown in~\cite{Pa1998}, the leading logarithmic correction
to the self energy can be obtained, in nonrelativistic approximation,
from second-order perturbation theory based on
nonrelativistic quantum electrodynamics~\cite{CaLe1986}
(see also~\cite{BB1984}). We will investigate here, in a
systematic way, the leading logarithms generated for 
S states by self energy and relativistic-recoil effects 
(the so-called Salpeter correction), and show that these
are spin-independent.

The quantized electromagnetic field 
is [see Eq.~(5) of~\cite{Pa1998}],
\begin{equation}
\bbox{A}(\bbox{r}) = \sum_{\lambda=1,2} 
\int \frac{{\mathrm d}^3 k}{\sqrt{(2\pi)^3 \, 2 k}}\,
\bbox{\epsilon}_{\lambda}(\bbox{k}) \,
\left[ a^{+}_{\bbox{k},\lambda} \exp(-{\mathrm i} \bbox{k}\cdot\bbox{r}) +
a_{\bbox{k},\lambda} \exp({\mathrm i} \bbox{k}\cdot\bbox{r})\right]\,,
\end{equation}
and the nonrelativistic interaction hamiltonian for an atomic
system with two spinless particles (charges $e_1$ and $e_2$ and
masses $m_1$ and $m_2$) reads
\begin{equation}
\label{HI}
H_{\mathrm I} = - \frac{e_1}{m_1} \, \bbox{p_1}\cdot \bbox{A}(\bbox{r}_1) + 
\frac{e_1^2}{2 m_1} \bbox{A}(\bbox{r}_1)^2 
- \frac{e_2}{m_2} \, \bbox{p_2}\cdot \bbox{A}(\bbox{r}_2) +
\frac{e_2^2}{2 m_2} \bbox{A}(\bbox{r}_2)^2 \,.
\end{equation}
For two spin-$1/2$ particles, the terms
\begin{equation}
\label{HIS}
- \frac{e_1}{m_1} \, \bbox{\sigma}_1 \cdot \bbox{B}(\bbox{r}_1)
- \frac{e_2}{m_2} \, \bbox{\sigma}_2 \cdot \bbox{B}(\bbox{r}_2)
\end{equation}
have to be added to $H_I$ [see Eq.~(7) of~\cite{Pa1998}]. 
We will carry out the calculations for the general case of one
particle of charge $e_1 = e$ and the other having a charge $e_2 = - Z e$
(we follow the convention of~\cite{ItZu1980}
that for hydrogen, $e$ is the physical charge of the electron,
i.e.~$e = - |e|$).
The unperturbed hamiltonian of the system of the
two particles and the electromagnetic
field reads [see e.g.~Eq.~(6) of~\cite{Pa1998}],
\begin{equation}
\label{H0}
H_0 = \frac{\bbox{p}_1^2}{2 m_1} +
\frac{\bbox{p}_2^2}{2 m_2} - \frac{Z\alpha}{r} +
\sum_{\lambda=1,2} \int {\mathrm d}^3 k \, k \,
a^{+}_{k,\lambda} \, a_{k,\lambda} \,.
\end{equation}
where $\bbox{r} = \bbox{r}_1 - \bbox{r}_2$.
The eigenstates of the ``atomic part'' $H_0^A$ of this hamiltonian 
in the center-of-mass system $\bbox{p}_1 + \bbox{p}_2 = 0$ are the 
nonrelativistic Schr\"{o}dinger--Coulomb wave functions for a reduced
mass $m_{\rm r} = m_1 \, m_2 / (m_1 + m_2)$
[here,  the ``atomic part'' $H_0^A$ excludes the photon field,
i.e.~the last term of~(\ref{H0})]. We denote
$\bbox{p} \equiv \bbox{p}_1 = -\bbox{p}_2$.

Given that the first-order perturbation
$\langle \phi_{\mathrm S} | H_{\mathrm I} | \phi_{\mathrm S}\rangle$
vanishes, the second-order perturbation yields the dominant
nonvanishing perturbation. When evaluated on an atomic state, it 
is given by
\begin{equation}
\label{2ndorder}
\delta E_{\mathrm{SE}} = \langle \phi_{\mathrm S} | H_{\mathrm I}
\frac{1}{H_0 - E_{\mathrm S}} H_{\mathrm I} | \phi_{\mathrm S} \rangle\,.
\end{equation}
The interaction hamiltonian (\ref{HI}) gives rise to 
QED corrections that involve both
particles (in the current context, these are
recoil corrections involving the product $e_1 \, e_2$), 
and also to terms which involve only a single particle and
are proportional to $e_1^2$ or $e_2^2$. The latter effects 
correspond to the self-energies of the two particles.

The low-energy part of the self energy in leading order~\cite{Pa1993} 
can be inferred directly from
(\ref{2ndorder}), and it can be seen
that the spin-dependent parts from (\ref{HIS}) vanish
in leading order in the $(Z\alpha)$-expansion~\cite{Pa1998}:
\begin{equation}
\label{NRELe1e2}
E_{\mathrm L} = -\frac{e_1^2}{6 \, \pi^2} \,
\int_0^\epsilon {\mathrm d}k \, k \,
  \left< \phi \left| \frac{\bbox{p}}{m_1} \,\,\, 
    \frac{1}{H_0^A - (E_{\mathrm S} - k)} \,\,\,
     \frac{\bbox{p}}{m_1} \right| \phi \right> +
(e_1 \leftrightarrow e_2, m_1 \leftrightarrow m_2)\,.
\end{equation}
where 
\begin{equation}
E_{\mathrm S} = -\frac{(Z\alpha)^2 m_{\mathrm r}}{2 n^2}
\end{equation}
is the Schr\"{o}dinger energy ($m_{\mathrm r}$ in the reduced mass
of the atomic system under investigation). 
Starting from the {\em spin-independent} expression (\ref{NRELe1e2}),
it is now relatively straightforward to show
that the leading ``self energy logarithm'' for S states is given by
\begin{equation}
\label{SEestimate}
\delta E_{\mathrm{SE}} \approx 
\frac{4 \, \ln(Z\alpha)^{-2}}{3\,\pi\,n^3} \, 
\delta_{l0}\, \left[\alpha \, (Z \alpha)^4 \, \frac{m^3_{\mathrm r}}{m_1^2} 
+ Z \, (Z \alpha)^5 \, \frac{m^3_{\mathrm r}}{m_2^2}\right] \,.
\end{equation}
This result is by consequence spin-independent.
The derivation is simplified when using the $\epsilon$-method developed
and used in various bound-state calculations~\cite{Pa1993,JePa1996,JeSoMo1997}.
The two terms in square brackets in (\ref{SEestimate}) 
correspond to the two self-energies of the two constituent particles
with charges $e_1 = e$ and $e_2 = -Z e$ and masses $m_1$ and $m_2$,
respectively. It has been pointed out~\cite{SaYe1990}
that in contrast to the self energy corrections,
the vacuum polarization corrections given in Eq.~(\ref{resVP}) 
must not be double-counted. The ``double-counting'' of 
self energy corrections (and lack of it in the vacuum-polarization
case) finds a natural explanation in our formalism: whereas the 
vacuum-polarization correction mainly leads to a modification of the
$1/r$-type Coulomb attraction in (\ref{H0}) within a nonrelativistic
effective theory, the structure of the interaction
hamiltonian (\ref{HI}) implies the existence of the 
{\em two} self-energies of the {\em two} constituent particles
of the atomic system.

It might be instructive to point out that
the formula (\ref{SEestimate}) is consistent 
with Welton's argument for estimating the self energy
of a bound particle
which is based on analyzing the influence of the fluctuating
electromagnetic field [a detailed discussion is given on pp.~80--82 
of~\cite{ItZu1980}]. 
For a system with two particles of equal mass
$m_1 = m_2 = m$, we have 
$m_{\mathrm r} = m/2$.

The leading-order recoil correction (Salpeter correction)
can also be inferred from the
interaction hamiltonian (\ref{HI}) via second-order perturbation
theory, by ``picking up'' terms that involve
products $e_1 \, e_2$.
It has been shown in~\cite{Pa1998} that the leading {\em logarithm}
(for S states) of the Salpeter correction 
is spin-independent (just like the leading logarithm of the 
self energy correction). The Salpeter correction
is usually referred to as a relativistic recoil (RR) correction.
By following~\cite{Pa1998}, we obtain for the leading logarithm of this
effect
\begin{equation}
\label{RRestimate}
\delta E_{\mathrm{RR}} \approx 
\frac{2\,(Z \alpha)^5}{3\,\pi\,n^3} \, \delta_{l0} \,
\ln\left(\frac{1}{Z\alpha}\right) \, \frac{m^3_{\mathrm r}}{m_1 m_2} \,.
\end{equation}
This correction involves only 
the products $e_1 \, e_2 = - 4 \pi Z\alpha$ and can therefore
be written as a function of $Z\alpha$ alone.
 
For pionium, we have $Z=1$, $m_1=m_2=m=m_\pi$, $m_{\mathrm r} = m_\pi/2$.
The leading logarithmic correction from scalar QED for 
pionium in the order of $\alpha^5 \ln\alpha$ is obtained
by adding the corrections (\ref{SEestimate}) and (\ref{RRestimate}),
\begin{equation}
\label{reslog}
\delta E_{\mathrm{log}} = \delta E_{\mathrm{SE}} + 
  \delta E_{\mathrm{RR}} 
  = \frac{3}{4} \, 
    \frac{\alpha^5}{\pi\,n^3} \, \ln\left(\frac{1}{\alpha}\right) \, m_\pi\,.
\end{equation}
The non-logarithmic term of order $\alpha^5$ is spin-dependent,
and its evaluation requires a relativistic treatment of 
the self energy
of a bound spinless particle; such a calculation
would be of considerable theoretical
interest, but the size of the effect for pionium, which is roughly 
two orders of $\alpha$ smaller than the leading vacuum polarization 
correction, precludes experimental verification in the near future.
However, we would like to point out here that a fully relativistic
treatment of this problem, including a detailed discussion of the 
renormalization of the self energy of the spinless particle, has 
not yet been accomplished. Scalar QED is a renormalizable 
theory~\cite{ItZu1980}.

The dominance of vacuum polarization over the self energy
in pionium is expressed, in particular, by the fact that even
{\em two-loop} vacuum polarization of order $\alpha^4$ has
a stronger effect on the spectrum of pionium than the leading
logarithm from Eq.~(\ref{reslog}) according to Ref.~\cite{JeSoIn2002}, 
and that the strong-interaction
correction of order $\alpha^3$~\cite{GaLyRuGa2001} has to be well
understood before any experimental verification of~(\ref{reslog})
appears feasible.
Finally, we remark that for a manifestly non-elementary particle
like the pion which has a finite charge radius, form-factor 
corrections have to be taken into account.

%
% QED Bound--State Calculations: A Summary
%
\chapter{QED Calculations: A Summary}
\label{QEDConclu}

Calculations in the area of bound-state quantum electrodynamics
have a long history, starting with Hans Bethe's first evaluation
of the hydrogen Lamb shift~\cite{Be1947}. Since then, the field
has developed considerably, and the immense progress in high-resolution
spectroscopy (for a recent culmination of the
activities, see~\cite{NiEtAl2000}) has necessitated the inclusion
of higher- and higher-order corrections into the theoretical
framework, leading to increased complexity of the theoretical
calculations. Why should such an effort be carried out?
A more accurate understanding of the QED corrections is of crucial
importance for the determination of the fundamental constants
and the test of quantum electrodynamics. The theory of the 
QED bound states can, at little risk to over-statement, be described
as one the most developed theories in all of physics.
It combines the intricacies of quantum field theory (divergences,
renormalization) with the experimental possibilities implied by the
accuracy of high-resolution spectroscopy.
 
\begin{figure}[htb]
\begin{center}
\begin{minipage}{15cm}
\centerline{\mbox{\epsfysize=10.0cm\epsffile{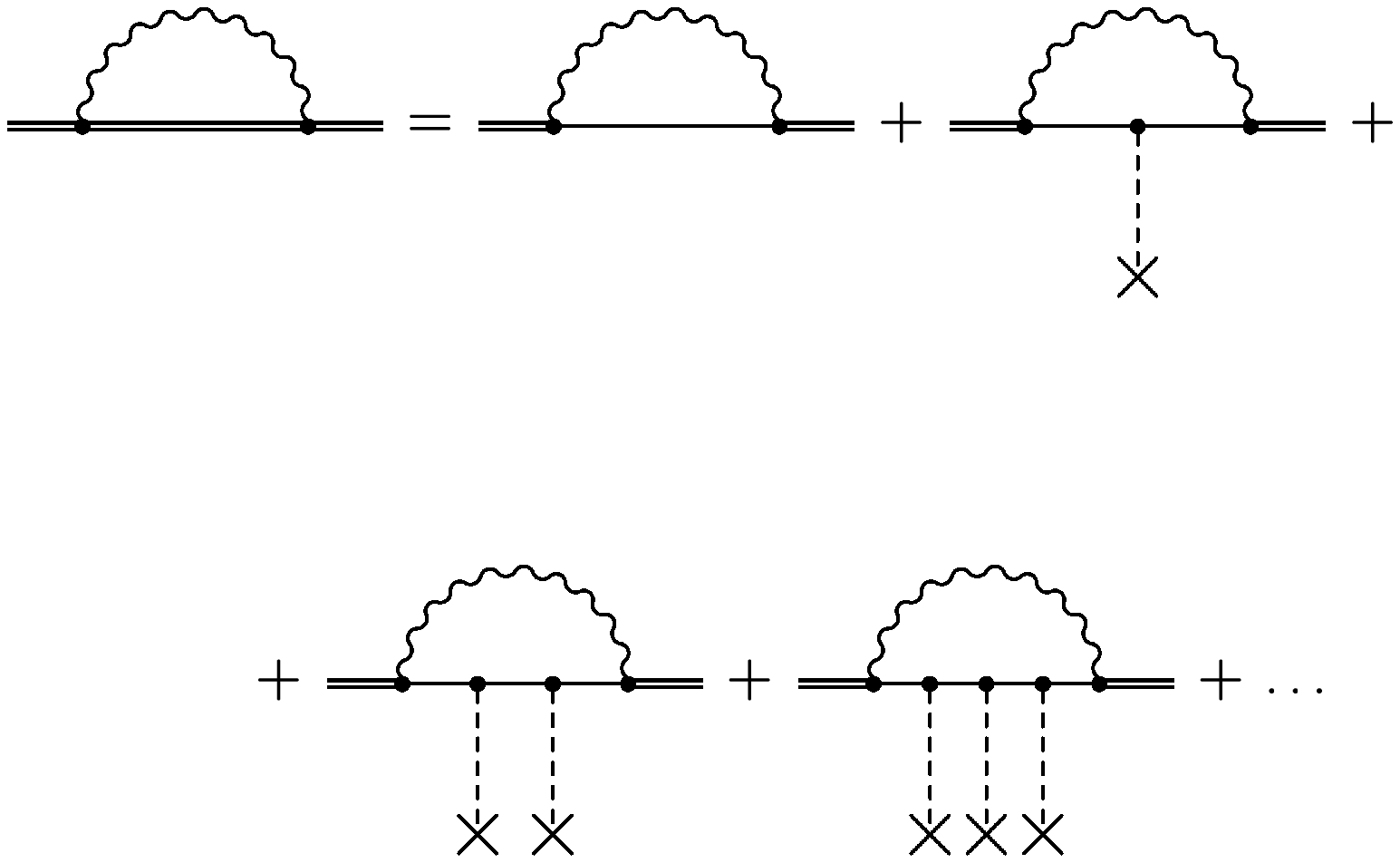}}}
\caption{\label{Vexp} Expansion of the bound electron self energy in
powers of the binding Coulomb field. The dashed lines denote Coulomb
photons, the crosses denote the interaction with the (external)
binding field. Each interaction with the binding Coulomb field
gives rise to a higher-order term in the $Z\alpha$-expansion.}
\end{minipage}
\end{center}
\end{figure}

\begin{figure}[htb]
\begin{center}
\begin{minipage}{14cm}
\centerline{\mbox{\epsfysize=10.0cm\epsffile{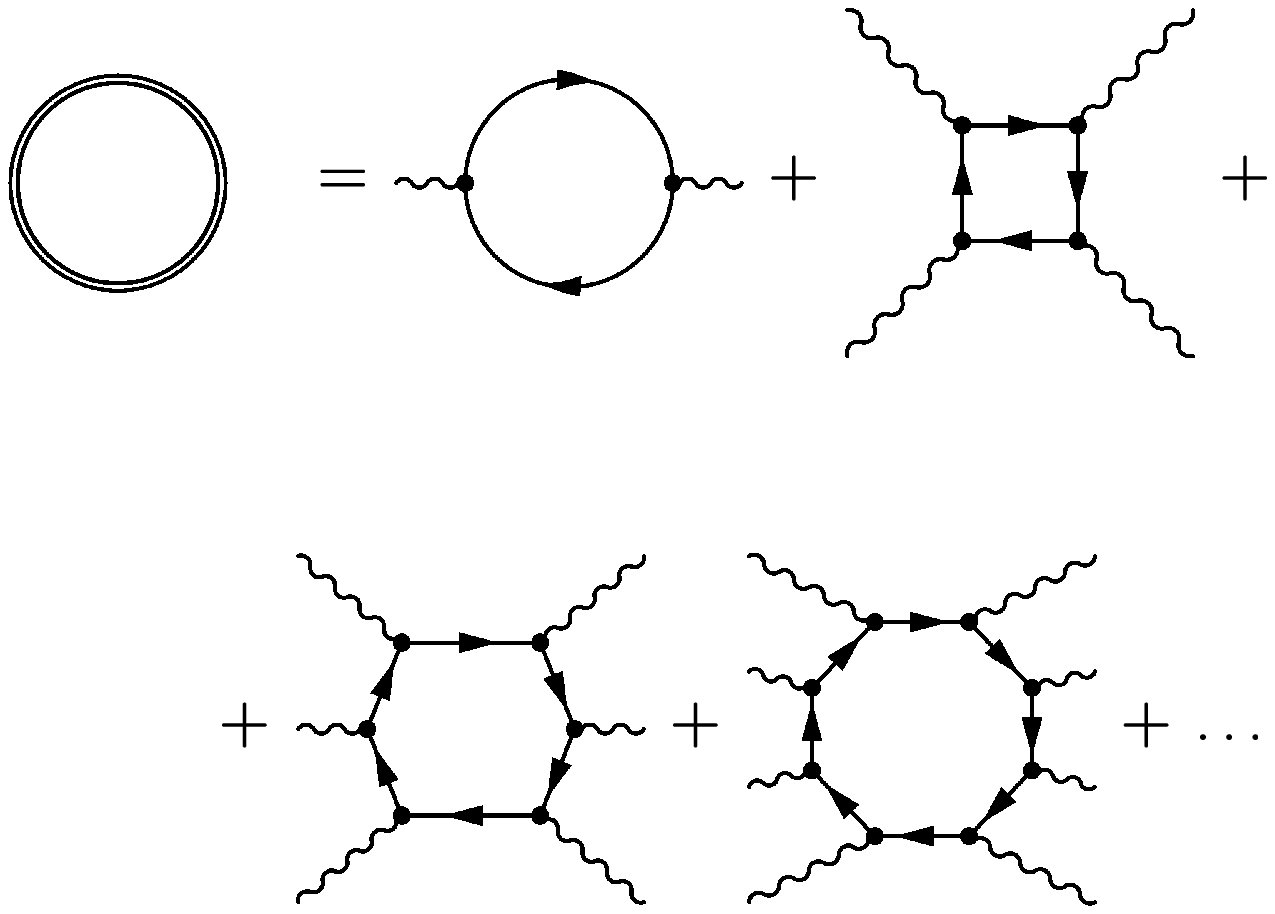}}}
\caption{\label{Eexp} Expansion of the QED effective
action in powers of the external field. As discussed in Sec.~\ref{DivQEDEff},
the resulting asymptotic series in powers of the coupling is divergent.}
\end{minipage}
\end{center}
\end{figure}

For decades, there has been a tremendous amount of work on analytic
self-energy calculations, notably on the coefficient 
$A_{60}$ for S states.
However, an accurate comparison of the analytic results 
to numerical data was impossible, simply because no numerical 
data were available in the region of low nuclear charge. The
semi-analytic expansion in $Z\alpha$ is in part 
problematic because there is evidence
that it may represent a divergent series. This argument
may be based in part on the analogy between Figs.~\ref{Vexp}
and~\ref{Eexp}, and on the observed rapid growth of the
analytic self energy coefficients ($A_{40}$ is of order one
for S states, $A_{50}$ is about $10$,
and the magnitude of $A_{60}$ is as large as $30$).
The QED effective action gives rise
to a divergent asymptotic expansion in appropriately defined
effective coupling parameters.
It is therefore {\em a priori} difficult to associate 
a finite remainder to a truncated series that involves all known
(but disregards the unknown) analytic coefficients. 

%
% Figure 4: V expansion
%

%
% Figure 4: V expansion
%

The results
reported here in Ch.~\ref{QEDSelOne} answer this conceptual question 
in the following way: the semi-analytic expansion in powers of $Z\alpha$
is in full agreement with the obtained numerical results.
We may therefore conclude that the semi-analytic expansion
probably gives rise to a generalized asymptotic series whose remainder is 
of the order of the first neglected semi-analytic term as
$Z\alpha \to 0$. Therefore, we may assume that a truncated
semi-analytic expansion in powers of $Z\alpha$ leads to 
reliable theoretical predictions in the realm 
of low nuclear charge numbers $Z$. 
However, although the numerical results confirm
the validity of the semi-analytic expansion in
this $Z$-region, they also show the limitations
of this approach: much more accurate results are obtainable
by numerical calculations. The accuracy on the level of 1~Hz 
for atomic hydrogen reported in Ch.~\ref{QEDSelOne}
would have involved analytic terms of an excessively high order in 
$Z\alpha$ [approximately $\alpha\,(Z\alpha)^9$)], and
the multitude and complexity of terms would have been a
severe obstacle for analytic calculations in this order of
$Z\alpha$. In general, the related activities have recently 
sparked a rather broad interest in the
numerical calculation of relativistic, QED self energy and
two-body corrections at low $Z$ and the comparison
of analytic and numerical results~\cite{ArShYe1995pra,ArShYe1995,%,
ShEtAl1998jpb,ShEtAl1998,PaGr1995,%,
Ye1998,BeSo1988,ScGrSo1993,Ka1993log,%,
MaSt2000,GoLaNePlSo1999,MaSa1998b,Ye2000}.

We have described in Ch.~\ref{QEDSelOne}
a nonperturbative evaluation of the one-photon
self energy in hydrogenlike ions with low nuclear charge
numbers $Z=1$ to $5$. The general outline of our approach is discussed
in Sec.~\ref{MethodOfEvaluation}. In Sec.~\ref{LowEnergyPart}, the
numerical evaluation of the low-energy part (generated by virtual
photons of low energy) is described. In Sec.~\ref{HighEnergyPart}, we
discuss the numerical evaluation of the high-energy part, which is
generated by high-energy virtual photons and contains the formally
infinite contributions, which are removed by the
renormalization. Sec.~\ref{HighEnergyPart} also contains a brief
discussion of the convergence acceleration methods as employed in the
current evaluation. We discuss in Sec.~\ref{ComparisonAnalytic} the
comparison of analytic and numerical data for K- and L-shell states in
the region of low $Z$. The main results of the current 
investigation are contained in
Table~\ref{tableFKL}: numerical data, nonperturbative in $Z\alpha$,
for the scaled function $F$ and the self energy remainder
function $G_{\rm SE}$ for K- and L-shell states at low nuclear charge.
The numerical accuracy of our data is 1~Hz or better in frequency
units for 1S, 2S and both 2P states in atomic hydrogen.

Still, calculations within
the semi-analytic expansion in addition to the 
numerical calculations (a kind of ``dual strategy'')
remain important. 
They serve as a confirmation and 
check for currently available and future numerical approaches.
In general, in view of the complexity of the calculations,
an independent check is always {\em highly} desirable.
For the two-loop effect, the $(Z\alpha)$-expansion
converges more rapidly than for the
one-loop effect in absolute frequency units because of the
additional radiative factor $\alpha/\pi$ which decreases the overall size
of the effect (see Ch.~\ref{QEDSelTwo}).

It is hoped that the analytic calculations reported
in Ch.~\ref{QEDSelTwo} for low nuclear
charge number $Z$ will be supplemented 
in the future by an accurate numerical treatment of the two-loop
self energy problem (see also related recent work in the 
high-$Z$ region, Refs.~\cite{Ye2000,YeSh2001,GoEtAl2001}). This presupposes 
that the considerable numerical 
problems in the domain of small nuclear charge
could be solved by adequate numerical methods, and that the further
problem of the increased computational demand of the two-loop effect
in comparison to the one-loop problem~\cite{JeMoSo1999,JeMoSo2001pra}
can be tackled -- possibly by massively parallel computer architectures.
Note, however, that the most accurate theoretical predictions could
only be reached in combining numerical and analytic results.
The reason is the following:
All numerical calculations are performed in the non-recoil limit
which is the limit of infinite nuclear mass. This is not quite
sufficient for an accurate theoretical treatment because
the self energy of a bound-state depends genuinely on the ratio of 
the orbiting particle to the nuclear mass -- an effect beyond the 
recoil correction. For example, the argument of the logarithms
in Eqs.~(\ref{defFLO}) and~(\ref{DefESE}) should be replaced according to
$\ln[(Z\alpha)^{-2}] \to \ln[\sigma\,(Z\alpha)^{-2}]$, 
where $\sigma = m/m_{\mathrm r}$ and $m_{\mathrm r}$ is the reduced
mass~\cite{SaYe1990}. The possibility to include these 
tiny, but important effects 
depends crucially on a reliable knowledge of the analytic 
coefficients {\em in combination} 
with an accurate numerical treatment of the problem
(see also the discussion in Sec.~\ref{Observations}).

The analytic results can be used
to obtain improved theoretical predictions 
for the hydrogenic fine structure
as compared to the previous 
order--$\alpha^7$--calculations~\cite{JePa1996,JeSoMo1997},
because they remove the principal theoretical uncertainty 
in the order of $\alpha^8$
due to the problematic two-loop self energy which is 
represented diagrammatically in Fig.~\ref{fig1}.
Our calculation in Ch.~\ref{QEDSelTwo}
illustrates the usefulness of the simplified
effective treatments of two-loop effects in the analytic approach
based on the modified Dirac hamiltonian (\ref{HDm}) and the 
``$\epsilon$ method''. This aspect highlights, as we believe, the 
need for systematic, simplified treatments of higher-order radiative
corrections in bound systems.  

In Ch.~\ref{QEDSelTwo}, we primarily address spin-dependent effects in
one-electron (hydrogenlike) systems. However, the same effects also
contribute to the fine-structure splitting in two-electron (heliumlike)
systems. There is currently remarkable interest in improved measurements
of the fine-structure splitting in helium and heliumlike atomic systems
with low nuclear charge~\cite{MyTa1999,MyEtAl1999,StGeHe2000,GeLoHe2001}. 
The effects addressed here 
contribute to the fine-structure splitting in helium on the level of
100~Hz, which is not much smaller than the current experimental accuracy
of about 1~kHz, and allows for an estimate of uncalculated 
higher-order contributions.

All analytic calculations rely on the separation of the virtual
photon energy into high- and low-energy energy regions
via a parameter $\epsilon$. This 
separation is illustrated in Sec.~\ref{EpsilonMethod}.
For two-photon problems, four energy regions result from such
a separation, but for the one-photon problem discussed in
Ch.~\ref{QEDSelExc}, one separation parameter $\epsilon$
is sufficient. We present in Ch.~\ref{QEDSelExc}  analytic
calculations for excited atomic states based on the 
semi-analytic $Z\alpha$-expansion. These may be complemented
in the near future by numerical results~\cite{LBInPriv2002}. 
We discuss some intricacies of the analytic treatment
of highly excited atomic states in Sec.~\ref{ORISelExc}.
Results for the individual contributions are presented 
in Subsections~\ref{resultsP12} --~\ref{resultsG92}
and summarized in Sec.~\ref{SummaryOfResults}.
Final values of the analytic coefficients can be found
in Tables~\ref{tableP12} --~\ref{tableG}.
The characteristic numerical cancellations between individual
contributions are described in Sec.~\ref{TypicalCancellations}.
Finally, in Sec.~\ref{Observations}, we describe a number
of observations regarding the dependence of the
$A_{61}$ and the $A_{60}$-coefficients on the quantum numbers of the 
atomic state.

All of the above calculations are related to the bound electron,
which is a spin-$1/2$ particle. In nature, particles with 
different spin quantum numbers exist, and these, too, can form
bound systems. The spin-dependence of the corrections is a 
conceptually interesting issue that forms the subject
of Ch.~\ref{QEDSpiZer}. This interest is sparked by the 
projected DIRAC experiment at CERN where the bound system 
of two oppositely charged pions (``pionium'') will be
studied. Notably, even the relativistic 
corrections for a bound two-body system consisting of 
spinless particles are different for spinless
particles than for particles with nonvanishing spin.
They are given by a generalized Breit hamiltonian that
we discuss in Sec.~\ref{BreitEffects}.
There, we present
a simplified derivation for the 
relativistic and recoil corrections of order $\alpha^4$
to a bound state of two spinless particles. The results
agree with previous calculations~\cite{Na1972}. As evident from 
equation~(\ref{hbreit}), the zitterbewegung term
is absent in a bound system of two spinless particles. 

The self energy is suppressed in systems with spinless
particles in comparison to the vacuum polarization effect as
discussed in Secs.~\ref{VacuumEffects}
and~\ref{SelfEnergyEffects}, because 
the lightest known spinless particle is much heavier than the 
electron, which implies that the vacuum polarization effect is larger 
by two orders of $Z \alpha$ than the self energy in 
bound systems of spinless particles. We provide a
complete result for the leading scalar-QED correction
of order $\alpha^5\,\ln\alpha$ in Sec.~\ref{SelfEnergyEffects}. 
A list of further QED corrections to the 1S level of pionium,
including two-loop vacuum polarization, finite-size corrections
and other effects, can be found in~\cite{JeSoIn2002}.

To conclude this Chapter, we would like to mention the 
complementarity of the high-precision calculations and 
experiments at low nuclear charge and the strong-field
calculations and related experiments at high nuclear charge
numbers. While an accuracy~\cite{NiEtAl2000} of one part in $10^{14}$ 
currently appears 
to be out of reach for hydrogenlike Uranium~\cite{StElAl2000},
the experiments on highly charged ions have meanwhile attained a
precision where a detailed understanding of the nuclear
structure and shape is necessary in order to describe
QED effects rigorously~\cite{MoSo1993}. 

Bound systems described by quantum electrodynamics belong to the
most accurately understood physical systems today. 
They convey a certain esthetic appeal because the apparent
simplicity of the two- and three-particle systems gives rise to a 
large number of conceptual as well as calculational difficulties when a
thorough understanding is sought in higher orders of perturbation
theory.

%
% Part 2: Large-Order Perturbation Theory
%
%
% Convergence Acceleration and Divergent Series
%
\part{Convergence Acceleration and Divergent Series}

\typeout{}
\typeout{Part 2:}
\typeout{=============================================}
\typeout{Convergence Acceleration and Divergent Series}
\typeout{=============================================}
\typeout{}

\label{part2}

%
% Overview of Convergence Acceleration and Divergent Series
%
\chapter{Introduction to Convergence Acceleration and
Divergent Series in Physics}

\typeout{========================================}
\typeout{Introduction to Convergence Acceleration}
\typeout{========================================}
\typeout{}

\label{ConDivInt}

The second part of this thesis will be concerned with 
convergence acceleration methods and with the summation of divergent
series. In theoretical physics, we often encounter sequences of numbers,
either as a result of perturbation theory or as partial sums of 
an infinite series defining a physical quantity, whose numerical
properties are disfavourable: in many cases, the sequences converge
so slowly that they are numerically useless. In other 
cases, the sequences do not converge at all, and yet we know -- or 
assume -- that the sequences are associated, nevertheless, to the physical
quantity in question, for example in the form of a divergent asymptotic
expansion.  

Let us consider the case where the sequence $\{s_n\}_{n=0}^{\infty}$
under investigation is given by partial sums of an infinite series,
\begin{equation}
\label{FundSeq}
s_n = \sum_{k=0}^n a_k\,.
\end{equation}
We assume, furthermore, that the $a_k$ are real (this condition
will be relaxed in the sequel). 

(i) Let us assume that the sequence
$\{s_n\}_{n=0}^{\infty}$ is {\em convergent}. The purpose of convergence
acceleration is to convert $\{s_n\}_{n=0}^{\infty}$ into a new
sequence $\{s'_n\}_{n=0}^{\infty}$ with hopefully better numerical
properties. If we denote the limit of the 
sequence by $s$, that is~$s \equiv \lim_{n \to \infty} s_n$, 
then the rate of convergence is obviously increased if
\begin{equation}
\label{ConvAccelDef1}
\lim_{n \to \infty} \frac{s'_n - s}{s_n - s} = 0\,.
\end{equation}   

(ii) Alternatively, let us assume that the sequence
$\{s_n\}_{n=0}^{\infty}$ is {\em divergent}. The purpose of 
the resummation is to determine the generalized limit
(or antilimit) of the sequence. The following question immediately
arises: how is it possible to associate a finite generalized
limit with a divergent series. L.~Euler wrote in a letter to 
Goldbach (1745): ``Summa cuiusque seriei est valor 
expressionis illius finitae, ex cuius
evolutione illa series oritur.'', which means: the generalized limit
of any divergent series is the value of the particular characteristic
finite expression, whose expansion gives rise to the divergent 
series. This is still the main notion of resummation today:
Given a divergent series,
to ``reconstruct'' {\em a posteriori} the function whose expansion
gave rise to the divergent series, with a suitable mathematical
resummation method.

We illustrate this idea with a trivial example:
A divergent series can be generated, e.g., by expanding the 
function $f(z) = 1/(1+z) \sim \sum_{j=0}^{\infty} (-z)^j$
at $z=1$ in powers of $z$. The partial sums of the resulting
divergent series read
\begin{equation}
s_n = \sum_{j=0}^n (-1)^j = \left\{ \begin{array}{cc} 
1 & \mbox{for $n$ even} \\[1ex] 
0 & \mbox{for $n$ odd} \end{array} \right.\,.
\end{equation}
The resulting sequence, which is $1, 0, 1, 0, \dots$,
can be identified as originating from the discussed function
by expansion at unit argument, and its antilimit is therefore $\lfrac{1}{2}$.
Since this argument may not sound convincing, let us briefly indicate
that the result may also be obtained by imposing the 
postulate that the antilimit of the sum of 
two divergent series should be the sum of the two antilimits each. 
Then, it is easy to derive (by inspection) that the antilimit $a$
of the series $\sum_{j=0}^\infty (-1)^j$ should fulfill
$1 - a = a$, and thus $a=\lfrac{1}{2}$ (again).
A more formal investigation of this series and its
resummation can be found in~\cite{Ha1949} [see Eq.~(1.1.1)
and the instructive discussion on the p.~6 {\em ibid.}]. 
It is probably a triviality to express that the most familiar
resummation methods like the Euler transformation,
Pad\'{e} approximants and recently developed 
nonlinear sequence transformations of the Levin--Weniger type
all convert the divergent sequence $1, 0, 1, 0, \dots$ to the 
convergent sequence $\lfrac{1}{2}, \lfrac{1}{2}, \lfrac{1}{2}, \lfrac{1}{2},
\dots$, in mutual agreement. Some of
these resummation methods are discussed in
the sequel, and we also refer the reader to
the seminal review~\cite{We1989}.
It has been stressed in J. E.~Littlewood's introduction to~\cite{Ha1949}
that the subject of divergent series is in no way ``mystical or unrigorous''.

Fundamental differences exist between nonalternating 
series (without loss of generality we can assume $a_k > 0$ $\forall k$ in 
that case), and strictly alternating series $a_k = (-1)^k \, b_k$
where $b_k > 0$ $\forall k$ (mixed forms will also be discussed in
the sequel). The acceleration of the 
convergence and the summation of divergent {\em nonalternating}
series has been historically problematic.

For example, it has been known for a long time that the acceleration
of nonalternating
slowly convergent series can be highly unstable numerically
(see~\cite{We1989} and references therein).
Likewise, the summation of nonalternating
factorially divergent series has represented a considerable
problem, because the series are typically non--Borel 
summable (see~\cite{FrGrSi1985,CaGrMa1986,CaGrMa1993,Ca2000}
for suitable generalizations of the concept of Borel summability
which provide solutions for large classes of problems
of practical interest). 

Recently, considerable progress
has been achieved for the nonalternating case, and 
we will describe some of the relevant applications. 
For the acceleration of convergence of nonalternating series, 
new powerful algorithms have been devised~\cite{JeMoSoWe1999},
and generalizations of the Borel summation process have led
to promising results in the case of nonalternating factorially 
divergent series~\cite{Je2000prd,JeSo2001,Je2001pra}.

One would intuitively assume that the perturbation
series, even if it is divergent, should be associated to the 
physical quantity of interest in a more or less unique way.
Unfortunately, many divergent series and the associated 
physical observables (e.g., energy levels as a function of the 
coupling strength) fail to fulfill the Carleman criterion~\cite{Ca1926} which 
guarantees that an asymptotic series can be uniquely associated 
to a given mathematical function. It appears that some ambiguity has 
to accepted as far as the connection between perturbation theory 
and physical quantities is concerned, although some attempts have
been made to resolve ambiguities in particular cases
(see e.g.~\cite{KnRe2001}).

For completeness, we give here the Carleman criterion
which guarantees that there is a one-to-one
correspondence between a function and its associated asymptotic series
(see for example~\cite{GrGrSi1970}, Theorems XII.17 -- XII.19 and the
definition on p.~43 in~\cite{ReSi1978}, p.~410 in~\cite{BeOr1978}, or
the comprehensive and elucidating
review~\cite{Fi1997}): 
\begin{center}
\begin{minipage}{12cm}
{\bf Carleman Criterion.} Let $f (z)$ be a function
which is analytic in the interior and continuous on a sectorial region
${\cal S} = \{ z \vert \vert \arg (z) \vert \le k\,\pi/2 + \epsilon, 0
< \vert z \vert < R \}$ of the complex plane for some $\epsilon >
0$. Let the function $f$ have an asymptotic expansion $f (z) \; \sim
\; \sum_{n=0}^{\infty} \, c_n \, z^n$ (for $z \to 0$). The function
$f$ obeys a strong asymptotic condition (of order $k$) if there are
suitable positive constants $C$ and $\sigma$ such that $| f (z) \, -
\, \sum_{n=0}^m \, c_n \, z^n | \le C \, \sigma^{m+1} \, [k\,(m+1)]!
\, |z|^{m+1}$ holds for all $m$ and for all $z \in {\cal S}$. The
validity of such a condition implies that the function $f (z)$ is
uniquely determined by its asymptotic series. 
\end{minipage}
\end{center}
Typically, nonalternating
factorially divergent series which entail nonperturbative
(imaginary) contributions do not fulfill the Carleman condition. 
This failure gives rise to the inevitable ambiguities.

Even the generalized Borel resummation methods introduced
in~\cite{FrGrSi1985,CaGrMa1986,CaGrMa1993,Ca2000}
cannot provide conclusive answers for all possible
physical scenarios. One highly problematic application 
is given by the quantum-mechanical double-well oscillator.
The perturbation series for the ground state (as well as
for any other unperturbed state) is divergent.
Yet it is known that the ground state splits into two 
states of opposite parity when the second minimum of the potential
approaches the first from infinity~\cite{LaLi1958,LaLi1979}.
The interesting observation is that two distinct quantum
states are described by one and the same perturbation series.
The energy difference is nonperturbative in the coupling and 
cannot be inferred {\em in principle} from perturbation
theory. The problem has been the subject of intensive
discussion~\cite{CaGrMa1988,CaGrMa1996,ZJ1981jmp,ZJ1981npb,ZJ1983npb,ZJ1984jmp}.
The resulting multi-instanton
expansion has recently found a more mathematically motivated
explanation in the theory of resurgent 
functions~\cite{Ph1988,DeDi1991,CaNoPh1993}.

We conclude this introduction by expressing that convergence
acceleration and resummation methods cannot be separated. The
connection can be seen as follows. We consider the remainders
$r_n \equiv s_n - s$, i.e.~the difference of the element $s_n$   
of the sequence and the limit $s$. The acceleration of convergence is
achieved by eliminating, as far as possible,
the remainders $r_n$ from a necessarily finite
set of input data $\{s_n\}_{n=0}^m$, in order to obtain a
better estimate $s'_m$ to $s$ than the obvious estimate
which would be provided by $s_m$. Clearly, the ``elimination
of remainders'' is a concept which can be transported to the 
case of a divergent series. The only difference is 
that for a convergent series, $r_n \to 0$ as $n \to \infty$,
whereas for a divergent series, $\lim_{n\to\infty} r_n$ does not
exist. Otherwise, the concept of the elimination of 
remainders is the same for both cases.

Returning to our (simple) example,
it is intuitively obvious that a method which helps to eliminate
the remainders 
\begin{equation}
r_n = \sum_{j=n+1}^{\infty} (-z)^j = \frac{(-z)^{n+1}}{1+z}
\end{equation}
{\em inside} the circle of convergence $|z| < 1$, could
be suitable to accomplish an analogous elimination of the $r_n$
if $z$ assumes values {\em outside} of the domain of convergence
of the series $\sum_{j=0}^{\infty} (-z)^j$ because the 
mathematical structure of the expression which defines
$r_n$ does not change. These entirely heuristic considerations
will be illustrated and explained in the sequel.

%
% Convergence Acceleration
%
\chapter{Convergence Acceleration}

\typeout{========================}
\typeout{Convergence Acceleration}
\typeout{========================}
\typeout{}

\label{ConvAccel}

%
% The Concept of Convergence Acceleration
%
\section{The Concept of Convergence Acceleration}
\label{ConvAccelConc}

%
% Difference Operator
%
\subsection{A Brief Survey}
\label{BriefSurvey}

We intend to give a brief overview of the convergence
acceleration methods used in this Thesis, without any
claim for completeness or comprehensiveness with regard to
currently known convergence acceleration techniques. 
Indeed, excellent textbooks (e.g.~\cite{Ba1975,BaGr1996,BrRZ1991})
and review articles~\cite{Ba1965,BaGa1970,Ba1972,We1989,Ba1990} 
on the subject are available.
Here, we focus on aspects of the methods which will become
useful for the applications discussed in later Chapters of the
Thesis.

This brief introduction is structured as follows:
After a discussion of basic concepts (Secs.~\ref{DiffOp} and~\ref{LinLogConv}),
we introduce Pad\'{e} approximants (one of {\em the} standard mathematical
tools for extrapolation and convergence acceleration) in 
Sec.~\ref{PadeApproximation}. Next, we discuss nonlinear sequence
transformations in Sec.~\ref{NonlinearST}. Both of these convergence
acceleration methods
are much more powerful and much more numerically stable in the case
of alternating input series than in the case of input series whose
terms have the same sign (nonalternating series). We conclude in
Sec.~\ref{CNCT} with a discussion of the recently proposed~\cite{JeMoSoWe1999}
combined nonlinear-condensation transformation (CNCT) which 
provides an efficient algorithm in the historically problematic case of 
{\em nonalternating} input series.

Before we start the discussion of Pad\'{e} approximants
and nonlinear sequence transformations,
we would like to mention that a large number of other convergence
acceleration methods exist which have all been documented
in the literature: examples include the Euler transformation,
Aitken's $\Delta^2$ process, the iterated 
$\Delta^2$ process, and Richardson extrapolation which can
be implemented using Wynn's rho algorithm, and its iterations.
Last, but not least, we mention the Euler--Maclaurin sum formula.
The interested reader will find related information
e.g.~in Chs.~5 and 6 of~\cite{We1989},
App.~E.2 of~\cite{Je1999}, and in Eq.~(3.6.28) ff.~of~\cite{AbSt1972}.

Concerning the mutual differences between the Pad\'{e} method
(Sec.~\ref{PadeApproximation} and nonlinear sequence transformations
(Sec.~\ref{NonlinearST}), we would like to expand: Given a formal 
series
\begin{equation} 
\label{formalperser1}
{\cal P}(z) \;=\; \sum_{j=0}^\infty \; c_j \, z^j
\end{equation}
as input, the Pad\'{e} method produces rational approximants
which depend exclusively on a finite subset of the coefficients 
$\{ c_j \}_{j=0}^\infty$, which in turn determine
the partial sums of (\ref{formalperser1}). Nonlinear sequence transformations
require, in addition to the partial sums of (\ref{formalperser1}),
also explicit remainder estimates, i.e.~estimates for the 
remainder $\sum_{j=n+1}^\infty \; c_j \, z^j$ which remains 
after the summation of the first $n$ terms of (\ref{formalperser1}).

Thus, the remainder estimates introduce additional
degrees of freedom in the construction of sequence
transformations as compared to Pad\'{e} approximants.
As pointed out in~\cite{JeWeSo2000},
one may draw an analogy between sequence
transformations and Pad\'{e} approximants on the
one hand and the Gaussian integration and
the Simpson rule on the other hand;
the variable integration nodes and weight factors
of the Gaussian integration
yield additional degrees of freedom which may be used in order
to construct a potentially much more powerful algorithm for
numerical integration.

Except for the additional degrees of freedom, nonlinear sequence
transformations have additional advantages.
For example, the delta transformation to be discussed in Sec.~\ref{NonlinearST}
gave very good results in the case of the sextic anharmonic oscillator, and
there is strong numerical evidence that it is able to sum the extremely
violently divergent perturbation series for the octic anharmonic
oscillator~\cite{WeCiVi1993}.
By contrast, as discussed in~\cite{WeCiVi1993},
Pad\'e approximants are apparently not powerful
enough to sum divergent series whose coefficients diverge like $(2 n)!$
or even $(3 n)!$. These observations stimulated research on sequence
transformations considerably. The rapid progress in this field is
convincingly demonstrated by the large number of monographs
\cite{Br1980,Br1977,Br1978,Br1997,BrRZ1991,De1988,LiLuSh1995,%
MaSh1983,Wa1996,Wi1981} and review articles
\cite{We1989,Br1985jcam,GaGu1974,Gu1989} which appeared in recent years.

%
% Difference Operator
%
\subsection{The Forward Difference Operator}
\label{DiffOp}

An essential ingredient in the construction of sequence
transformations, both for convergence acceleration
and for the resummation of divergent series,
is the (forward) difference operator.
We define the difference operator $\Delta$
acting on the sequence $\{s_n\}_{n=0}^{\infty}$:
\begin{equation}
\label{defOpDelta}
\Delta \; : \;
\{s_n\}_{n=0}^{\infty} \to \{(\Delta s)_n\}_{n=0}^{\infty} \; , \;
(\Delta s)_n \equiv s_{n+1} - s_n\,,
\end{equation}
The $n$th element of the sequence $\{(\Delta s)_n\}_{n=0}^{\infty}$ is
given by $(\Delta s)_n = s_{n+1} - s_n$.
This way of writing the $n$th element of the {\em transformed}
sequence stresses the fact that $(\Delta s)_n$ is to be regarded
as an element of some new sequence. However, the brackets are often left out,
and we write
\begin{equation}
\Delta s_n \equiv s_{n+1} - s_n = (\Delta s)_n\,.
\end{equation}
An important relation is,
\begin{equation}
\label{DeltaToTheK}
(\Delta^k s)_n = (-1)^k \,
\sum_{j = 0}^{k} (-1)^j \, {{k} \choose {j}} \, s_{n+j}\,.
\end{equation}
Again, we identify
\begin{equation}
\Delta^k s_n \equiv (\Delta^k s)_n\,.
\end{equation}
The $k$th power of the difference operator plays a crucial rule in constructing
sequence transformations.

%
% Linear and Logarithmic Convergence
%
\subsection{Linear and Logarithmic Convergence}
\label{LinLogConv}

As already stressed in Ch.~\ref{ConDivInt},
the main notion of convergence acceleration is to
extract information ``hidden in trailing digits''
from a necessarily finite number of sequence elements,
in order to convert a sequence $\{s_n\}_{n=0}^{\infty}$ 
into a new sequence $\{s'_n\}_{n=0}^{\infty}$ with hopefully
better numerical properties.

We briefly recall the notion of logarithmic convergence and
the difficulties associated with the 
acceleration of the convergence of nonalternating series.
Let a sequence $\{ s_n\}_{n=0}^{\infty}$  
fulfill the asymptotic condition
\begin{equation}
\label{LimRatSeq}
\lim_{n \to \infty} \frac {s_{n+1} - s} {s_n - s} \; = \; \rho \,,
\end{equation}
where $s = s_{\infty}$ is the limit of the sequence as $n \to
\infty$. If $\rho > 1$, then the sequence 
$\{ s_n\}_{n=0}^{\infty}$ is divergent.
For $\rho = 1$, the sequence may either be convergent or divergent.
A convergent sequence with $\rho = 1$ is called 
{\em logarithmically convergent}
(if $\rho < 1$, the series
is called {\em linearly convergent}).
Let us further assume that the
elements of the sequence $\{ s_n \}_{n=0}^{\infty}$ in Eq.\
(\ref{LimRatSeq}) represent partial sums 
\begin{equation}
\label{defsn}
s_n = \sum_{k=0}^{n} a(k)
\end{equation}
of an infinite series. Here, we will almost exclusively
investigate slowly convergent nonalternating
sequences $\{ s_n\}_{n=0}^{\infty}$ 
whose elements are all real and positive 
(for these sequences, $0 < \rho \leq 1$). 
In the case of slow convergence, $\rho$ is either very close 
or equal to unity.

As observed by many authors~(e.g.~\cite{We1989}), the 
acceleration of the convergence of nonalternating sequences
is a potentially unstable numerical process.
The reason is the following: A sequence transformation can only
accelerate convergence if it
succeeds in extracting additional information about the
index-dependence of the truncation errors 
\begin{equation}
\label{defrn}
r_n = s_n - s 
\end{equation} 
from a necessarily finite set of partial sums 
$\{ s_n \}_{n=0}^k$ of the input series. Normally, this is
done by forming arithmetic expressions involving higher weighted
differences of the $s_n$. The calculation of
higher weighted differences is a
potentially unstable process which can easily lead to a serious loss
of numerical significance due to cancellation if the input data
all have the same sign.

We had mentioned in Ch.~\ref{ConDivInt} that the rate of convergence
can be said to be increased if the transformed sequence elements
$\{ s'_n \}_{n=0}^\infty$ and the input data $\{ s_n \}_{n=0}^\infty$
fulfill [see also Eq.~(\ref{ConvAccelDef1})]
\begin{equation}
\label{ConvAccelDef}
\lim_{n \to \infty} \tau_n \equiv
\lim_{n \to \infty} \frac{s'_n - s}{s_n - s} = 0\,,
\end{equation}
where we implicitly define the $\tau_n$. Let us assume
that the input sequence $\{ s_n \}_{n=0}^\infty$
is logarithmically convergent and that the transformed sequence
$\{ s'_n \}_{n=0}^\infty$ is linearly convergent with 
$\lim_{n \to \infty} (s'_{n+1} - s)/(s'_n - s) \; = \; \rho' < 1$
[see Eq.~(\ref{LimRatSeq}) and the definition of logarithmic convergence]. 
Then
\begin{equation}
\lim_{n \to \infty} \frac{\tau_{n+1}}{\tau_n} \equiv
\lim_{n \to \infty} \frac{s'_{n+1} - s}{s'_n - s} \,
\frac{s_n - s}{s_{n+1} - s} = \rho' < 1\,,
\end{equation}
so that $\lim_{n \to \infty} \tau_n = 0$, and the rate of
convergence is increased [according to the definition
of convergence acceleration in Eq.~(\ref{ConvAccelDef})].

%
% Pade approximants
%
\subsection{The Standard Tool: Pad\'{e} Approximants}
\label{PadeApproximation}

The  Pad\'e approximation produces a rational approximation to the
original input series; this approximation is given by the ratio of two
polynomials.  The $[ l / m ]$ Pad\'e approximant to a series ${\cal P}(z)$,
\begin{equation}
\label{formalperser}
{\cal P}(z)\;=\; \sum_{j=0}^\infty \; c_j \, z^j\,,
\end{equation}
is given by the ratio of two polynomials $P_l(z)$ and $Q_m(z)$ of
degree $l$ and $m$, respectively:
\begin{equation}
\label{pade}
[ l / m ]_{\cal P} (z) \;=\; \frac{P_l(z)}{Q_m(z)}\;=\;
\frac{p_0 + p_1 \, z + \ldots + p_{l} \,  z^{l}}
{q_0 + q_1 \, z + \ldots + q_{m} \, z^{m}} \;.
\end{equation}
The polynomials $P_l(z)$ and $Q_m(z)$ are constructed so that the
Taylor expansion of the Pad\'e approximation Eq.~(\ref{pade}) agrees
with the original input series Eq.~(\ref{formalperser}) up to the
order of $l+m$
\begin{equation}
\label{padedef}
{\cal P}(z) \;-\; [ l / m ]_{\cal P}(z) \; = \; {\rm O}(z^{l+m+1})\,.
\end{equation}
In other words, the Pad\'e approximant $[ l/m ]_{\cal P}(z)$ reproduces upon
re-expansion the first $l+m+1$ ($c_0,\dots,c_{l+m}$) coefficients of
the input series.  

Usually the normalization condition is choosen such as $q_0$ is one.
Hence there are $l+m+1$ equations to determine the coefficients of
$P_l(z)$ and $Q_m(z)$.  In most cases it is quite unattractive
to solve this system of equations.  A more
attractive {\em recursive} prescription for the evaluation of Pad\'e
approximants is the $\epsilon$-algorithm given by the
formulas~\cite{Wy1956a}
\begin{eqnarray}
\label{epsalg}
&& \epsilon_{-1}^{(n)} \; = \; 0\,,\;\;\;\;\;
\epsilon_0^{(n)} \;=\; s_n\,, \nonumber \\
&& \epsilon_{k+1}^{(n)} \; = \; 
\epsilon_{k-1}^{(n+1)} \;+\; 1/ [ \epsilon_k^{(n+1)} - \epsilon_k^{(n)} ]
\end{eqnarray}
for $k$ and $n$ integer. 
The $\epsilon$-algorithm is discussed in~\cite{Wy1956a}, and
a detailed discussion of its implementation can be found
in~\cite{We1989}. If the input data $\epsilon_0^{(n)} = s_n$
are the partial sums
\begin{equation}
s_n \;=\; \sum_{j=0}^n \; c_j \; z^j
\end{equation}
of the power series Eq.~(\ref{formalperser}), then the elements
$\epsilon_{2k}^{(n)}$ of the two-dimensional table of transformations
produce Pad\'e approximants according to
\begin{equation}
\label{EpstoPade}
\epsilon_{2k}^{(n)} \;=\; [ n+k / k ]_P (z) \;.
\end{equation}
The elements with odd lower subscripts $\epsilon_{2k+1}^{(n)}$
of the table of transformations are only auxiliary quantities.

In practical applications (see e.g. Tables 1 -- 3 of~\cite{JeBeWeSo2000}), 
the epsilon algorithm can be used for evaluation
of the following staircase sequence in the Pad\'e table:
\begin{equation}
\label{PadeSeq}
[0/0],\, [1/0],\, [1/1],\, [2/1], \ldots,
[\nu/\nu],\, [\nu+1/\nu],\, [\nu+1/\nu+1],\, \ldots \;.
\end{equation}
Because of Eq.~(\ref{EpstoPade}) this corresponds to the sequence
\begin{equation}
\epsilon_0^{(0)},\epsilon_0^{(1)},\epsilon_2^{(0)},\epsilon_2^{(1)},
\epsilon_4^{(0)},\epsilon_4^{(1)},\epsilon_6^{(0)},\ldots
\epsilon_{2 \nu}^{(0)},\epsilon_{2 \nu}^{(1)},\epsilon_{2 \nu+2}^{(0)},\ldots\;,
\end{equation}
which can be written as 
$\{ \epsilon_{2 \Ent {\nu/2}}^{(\nu - 2 \Ent {\nu/2} )} \}_{\nu=0}^\infty$.
Here $\Ent {\nu/2}$ denotes the integral part of $\nu/2$, which is the largest
integer $\mu$ satisfying $\mu \leq \nu/2$. With this notation the Pad\'e
sequence in Eq.~(\ref{PadeSeq}) can also be written as
$\{ \lbrack \nu - \Ent {\nu/2} / \Ent {\nu/2} \rbrack_P(z) \}_{\nu=0}^\infty$.

Modern computer algebra packages~\cite{Wo1988} contain built-in
routines for the evaluation of Pad\'{e} approximants. While these
are computationally less effective than the epsilon algorithm,
they are more flexible with regard to the degree of the numerator
and denominator polynomials. In particular, the sequence allow 
for the evaluation of the sequence of lower-diagonal Pad\'{e} approximants
of the form 
$\{ \lbrack \Ent {\nu/2} / \nu - \Ent {\nu/2} \rbrack_P(z) \}_{\nu=0}^\infty$.

As already mentioned in Sec.~\ref{BriefSurvey},
it is a typical feature of all algorithms for Pad\'{e} approximants that
the coefficients $c_0$, $c_1$, $\ldots$ , $c_{l+m}$ of the formal power
series (\ref{formalperser}) suffice to determine $[l / m]_{\cal P} (g)$
completely. No additional information is needed.
At first sight, this seems to be very advantageous. However, there are
situations in which this apparent advantage becomes a
disadvantage. For instance, the truncation error of an asymptotic
power series which has the Stieltjes property is for 
real argument bounded in magnitude by the first
term not included in the partial sum (see Example 9.3
on p.~92 of~\cite{Ol1974}). Accordingly, the
truncation error of a strictly alternating divergent hypergeometric
series ${}_2 F_0 (\alpha, \beta; - x)$ with $\alpha, \beta, x > 0$,
which provides a good mathematical model for many divergent perturbation
expansions, is bounded in magnitude by the first term not included in
the partial sum. In addition, the truncation error and the first term
not included in the partial sum have the same sign (see Theorem 5.12-5
of~\cite{Ca1977}). In principle, this information should be very
helpful in summation processes. Unfortunately, Pad\'{e} approximants are
not able to profit from it. This greatly reduces their efficiency if a
divergent hypergeometric series ${}_2 F_0$ of that kind or a divergent
perturbation series of a similar type is to be summed.

%
% Nonlinear Sequence Transformations
%
\subsection{Nonlinear Sequence Transformations}
\label{NonlinearST}

A sequence transformation ${\cal T}$ 
is a rule which converts a slowly convergent
or divergent sequence $\{ s_n \}_{n=0}^{\infty}$ whose elements may be
the partial sums
$s_n \; = \; \sum_{k=0}^{n} \, a_k$ [see Eq.~(\ref{FundSeq})]
of an infinite series, into a new sequence $\{ s'_n \}_{n=0}^{\infty}$
with hopefully better numerical properties. Pad\'{e} approximants,
which were discussed in Section \ref{PadeApproximation}, are special
sequence transformations since they transform the partial sums of a
formal power series (\ref{formalperser}) 
into a doubly indexed sequence of rational
approximants (\ref{PadeSeq}). In the current Section, alternative transformations
will be discussed.

Let us now assume that the sequence $\{ s_n \}_{n=0}^{\infty}$ of
partial sums of an infinite series diverges but can be summed to its
generalized limit $s$. Then the sequence elements $s_n$ can for all
integers $n \ge 0$ be partitioned into the generalized limit $s$ and
remainders $r_n$:
\begin{equation}
s_n \; = \; s \, + \, r_n \, .
% \label{def_Rem_Est}
\end{equation}
If $\{ s_n \}_{n=0}^{\infty}$ diverges, then the remainders
$r_n$ do not vanish as $n \to \infty$.
A divergent sequence would be summed to its generalized limit $s$ if the
remainders $r_n$ could somehow be eliminated from the sequence
elements $s_n$. For a slowly convergent sequence, one can hope to accelerate
convergence by eliminating approximations to the $r_n$ obtained from a finite
set of sequence elements which are used as input data for the 
transformation.

The sequence $\{ s_n \}_{n=0}^{\infty}$ is transformed into a
new sequence $\{ s^{\prime}_n \}_{n=0}^{\infty}$, whose elements can for
all $n \ge 0$ be partitioned into the generalized limit $s$ of the
original sequence and a transformed remainder $r^{\prime}_n$:
\begin{equation}
s^{\prime}_n \; = \; s \, + \, r^{\prime}_n \, .
% \label{}
\end{equation}
This approximative elimination process can be termed successful 
if the transformed
remainders $r^{\prime}_n$ either vanish as $n \to \infty$ 
(in the case of a divergent input sequence where the
$r_n$ do not vanish at all as $n \to \infty$), 
or if the $r^{\prime}_n$ vanish faster than the original remainders
$r_n$ as $n \to \infty$ [in the case of a slowly convergent 
input sequence, see also Eq.~(\ref{ConvAccelDef})]. 
A {\it sequence transformation\/} is a rule which
transforms a slowly convergent or divergent sequence $\{ s_n
\}_{n=0}^{\infty}$ into a new sequence $\{ s^{\prime}_n
\}_{n=0}^{\infty}$ with hopefully better numerical properties.

However, the truncation errors of a divergent series
can depend on $n$ in a very complicated way. Consequently, the direct
elimination of approximations to $r_n$ from $s_n$ can be very
difficult. A considerable simplification can often be achieved by means
of a suitable reformulation. Let us consider the following model
sequence (see Section 3.2 of \cite{We1989}):
\begin{equation}
\tilde{s}_n \; = \; \tilde{s} \, + \, \omega_n \, z_n \, .
\label{Mod_Seq_Om}
\end{equation}
Here, $\omega_n$ is a {\em remainder estimate\/}, which has to be chosen
according to some rule and which may depend on $n$ in a very complicated
way, and $z_n$ is a {\em correction term\/}, which should be chosen in
such a way that it depends on $n$ in a relatively smooth way. Moreover,
the products $\omega_n z_n$ should be capable of producing sufficiently
accurate approximations to the actual remainders $r_n$ of the sequence
to be transformed.

The principal advantage of this approach is that now only the correction
terms $\{ z_n \}_{n=0}^{\infty}$ have to be determined. The subsequent
elimination of $\omega_n z_n$ from $s_n$ is often much easier than the
direct elimination of approximations to $r_n$ from $s_n$. The use of
remainder estimates $\{ \omega_n \}_{n=0}^{\infty}$ is also an efficient
way of incorporating {\it additional\/} information about the remainders
into the summation scheme.

The model sequence (\ref{Mod_Seq_Om}) has another undisputable
advantage: There is a {\em systematic\/} way of constructing a sequence
transformation which is exact for this model sequence. Let us assume
that a {\em linear\/} operator ${\hat T}$ can be found which annihilates
the correction term $z_n$. Then, a sequence transformation, which is
exact for the model sequence (\ref{Mod_Seq_Om}), can be constructed
quite easily. Just apply ${\hat T}$ to $[\tilde{s}_n - \tilde{s}] /
\omega_n = z_n$. Since ${\hat T}$ annihilates $z_n$ according to ${\hat
T} (z_n) = 0$ and is by assumption linear, we find that the following
sequence transformation ${\cal T}$ is {\it exact\/} for the model
sequence (\ref{Mod_Seq_Om}) (see Eq.\ (3.2-11) of \cite{We1989}):
\begin{equation}
{\cal T} (\tilde{s}_n, \omega_n) \; = \; \frac
{{\hat T} (\tilde{s}_n / \omega_n )} {{\hat T} (1 / \omega_n )}
\; = \; \tilde{s} \, .
\label{GenSeqTr}
\end{equation}
Simple and yet very powerful sequence transformations are obtained if
the annihilation operators are based upon the finite difference operator
$\Delta$ defined in Sec.~\ref{DiffOp}. A polynomial $P_{k-1} (n)$ of
degree $k - 1$ in $n$ is annihilated by the $k$-th power of
$\Delta$. Thus, we now assume that the correction terms $\{ z_n
\}_{n=0}^{\infty}$ can be chosen in such a way that multiplication of
$z_n$ by some suitable quantity $w_k (n)$ yields a polynomial $P_{k-1}
(n)$ of degree $k-1$ in $n$:
\begin{equation}
w_k (n) \, z_n \; = \; P_{k-1} (n) \, .
% \label{}
\end{equation}
Since $\Delta^k P_{k-1} (n) = 0$, the weighted difference operator
${\hat T} = \Delta^k w_k (n)$ annihilates $z_n$, and the corresponding
sequence transformation (\ref{GenSeqTr}) is given by the ratio
\begin{equation}
{\cal T}_k^{(n)} \bigl( w_k (n) \big\vert s_n, \omega_n \bigr)
\; = \; \frac
{\Delta^k \{ w_k (n) s_n / \omega_n \}}
{\Delta^k \{ w_k (n) / \omega_n \}} \, .
\label{Seq_w_k}
\end{equation}
A number of sequence transformations can be obtained by 
specializing $w_k(n)$. For instance, 
$w_k(n) = (n + \beta)^{k-1}$ with $\beta > 0$
yields Levin's sequence transformation \cite{Le1973}:
\begin{eqnarray}
{\cal L}_{k}^{(n)} (\beta, s_n, \omega_n) & = & \frac
{ \Delta^k \, \{ (n + \beta)^{k-1} \> s_n / \omega_n\} }
{ \Delta^k \, \{ (n + \beta)^{k-1}  / \omega_n \} }
\nonumber \\
& = & \frac
{\displaystyle
\sum_{j=0}^{k} \; ( - 1)^{j} \; {{k} \choose {j}} \;
\frac
{(\beta + n +j )^{k-1}} {(\beta + n + k )^{k-1}} \;
\frac {s_{n+j}} {\omega_{n+j}} }
{\displaystyle
\sum_{j=0}^{k} \; ( - 1)^{j} \; {{k} \choose {j}} \;
\frac
{(\beta + n +j )^{k-1}} {(\beta + n + k )^{k-1}} \;
\frac {1} {\omega_{n+j}} } \, .
\label{LevTr}
\end{eqnarray}
We follow the notation of~\cite{We1989}. The subscript $k$
denotes the {\em order\/} of the transformation, and $n$ denotes the
starting index (the initial element)
of the input data. Observe that the $k+1$ sequence elements $s_{n}$, \dots
$s_{n+k}$ and the $k+1$ remainder 
estimates $\omega_{n}$ \ldots $\omega_{n+k}$ are
needed for the computation of the transformation
${\cal L}_{k}^{(n)}$. The shift parameter
$\beta$ has to be positive in order to admit $n = 0$ in (\ref{LevTr}).
In most practical applications, it is observed that any choice
other than $\beta = 1$ does not lead to a significant improvement
of the rate of convergence, and therefore the choice 
$\beta = 1$ is standard~\cite{We1989}.

The notation ${\cal L}_{k}^{(n)} (\beta, s_n, \omega_n)$ is in need
of a certain further explanation. As explained above, $n$ represents
the initial element of the sequence $\{ s_n \}_{n=0}^{\infty}$
used in the evaluation of the transform. The specification of $s_n$
as the second argument of ${\cal L}_{k}^{(n)} (\beta, s_n, \omega_n)$
is therefore redundant as far as the index $n$ is concerned,
because $n$ already appears as an upper index in ${\cal L}_{k}^{(n)}$.
The arguments $s_n$ (and $\omega_n$) are to be interpreted
as follows: they rather specify the
initial element $s_n$ and the initial remainder estimate
$\omega_n$ which have to be used in the calculation
of the right-hand side of (\ref{LevTr}). Of course,
the evaluation of ${\cal L}_{k}^{(n)} (\beta, s_n, \omega_n)$
requires the $k+1$ sequence elements $\{ s_n,\dots, s_{n+k}, s_{n+k} \}$
as input. The knowledge of the ``post-initial''
elements $\{ s_{n+1},\dots, s_{n+k} \}$ is implicitly assumed
in writing the definition (\ref{LevTr}).

According to Smith and Ford, who had investigated the performance of
sequence transformations for a wide range of test problems
\cite{SmFo1979,SmFo1982}, Levin's transformation is among the most
powerful and most versatile sequence transformations that are currently
known.

We can also assume that the unspecified weights $w_k (n)$ in
(\ref{Seq_w_k}) are Pochhammer symbols according to $w_k (n) = (n +
\beta)_{k-1} = \Gamma (n+\beta+k-1)/\Gamma (n+\beta)$ with $\beta >
0$. This yields the delta transformation (see Eq.\ (8.2-7) of
\cite{We1989}):
\begin{eqnarray}
{\cal S}_{k}^{(n)} (\beta , s_n, \omega_n) & = & \frac
{ \Delta^k \, \{ (n + \beta)_{k-1} \> s_n / \omega_n\} }
{ \Delta^k \, \{ (n + \beta)_{k-1}  / \omega_n \} }
\nonumber\\
& = & \frac
{\displaystyle
\sum_{j=0}^{k} \; ( - 1)^{j} \; {{k} \choose {j}} \;
\frac {(\beta + n +j )_{k-1}} {(\beta + n + k )_{k-1}} \;
\frac {s_{n+j}} {\omega_{n+j}} }
{\displaystyle
\sum_{j=0}^{k} \; ( - 1)^{j} \; {{k} \choose {j}} \;
\frac {(\beta + n +j )_{k-1}} {(\beta + n + k )_{k-1}} \;
\frac {1} {\omega_{n+j}} } \, .
\label{WenTr}
\end{eqnarray}
As in the case of Levin's transformation, the most obvious choice for
the shift parameter is $\beta = 1$.

The numerator and denominator sums in (\ref{LevTr}) and (\ref{WenTr})
can be computed more effectively with the help of the three-term
recursions [see Eq.~(7.2-8) of \cite{We1989}]:
\begin{equation}
L_{k+1}^{(n)} (\beta) = L_k^{(n+1) (\beta)}
- \, {\displaystyle \frac
{ (\beta + n + k ) (\beta + n + k)^{k-1} }
{ (\beta + n + k + 1)^k }} \, L_k^{(n)} (\beta)\,,
\label{LevRec}
\end{equation}
and (see Eq.\ (8.3-7) of \cite{We1989})
\begin{equation}
S_{k+1}^{(n)} (\beta) = S_k^{(n+1)} (\beta)
- \, {\displaystyle \frac
{ (\beta + n + k ) (\beta + n + k - 1) }
{ (\beta + n + 2 k ) (\beta + n + 2 k - 1) }} \, S_k^{(n)} (\beta) \, .
\label{WenRec}
\end{equation}
The initial values $L_0^{(n)} (\beta) = S_0^{(n)} (\beta) = s_n /
\omega_n$ and $L_0^{(n)} (\beta) = S_0^{(n)} (\beta) = 1 / \omega_n$
produce the numerator and denominator sums, respectively, of ${\cal
L}_{k}^{(n)} (\beta , s_n, \omega_n)$ and ${\cal S}_{k}^{(n)} (\beta ,
s_n, \omega_n)$.

Let us assume that the terms $a_k$ in (\ref{FundSeq}) are 
strictly alternating, i.e.~$a_k = (-1)^k b_k$ with $b_k > 0$.
In this case, the first term $(-1)^{n+1} b_{n+1}$ not included in the partial
sum $s_n = \sum_{k=0}^{n} (-1)^k b_k$ is the best {\em simple\/}
estimate for the truncation error. This corresponds to the
remainder estimate
\begin{equation}
\omega_n \; = \; (-1)^{n+1} b_{n+1} \; = \; \Delta s_n \, ,
\label{d_Est}
\end{equation}
which was proposed by Smith and Ford \cite{SmFo1979}. The use of this
remainder estimate in (\ref{LevTr}) and (\ref{WenTr}) yields the
following sequence transformations [Eqs.~(7.3-9) and (8.4-4) of
\cite{We1989}]:
\begin{eqnarray}
d_k^{(n)} (\beta, s_n) & = &
{\cal L}_{k}^{(n)} (\beta , s_n, \Delta s_n)
\nonumber \\
& = & \frac
{\displaystyle
\sum_{j=0}^{k} \; (-1)^{j} \; {{k} \choose {j}} \;
\frac
{(\beta + n +j )^{k-1}} {(\beta + n + k )^{k-1}} \;
\frac {s_{n+j}} {\Delta s_{n+j}} }
{\displaystyle
\sum_{j=0}^{k} \; ( - 1)^{j} \; {{k} \choose {j}} \;
\frac
{(\beta + n +j )^{k-1}} {(\beta + n + k )^{k-1}} \;
\frac {1} {\Delta s_{n+j}} } \, ,
\label{dLevTr} \\
{\delta}_k^{(n)} (\beta, s_n) & = &
{\cal S}_{k}^{(n)} (\beta , s_n, \Delta s_n)
\nonumber \\
& = & \frac
{\displaystyle
\sum_{j=0}^{k} \; ( - 1)^{j} \; {{k} \choose {j}} \;
\frac {(\beta + n +j )_{k-1}} {(\beta + n + k )_{k-1}} \;
\frac {s_{n+j}} {\Delta s_{n+j}} }
{\displaystyle
\sum_{j=0}^{k} \; ( - 1)^{j} \; {{k} \choose {j}} \;
\frac {(\beta + n +j )_{k-1}} {(\beta + n + k )_{k-1}} \;
\frac {1} {\Delta s_{n+j}} } \, .
\label{dWenTr}
\end{eqnarray}
Alternative remainder estimates for the 
sequence transformations (\ref{LevTr}) and
(\ref{WenTr}) were discussed in Sections 7 and 8 of \cite{We1989} or in
\cite{We1994a,HoWe1995}.

A final word on sequence transformations:
It is observed that sequence transformations constructed according to
(\ref{dWenTr}) often lead to much better numerical results than
better known convergence accelerators like the Euler transformation,
Wynn's epsilon or rho algorithms, or Aitken's $\Delta^2$ process~\cite{We1989}.
In particular, there is overwhelming empirical
evidence that $d_{k}^{(n)}$ and ${\delta}_{k}^{(n)}$ work very well in
the case of convergent or divergent alternating series for instance as
they occur in special function theory
\cite{Le1973,We1989,We1996c,JeMoSoWe1999,BrRZ1991,SmFo1979,SmFo1982,%
We1990,WeCi1990,RoBhBh1996,BhRoBh1997} or in
perturbation theory \cite{We1997,WeCiVi1993,We1992,We1990,WeCi1990,%
WeCiVi1991,We1996a,We1996b,We1996d,RoBhBh1998}.
Only for some special model problems could rigorous convergence
proofs be obtained (see~\cite{Si1981} or Sections 13
and 14 of~\cite{We1989}).
No general convergence proof is known for delta transformation 
${\delta}_{k}^{(n)}$ as
well as their parent transformations ${\cal L}_{k}^{(n)}$ and ${\cal
S}_{k}^{(n)}$. In view of the partially unsatisfactory situation concerning
the availability of mathematical proof, we have carried out
high-precision studies of the convergence of sequence transformations
of the type (\ref{dWenTr}), in part using multi-precision
arithmetic (see Sec.~\ref{sec_math}). In the absence of a rigorous proof,
one of the concerns which may be raised
against the convergence of the transforms
(\ref{dWenTr}) is a plausible asymptotic nature of the
sequence of transforms. In our numerical experiments, we found no
indication of asymptotic behaviour. The rate of convergence
of the transforms remained constant, and we observed
apparent convergence up
to 430 decimal figures (see Sec.~\ref{sec_math} below).
In the absence of rigorous proof and the presence of considerable
numerical evidence for the computational usefulness
of sequence transformations of the type (\ref{dWenTr}),
we believe that experimental mathematics may give some hints
at the theoretical soundness of the concepts involved.

%
% The Combined Nonlinear--Condensation Transformation
%
\subsection{The Combined Nonlinear--Condensation Transformation (CNCT)}
\label{CNCT}

We assume the series $\sum_{k=0}^\infty a_k$ is slowly convergent and
nonalternating, i.e.~$a_k > 0$ $\forall k$. In order to 
make the notation of involved subscripts more clear, we
temporarily identify $a_k \equiv a(k)$.
Following Van Wijngaarden \cite{vW1965}, we transform the 
nonalternating input series 
\begin{equation}
\sum_{k=0}^{\infty} a(k)\,,
\quad a(k) \geq 0\,,
\end{equation}
whose partial sums are given by (\ref{defsn}),
into an alternating series 
$\sum_{j=0}^{\infty} (-1)^j {\bf A}_j$.
After the first step of the transformation,
the limit of the input series is recovered according to
\begin{equation}
\label{VWSerTran}
\sum_{k=0}^{\infty} \, a (k) =
\sum_{j=0}^{\infty} \, (-1)^j \, {\bf A}_j \,.
\end{equation}
The quantities ${\bf A}_j$ are defined according to 
\begin{equation}
\label{A2B}
{\bf A}_j = \sum_{k=0}^{\infty} \, {\bf b}_{k}^{(j)} \, ,
\end{equation}
where
\begin{equation}
\label{B2a}
{\bf b}_{k}^{(j)} = 2^k \, a(2^k\,(j+1)-1) \, .
\end{equation}
Obviously, the terms ${\bf A}_j$ defined in Eq.~(\ref{A2B}) are all
positive if the terms $a (k)$ of the original series are all
positive. The ${\bf A}_j$ are referred
to as the {\em condensed series}~\cite{JeMoSoWe1999}, and the series
$\sum_{j=0}^{\infty} (-1)^j {\bf A}_j$ is referred to as the {\em
transformed alternating series}, or alternatively as the {\em Van
Wijngaarden transformed series}.

The construction of the condensed series reminds one of
Cauchy's condensation theorem (see e.g.~p.~28 of
Ref.~\cite{Bw1991} or p.~121 of Ref.~\cite{Kn1964}).
Given a nonalternating series $\sum_{k=0}^{\infty} a (k)$ 
with terms that satisfy $\vert a(k+1)
\vert < \vert a(k) \vert$, Cauchy's condensation theorem states that
$\sum_{k=0}^{\infty} a (k)$ converges if and only if
the first condensed sum ${\bf A}_0$ defined according to 
Eq.~(\ref{A2B}) converges.

The summation over $k$ in Eq.~(\ref{A2B}) dose not pose
numerical problems. Specifically, it can be easily 
shown in many cases of practical importance that the 
convergence of $\sum_{k=0}^{\infty} \, {\bf b}_{k}^{(j)}$ (in $k$)
is linear even if the convergence of $\sum_{k=0}^{\infty} a(k)$ is
only logarithmic. We will illustrate this statement
by way of two examples. 

\underline{Example 1:} a logarithmically
convergent input series whose terms behave asymptotically as
$a (k) \sim k^{-1 - \epsilon}$ with $\epsilon > 0$.
In this case, the partial sums
\begin{equation}
{\bf A}^{(n)}_j = \sum_{k=0}^{n} \, {\bf b}_{k}^{(j)}
\end{equation}
converge linearly with
\begin{equation}
\lim_{n \to \infty} \frac{{\bf A}^{(n+1)}_j - {\bf A}_j} 
  {{\bf A}^{(n)}_j - {\bf A}_j} \; = \; 
\frac{1}{2^\epsilon \, (j+1)^{1+\epsilon}} < 1 \,, \qquad
a(k) \sim k^{- 1 - \epsilon}\,, \quad k \to \infty.
\end{equation}
 
\underline{Example 2:} a series with $a (k) \sim k^{\beta} r^k$ 
where $0 < r < 1$ and $\beta$ real. Here, we have
$\rho = r < 1$, and the series
is (formally) linearly convergent. However, slow convergence may result 
if $\rho$ is close to one. In this case, the condensed series are
very rapidly convergent,  
\begin{equation}
\lim_{n \to \infty} \frac {{\bf A}^{(n+1)}_j - {\bf A}_j} 
  {{\bf A}^{(n)}_j - {\bf A}_j} \; = \; 0 \,, \qquad
a(k) \sim k^{\beta} r^k \,, \quad k \to \infty.
\end{equation}
Therefore, when summing over $k$ in evaluating the condensed series
according to Eq.~(\ref{A2B}), it is in many cases 
sufficient to evaluate the condensed series by adding the terms successively,
and no further acceleration of the convergence is required.

As shown in~\cite{Da1969,JeMoSoWe1999}, the condensation transformation
defined according to Eqs~(\ref{VWSerTran})-- (\ref{B2a}) is 
essentially a reordering
of the terms of the input series $\sum_{k=0}^{\infty} a(k)$.
Furthermore, Daniel was able to show 
(see the Corollary on p.\ 92 of Ref.\ \cite{Da1969}) that for 
nonalternating convergent series whose terms decrease in magnitude
($\vert a(k) \vert > \vert a(k+1) \vert$), the equality 
(\ref{VWSerTran}) holds. This formally justifies the correctness
of the condensation transformation defined according to
Eqs.~(\ref{VWSerTran}) -- (\ref{B2a}). 

Note that the property
\begin{equation}
{\bf A}_{2\,j-1} = \frac{1}{2} \,
\left[ {\bf A}_{j-1} - a_{j-1} \right]\,, \qquad
\mbox{($j = 1,2,\dots$)}\,,
\end{equation}
derived in~\cite{Da1969}, facilitates the numerical evaluation of a 
set of condensed series, by reducing the evaluation of condensed series 
of odd index to a trivial computation.

In the second step of the CNCT, the convergence of
the Van Wijngaarden transformed series 
$\sum_{j=0}^{\infty} \, (-1)^j \, {\bf A}_j$
on the right-hand side of
Eq.~(\ref{VWSerTran}) is accelerated by
a suitable nonlinear sequence transformation.
We start from the partial sums 
\begin{equation}
\label{PsumS}
{\bf S}_n \; = \; \sum_{j=0}^{n} \, (-1)^j \, {\bf A}_j
\end{equation}
of the Van Wijngaarden transformed series.

We use the partial sums ${\bf S}_n$ defined
in Eq.~(\ref{PsumS}) as input data for the delta 
transformation~(\ref{dWenTr}).
This leads to the CNC transforms
\begin{equation}
\label{defcnc}
{\cal T}_{\rm CNC}(n) = \delta_{n}^{(0)} \left( 1, {\bf S}_0 \right)\,.
\end{equation}
which require as input the elements
$\{ {\bf S}_0, \dots, {\bf S}_{n}, {\bf S}_{n+1} \}$
of the Van Wijngaarden transformed series. 

%
% Applications of Convergence Acceleration Methods
%
\section{Applications of Convergence Acceleration Methods}
\label{AppConvAccel}

%
% Special functions in statistical theory
%
\subsection{Applications in Statistics and Applied Biophysics}
\label{sec_stat}

Several slowly convergent series of the type considered in this
Section of the Thesis define special functions that have important uses in
statistics, namely the Riemann zeta, generalized zeta, Jonqui\`{e}re
functions and Lerch's transcendent (see~\cite{Ba1953vol1} for
definitions and some properties of these functions).  Discrete
distributions are related to these functions by associating
probability mass functions (p.m.f.s) with the terms of 
the particular infinite series defining the special functions.
In turn, the normalization constants of the p.m.f.s are associated
with the sums of the series.

Consider a discrete distribution with the p.m.f. 
\begin{equation} 
\label{discrete}
{\rm Pr}[X=n] = c\,p(n;\theta)
\end{equation}
and with support of all nonnegative integers $n=0,1,\dots$, where
$\theta$ is a (vector of) parameter(s) and $c$ is a constant.  
The probability over the set of all
outcomes is unity:
\begin{equation} 
\label{proper}
\sum_{n=0}^{\infty} {\rm Pr}[X=n] = 1\,,
\end{equation}
from which we obtain the normalization constant
\begin{equation} 
\label{normc}
c^{-1} = \sum_{n=0}^{\infty} p(n;\theta)
\end{equation}
(in statistics, the random variable is usually denoted by the symbol $n$,
whether continuous or discrete, but we will denote it here by the
symbol $n$ in order to follow the usual notation used in the treatment
of special functions).
Thus we expect the sum in Eq.~(\ref{normc}) to exist and to converge
to a finite value. In the sequel,
we consider several related discrete distributions.
The Zipf distribution has the p.m.f.
\begin{equation}
\label{zipf}
p_x = c\,\frac1{n^s}
\end{equation}
with support of all positive integers $n=1,2,\dots$.  The
normalization constant~(\ref{normc}) is then
\begin{equation} 
\label{zeta}
c^{-1} = \zeta(s) = \sum_{n=1}^\infty \frac{1}{n^s}\,,
\end{equation}
where $s>1$, which is immediately recognized as the Dirichlet series
for the Riemann zeta function [Eq.~(1) on p.~32 of
Ref.~\cite{Ba1953vol1}].  Basic properties of the Zipf distribution
can be found in~\cite{JoKoKe1992}. 

The Zipf--Mandelbrot distribution~\cite{Ma1983} is the generalization of the Zipf
distribution that adds a constant to the ranks $n$ and has the p.m.f. 
\begin{equation}
\label{zipfman}
p_x=c\,\frac1{(n+v)^s}
\end{equation}
with the support of all nonnegative integers $n=0,1,\dots$. The
normalization constant~(\ref{normc}) is then
\begin{equation} 
\label{genzeta}
c^{-1} = \zeta(s,v) = \sum_{n=0}^\infty \frac1{(n+v)^s}\,,
\end{equation}
where $v \neq 0,-1,\dots$. The sum~(\ref{genzeta}) is recognized as
the generalized zeta function [Eq.~(1) on p.~24 of
Ref.~\cite{Ba1953vol1}].

Further, the p.m.f.
\begin{equation}
\label{good}
p_x=c\,\frac{z^n}{n^s}
\end{equation}
with the support of all positive integers $n=1,2,\dots$ defines the
Good distribution~\cite{Go1953}.  The constant~(\ref{normc}) is then
\begin{equation}
\label{jonquiere}
c^{-1} = F(z,s) = \sum_{n=1}^\infty \frac{z^n}{n^s}\,,
\end{equation}
where $\vert z \vert <1$, which is recognized as the Jonqui\`{e}re's
function [Eq.~(14) on p.~30 of Ref.~\cite{Ba1953vol1}; 
see also~\cite{Tr1945}]. Important properties of the 
Good distribution were derived in~\cite{KuTo1992,DoLu1997}.

In biophysics, one is interested in establishing the statistical
structure of DNA and RNA sequences.
Indeed, an interest in statistical compositions of DNA and protein sequences
originated shortly after the discovery that the nucleotide sequences
of RNA uniquely determines the amino-acid sequence of the proteins. 
An early analysis was performed in~\cite{GaYc1955}, 
where it was observed that the distributions of
relative abundances of amino-acids and nucleotides were nonrandom,
i.e. deviated from models assuming uniform distribution of different
words in any given text. Zipf distributions and related generalizations
proved to be very useful in further studies~\cite{Maetal1994,KoMa1995,MaKo1996}.

A formal generalization of the Zipf, the Zipf--Mandelbrot and the
Good distributions is possible upon realizing that the Riemann zeta,
the generalized zeta and the Jonqui\`{e}re functions constitute
special cases of the Lerch transcendent which is defined by the
following series
\begin{equation}
\label{lerch}
\Phi(z,s,v) = \sum_{n=0}^\infty \frac{z^n}{(n+v)^s}\,,
\end{equation}
where $|z|<1$ and $v \neq 0,-1,\dots$ [Eq.~(1) on p.~27 of
Ref.~\cite{Ba1953vol1}]. We use here the the nonstandard notation
of a dummy index $n$ in order to highlight the connection
with statistical theory. The relations between $\Phi(z,s,v)$ and
$\zeta(s)$, $\zeta(s,v)$ and $F(z,s)$ can be obtained by considering 
appropriate limits if $\Phi(z,s,v)$ (e.g., by letting $z$
and/or $v$ to 0 and/or 1), and by making use of the following functional
relation [Eq.~(2) on p.~27 of Ref.~\cite{Ba1953vol1}]
\begin{equation}
\label{lerchrel}
\Phi(z,s,v) = z^m \Phi(z,s,m+v) + \sum_{n=0}^{m-1} \frac{z^n}{(n+v)^s}\,.
\end{equation}
In particular, Eq.~(\ref{lerchrel}) with $m=1$ can be used
to evaluate sums that start from $n=1$ rather than from $n=0$:
\begin{equation}
\sum_{n=1}^\infty \frac{z^n}{(n+v)^s} = z \, \Phi(z,s,1+v)
\end{equation}

\begin{table}[htb]
\begin{center}
\begin{minipage}{14cm}
\begin{center}
\begin{tabular}{llll}
\hline
\hline\\[0.3ex]
Distribution & Support & p.m.f. & $c^{-1}$ \\
(related function) & & & \\[2ex]
\hline\\[0.3ex]
Zipf & 
$1,2,\dots$ & 
$c\,n^{-s}$ & 
$\zeta(s) = \Phi(1,s,1)$\\
(zeta function) & & & \\[2ex]
Zipf--Mandelbrot & 
$0,1,\dots$ & 
$c\,(n+v)^{-s}$ & 
$\zeta(s,v) = \Phi(1,s,v)$\\
(generalized zeta) & & & \\[2ex]
Good & 
$1,2,\dots$ & 
$c\,z^n\,n^{-s}$ & 
$F(z,s) = z \, \Phi(z,s,1)$\\
(Jonqui\`{e}re function) & & & \\[2ex]
Lerch & 
$0,1,\dots$ & 
$c\,z^n\,(n+v)^{-s}$ & 
$\Phi(z,s,v)$\\
(Lerch's transcendent) & & & \\[0.3ex]
\hline
\hline
\end{tabular}
\end{center}
\caption{Interrelationships between distributions defined by 
Riemann zeta, generalized zeta, Jonqui\`{e}re and Lerch functions.}
\label{speccases}
\end{minipage}
\end{center}
\end{table}

In Tab.~\ref{speccases}, we have summarized the relations between
the different statistical distributions and their defining special
functions. The table illustrates the fact that
the three distributions, Zipf (Zeta), Zipf--Mandelbrot
(Generalized zeta) and Good (Jonqui\`{e}re), are special cases of
the Lerch distribution, and that their properties can be expressed in
terms of Lerch's transcendent with special values of parameters.

We have found that the application of the 
combined nonlinear-condensation transformation (CNCT)
described in detail in Sec.~\ref{CNCT} is very suitable
for the accelerated evaluation of the Lerch transcendent
(\ref{lerch}) in the nonalternating case ($z > 0$).
Details of the implementation,
both in C and {\em Mathematica}~\cite{Wo1988},
are discussed in~\cite{JeAkMoSaSo2003,AkSaJeBeSoMo2003}.
For $z < 0$, the series (\ref{lerch}) is alternating.
In this case, the direct application of the 
delta transformation (\ref{dWenTr}) leads to 
an efficient calculational scheme~\cite{AkSaJeBeSoMo2003}.
A C code for the Lerch transcendent is available~\cite{AkHome,JeHome}.

%
% An Application in Experimental Mathematics
%
\subsection{An Application in Experimental Mathematics}
\label{sec_math}

We would like to begin this section by 
quoting~\cite{Ba1994tech}: ``In
April 1993, Enrico Au--Yeung, an undergraduate at the University of
Waterloo, brought to the attention of the author's [David Bailey's]
colleague Jonathan Borwein the curious fact that 
\begin{equation}
\label{SumBailey} 
\sum_{k=1}^\infty \;  \left( 1 + \frac{1}{2} + \cdots +
\frac{1}{k} \right)^2 \, k^{-2} \;=\; 4.59987\cdots \; \approx \;  
\frac{17}{4} \, \zeta(4) \;=\;  \frac{17 \pi^2}{360} \,, 
\end{equation} 
based on a computation of 500,000 terms. Borwein's reaction was to
compute the value of this constant to a higher level of precision in order
to dispel this conjecture.  Surprisingly, his computation to 30 digits
affirmed it. [David Bailey] then computed this constant to 100 decimal
digits, and the above equality was still affirmed.''
Many formulas similar to (\ref{SumBailey}) have subsequently been established
by rigorous proof~\cite{BaBoGi1994}. 

With the help of a multiprecision system~\cite{Ba1990tech,Ba1993,Ba1994tech} 
and the CNCT, we have verified (\ref{SumBailey}) ``experimentally''
to a couple of hundred decimals. The calculation will be
sketched in the following. Using the definition
\begin{equation}
{\bar b}(k) \;=\; 
\left( 1 + \frac{1}{2} + \ldots + \frac{1}{k} \right)^2 \, k^{-2}\,,
\end{equation}
we rewrite (\ref{SumBailey}) as follows,
\begin{eqnarray}
\sum_{k=0}^\infty \; {\bar b}(k)
& = & 
\sum_{k=0}^\infty \; 
\left( \sum_{j=0}^{k} \; \frac{1}{j+1} \right)^2 \, (k+1)^{-2}
\nonumber \\[2ex]
& = &
\sum_{k=0}^\infty \; \left( \frac{\psi(k+2) + C}{k+1} \right)^2 \,,
\end{eqnarray}
where $C$ is the {\em Euler-Mascheroni} constant
$C = 0.577~215\ldots$, and $\psi(z)$ is the 
logarithmic derivative of the 
Gamma function~\cite{Ol1974,Ba1953vol1}\,,
\begin{equation}
\psi(z) \;=\; \frac{\mathrm d}{{\mathrm d}z} \ln \Gamma(z)\,.
\end{equation}
With the help of the relation
\begin{equation}
\sum_{k=0}^\infty \; \frac{\psi(k+2)}{(k+1)^2} =
  2 \, \zeta(3) - C \, \zeta(2)\,,
\end{equation}
the conjecture (\ref{SumBailey}) can be rewritten as
\begin{equation}
\label{SimplifySumBailey}
\sum_{k=0}^\infty \; \left( \frac{\psi(k+2)}{k+1} \right)^2 =
\frac{17}{4} \, \zeta(4) - 4 C\,\zeta(3) + C^2 \, \zeta(2)\,.
\end{equation}
We proceed to calculate numerically, to high
precision, the infinite sum
\begin{equation}
\label{SumTodo}
\sum_{k=0}^\infty \; {\bar a}(k) \,, \qquad
{\bar a}(k) = \left( \frac{\psi(k+2)}{k+1} \right)^2\,,
\end{equation}
using the CNC transformation.

In order to establish the rate of convergence
of (\ref{SumTodo}), we investigate the asymptotic
behaviour of the ${\bar a}(k)$ as $k \to \infty$.
The logarithm of the Gamma function can be expanded into an 
asymptotic series [see Eq.~(4.03) on p.~294 of~\cite{Ol1974}]:
\begin{equation}
\label{LogGamma2} 
\ln \Gamma(z) \;=\; \left(z - \frac{1}{2} \right) \, \ln z
\;-\; z \;+\; 
\frac{1}{2} \, \ln (2 \pi) \;+\; 
\sum_{s=1}^{m-1} \; 
\frac{{\cal B}_{2s}}{2s(2s-1) z^{2s-1} } \;+\; R_m(z) \,, 
\end{equation}
where
\begin{equation}
R_m(z) \;=\; 
\int_0^\infty \frac{ {\mathcal B}_{2m} -
  {\mathcal B}_{2m}(x - \Ent{x})} 
    {2 m (x + z)^{2m}} {\mathrm d}x \,. 
\end{equation}
Here, $\Ent{x}$ is the integral part of
$x$, i.e., the largest integer $m$ satisfying $m \le x$, 
$B_k (x)$ is a Bernoulli polynomial defined by the 
generating function 
[see Eq.~(1.06) on p.~281 of Ref.~\cite{Ol1974}]:
\begin{equation}
\frac {t \exp(x t)}{\exp(t) - 1} \; = \;
\sum_{m=0}^{\infty} \, B_m(x) \, \frac {t^m} {m!} \, ,
\qquad \vert t \vert < 2 \pi \, ,
\end{equation}
and
\begin{equation}
B_m \; = \; B_m (0)
\label{BernNum}
\end{equation}
is a Bernoulli number (p.\ 281 of Ref.~\cite{Ol1974}).
The following asymptotic relation for $\psi(z)$ follows:
\begin{equation}
\label{PolyAsym}
\psi(z) = \ln z \;-\; 
\frac{1}{2z} \;-\; \sum_{s=1}^{m-1} \; 
\frac{{\mathcal B}_{2s}}{2s z^{2s} } \;+\; 
{\mathcal O}\left(\frac{1}{z^{2m}} \right)\,.
\end{equation}
The leading asymptotics of
the remainder of the 
sum (\ref{SumTodo}) after adding $N-1$ terms can thus be derived easily.
We have for large $k$,
\begin{equation} 
\label{asymp_a}
{\bar a}(k) \;  \sim \; 
\frac{\ln(k + 2)^2}{(k+1)^2} -
\frac{\ln(k + 2)}{(k + 1)^2 \, (k + 2)} -
\frac{\ln(k + 2)}{6\,(k + 1)^2 \, (k + 2)^2}
+ {\mathcal O}\left(\frac{1}{k^4} \right)  \,,
\qquad k \to \infty\,.
\end{equation}
Based on these formulas, 
the remainder of the sum (\ref{SumTodo}), for large $N$, 
can be written as
\begin{equation}
\label{emestimate}
\sum_{k=N}^\infty \; {\bar a}(k) \; \sim \;
\frac{\ln^2 N}{N} \;+\; \frac{\ln N}{N} \;+\; \frac{1}{N} \;+\;
{\mathcal O}\left( \frac{ \ln^2 N}{N^2} \right) \,. 
\end{equation}
Here, the Euler-Maclaurin formula [Eqs.~(2.01)
and~(2.02) on p.~285 of Ref.~\cite{Ol1974}] has been used in order
to convert the sum over the ${\bar a}(k)$ in the
asymptotic regime of large $k$ [see Eq.~(\ref{asymp_a})]
into an integral plus correction terms.
In order to calculate (\ref{SimplifySumBailey}) to an accuracy of
200 decimals, Eq.~(\ref{emestimate}) says
that we would be required to add
on the order of $10^{205}$ terms. Without the use 
of convergence acceleration methods, this would represent
a formidable computational task. 

Using the CNCT, it is easy to
calculate the sum (\ref{SimplifySumBailey}) to 200 digits, 
based on multiprecision
arithmetic~\cite{Ba1994tech} and a Linux personal computer, within a few
hours. We obtain for the 246th and the 247th CNC transform defined
according to Eq.~(\ref{defcnc}),
\begin{eqnarray}
\label{t246}
\!\!\! {\mathcal T}_{\mathrm{CNC}}(246) \!\!\! 
&=& \!\!\! 2.37254~51620~38445~67035~68130~69148~85258~25756~18499~54254
\nonumber\\[1ex]
& & \; 97013~57806~20011~72404~62937~46020~32218~23862~67095~00004
\nonumber\\[1ex]
& & \; 69194~36541~28946~10390~15116~52595~90270~23975~58737~74256
\nonumber\\[1ex]
& & \; 23420~48480~95165~00802~19816~35378~76591~98589~60393~32103,
\end{eqnarray}
and
\begin{eqnarray}
\label{t247}
\!\!\! {\mathcal T}_{\mathrm{CNC}}(247)  \!\!\! 
&=& \!\!\! 2.37254~51620~38445~67035~68130~69148~85258~25756~18499~54254
\nonumber\\[1ex]
& & \;  97013~57806~20011~72404~62937~46020~32218~23862~67095~00004
\nonumber\\[1ex]
& & \; 69194~36541~28946~10390~15116~52595~90270~23975~58737~74256
\nonumber\\[1ex]
& & \; 23420~48480~95165~00802~19816~35378~76591~98589~60393~32112.
\end{eqnarray}
The apparent convergence to 200 decimals can be verified
against the right-hand side of~Eq.~(\ref{SimplifySumBailey}).
Of course, the right-hand side of~Eq.~(\ref{SimplifySumBailey}),
which involves only rationals, zeta functions and
the Euler--Mascheroni constant,
\[
\frac{17}{4} \, \zeta(4) - 4 C\,\zeta(3) + C^2 \, \zeta(2)\,,
\]
can be easily evaluated to 200 decimals
using known algorithms which are included in 
computer algebra systems~\cite{Wo1988}.

The evaluation of the terms ${\bar a}(k)$ proceeds as follows.
For {\em small} index $k$, it is easy to write a recursion
relation relating ${\bar a}(k)$ and ${\bar a}(k+1)$ based on the 
(trivial) recursion for the $\psi$ function,
\begin{equation}
\psi(k+1) = \psi(k) + \frac{1}{k}\,.
\end{equation}
For {\em large} index $k$, the asymptotic formula (\ref{PolyAsym})
can be used in order to calculate the $\psi$ function to
high precision. The point at which index one may switch
from the recursion to the asymptotic method depends on how many
explicit values for Bernoulli numbers are available to the 
machine. We use the first values for the first 60 Bernoulli numbers,
to 250 decimals, for our calculation.
We switch from one method to the other when the index $k$
of ${\bar a}(k)$ has reached a value of 500.

With 84~308 ${\bar a}(k)$ terms evaluated (out of which 
1364 by recursion and 82944 by the asymptotic method),
we evaluate the first 247 transforms with the results
presented above in Eqs.~(\ref{t246}) and (\ref{t247}).
If the terms of the series (\ref{SumBailey}) were
added on a term-by-term basis, then about $10^{205}$
would be required for an accuracy of 200 decimals in the 
final result. The reduction of this number to
roughly 84~000 corresponds to an acceleration of the convergence by
roughly 200 orders of magnitude.
We have also carried out, using enhanced precision, 
a calculation to 430 decimals,
involving about 500 CNC transformations and
arithmetic with 600 decimal figures. These evaluations
not only confirm the conjecture (\ref{SimplifySumBailey}) to high 
precision,
but they also represent one of the most accurate experimental
verifications of the convergence properties of the 
delta transformation (\ref{dWenTr}) today.
Moreover, it is observed that the rate of convergence of the
CNC transforms, which is approximately one decimal figure per
transformation, remains constant over a wide range of transformation
orders. This observed behaviour is consistent with 
linear convergence [$\rho \approx 0.1$ in Eq.~(\ref{LimRatSeq})]
in higher transformation orders. By contrast,
the series (\ref{SimplifySumBailey}) is only logarithmically convergent.
This corresponds to convergence
acceleration according to the definition~(\ref{ConvAccelDef}). 

%
% Special Functions
%
\subsection{Other Applications of the CNCT}
\label{sec_other}

We would like to mention the existing applications in the domain
of quantum electrodynamic bound-state 
calculations~(see~\cite{JeMoSo1999,Je1999,JeMoSo2001pra} and
Ch.~\ref{QEDSelOne} of this Thesis). Another existing application
concerns the quantum electrodynamic effective action (see
Ref.~\cite{JeGiVaLaWe2002}).

J.~Becher~\cite{Be1999master} has investigated the applicability of the 
combined nonlinear-condensation transformation to
series of the form
\begin{equation}
\label{RpSer} 
R_p(x) \;=\; 
\sum_{k=0}^\infty \; \frac{x^{2k+1}}{(2k+1)^p} \,,
\end{equation}
as well as
\begin{equation}
\label{TpSer} 
T_p(x,b) \;=\; 
\sum_{k=0}^\infty \; \frac{1} {(2k+1)^p} \frac{ \cosh(2k+1)x}{\cosh(2k+1)b} \,,
\end{equation}
and
\begin{equation}
\label{UpSer} 
U_p(x,b) \;=\; 
\sum_{k=0}^\infty \; \frac{1}{(2k+1)^p}\frac{\cosh(2k+1)x}{\sinh(2k+1)b} \,. 
\end{equation}
Series of this type occur naturally in the context of plate
contact problems with periodic boundary conditions~\cite{DeKePaGl1984}.
The arguments $p$, $x$ and $b$ are real and positive for cases
of practical relevance. For $x \approx b$ and $p \approx 1$,
the series $T_p$ and $U_p$ are
very slowly convergent. In App.~A.2 of~\cite{Be1999master}, it
is demonstrated by way of numerical example that the CNCT is
able to efficiently accelerate the convergence of these series
in problematic parameter regions. 

In the numerical calculations, it is necessary to evaluate
terms with large index $k$. This can lead to numerical overflow 
because of the large arguments of the hyperbolic functions.
Clearly, representations like
\begin{equation}
T_p(x,b) \;=\; 
\sum_{k=0}^\infty \; \frac{e^{(2k+1)(x-b)}}{(2k+1)^p} \;
\frac{1+e^{-2x(2k+1)}}{1+e^{-2b(2k+1)}}
\end{equation}
provide a solution for this problem.

Let us recall that considerable effort has been invested in the 
development of efficient numerical methods for the evaluation
of the series (\ref{RpSer}) -- 
(\ref{UpSer})~\cite{Ma1997,BaGa1995,BoDe1992,Ga1991,DeLiDe1990}.
These alternative methods make intensive use of special properties
of the series. They involve integral transformations and
infinite series over numerical integral~\cite{Ga1991}, and
they make use of
special properties of Legendre's chi-function~\cite{BoDe1992}
which is related to the functions (\ref{RpSer}) --
(\ref{UpSer}).

A further application of convergence acceleration methods
discussed by J.~Becher~\cite{Be1999master} is the calculation
of the incomplete Gamma function $\Gamma(0,x)$ whose
asymptotic expansion for large argument reads
\begin{equation}
\Gamma(0,x) \;\sim\; 
\frac{e^{-x}}{x} \; \sum_{n=0}^\infty \; (-1)^n \; \frac{n!}{x^n}\,,
\qquad x \to \infty \,.
\;. \end{equation}
It is observed (see App.~A.3 
{\em ibid.}) that the divergent asymptotic series of this
function for large positive argument can be resummed effectively
using the delta transformation, without any recourse to the 
CNCT.  

%
% Conclusions and Outlook for Convergence Acceleration
%
\section{Conclusions and Outlook for Convergence Acceleration}

We have discussed several applications of the
convergence acceleration methods introduced in Sec.~\ref{ConvAccelConc}:
in statistical physics (Sec.~\ref{sec_stat}),
in experimental mathematics (Sec.~\ref{sec_math}),
and other applications, mainly in the evaluation of 
special functions (Sec.~\ref{sec_other}).  
Specifically, it is observed that a combination of 
the methods introduced in Secs.~\ref{NonlinearST}
and~\ref{CNCT} leads to an efficient calculational
scheme for the Lerch transcendent $\Phi$ defined in
Eq.~(\ref{lerch}). This special function provides a generalization
of several kinds of probability density functions which 
are of significance for the statistical analysis 
of DNA sequences (see Table~\ref{speccases}).

Other applications can be envisaged:
For example, the evaluation
of generalized hypergeometric functions --
notably of two-variable hypergeometric functions (Appell functions) $F_1$ and
$F_2$~\cite{ApKa1926}  -- has been historically problematic.
In the recent investigation~\cite{CoGaMi2001},
the evaluation of  $F_1(\alpha,\beta_1,\beta_2,\gamma;x,y)$
is considered: The algorithm consists in two steps: (i) the sophisticated use
of analytic continuations to map the point $(x,y)$ into the
convergence region of the defining hypergeometric series of the
Appell function, and (ii) the computation of the Appell function
within the region of convergence.
The analytic continuations give rise -- in addition
to $F_1$ functions of transformed argument -- to the $G_2$ function,
which can in turn be written in terms of Appell's $F_2$ function.
This implies that numerical methods also have to be devised for the
$F_2$ function.
The second step of the algorithm described in~\cite{CoGaMi2001} is the actual
numerical calculation of the Appell
function(s) ($F_1$ and $F_2$) within the region of convergence.
This calculation proceeds by investigating
the criterion $t_0 = {\rm max}(|x|,|y|) < 0.5$.
If the criterion is fulfilled,
then both the $F_1$ and the $F_2$ are evaluated
by single-index series expansions
in terms of ${}_2 F_1$'s, for example employing
the formula~\cite{BuCh1940,BuCh1941}
\begin{eqnarray}
\label{Burchnall}
\lefteqn{F_1(\alpha,\beta_1,\beta_2,\gamma;x,y) = \sum_{r = 0}^\infty
\frac{(\alpha)_r \, (\beta_1)_r \, (\beta_2)_r \, (\gamma - \alpha)_r}
  {(\gamma + r - 1)_r \, (\gamma)_{2r} \, r!}} \nonumber\\[2ex]
& & \;\;\; \times \;\; {}_2 F_1 (\alpha+r, \beta_1+r,\gamma+2 r, x) \,
{}_2 F_1 (\alpha+r, \beta_2+r,\gamma+2 r, y) \,.
\end{eqnarray}
If the criterion is not fulfilled,
the authors resort to the solution of a third-order
differential equation for the evaluation of the
Appell function [starting from the point $(x,y) = (0,0)$
where $F_1$ vanishes]. It appears possible that 
an application of convergence acceleration algorithms to the
series (\ref{Burchnall}) could lead to an efficient
calculational scheme in the problematic case $t_0 > 0.5$
where the authors of~\cite{CoGaMi2001} had to resort
to an integration of a third-order
differential equation. 

Possible improvements over existing algorithms concern 
``multi-stage transformations''. For example,
it was observed in~\cite{JeWeSo2000} that the rate of 
convergence of the output data
resulting from Borel summation (see Sec.~\ref{sec_bp})
of a specific divergent series describing the quantum 
electrodynamic effective action
could be further accelerated by using two iterations
of the epsilon algorithm [defined in Eq.~\ref{epsalg})],
i.e.~by evaluating Pad\'{e} approximants to the 
Pad\'{e} approximants of the Borel transforms (in this
sequence). Iterated forms of a number of convergence
accelerators have been discussed in~\cite{We1989}.
In general, the iteration of suitable convergence
accelerators depends on a thorough analysis of the 
remainder structure resulting from previous ``stages''
of the transformation, and it is not {\em a priori}
guaranteed that a ``multi-stage transformation'' with only
{\em one} iterated algorithm will help. In principle,
the use of iterated
algorithms can only be helpful when the evaluation of the sequence
transforms or Pad\'{e} approximants is computationally
cheap in comparison to the evaluation of further terms
of the input sequence which could potentially be used to
calculate higher-order terms in a single-stage
transformation process.

%
% Divergent Series
%
\chapter{Divergent Series}

\typeout{================}
\typeout{Divergent Series}
\typeout{================}
\typeout{}

\label{DivIntrod}

%
% Introduction
%
\section{Introduction to Divergent Series in Physics}
\label{IntroDivSer}

The observation that many perturbation series encountered
in quantum mechanics and quantum field theory are divergent,
sparked the development of large-order perturbation theory~\cite{LGZJ1990}.
This subject is inherently connected to the resummation
of divergent series, and with the development of appropriate 
numerical methods which exploit all available information
contained in a necessarily finite number of perturbative coefficients.
Of prime importance in this context is the Borel method
to be discussed in Sec.~\ref{sec_bp}.

As already stressed in Ch.~\ref{DivIntrod}, there are connections
between the convergence acceleration methods discussed in
Ch.~\ref{ConvAccel} and the resummation methods discussed in
the current Chapter. Specifically, the methods introduced
in Secs.~\ref{PadeApproximation} and~\ref{NonlinearST} will
be used below in Sec.~\ref{DivQEDEff} in the context of the
resummation of the divergent perturbation series for the 
QED effective action.

Essentially, we will consider four physical applications
of resummation methods: (i) the energy displacement of a hydrogen
atom in a background electric field, including the autoionization
width (Sec.~\ref{sec_stark}), 
(ii) zero-dimensional field theories (Sec.~\ref{degen}),
(iii) the QED effective Lagrangian for background magnetic
and electric fields (Sec.~\ref{DivQEDEff}), 
and (iv) the energy levels of the double-well
potential which receive nonperturbative contributions from
multi-instanton effects (Sec.~\ref{DivMATIns}).

Several important aspects and basic concepts connected with 
divergent series and the large-order
behavior of perturbation theory can be illustrated
by investigating the (divergent) perturbation series
of the Stark effect~\cite{St1913}. This effect, which is also known as the 
LoSurdo--Stark effect in view of the 
existence of the investigation~\cite{Su1913}), describes the energy
displacement of a hydrogen atom in a static external
electric field and can be expressed as a
perturbation series (power series) in the electric
field strength.

The first nonvanishing perturbation (in atomic units) is the second-order effect
\[
F^2 \, \sum_{m \neq \mathrm{1S}}
\frac{\langle \phi_{\mathrm{1S}} | z | \phi_m \rangle \,
\langle \phi_m | z | \phi_{\mathrm{1S}} \rangle}{E_{\mathrm{1S}} - E_m}\,,
\]
where the sum over $m$ runs over the entire spectrum,
including the continuum, but excluding the 1S ground state,
and $E_m$ is the nonrelativistic (Schr\"{o}dinger) energy of the $m$th state.
A well known, but perhaps surprising result says that
the coefficients of the terms of order $F^4$, $F^6$, $F^8,\dots$
grow so rapidly that the series in $F$ ultimately diverges,
irrespective of how small the field strength is.
The convergence radius of the factorially divergent perturbation
series is zero.
The perturbation series for the
Stark effect~\cite{Su1913,St1913} can be formulated to
arbitrarily high order~\cite{Si1978}. The perturbative coefficients
grow factorially in absolute magnitude~\cite{BeGrHaSi1979}.
The perturbation series is a divergent, asymptotic expansion in
the electric field strength $F$, i.e.~about zero electric field.  This
means that the perturbative terms at small coupling first decrease in
absolute magnitude up to some minimal term.  After passing through the
minimal term, the perturbative terms increase again in magnitude, and
the series ultimately diverges.

By the use of a {\em resummation} method, it is possible to assign a
finite value to an otherwise divergent series, and various
applications of resummation methods in mathematics and physics have
been given, e.g., in~\cite{BeOr1978,We1989,BaGr1996,DuHa1999,JeBeWeSo2000}.
When a divergent series is resummed (cf. Ch.~\ref{ConDivInt}),
the superficial precision limit set by the
minimal term can be overcome,
and more accurate results can be obtained as compared to the
optimal truncation of the perturbation series.
However, the resummation of the relevant divergent
series is problematic in the considered case,
because the Borel transform, from which the
physically correct, finite
result is obtained by evaluating a {\em generalized} Borel
integral -- see Eq.~(\ref{ContourIntegral}) in Sec.~\ref{sec_bp} below --,
exhibits a rich singularity structure in the complex plane.
The perturbation series is formally not Borel summable.

The Stark effect and its associated divergent
perturbative expansion including 
the nonperturbative, nonanalytic imaginary contributions have attracted
considerable attention, both theoretically and
experimentally~\cite{Si1978,BeGrHaSi1979,SiKo1979,Op1928,Al1969,%,
HeInBr1974,GuNi1975,FrBr1975,DaKo1976,YaTaSi1977,%,
DaKo1978,GrGr1978b,HeSi1978,%,
BeGrHaSi1979err,DaKo1979,SiAdCiOt1979,%,
AvEtAl1979,AdEtAl1980,BeGr1980,LKBa1980a,LKBa1980b,SiHaGr1981,%,
MaChRe1983,FaRe1983,%,
FrGrSi1985,GlNgYaNa1985,Ko1987,Ko1989,AlSi1989,%,
AlDaSi1991,Fe1992,NiTh1992,AlSi1994,ZaCiSk2000}. Experiments have
been performed in field strengths up to a couple of
MV/cm~\cite{Ko1978,BeEtAl1984,GlNa1985,RoWe1986}.

Rather mathematically motivated investigations regarding the Borel summability
of the divergent perturbation series for the Stark effect were
performed in~\cite{GrGr1978b} and~\cite{CaGrMa1993}, and it
was established that
the perturbation series of the Stark effect
is Borel summable in the distributional
sense (for the definition of ``distributional Borel summability''
we refer to~\cite{CaGrMa1986}).

In comparison to the previous investigation~\cite{FrGrSi1985}, 
we use here an optimized resummation
method that combines an analytic continuation of the Borel transform 
with Pad\'{e} approximants in the conformal variable and 
the explicit use of the ``leading renormalon pole''. All of
these methods will be described in Secs.~\ref{sec_bp} and~\ref{sec_cm}.
These optimized Borel transforms represent
an alternative to so-called order-dependent mappings which have
been discussed intensively in the literature~\cite{SeZJ1979,LGZJ1983,%,
GuKoSu1995,GuKoSu1996}.

After the discussion of the Stark effect (Secs.~\ref{sec_bp} --
\ref{sec_nc}), we turn our attention
to the resummation of the divergent perturbation series
for zero-dimensional field theories (Sec.~\ref{degen}),
to the QED effective action (Sec.~\ref{DivQEDEff}),
and to the double-well potential (Sec.~\ref{DivMATIns}).

%
% The Stark Effect: A Paradigmatic Example
%
\section{The Stark Effect: A Paradigmatic Example}
\label{sec_stark}

%
% PERTURBATION SERIES FOR THE LoSURDO--STARK EFFECT
%
\subsection{Perturbation Series for the Stark Effect}
\label{sec_ps}

In the presence of an electric field, the ${\rm SO}(4)$ symmetry of
the hydrogen atom is broken, and parabolic quantum numbers $n_1$,
$n_2$ and $m$ are used for the classification of the atomic
states~\cite{LaLi1979}. For the Stark effect, the perturbative
expansion of the energy eigenvalue $E(n_1,n_2,m,F)$ reads [see
Eq.~(59) of~\cite{Si1978}],
\begin{equation}
\label{PertSer}
E(n_1,n_2,m,F) \sim \sum_{N=0}^{\infty}
E^{(N)}_{n_1 n_2 m} \, F^N\,,
\end{equation}
where $F$ is the electric field strength.  For $N \to \infty$, the leading
large-order factorial asymptotics of the perturbative coefficients have been
derived in~\cite{SiAdCiOt1979} as
\begin{equation}
\label{LargeN}
E^{(N)}_{n_1 n_2 m} \sim A^{(N)}_{n_1 n_2 m} +
(-1)^N \, A^{(N)}_{n_2 n_1 m}\,, \quad N \to \infty\,,
\end{equation}
where $A^{(N)}_{n_i n_j m}$ is given as an asymptotic series,
\begin{equation}
\label{DefAN}
A^{(N)}_{n_i n_j m} \sim K(n_i,n_j,m,N) \,
\sum_{k=0}^{\infty} a^{n_i n_j m}_k \, (2\,n_j + m + N - k)!\,.
\end{equation}
The quantities $a^{n_i n_j m}_k$ are constants.
The $K$-coefficients in Eq.~(\ref{DefAN}) are given by
\begin{equation}
\label{Prefactor}
K(n_i,n_j,m,N) = 
- \left[ 2\pi n^3 n_j! \, (n_j + m)! \right]^{-1}\,
\exp\left\{ 3 \, (n_i - n_j) \right\} \,
6^{2 \, n_j + m + 1} \, (3 n^3/2)^N \,.
\end{equation}
Here, the principal quantum number $n$ as a function of the parabolic
quantum numbers $n_1$, $n_2$ and $m$ is given by [see Eq.~(65)
in~\cite{Si1978}]
\begin{equation}
\label{PrincQuant}
n = n_1 + n_2 + |m| + 1\,.
\end{equation}
According to Eq.~(\ref{LargeN}), the perturbative coefficients
$E^{(N)}_{n_1 n_2 m}$, for large order $N \to \infty$ of perturbation
theory, can be written as a sum of a nonalternating factorially
divergent series [first term in Eq.~(\ref{LargeN})] and of an
alternating factorially divergent series [second term in
Eq.~(\ref{LargeN})].  Because the $a^{n_i n_j m}_k$ in
Eq.~(\ref{DefAN}) are multiplied by the factorial $(2\,n_i + m + N -
k)!$, we infer that for large perturbation theory order $N$, the term
related to the $a^{n_i n_j m}_0$ coefficient ($k = 0$) dominates.
Terms with $k \geq 1$ are suppressed in relation to the leading term
by a relative factor of $1/N^k$ according to the asymptotics
\begin{equation}
\label{AsyN}
\frac{(2\,n_j + m + N - k)!}{(2\,n_j + m + N)!} \sim
\frac{1}{N^k} \, \left[ 1 + {\mathcal O}\left(\frac{1}{N}\right)\right]\,
\end{equation}
for $N \to \infty$. The leading ($k=0$)--coefficient has been evaluated
in~\cite{BeGrHaSi1979} as
\begin{equation}
\label{ninjm0}
a^{n_i n_j m}_0 = 1\,.
\end{equation}
According to Eqs.~(\ref{LargeN}), (\ref{DefAN}) and (\ref{ninjm0}),
for states with $n_1 < n_2$, the nonalternating component of the
perturbation series dominates in large order of perturbation theory,
whereas for states with $n_1 > n_2$, the alternating component is
dominant as $N \to \infty$. For states with $n_1 = n_2$, the odd-$N$
perturbative coefficients vanish~\cite{SiAdCiOt1979}, and the even-$N$
coefficients necessarily have the same sign in large order [see
Eq.~(\ref{LargeN})]. According to Eq.~(\ref{LargeN}), there are
subleading divergent nonalternating contributions for states with $n_1
> n_2$, and there exist subleading divergent alternating contributions
for states with $n_1 < n_2$. This complicates the resummation of the
perturbation series.

%
% figure 1
%
\begin{figure}[htb!]
\begin{center}
\begin{minipage}{14cm}
\centerline{\mbox{\epsfysize=10.0cm\epsffile{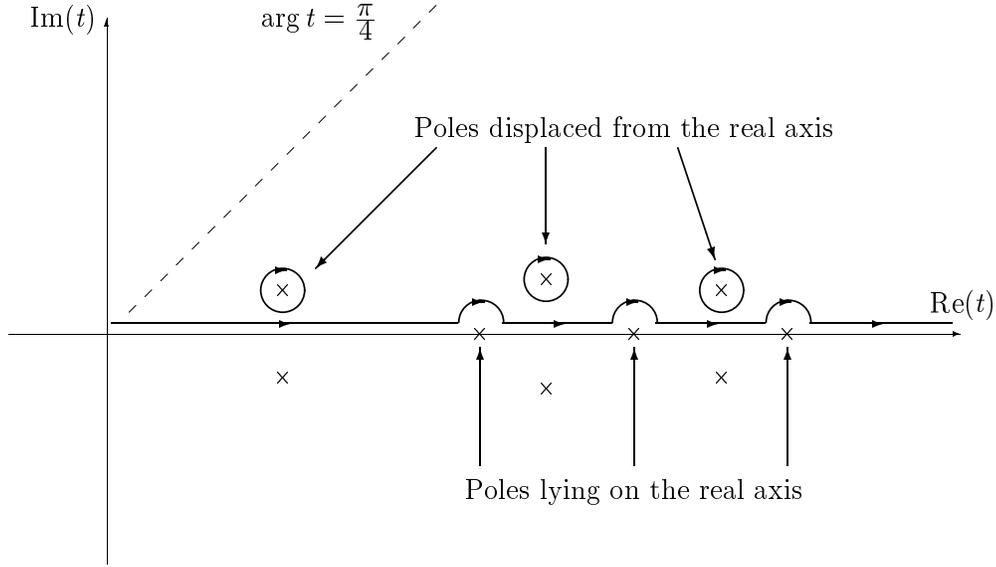}}}
\caption{\label{figg1} Integration contour
$C_{+1}$ for the evaluation of the
generalized Borel integral defined
in Eq.~\protect{(\ref{ContourIntegral})}.
Poles displaced from the real axis are evaluated
as full poles, whereas those poles which lie on the
real axis are treated as half poles.}
\end{minipage}
\end{center}
\end{figure}

%
% BOREL--PAD\'{E} RESUMMATION
%
\subsection{Borel--Pad\'{e} Resummation}
\label{sec_bp}

The resummation of the perturbation series~(\ref{PertSer}) 
by a combination of the Borel transformation 
and Pad\'{e} approximants proceeds as
follows. First we define the parameter
\begin{equation}
 \label{deflambda}
\lambda = 2 \, {\rm max}(n_1,n_2) + m + 1\,.
\end{equation}
The large-order growth of the perturbative coefficients [see
Eqs.~(\ref{LargeN}) and (\ref{DefAN})] suggests the definition of the
(generalized) Borel transform [see Eq.~(4) in~\cite{JeWeSo2000}]
\begin{eqnarray}
\label{BorelTrans}
E_{\rm B}(z) & \equiv & E_{\rm B}(n_1,n_2,m,z) \nonumber\\[2ex]
& = & {\mathcal B}^{(1,\lambda)}\left[E(n_1,n_2,m); \, z\right] 
\nonumber\\[1ex]
& = & \sum_{N=0}^{\infty}
\frac{E^{(N)}_{n_1 n_2 m}}{\Gamma(N + \lambda)} \, z^N\,,
\end{eqnarray}
where we consider the argument $z$ of $E_{\rm B}(z)$ as a complex
variable and $\lambda$ is defined in Eq.~(\ref{deflambda}). 
The additive constant (in this case $\lambda$)
in the argument of the Gamma function
is chosen in accordance with the notion of an ``asymptotically
improved'' resummation (see also~\cite{JeWeSo2000}).
It is observed that the additive constant $\lambda$
can be shifted by a small integer without 
affecting the convergence of the Borel resummed series.
Because the perturbative coefficients $E^{(N)}_{n_1 n_2 m}$
diverge factorially in absolute magnitude, the 
Borel transform $E_{\rm B}(z)$ has a finite radius of 
convergence about the origin.  The
evaluation of the (generalized)
Laplace--Borel integral~[see Eq.~(\ref{ContourIntegral})
below] therefore requires an analytic continuation of $E_{\rm B}(z)$
beyond the radius of convergence.  The ``classical'' Borel integral is
performed in the $z$-range $z \in (0,\infty)$, i.e.~along the positive
real axis [see e.g.~Eqs.~(8.2.3) and~(8.2.4) of~\cite{BeOr1978}].  It
has been suggested in~\cite{Pi1999} that the analytic continuation
of~(\ref{BorelTrans}) into regions where $F$ retains a nonvanishing,
albeit infinitesimal, imaginary part can be achieved by evaluating
Pad\'{e} approximants.  Using the first $M+1$ terms in the power
expansion of the Borel transform $E_{\rm B}(z)$, we construct the
Pad\'{e} approximant~[we follow the notation of~\cite{BaGr1996},
see also Eq.~(\ref{pade})]
\begin{equation}
\label{padeapp}
{\mathcal P}_M(z) = \bigg[ [\mkern - 2.5 mu [M/2] \mkern - 2.5 mu ] \bigg/
[\mkern - 2.5 mu [(M+1)/2] \mkern - 2.5 mu ]
\bigg]_{E_{\rm B}}\!\!\!\left(z\right)\,,
\end{equation}
where $[\mkern - 2.5 mu [x] \mkern - 2.5 mu ]$ denotes the largest
positive integer smaller than $x$.  We then evaluate the (modified)
Borel integral along the integration contour $C_{+1}$ shown in
Fig.~\ref{figg1} in order to construct the transform ${\mathcal T}\!E_M(F)$
where
\begin{equation}
\label{ContourIntegral}
{\mathcal T}\!E_M(F) = \int_{C_{+1}} {\rm d}t \, t^{\lambda - 1}\,
\exp(-t)\,{\mathcal P}_M(F\,t) \,.
\end{equation}
The successive evaluation of transforms ${\mathcal T}\!E_M(F)$ in increasing
transformation order $M$ is performed, and the apparent convergence of
the transforms is examined. This procedure is illustrated in Tables~I
and II of~\cite{Je2000prd}.  

The contour
$C_{+1}$ is supposed to 
encircle all poles at $t = z_i$ in the upper right quadrant 
of the complex plane 
with $\arg \, z_i < \pi / 4$ in the mathematically negative sense. 
That is to say,
the contribution of all poles $z_i$ with ${\rm Re}\,z_i > 0$,
${\rm Im}\,z_i > 0$ and ${\rm Re}\,z_i > {\rm Im}\,z_i$,
\[
- 2\,\pi\,{\rm i}\, \sum_i \,\Res{t=z_i} t^{\lambda - 1}\,
\exp(-t)\,{\mathcal P}_M(F \, t)\,,
\]
is added to the principal value (P.V.)
of the integral (\ref{ContourIntegral})
carried out in the range $t \in (0,\infty)$. 
We have,
\begin{eqnarray}
\label{ContourIntegral2}
{\mathcal T}\!E_M(F) &=& {\rm (P.V.)}
\int_0^\infty {\rm d}t \, t^{\lambda - 1}\, \exp(-t)\,{\mathcal P}_M(F\,t)
\nonumber\\[1ex]
& & - 2\,\pi\,{\rm i}\, \sum_i \,\Res{t=z_i} t^{\lambda - 1}\,
\exp(-t)\,{\mathcal P}_M(F \, t) \,.
\end{eqnarray}
The principal-value prescription [first term in 
Eq.~(\ref{ContourIntegral2})]
for the evaluation of the Laplace--Borel integral
has been recommended in~\cite{Pi1999,Raczka1991}.
This prescription leads to a real (rather than complex) result for the
energy shift and cannot account for the width of the 
quasistationary state. 
The additional pole contributions [second term in
Eq.~(\ref{ContourIntegral2})]
are responsible for complex-valued (imaginary) corrections which lead, in
particular, to the decay width.  

By contour integration (Cauchy Theorem) and Jordan's Lemma,
one can show that the result obtained along $C_{+1}$ is 
equivalent to an integration along the straight line with
$\arg z = \pi/4$,
\begin{equation}
\label{ContourIntegral3}
{\mathcal T}\!E_M(F) = c^\lambda \, \int_0^\infty 
{\rm d}t \, t^{\lambda - 1}\, \exp(-c\,t)\,{\mathcal P}_M(F \, c\, t)
\end{equation}
where $c = \exp(i\,\pi/4)$. This contour has been used 
in~\cite{FrGrSi1985} (see also p.~815 in~\cite{ZJ1996}).
The exponential factor $\exp(-c\,t)$ and the asymptotic behavior of the
Pad\'{e} approximant ${\mathcal P}_M(F \, c\, t)$ as $t \to \infty$
together ensure that the integrand falls off sufficiently rapidly so that
the Cauchy Theorem and
Jordan's Lemma can be applied to show the equivalence of the 
representations (\ref{ContourIntegral2}) and (\ref{ContourIntegral3}).

The representation (\ref{ContourIntegral3}) illustrates the fact
that the integration in the complex plane along $C_{+1}$ analytically
continues the resummed result in those cases where the evaluation of the
standard Laplace--Borel integral is not feasible due to poles on the
real axis. The representations (\ref{ContourIntegral}) and
(\ref{ContourIntegral2}) serve to clarify the r\^{o}le of the additional
terms which have to be added to the result obtained by the principal-value
prescription in order to obtain the full physical result, including
the nonperturbative, nonanalytic contributions. Note that, as 
stressed in~\cite{Je2000prd}, the pole contributions in general do not
only modify the imaginary, but also the real part of the resummed
value for the perturbation series.

Formally, the limit of the sequence of the
${\mathcal T}\!E_M(F)$ as $M\to\infty$, provided it exists,
yields the nonperturbative result
inferred from the perturbative expansion~(\ref{PertSer}),
\begin{equation}
\label{Limit}
\lim_{M \to \infty} {\mathcal T}\!E_M(F) = E(F) \equiv 
E(n_1,n_2,m,F)\,.
\end{equation}
Because the contour $C_{+1}$ shown in Fig.~\ref{figg1} extends into
the complex plane, the transforms ${\mathcal T}\!E_M(F)$ acquire an
imaginary part even though the perturbative coefficients in
Eq.~(\ref{PertSer}) are real.

In the context of numerical analysis, the concept of
incredulity~\cite{Wi1966} may be used for the analysis of the
convergence of the transforms ${\mathcal T}\!E_M(F)$ of increasing order
$M$. If a certain number of subsequent transforms exhibit apparent
numerical
convergence within a specified relative accuracy, then the calculation
of transforms is stopped, and the result of the last calculated
transformation is taken as the numerical limit of the series under
investigation.  It has been observed in~\cite{Je2000prd,Pi1999} that for
a number of physically relevant perturbation series, the apparent
numerical convergence of the transforms (\ref{ContourIntegral}), with
increasing transformation order, leads to the physically correct
results. 

It is observed that the rate of convergence of the transforms 
(\ref{ContourIntegral}) can be enhanced if instead of the 
unmodified Pad\'{e} approximants (\ref{padeapp}) leading 
``renormalon'' poles are explicit used for the construction
of modified approximants. For the ground state,
this entails the following replacement in
Eq.~(\ref{ContourIntegral}):
\[
{\mathcal P}_M(z) \to {\mathcal P'}_M(z)\,,
\]
where
\begin{equation}
\label{replacement000}
{\mathcal P'}_M(z) = \frac{1}{1-z^2}\,
\bigg[ [\mkern - 2.5 mu [\frac{M+4}{2}] \mkern - 2.5 mu ] \bigg/
[\mkern - 2.5 mu [\frac{M-3}{2}] \mkern - 2.5 mu ]
\bigg]_{E'_{\rm B}(\zeta)}\!\!\!\left(z\right)\,,
\end{equation}
where $E'_{\rm B}(\zeta) = (1-\zeta^2)\,E_{\rm B}(\zeta)$.
For the excited state with quantum numbers $n_1 = 3$,
$n_2 = 0$ and $m = 1$, we replace
\[
{\mathcal P}_M(z) \to {\mathcal P''}_M(z)\,,
\]
where
\begin{equation}
\label{replacement301}
{\mathcal P''}_M(z) = \frac{1}{1-z}\,
\bigg[ [\mkern - 2.5 mu [\frac{M+2}{2}] \mkern - 2.5 mu ] \bigg/
[\mkern - 2.5 mu [\frac{M-1}{2}] \mkern - 2.5 mu ]
\bigg]_{E''_{\rm B}}\!\!\!\left(z\right)\,,
\end{equation}
where $E''_{\rm B}(\zeta) = (1-\zeta^2)\,E_{\rm B}(\zeta)$.
The resummation method by a combination of Borel and Pad\'{e}
techniques -- current Section -- will be
referred to as ``method I'' in the sequel.

%
% DOUBLY--CUT BOREL PLANE AND RESUMMATION BY CONFORMAL MAPPING
%
\subsection{Doubly--Cut Borel Plane}
\label{sec_cm}

According to Eqs.~(\ref{LargeN}) and (\ref{DefAN}), the perturbative
coefficient $E^{(N)}_{n_1 n_2 m}$, for large $N$, can be written as
the sum of an alternating and of a nonalternating divergent series. In
view of Eqs.~(\ref{Prefactor}) and (\ref{ninjm0}), we conclude
that the series defined in~Eq.~(\ref{BorelTrans}),
\[          
E_{\rm B}(z) = \sum_{N=0}^{\infty}    
\frac{E^{(N)}_{n_1 n_2 m}}{\Gamma(N + \lambda)} \, z^N\,,
\]
has a radius of convergence 
\begin{equation}   
\label{defS}
s = \frac{2}{3 \, n^3}
\end{equation}  
about the origin, where $n$ is the principal quantum number
[see Eq.~(\ref{PrincQuant})]. Therefore, the function
\begin{equation}      
\label{calEB}
{\mathcal E}_{\rm B}(w) = \sum_{N=0}^{\infty}
\frac{E^{(N)}_{n_1 n_2 m}\,\,s^N}{\Gamma(N + \lambda)} \, w^N\,, 
\end{equation}
has a unit radius of convergence about the origin.  It is not {\em a
  priori} obvious if the points $w = -1$ and $w = +1$ represent
isolated singularities or branch points.  The asymptotic properties
(\ref{LargeN}) and (\ref{DefAN}) together with Eq.~(\ref{AsyN})
suggest that the points $w = -1$ and $w = +1$ do not constitute poles
of finite order.  We observe that the leading factorial growth of the
perturbative coefficients in large perturbation order $N$ is divided
out in the Borel transform (\ref{calEB}), which is a sum over $N$. The
perturbative coefficient $E^{(N)}_{n_1 n_2 m}$ can be written as an
asymptotic series over $k$ [see Eq.~(\ref{DefAN})].  We interchange
the order of the summations over $N$ and $k$, we use Eq.~(\ref{AsyN})
and take advantage of the identity
\begin{equation}
\label{polylog}
\sum_{N=0}^{\infty} \frac{w^N}{N^k} = {\rm Li}_k (w)\,.
\end{equation}
The Borel transform ${\mathcal E}_{\rm B}(w)$ can then be written
as a sum over terms of the form $T_k(w)$ where for $k \to \infty$,
\begin{equation}
T_k(w) \sim C(n_i,n_j,m) \, a_k^{n_i n_j m} \, {\rm Li}_k (w)\,.
\end{equation}
The coefficient $C(n_i,n_j,m)$ is given by
\begin{equation}
C(n_i,n_j,m) = - \left[ 2\pi n^3 n_j! \, (n_j + m)! \right]^{-1}
\exp\left\{ 3 \, (n_i - n_j) \right\} \, 6^{2 \, n_j + m + 1} \,.
\end{equation}
These considerations suggest that the points $w = -1$ and $w = +1$
represent essential singularities (in this case, branch points) of the
Borel transform ${\mathcal E}_{\rm B}(w)$ defined in Eq.~(\ref{calEB}).
For the analytic continuation of ${\mathcal E}_{\rm B}(w)$ by conformal
mapping, we write $w$ as
\begin{equation}      
\label{conformal1}
w = \frac{2\,y}{1 + y^2}
\end{equation}
(this conformal mapping preserves the origin of the complex plane).
Here, we refer to $w$ as the Borel variable, and we call $y$ the
conformal variable.  We then express the $M$th partial sum of
(\ref{calEB}) as
\begin{eqnarray}
\label{calEBM}
\lefteqn{{\mathcal E}^{M}_{\rm B}(w) = \sum_{N=0}^{M}
\frac{E^{(N)}_{n_1 n_2 m}\,s^N}{\Gamma(N + \lambda)} \, w^N}
\nonumber\\[1ex]
& & \quad = \sum_{N=0}^{M} C_N \, y^N + {\mathcal O}(y^{M+1})\,,
\end{eqnarray}
where the coefficients $C_N$ are uniquely determined [see, e.g.,
Eqs.~(36) and (37) of \cite{CaFi1999}].  We define the partial sum of the
Borel transform, re-expanded in terms of the conformal variable
$y$, as
\begin{equation}
\label{PartialConformal}
{{\mathcal E}'}^{M}_{\rm B}(y) = \sum_{N=0}^{M} C_N\,y^N\,.
\end{equation}
We then evaluate (lower-diagonal) Pad\'{e} approximants to the
function ${{\mathcal E}'}^{M}_{\rm B}(y)$,
\begin{equation}
\label{ConformalPade}
{{\mathcal E}''}^{M}_{\rm B}(y) =
\bigg[ [\mkern - 2.5 mu [M/2] \mkern - 2.5 mu ] \bigg/
[\mkern - 2.5 mu [(M+1)/2] \mkern - 2.5 mu ]
\bigg]_{{{\mathcal E}'}^{M}_{\rm B}}\!\!\!\left(y\right)\,.
\end{equation}
We define the following transforms,
\begin{equation}
\label{AccelTrans}
{{\mathcal T}''}\!E_M(F) = s^\lambda\,
\int_{C_{+1}} {\rm d}w \,w^{\lambda - 1} \,\exp\bigl(-w\bigr)\,
{{\mathcal E}''}^{M}_{\rm B}\bigl(y(w)\bigr)\,.
\end{equation}
At increasing $M$, the limit as $M\to\infty$, provided it exists, is
then again assumed to represent the complete, physically relevant
solution,
\begin{equation}
E(F) = \lim_{M\to\infty} {{\mathcal T}''}\!E_M(F)\,.
\end{equation}
We do not consider the question of the existence of this limit here
(for an outline of questions related to these issues we refer
to~\cite{CaFi2000}; potential problems at excessively strong
coupling are discussed in Sec. IIC of~\cite{LGZJ1980}).

Inverting Eq.~(\ref{conformal1}) yields [see Eq.~(\ref{AccelTrans})]
\begin{equation}
\label{conformal2}
y(w) = \frac{\sqrt{1+w}-\sqrt{1-w}}{\sqrt{1+w}+\sqrt{1-w}}\,.
\end{equation}
The conformal mapping given by Eqs.~(\ref{conformal1}) and
(\ref{conformal2}) maps the doubly cut $w$-plane with cuts running
from $w = 1$ to $w = \infty$ and $w = -1$ to $w = -\infty$ unto the
unit circle in the complex $y$-plane (i.e., it is a surjective mapping). 
The cuts themselves are mapped
to the edge of the unit circle in the $y$-plane.  

In comparison to the investigations~\cite{CaFi1999} and~\cite{CaFi2000}, 
we use here a
different conformal mapping defined in Eqs.~(\ref{conformal1})
and~(\ref{conformal2}) which reflects the different singularity
structure in the complex plane [cf.~Eq.~(27) in~\cite{CaFi1999}]. We
also mention the application of Pad\'{e} approximants for the
numerical improvement of the conformal mapping performed according to
Eq.~(\ref{ConformalPade}).  In comparison 
to~\cite{JeSo2001}, where the additional
Pad\'{e}--improvement in the conformal variable is also used, we
perform here the analytic continuation by a mapping whose structure
reflects the {\em double} cuts suggested by the asymptotic properties of the
perturbative coefficients given in Eqs.~(\ref{LargeN}), (\ref{DefAN})
and (\ref{AsyN}) [cf.~Eq.~(5) in~\cite{JeSo2001}].

The method introduced in this Section will be referred to as ``method
II''. One goal of the current investigation is to
contrast and compare the two methods I and II.  A comparison of
different approaches to the resummation problem for series with both
alternating and nonalternating divergent components appears useful, in
part because the conformal mapping (without further Pad\'{e}
improvement) has been recommended for the resummation of quantum
chromodynamic perturbation series~\cite{CaFi1999,CaFi2000}.

%
% NUMERICAL CALCULATIONS
%
\subsection{Numerical Calculations for the Stark Effect}
\label{sec_nc}

In this section, the numerical results based on the resummation
methods introduced in Secs.~\ref{sec_bp} and~\ref{sec_cm} are
presented. Before we describe the calculation in detail, we should
note that relativistic corrections to both the real and the imaginary
part of the energy contribute at a relative order of $(Z\alpha)^2$
compared to the leading nonrelativistic effect which is treated here 
(and in the previous work on the subject, see
e.g.~\cite{SiAdCiOt1979,FrGrSi1985}). Therefore, the theoretical
uncertainty due to relativistic effects
can be estimated to be, at best, 1 part in $10^4$ (for an outline of
the relativistic and quantum electrodynamic corrections in hydrogen
see~\cite{BeLiPi1973,JoSo1985,Mo1996,JePa1996,JeSoMo1997,%,
MoPlSo1998,EiGrSh2001}).
Measurements in very high fields are difficult~\cite{Ko1978}. At the
achievable field strengths to date (less than $0.001\,{\rm a.u.}$ or
about $5\,{\rm MV}/{\rm cm}$), the accuracy of the theoretical
prediction exceeds the experimental precision, and relativistic
effects do not need to be taken into account.

The perturbative coefficients $E^{(N)}_{n_1 n_2 m}$ defined in
Eq.~(\ref{PertSer}) for the energy shift can be inferred, to
arbitrarily high order, from the Eqs.~(9), (13--15), (28--33),
(59--67) and (73) in~\cite{Si1978}.  The atomic unit system is used in
the sequel, as is customary for this type of
calculation~\cite{Si1978,HeInBr1974,DaKo1976,DaKo1978}.  The unit of
energy is $\alpha^2\,m_{\rm e}\,c^2 = 27.211\,{\rm eV}$ where $\alpha$
is the fine structure constant, and the unit of the electric field is
the field strength felt by an electron at a distance of one Bohr
radius $a_{\rm Bohr}$ to a nucleus of elementary charge, which is
$1/(4\,\pi\,\epsilon_0)\,(e/a_{\rm Bohr}^2) = 5.142\times 10^{3}\,
{\rm MV}/{\rm cm}$ (here, $\epsilon_0$ is the permittivity of the
vacuum).

We consider the resummation of the divergent perturbative
expansion~(\ref{PertSer}) for two states of atomic hydrogen. These are
the ground state ($n_1 = n_2 = m = 0$) and an excited state with
parabolic quantum numbers $n_1 = 3$, $n_2=0$, $m=1$. 
We list here the first few perturbative
coefficients for the states under investigation. For the ground state,
we have (in atomic units),
\begin{equation}
\label{Ser000}
E_{000}(F) = -\frac{1}{2} - \frac{9}{4}\,F^2 -
\frac{3\,555}{64} \, F^4 - \frac{2\,512\,779}{512} \, F^6 -
\frac{13\,012\,777\,803}{16~384} \, F^8 + \dots
\end{equation}
The perturbation series for the state $n_1 = 3$, $n_2 = 0$, $m = 1$ is
alternating, but has a subleading nonalternating component [see
Eq.~(\ref{LargeN})]. The first perturbative terms read
\begin{equation}
\label{Ser301}
E_{301}(F) = -\frac{1}{50} + \frac{45}{2}\,F -
\frac{31875}{2} \, F^2 + \frac{54\,140\,625}{4} \, F^3 -
\frac{715\,751\,953\,125}{16} \, F^4 + \dots
\end{equation}
Note that for $F = 0$, the unperturbed nonrelativistic energy is
recovered, which is $-1/(2\,n^2)$ in atomic units.  In contrast to the
real perturbative coefficients, the energy pseudoeigenvalue (resonance)
$E(n_1,n_2,m,F)$ has a real and an imaginary component,
\begin{equation}
\label{ComplexEnergy}
E(n_1,n_2,m,F) = {\rm Re}\,E_{n_1 n_2 m}(F) -
\frac{i}{2} \, \Gamma_{n_1 n_2 m}(F)\,,
\end{equation}
where $\Gamma_{n_1 n_2 m}(F)$ is the autoionization width.

Using a computer algebra system~\cite{Wo1988}, the first
50 nonvanishing perturbative
coefficients are evaluated for the ground state,
and for the state with parabolic quantum numbers
$n_1 = 3$, $n_2 = 0$, $m = 1$, we evaluate the first
70 nonvanishing perturbative
coefficients.
The apparent convergence of the
transforms defined in Eqs.~(\ref{ContourIntegral})
and~(\ref{AccelTrans}) in higher order is examined. In the case
of the Borel--Pad\'{e} transforms defined in Eq.~(\ref{ContourIntegral}),
use is made of the replacements
in Eqs.~(\ref{replacement000}) and~(\ref{replacement301})
[``leading renormalon poles are being put in by hand''].
This procedure leads to the numerical results listed in 
Tables~\ref{tablestark1} and~\ref{tablestark2}.  
The numerical error of our results is estimated on the basis
of the highest and lowest value of the four
highest-order transforms.

An important result of the comparison of the methods introduced in
Secs.~\ref{sec_bp} and~\ref{sec_cm} is the following: Both methods
appear to accomplish a resummation of the perturbation series to the
physically correct result.  Method I (Borel$+$Pad\'{e} with 
leading renormalon poles, see
Sec.~\ref{sec_bp}) and method II (Borel$+$Pad\'{e}-improved
conformal mapping, see Sec.~\ref{sec_cm}) appear to lead to results
of comparable accuracy.

To date, a rigorous theory of the performance of the resummation
methods for divergent series of the type discussed in this work (with
alternating and nonalternating components) does not exist.
The {\em logarithmic} singularities introduced by
the branch points of higher-order polylogarithms 
[see the index $k$ in Eq.~(\ref{polylog})] are difficult to 
approximate with the rational functions employed in the construction of 
Pad\'{e} approximants. A solution to the problem of approximating
the logarithmic singularities, based on finite number of perturbative
coefficients, would probably lead to further optimizimation of the rate of
convergence of the transformed series. Within
the current scheme of evaluation, the problematic logarithmic 
singularities may be responsible, at least in part,
for certain numerical instabilities
at higher transformation order, e.g.~in the result for
${\mathcal T}''\!E_{70}(F = 2.1393\times10^{-4})$ in 
Eq.~(\ref{method1}) below.

For the atomic state with quantum numbers $n_1 = 3$, $n_2=0$ and $m=1$, the
evaluation of the transforms ${\mathcal T}\!E_M(F)$ defined in
Eq.~(\ref{ContourIntegral}) (method I) and of the transforms ${\mathcal
  T}''\!E_M(F)$ defined in Eq.~(\ref{AccelTrans}) (method II) in
transformation order $M=67, 68, 69, 70$ for a field strength of
$F=2.1393\times10^{-4}$.  Method I leads to the following results,
\begin{eqnarray}
\label{method1}
& & {\mathcal T}\!E_{67}(F = 2.1393\times10^{-4}) =
-0.015\,860\,468\,199~2
- {\rm i} \; 0.529~048 \times 10^{-6} \,, \nonumber\\[2ex]
& & {\mathcal T}\!E_{68}(F = 2.1393\times10^{-4}) =
-0.015\,860\,468\,200~9
- {\rm i} \; 0.529~047 \times 10^{-6} \,, \nonumber\\[2ex]
& & {\mathcal T}\!E_{69}(F = 2.1393\times10^{-4}) =
-0.015\,860\,468\,198~9
- {\rm i} \; 0.529~048 \times 10^{-6} \,, \nonumber\\[2ex]
& & {\mathcal T}\!E_{70}(F = 2.1393\times10^{-4}) =
-0.015\,860\,468\,194~5
- {\rm i} \; 0.529~015 \times 10^{-6} \,.
\end{eqnarray}
Method II yields the following data,
\begin{eqnarray}
\label{method2}
& & {\mathcal T}''\!E_{67}(F = 2.1393\times10^{-4}) =
-0.015\,860\,468\,200~4
- {\rm i} \; 0.529~047 \times 10^{-6} \,, \nonumber\\[2ex]
& & {\mathcal T}''\!E_{68}(F = 2.1393\times10^{-4}) =
-0.015\,860\,468\,200~3
- {\rm i} \; 0.529~047 \times 10^{-6} \,, \nonumber\\[2ex]
& & {\mathcal T}''\!E_{69}(F = 2.1393\times10^{-4}) =
-0.015\,860\,468\,200~4
- {\rm i} \; 0.529~047 \times 10^{-6} \,, \nonumber\\[2ex]
& & {\mathcal T}''\!E_{70}(F = 2.1393\times10^{-4}) =
-0.015\,860\,468\,203~3
- {\rm i} \; 0.529~046 \times 10^{-6} \,. 
\end{eqnarray}
Numerical results obtained by resummation are presented
in Tables~\ref{tablestark1} and~\ref{tablestark2} for a variety of 
field strengths and for the two atomic states under investigation here. 

%
% table1
%
\begin{center}
\begin{table}[htb!]
\begin{minipage}{14cm}
\begin{center}
\begin{tabular}{lll}
\hline
\hline
\multicolumn{1}{c}{\rule[-3mm]{0mm}{8mm} $F$ (a.u.)} &
\multicolumn{1}{c}{\rule[-3mm]{0mm}{8mm} ${\rm Re}\,E_{000}(F)$} &
\multicolumn{1}{c}{\rule[-3mm]{0mm}{8mm} $\Gamma_{000}(F)$} \\
\hline
\rule[-3mm]{0mm}{8mm}
%
% F = 0.04
%
 $0.04$ &
 $-0.503~771~591~013~654~2(5)$ &
 $3.892~699~990(1) \times 10^{-6}$ \\
%
% F = 0.06
%
\rule[-3mm]{0mm}{8mm}
 $0.06$ &
 $-0.509~203~451~088(2)$ &
 $5.150~775~0(5) \times 10^{-4}$ \\
%
% F = 0.08
%
\rule[-3mm]{0mm}{8mm}
 $0.08$ &
 $-0.517~560~50(5)$ &
 $4.539~63(5) \times 10^{-3}$ \\
%
% F = 0.10
%
\rule[-3mm]{0mm}{8mm}
 $0.10$ &
 $-0.527~419~3(5)$ &
 $1.453~8(5) \times 10^{-2}$ \\
%
% F = 0.12
%
\rule[-3mm]{0mm}{8mm}
 $0.12$ &
 $-0.537~334(5)$ &
 $2.992~7(5) \times 10^{-2}$ \\
%
% F = 0.16
%
\rule[-3mm]{0mm}{8mm}
 $0.16$ &
 $-0.555~24(5)$ &
 $7.131(5) \times 10^{-2}$ \\
%
% F = 0.20
%
\rule[-3mm]{0mm}{8mm}
 $0.20$ &
 $-0.570~3(5)$ &
 $1.212(5) \times 10^{-1}$ \\
%
% F = 0.24
%
\rule[-3mm]{0mm}{8mm}
 $0.24$ &
 $-0.582~6(1)$ &
 $1.767(5) \times 10^{-1}$ \\
%
% F = 0.28
%
\rule[-3mm]{0mm}{8mm}
 $0.28$ &
 $-0.591~7(5)$ &
 $2.32(3) \times 10^{-1}$ \\
%
% F = 0.32
%
\rule[-3mm]{0mm}{8mm}
 $0.32$ &
 $-0.600(5)$ &
 $2.92(3) \times 10^{-1}$ \\
%
% F = 0.36
%
\rule[-3mm]{0mm}{8mm}
 $0.36$ &
 $-0.604(5)$ &
 $3.46(3) \times 10^{-1}$ \\
%
% F = 0.40
%
\rule[-3mm]{0mm}{8mm}
 $0.40$ &
 $-0.608(5)$ &
 $4.00(5) \times 10^{-1}$ \\
\hline
\hline
\end{tabular}
\caption{\label{tablestark1}
Real and imaginary
part of the energy pseudoeigenvalue $E_{000}(F)$
for the ground state of atomic hydrogen
(parabolic quantum numbers
$n_1 = 0, n_2 = 0, m = 0$).}
\end{center}
\end{minipage}
\end{table}
\end{center}

%
% table2
%
\begin{table}[tbh!]
\begin{center}
\begin{minipage}{16cm}
\begin{center}
\begin{tabular}{l@{\hspace*{1.5cm}}l@{\hspace*{0.5cm}}%
l@{\hspace*{1.5cm}}}
\hline
\hline
\multicolumn{1}{c}{\rule[-3mm]{0mm}{8mm} $F$ (a.u.)} &
\multicolumn{1}{c}{\rule[-3mm]{0mm}{8mm} ${\rm Re}\,E_{301}(F)$} &
\multicolumn{1}{c}{\rule[-3mm]{0mm}{8mm} $\Gamma_{301}(F)$} \\
\hline
%
% F = 1.5560 {-4}
%
\rule[-3mm]{0mm}{8mm}
 $1.5560 \times 10^{-4}$ &
 $-0.016~855~237~140~761~7(5)$ &
 $0.421~683(5)\times 10^{-9}$ \\
%
% F = 1.9448 {-4}
%
\rule[-3mm]{0mm}{8mm}
 $1.9448 \times 10^{-4}$ &
 $-0.016~179~388~257~0(5)$ &
 $0.143~773(5) \times 10^{-6}$ \\
%
% F = 2.1393 {-4}
%
\rule[-3mm]{0mm}{8mm}
 $2.1393 \times 10^{-4}$ &
 $-0.015~860~468~20(1)$ &
 $0.105~09(5) \times 10^{-5}$ \\
%
% F = 2.5282 {-4}
% 
\rule[-3mm]{0mm}{8mm}
 $2.5282 \times 10^{-4}$ &
 $-0.015~269~293(1)$ &
 $0.176~39(5) \times 10^{-4}$ \\
%
% F = 2.9172 {-4}
%
\rule[-3mm]{0mm}{8mm}
 $2.9172 \times 10^{-4}$ &
 $-0.014~742~60(3)$ &
 $0.999~96(9)\times 10^{-4}$ \\
%
% F = 3.3061 {-4}
%
\rule[-3mm]{0mm}{8mm}
 $3.3061 \times 10^{-4}$ &
 $-0.014~260~2(3)$ &
 $0.295~4(2) \times 10^{-3}$ \\
\hline
\hline
\end{tabular}
\caption{\label{tablestark2} Real part and imaginary part
of the energy pseudoeigenvalue $E_{301}(F)$
for the excited state with parabolic quantum numbers
$n_1 = 3, n_2 = 0, m = 1$.
The field strength $F$ is given
in atomic units.}
\end{center}
\end{minipage}
\end{center}
\end{table}
%
% Table of a zero-dimensional theory 
%
\begin{table}[htb!]
\begin{center}
\begin{minipage}{14cm}
\begin{center}
\label{TableZeroDimensional}
\begin{tabular}{lrrr}
\hline
\hline
\multicolumn{1}{c}{$n$}
& \multicolumn{1}{c}{$s_n $}
& \multicolumn{1}{c}{$\lbrack \nu - \Ent {\nu/2} / \Ent {\nu/2} \rbrack$}
& \multicolumn{1}{c}{${\delta}_{n}^{(0)} \bigl(1, s_0 \bigr)$}   \\
\hline
$ 1 $&
  $-2$&
  $-0.455~882~352~941$&
  $\underline{ 0}.837~837~837~838$\\
$ 2 $&
  $ 50.5$&
  $ 0.011~324~180~660$&
  $\underline{ 0.6}52~631~578~947$\\
$ 3 $&
  $-1~682$&
  $ 0.225~195~251~591$&
  $\underline{ 0.6}13~275~696~169$\\
$ 4 $&
  $ 82~777.375$&
  $ 343~344~968~231$&
  $\underline{0.6}09~400~007~774$\\
$ 5 $&
  $ -5.373~298 \times 10^{6}$&
  $ 0.416~165~471~150$&
  $\underline{0.6}12~698~290~875$\\
$\dots$ &
  $\dots$ &
  $\dots$ &
  $\dots$ \\
$ 60 $&
  $ 5.492~129 \times 10^{151}$&
  $\underline{0.619}~816~798~266$&
  $\underline{0.620~282~559~59}2$\\
$ 61 $&
  $-5.272~807 \times 10^{154}$&
  $ 0.619~846~730~800$&
  $\underline{0.620~282~559~59}3$\\
$ 62 $&
  $ 5.146~602 \times 10^{157}$&
  $ \underline{0.619}~874~504~158$&
  $\underline{0.620~282~559~59}4$\\
$ 63 $&
  $ -5.105~759 \times 10^{160}$&
  $\underline{0.619}~900~292~634$&
  $\underline{0.620~282~559~59}4$\\
$ 64 $&
  $ 5.146~926 \times 10^{163}$&
  $\underline{0.619}~924~254~847$&
  $\underline{0.620~282~559~595}$\\
$ 65 $&
  $ -5.270~772 \times 10^{166}$&
  $\underline{0.619}~946~535~301$&
  $\underline{ 0.620~282~559~595}$\\
$ 66 $&
  $ 5.481~925 \times 10^{169}$&
  $\underline{0.619}~967~265~778$&
  $\underline{0.620~282~559~595}$\\
$ 67 $&
  $ -5.789~242 \times 10^{172}$&
  $\underline{0.619}~986~566~577$&
  $\underline{ 0.620~282~559~595}$\\
$ 68 $&
  $ 6.206~411 \times 10^{175}$&
  $\underline{0.620}~004~547~622$&
  $\underline{0.620~282~559~595}$\\
$ 69 $&
  $-6.752~937 \times 10^{178}$&
  $\underline{0.620}~004~547~622$&
  $\underline{0.620~282~559~595}$\\
$ 70 $&
  $ 7.455~631 \times 10^{181}$&
  $\underline{0.619}~986~566~577$&
  $\underline{0.620~282~559~595}$\\
\hline
exact&
  $0.620~282~559~595$&
  $0.620~282~559~595$&
  $0.620~282~559~595$\\
\hline
\hline
\end{tabular}
\caption{Evaluation of the perturbation series for $Z(\Phi)$
[zero-dimensional $\phi^4$-theory,
see Eq.~(\ref{ZeroDimGenerating})] for $g=1$.
The $s_n$ are the partial sums of the divergent asymptotic series
(\ref{ZeroDimAsymptotic}), the Pad\'{e} approximants
$\lbrack \nu - \Ent {\nu/2} / \Ent {\nu/2} \rbrack$ are
calculated according to Eq.~(\ref{padedef}), and the delta transforms
are evaluated according to Eq.~(\ref{dWenTr}).}
\end{center}
\end{minipage}
\end{center}
\end{table}

%
% Further Applications of Resummation Methods}
%
\section{Further Applications of Resummation Methods}
\label{FurtherApp}

%
% Zero--Dimensional Theories with Degenerate Minima
%
\subsection{Zero--Dimensional Theories with Degenerate Minima} 
\label{degen}

We consider the generating functional in a zero-dimensional
theory (in this case, the 
usual path integral reduces to an ordinary integral).
First, we briefly consider the $\Phi^4$-theory in zero
dimensions [see Eq.~(9-177) ff.~in~\cite{ItZu1980}];
the generating functional reads 
\begin{equation}
\label{ZeroDimGenerating}
Z(\Phi) = \int\limits_{-\infty}^{\infty}
\frac{{\rm d} \Phi}{\sqrt{2 \pi}} \, \exp\left[ - \frac{1}{2} \, \Phi^2 \, 
- g \, \Phi^4 \right] \,.
\end{equation}
The strictly alternating
divergent asymptotic expansion in powers of $g$ for
$g \to 0$ reads,
\begin{equation}
\label{ZeroDimAsymptotic}
Z(\Phi) \sim \sum_{N=0}^{\infty} 
\frac{4^{N} \, \Gamma(2 N + 1/2)}
{\sqrt{\pi} \, \Gamma(N + 1)} \, (-g)^N\,.
\end{equation}
On using the known asymptotics valid 
for $N \to \infty$, which in this case 
yield the ``large-order'' asymptotics
of the perturbative coefficients,
\begin{equation}
\label{asympgamma}
\frac{\Gamma(2 N + 1/2)}{\Gamma(N + 1)} \sim
\frac{4^N}{\sqrt{2 \pi}} \, \Gamma(N) \, \left[ 1 + 
{\mathcal O}\left(\frac{1}{N}\right) \right] \,
\end{equation}
it is easy to explicitly establish the factorial divergence
of the series~(see also p.~888 of~\cite{ZJ1996}).
The generating functional in zero dimensions has been
proposed as a paradigmatic example for the divergence of perturbation
theory in higher order. It can be resummed easily to the nonperturbative
result; in particular it is manifestly Borel summable, and no
singularities are present on the positive real axis.
Specifically, in Table~\ref{TableZeroDimensional}, we consider the 
resummation of the series (\ref{ZeroDimAsymptotic}) by 
Pad\'{e} approximants~(Sec.~\ref{PadeApproximation}) 
and delta transformations~(Sec.~\ref{NonlinearST}).
The rapid rate of convergence due to the delta transformations
in comparison to the Pad\'{e} approximants is obvious; this
finding is consistent with the results of Ref.~\cite{JeBeWeSo2000}.

Complications are introduced
by degenerate minima. As a second example, we consider the 
modified generating functional [compare with
Eq.~(2.6) on p.~15 of~\cite{LGZJ1990} and with
Eq.~(40.1) on p.~854 of~\cite{ZJ1996}]:
\begin{eqnarray}
\label{zp}
Z'(\Phi) &=& \int\limits_{-\infty}^{\infty}
\frac{{\rm d} \Phi}{\sqrt{2 \pi}} \, \exp\left[ - \frac{1}{2} \, \Phi^2 \, 
(1 - \sqrt{g} \, \Phi)^2 \right] \nonumber\\[2ex]
&=& \int\limits_{-\infty}^{\infty}
\frac{{\rm d} \Phi}{\sqrt{2 \pi}} \, \exp\left[ - \frac{1}{2} \, \Phi^2 + 
\sqrt{g} \, \Phi^3  - \frac{1}{2} \, g \, \Phi^4 \right] \,.
\end{eqnarray}
The expansion of the exponential in powers of the coupling
$g$ leads to a divergent asymptotic series,
\begin{eqnarray}
\label{powerg}
Z'(\Phi) &=& \sum_{N=0}^{\infty} \frac{1}{N!} \,
\int\limits_{-\infty}^{\infty}
\frac{{\rm d} \Phi}{\sqrt{2 \pi}} \,\, {\rm e}^{- 1/2 \, \Phi^2} \,
\left(\sqrt{g} \, \Phi^3  - \frac{1}{2} \, g \, \Phi^4\right)^N
\nonumber\\[3ex]
&=& \sum_{N=0}^{\infty} \,\,
\int\limits_{-\infty}^{\infty}
\frac{{\rm d} \Phi}{\sqrt{2 \pi}} \,\, {\rm e}^{- 1/2 \, \Phi^2} 
\sum_{j=0}^{N} \frac{(-1)^j}{\Gamma(2 N - j + 1)} 
\nonumber\\
& & \quad \times {2 N - j \choose j} \,
\left( \sqrt{g} \, \phi^3 \right)^{2 (N - j)} \,
\left( \frac{g \, \phi^4}{2} \right)^j 
\nonumber\\[3ex]
&=& \sum_{N=0}^{\infty} \, 2 \sqrt{\pi} \,
\frac{(-1)^N \, C_{2 N}^{N + 1/2}(1)}{\Gamma(N - 1/2)} \, g^N
\nonumber\\[3ex]
&=& \sum_{N=0}^{\infty} 
\frac{8^N \, \Gamma(2 N + 1/2)}{\sqrt{\pi} \, \Gamma(N + 1)} \, g^N\,,
\end{eqnarray}
where $C_M^N(x)$ denotes a Gegenbauer (ultraspherical) polynomial. 
Note that terms of half-integer
power of $g$ entail an odd power of the 
field and vanish after integration. The first few terms
of the asymptotic expansion read,
\begin{eqnarray}
\label{divexp}
Z'(\Phi) &=& 1 + 6 \, g +
210 \, g^2 + 13860 \, g^3 \nonumber\\[2ex]
& & + 1351350 \, g^4 +
174594420 \, g^5 \nonumber\\[2ex]       
& & + 28109701620 \, g^6    
+ 5421156741000 \, g^7 \nonumber\\[2ex]
& & + 1218404977539750 \, g^8 + 
{\mathcal O}(g^9) \,. 
\end{eqnarray}
For the perturbative coefficients
\begin{equation}
C_N = \frac{8^N \, \Gamma(2 N + 1/2)}{\sqrt{\pi} \, \Gamma(N + 1)} \,,
\end{equation}
we establish the following asymptotics,
\begin{equation} 
\label{asympg}
C_N \sim \frac{1}{\pi \sqrt{2}} \, N^{-1} \, 32^N \, \Gamma(N+1) \,.
\end{equation}
Due to the nonalternating character of the expansion (\ref{powerg}),
it is not Borel summable in the ordinary sense. 
Rather, it is Borel summable in the 
distributional sense~\cite{CaGrMa1986,CaGrMa1993}.
Here, we present numerical evidence supporting the 
summability of the divergent expansion~(\ref{divexp}) 
based on a finite number of perturbative coefficients. 
The final integration is carried out along the 
contour $C_0$ introduced in~\cite{Je2000prd}
[see also Eq.~(\ref{stenmark}) below].
The same contour has also been used for the resummation of
divergent perturbation series describing renormalization
group (anomalous dimension) $\gamma$ functions~\cite{JeSo2001}.
As explained in~\cite{Je2000prd}, the integration along
$C_0$, which is based on the mean value of the results
obtained above and below the real axis,
leads to a {\em real} final result if all perturbative
coefficients are real.

\begin{table}[htb!]
\begin{center}
\begin{minipage}{14cm}
\begin{center}
\begin{tabular}{cll}
\hline
\hline
\multicolumn{1}{c}{\rule[-3mm]{0mm}{8mm}{$M$}} &
\multicolumn{1}{c}{\rule[-3mm]{0mm}{8mm}{partial sum}} &
\multicolumn{1}{c}{\rule[-3mm]{0mm}{8mm}{${\mathcal T}Z_M(g = 0.01)$}} \\
\hline
2  &
$1.081~000$ &
$1.102~326$ \\
3  &
$1.094~860$ &
$1.096~141$ \\
4  &
$1.108~373$ &
$1.089~875$ \\
5  &
$1.125~832$ &
$1.090~695$ \\
6  &
$1.153~942$ &
$1.092~000$ \\
7  &
$1.208~154$ &
$1.091~596$ \\
8  &
$1.329~994$ &
$1.091~389$ \\
9  &
$1.642~718$ &
$1.091~553$ \\
10 &
$2.545~239$ &
$1.091~545$ \\
11 &
$5.438~230$ &
$1.091~503$ \\
12 &
$1.5 \times 10^1$ &
$1.091~525$ \\
13 &
$5.5 \times 10^1$ &
$1.091~527$ \\
14 &
$2.2 \times 10^2$ &
$1.091~519$ \\
15 &
$9.5 \times 10^2$ &
$1.091~523$ \\
16 &
$4.5 \times 10^3$ &
$1.091~523$ \\
17 &
$2.2 \times 10^4$ &
$1.091~521$ \\
18 &
$1.2 \times 10^5$ &
$1.091~522$ \\
19 &
$6.9 \times 10^5$ &
$1.091~522$ \\
20 &
$4.1 \times 10^6$ &
$1.091~522$ \\
\hline
\multicolumn{1}{l}{exact} &
$1.091~522$ &
$1.091~522$ \\
\hline
\hline
\end{tabular}
\caption{\label{sailer} Resummation of the asymptotic series for the
generating functional of a zero-dimensional theory with degenerate
minima given in Eqs.~(\ref{powerg}) and~(\ref{divexp}).
We have $g = 0.01$. 
Results in the third column are obtained by the method indicated in
Eq.~(\ref{stenmark}) along the integration contour $C_{0}$
(see~\protect\cite{Je2000prd}). The
partial sums in the second column are obtained from the asymptotic
series~(\ref{powerg}).}
\end{center}
\end{minipage}
\end{center}
\end{table}

In particular, the resummation of the divergent expansion 
(\ref{divexp}) is accomplished as follows.
We first define the Borel transform of the 
generating functional by [see Eq.~(4) in~\cite{JeWeSo2000}
and the discussion after Eq.~(\ref{BorelTrans})]
\begin{eqnarray}
\label{ZBorel}
Z'_{\rm B}(z) & \equiv &  
{\mathcal B}^{(1,1)}\left[Z'; \, z\right] 
\nonumber\\[1ex]
& = & \sum_{N=0}^{\infty}
\frac{C_N}{\Gamma(N + 1)} \, z^N\,.
\end{eqnarray}
Pad\'{e} approximants to this Borel transform are evaluated, 
\begin{equation}
{\mathcal P'}_M(z) = \bigg[ [\mkern - 2.5 mu [M/2] \mkern - 2.5 mu ] \bigg/
[\mkern - 2.5 mu [(M+1)/2] \mkern - 2.5 mu ]
\bigg]_{Z'_{\rm B}}\!\!\!\left(z\right)\,,
\end{equation}
where $[\mkern - 2.5 mu [x] \mkern - 2.5 mu ]$ denotes the largest
positive integer smaller than $x$.  We then evaluate the (modified)
Borel integral along the integration contour $C_{0}$ 
introduced in~\cite{Je2000prd}; specifically we define
the transform ${\mathcal T}\!Z_M(g)$
\begin{equation}
\label{stenmark}
{\mathcal T}\!Z_M(g) = \int_{C_{0}} {\rm d}t \, 
\exp(-t)\,{\mathcal P'}_M(g\,t) \,.
\end{equation}
In this case, poles above and below the real axis must be considered,
and the final result involves no imaginary part. 
The particular case of $g = 0.01$ is considered. Values for
the partial sums of the perturbation series (\ref{divexp}) and the
transforms defined in Eq.~(\ref{stenmark}) are 
shown in Tab.~\ref{sailer}. The transforms exhibit apparent
convergence to 6 decimal places in 20th order, whereas the 
partial sums of the perturbation series diverge.
Between the second and forth term of the perturbation series,
(the forth term constitutes the minimal term), the partial sums
provide approximations to the exact result. 
It might seem surprising that the minimal term in the perturbative
expansion is reached already in forth order, although the 
coupling assumes the small value $g = 0.01$. This behavior
immediately follows from the large geometric factor 
in Eq.~(\ref{asympg}) which leads to a 
``resultative coupling strength parameter''
of $g_{\rm res} = 0.32$. ``Nonperturbative effects'' of the order
of $\exp(-1/g_{\rm res})$ provide a fundamental limit to the accuracy
obtainable by optimal truncation of the perturbation series;
this is consistent with the numerical data in Table~\ref{sailer}.

We have also investigated the 
resummation of the divergent series (\ref{divexp}) via a
combination of a conformal mapping and Pad\'{e} approximants
in the conformal variable. The situation is analogous to the
Stark effect: Results are consistent than those
presented in Table~\ref{sailer} obtained by the
``pure'' Borel--Pad\'{e} and in this case slightly
more accurate.
The radius of convergence of the Borel transform $Z'_{\rm B}(z)$
defined in Eq.~(\ref{ZBorel}) is $s = 1/32$
[cf.~Eq.~(\ref{defS}) for the Stark effect], and the 
appropriate conformal mapping in this case reads
\begin{equation}
\label{conformal3}
w = \frac{4\,y}{(1 + y)^2}
\end{equation}
[cf.~Eq.~(\ref{conformal1})]. The inverse reads
\[
y(w) = \frac{1 - \sqrt{1 - w}}{1 + \sqrt{1 - w}}
\]
[cf.~Eq.~(\ref{conformal2})]. The conformal mapping (\ref{conformal3})
maps the complex $w$-plane with a cut along $(1,\infty)$
unto the unit circle in the complex $y$-plane.
While the zero-dimensional model example given in Eq.~(\ref{zp}) does
not exhibit all problematic features of
degenerate anharmonic double-well oscillators,
certain analogies can be established; these comprise
in particular the need to evaluate the mean value of Borel
transforms above and below the real axis
(see also~\cite{JeSo2001}).

%
% The QED Effective Action as a Divergent Series
%
\subsection{The Effective Action as a Divergent Series}
\label{DivQEDEff}

Maxwell's equations receive corrections from virtual excitations
of the charged quantum fields (notably electrons and positrons).
This leads to interesting effects~\cite{GrRe1992,DiGi2000}:
light-by-light scattering, photon splitting, modification
of the speed of light in the presence of strong electromagnetic fields,
and -- last, but not least -- pair production.

When the heavy degrees of freedom are integrated out (in this case,
the ``heavy particles'' are the electrons and positrons), an effective
theory results. The corrections can be described by an effective interaction,
the so-called quantum electrodynamic (QED) effective Lagrangian.
The dominant effect for electromagnetic fields that
vary slowly with respect to the Compton wavelength
(frequencies $\omega \ll 2\,m\,c^2/\hbar$) is described by
the one--loop quantum electrodynamic effective
(so-called ``Heisenberg--Euler'') Lagrangian which is known to
all orders in the electromagnetic
field~\cite{HeEu1936,We1936,Sc1951,BBBB1970,DiRe1985,DiGi2000}.

The Heisenberg--Euler Lagrangian $\Delta {\cal L}$, which
constitutes a quantum correction to the
Maxwell Lagrangian, is usually expressed as a one-dimensional
proper-time integral [see e.g. Eq.~(3.43) in~\cite{DiGi2000},
\begin{equation}
\label{EffAct}
\Delta {\mathcal L} = - \frac{e^2}{8 \pi^2} \,
\lim_{\epsilon,\eta \to 0^{+}} \,
\int_{{\mathrm i}\,\eta}^{\infty + {\mathrm i}\,\eta} \,
\frac{{\rm d} s}{s} \,
{\rm e}^{- (m^2 - {\rm i} \epsilon)\, s}\,
\biggl[  \, a b \, \coth(e a s) \, \cot(e b s) \, -
\frac{a^2 - b^2}{3} - \frac{1}{(e s)^2} \biggr] \,.
\end{equation}
To clarify the notation,
we introduce the well-known Lorentz invariants ${\mathcal F}$ and 
${\mathcal G}$ which are given by
\begin{eqnarray} 
\label{DefF}
{\mathcal F} & = 
\lfrac{1}{4} \, F_{\mu\nu} \, F^{\mu\nu} =
\lfrac{1}{2} \, 
\left(\bm{B}^2 - \bm{E}^2\right) = &
\lfrac{1}{2} \, \left(a^2 - b^2\right)\,, \\[2ex]
\label{DefG}
{\mathcal G} & = 
\lfrac{1}{4} \, F_{\mu\nu} \, (*F)^{\mu\nu} \;\; =
\;\; -\bm{E}\cdot\bm{B} \;\; = & \pm a b\,,
\end{eqnarray}
where $\bm{E}$ and $\bm{B}$ are the 
electric and magnetic field strengths, 
$F_{\mu\nu}$ is the field strength tensor,
and $(*F)^{\mu\nu}$ denotes the 
dual field strength tensor
$(*F)^{\mu\nu} = (1/2) \, \epsilon^{\mu\nu\rho\sigma}\,F_{\rho\sigma}$.
By $a$ and $b$ we denote the {\em secular invariants},
\begin{eqnarray}
\label{Defab}
a &=& \sqrt{\sqrt{{\mathcal F}^2 +
{\mathcal G}^2} + {\mathcal F}}\,,\nonumber\\
b &=& \sqrt{\sqrt{{\mathcal F}^2 +
{\mathcal G}^2} - {\mathcal F}}\,.
\end{eqnarray}
These Lorentz invariants are referred to as secular invariants
because they emerge naturally as eigenvalues of the field strength
tensor; these eigenvalues are conserved under proper 
Lorentz transformations of the field strength tensor.
There are connections between the different representations~\cite{DiGi2000}:
If the relativistic invariant ${\mathcal G}$ is positive,
then it is possible to transform to a Lorentz frame in which $\bm{E}$ and
$\bm{B}$ are {\em antiparallel}. In the case ${\mathcal G} < 0$,
it is possible to choose a Lorentz frame in which  $\bm{E}$ and
$\bm{B}$ are {\em parallel}. Irrespective of the sign
of ${\mathcal G}$ we have in the specified frame
\begin{equation}
a = |\bm{B}| \quad \mbox{and} \quad b = |\bm{E}| 
\quad \mbox{if and only if} \quad 
\mbox{$\bm{B}$ is (anti-)parallel to $\bm{E}$}\,.
\end{equation}
In any case, because $a$ and $b$ are positive definite, 
we have
\begin{equation}
a \, b = |\bm{E} \cdot \bm{B}| > 0 \quad \mbox{for} \quad 
\mbox{any Lorentz frame and ${\mathcal G} \neq 0$}\,,
\end{equation}
which clarifies the sign ambiguity in (\ref{DefG}).
We give in (\ref{DefF}) and (\ref{DefG}) seemingly redundant definitions,
but it will soon become apparent that each of the alternative
``points of view'' has its applications.
The Maxwell Lagrangian is given by
\begin{equation}
{\cal L}_{\rm{cl}} = - {\mathcal F} = 
-\lfrac{1}{4} \, F_{\mu\nu} \,
F^{\mu\nu} = 
\lfrac{1}{2}\,
\left(\bm{E}^2 - \bm{B}^2\right) =
\lfrac{1}{2} \, \left(b^2 - a^2\right)\,.
\end{equation}
As it is obvious from Eq.~(\ref{EffAct}),
the correction $\Delta {\cal L}$ to the
Maxwell Lagrangian is conveniently written 
in terms of the secular invariants $a$ and $b$.

It was observed as early as 1956~\cite{Og1956}
that the quantum correction~(\ref{EffAct}), when expressed as 
a perturbation series in the usual QED perturbation theory parameter
$\alpha = e^2/(4\pi)$, constitutes a divergent series.
This divergent character of the QED perturbative expansion
is supported by Lipatov's argument based on a saddle-point
expansion of the generating functional~\cite{Li1976lett,Li1976,Li1977},
and explicit estimates for the large-order (factorially
divergent) behaviour of the QED perturbative coefficients
have been obtained in~\cite{ItPaZu1977,BaItZuPa1978}.
Recently, the divergent character of the large-order 
behaviour of perturbative expansions in quantum field theory has 
found an exquisite confirmation in explicit 30-loop calculations
of renormalization group $\gamma$ functions in a six-dimensional
$\phi^3$ theory, and in a Yukawa 
theory~\cite{BrKr2000,BrKr2001}.

Here, we will study the divergent series generated by 
expanding the effective action (\ref{EffAct}) in powers
of $e^2$. We will distinguish the cases 
of a magnetic and an electric background fields.
We start with a magnetic field of strength $B$
in which case the effective action reads 
\begin{equation}
\label{SRB}
S_{\rm B} \equiv
\Delta {\mathcal L} (E = 0, B) = - \frac{e^2 B^2}{8 \pi^2} \,
\int_0^\infty \frac{ds}{s^2}
\left\{\coth s - \frac{1}{s} - \frac{s}{3} \right\}
\exp\left(-\frac{m_{\rm e}^2}{e\,B}\,s\right)
\end{equation}
[put $a = B$, $b = 0$ in Eq.~(\ref{EffAct})]. 
The nonperturbative results in Eq.~(\ref{SRB}) can be expanded in
powers of the effective coupling
\begin{equation}
\label{gbdef}
g_{\rm B} = \frac{e^2\,B^2}{m_{\rm e}^4} \,.
\end{equation}
This results in the divergent asymptotic series 
\begin{eqnarray}
\label{Bperser}
S_{\rm B} & \sim &
- \frac{2 e^2 B^2 }{\pi^2} \, g_{\rm B} \,
\sum_{n=0}^\infty \;
\frac{(-1)^{n+1} 4^n \left|{\cal B}_{2 n+4}\right| }{(2n+4)(2n+3)(2n+2)}
\; g_{\rm B}^n \nonumber\\[2ex]
& = & - \frac{2 e^2 B^2 }{\pi^2} \, g_{\rm B} \,
\left[
- \frac{1}{720}
+ \frac{1}{1260} \, g_{\rm B}
- \frac{1}{630} \, g_{\rm B}^2 \right. \nonumber\\[2ex]
& & \;\;\;\;\;
\left. + \ldots -2.33 \times 10^{107} \, g_{\rm B}^{50} + \ldots \right]
\end{eqnarray}
for $g_{\rm B} \to 0$. Here, $\left|{\cal B}_{2\,n+4}\right|$ denotes
the {\em modulus} of the $(2\,n+4)$th Bernoulli number. 
The Bernoulli numbers alternate in sign,
\begin{equation}
{\rm sign}\left({\cal B}_{2 n + 4}\right) = (-1)^{n + 1}\,.
\end{equation}
The expansion coefficients
\begin{equation}
c_n =
\frac{(-1)^{n + 1} \; 4^n \;
\left|{\cal B}_{2n+4}\right| }{(2n+4)(2n+3)(2n+2)}
\end{equation}
obviously display an alternating sign pattern
and grow factorially in absolute magnitude,
\begin{equation}
\label{cnasymptotic}
c_n \; \sim \; \frac{(-1)^{n+1}}{8} \; \frac{\Gamma(2n+2)}{\pi^{2n+4}}
\left( 1 + {\rm O}({2^{-(2n+4)}}) \right)
\end{equation}
for $n \to \infty$.

The series~(\ref{Bperser}) is an alternating series.
In~\cite{Og1956,GrGrSi1970,DuHa1999}, it was shown that
the series~(\ref{Bperser}) is Borel summable. 
A further expansion of the perturbative
coefficients $c_n$ for large $n$ is carried out in~\cite{DuHa1999}
in order to perform the Borel summation; this analysis follows the
asymptotic expansion in Eq.~(\ref{cnasymptotic}) above, and
higher-order terms in the asymptotics of the coefficients for
large $n$ are also used.
Here we consider the resummation of the divergent
series~(\ref{Bperser}) with the help of rational
approximants which
use as input data only a {\em finite} number of perturbation theory
coefficients in {\em numerical} form. 
Specifically, we
use Pad\'{e} approximants defined in Eq.~(\ref{pade}) and discussed in
Sec.~\ref{PadeApproximation} and the nonlinear sequence
transformations $d$ and $\delta$ defined in Eqs.~(\ref{dLevTr}) and
(\ref{dWenTr}), respectively.  The nonlinear transformations
were discussed in Sec.~\ref{NonlinearST}.

The transformations are applied to the divergent perturbation series
Eq.~(\ref{Bperser}), the transforms are calculated with the help of the
recurrence relations~(\ref{LevRec}) and (\ref{WenRec}). The
evaluations presented in Tables~\ref{TableB1} and~\ref{TableB2}
are carried out for values $g_{\rm B}=1/10$,
$g_{\rm B}=1$ and $g_{\rm B}=10$ of the expansion parameter $g_{\rm
B}$ defined in Eq.~(\ref{gbdef}).  The first column in the tables
contains the index $n$, the second column contains the partial sums of
the input series and in the last three columns are the results of the
Pad\'{e} Approximation, the d-transformation and the
delta-transformation.

%
% Table 1
%
\begin{table}[htb!]
\begin{center}
\begin{minipage}{16cm}
\begin{center}
\begin{tabular}{lrrrr}
\hline
\hline
\multicolumn{1}{c}{$n$}%
& \multicolumn{1}{c}{$s_n$}%
& \multicolumn{1}{c}{$\lbrack \nu - \Ent {\nu/2} / \Ent {\nu/2} \rbrack$ }%
& \multicolumn{1}{c}{$d_{n}^{(0)} \bigl(1, s_0 \bigr)$ }%
& \multicolumn{1}{c}{$\delta_{n}^{(0)} \bigl(1, s_0 \bigr)$ } \\
\hline
$ 1 $ &
  $\underline{-0.2}09~523~809$&
%  $\underline{-0.211~3}06~990$&
  $\underline{-0.2}09~523~809$&
  $\underline{-0.211}~640~211$&
  $\underline{-0.211}~640~211$\\
$ 2 $ &
  $\underline{-0.21}2~063~492$&
%  $\underline{-0.211~3}82~773$&
  $\underline{-0.211}~640~211$&
  $\underline{-0.211~3}42~466$&
  $\underline{-0.211~3}42~466$\\
$ 3 $ &
  $\underline{-0.210}~986~050$&
%  $\underline{-0.211~39}1~398$&
  $\underline{-0.211~3}06~990$&
  $\underline{-0.211~4}03~608$&
  $\underline{-0.211~4}00~530$\\
$ 4 $ &
  $\underline{-0.211}~771~470$&
%  $\underline{-0.211~393}~026$&
  $\underline{-0.211~4}15~931$&
  $\underline{-0.211~392}~250$&
  $\underline{-0.211~393}~488$\\
$ 5 $ &
  $\underline{-0.210}~896~256$&
%  $\underline{-0.211~393}~446$&
  $\underline{-0.211~3}82~773$&
  $\underline{-0.211~393~6}52$&
  $\underline{-0.211~393~5}13$\\
$ 6 $ &
  $\underline{-0.212}~279~564$&
%  $\underline{-0.211~393~5}80$&
  $\underline{-0.211~39}7~516$&
  $\underline{-0.211~393~7}76$&
  $\underline{-0.211~393~6}95$\\
$ 7 $ &
  $\underline{-0.209}~336~271$&
%  $\underline{-0.211~393~6}30$&
  $\underline{-0.211~39}1~398$&
  $\underline{-0.211~393~6}31$&
  $\underline{-0.211~393~6}75$\\
 $ 8 $ &
  $\underline{-0.21}7~447~776$&
%  $\underline{-0.211~393~6}51$&
  $\underline{-0.211~39}4~638$&
  $\underline{-0.211~393~67}7$&
  $\underline{-0.211~393~67}0$\\
 $ 9 $ &
  $\underline{-0}.189~339~933$&
%  $\underline{-0.211~393~6}60$&
  $\underline{-0.211~393}~026$&
  $\underline{-0.211~393~67}3$&
  $\underline{-0.211~393~671}$\\
 $ 10$ &
  $\underline{-0}.308~952~549$&
%  $\underline{-0.211~393~66}5$&
  $\underline{-0.211~393}~982$&
  $\underline{-0.211~393~67}0$&
  $\underline{-0.211~393~671}$\\
\hline
exact&
  ${-0.211~393~671}$&
  ${-0.211~393~671}$&
  ${-0.211~393~671}$&
  ${-0.211~393~671}$\\
\hline
\hline
\end{tabular}
\caption{\label{TableB1}
Evaluation of the perturbation series for $S_{\rm B}$
given in Eq.~(\ref{SRB}) for $g_{\rm B}=1/10$. Results are given
in terms of the scaled function ${\bar S}_{\rm B}$ given by
$S_{\rm B} = 10^{-2} \times (-e^2 \, B^2)/(8 \, \pi^2)
\times {\bar S}_{\rm B}$. The performance of three different resummation
methods is compared: in the second column, Pad\'{e} approximants are used.
In the third column, we list results obtained using the Levin 
transformation defined in Eq.~(\ref{dLevTr}),
whereas in the last column, the delta transformation 
(\ref{dWenTr}) is employed. The apparent convergence is indicated by
underlining the decimal figures that have stabilized in increasing
transformation order.}
\end{center}
\end{minipage}
\end{center}
\end{table}

%
% Table 2
%
\begin{table}[htb!]
\begin{center}
\begin{minipage}{16cm}
\begin{center}
\begin{tabular}{lrrrr}
\hline
\hline
\multicolumn{1}{c}{$n$}%
& \multicolumn{1}{c}{$s_n$}%
& \multicolumn{1}{c}{$\lbrack \nu - \Ent {\nu/2} / \Ent {\nu/2} \rbrack$}%
& \multicolumn{1}{c}{$d_{n}^{(0)} \bigl(1, s_0 \bigr)$ }%
& \multicolumn{1}{c}{$\delta_{n}^{(0)} \bigl(1, s_0 \bigr)$ } \\
\hline
$ 1 $ &
  $\underline{-0}.095~238~095$&
%  $\underline{-0.1}43~682~906$&
  $\underline{-0}.095~238~095$&
  $\underline{-0.1}79~894~179$&
  $\underline{-0.1}79~894~179$\\
$ 2 $ &
  $\underline{-0}.349~206~349$&
%  $\underline{-0.1}54~744~471$&
  $\underline{-0.1}79~894~179$&
  $\underline{-0.1}59~879~642$&
  $\underline{-0.1}59~879~642$\\
$ 3 $ &
  $\underline{-0.}728~234~728$&
%  $\underline{-0.1}58~920~070$&
  $\underline{-0.1}43~682~906$&
  $\underline{-0.164}~402~059$&
  $\underline{-0.16}3~671~330$\\
$ 4 $ &
  $\underline{-}7.125~955~525$&
%  $\underline{-0.16}0~923~924$&
  $\underline{-0.1}71~471~152$&
  $\underline{-0.16}5~580~804$&
  $\underline{-0.164}~993~772$\\
$ 5 $ &
  $\underline{}80.395~411~995$&
%  $\underline{-0.16}2~033~744$&
  $\underline{-0.1}54~744~471$&
  $\underline{-0.164}~224~936$&
  $\underline{-0.164}~738~292$\\
$\dots$ &
  $\dots$ &
  $\dots$ &
  $\dots$ &
  $\dots$ \\
$ 15 $ &
  $\underline{}2.034~128 \times 10^{18}$&
%  $\underline{-0.164}~242~255$&
  $\underline{-0.16}3~152~747$&
  $\underline{-0.164~59}9~332$&
  $\underline{-0.164~598~9}12$\\
$ 16 $ &
  $\underline{}-2.179~139 \times 10^{20}$&
%  $\underline{-0.164}~283~614$&
  $\underline{-0.16}5~362~567$&
  $\underline{-0.164~598}~614$&
  $\underline{-0.164~598~9}19$\\
$ 17 $ &
  $\underline{}2.630~171 \times 10^{22}$&
%  $\underline{-0.164}~318~228$&
  $\underline{-0.16}3~457~086$&
  $\underline{-0.164~599}~763$&
  $\underline{-0.164~598~9}28$\\
$ 18 $ &
  $\underline{}-3.552~782 \times 10^{24}$&
%  $\underline{-0.164}~347~482$&
  $\underline{-0.165}~217~014$&
  $\underline{-0.164~59}9~218$&
  $\underline{-0.164~598~93}5$\\
$ 19 $&
  $\underline{} 5.338~753 \times 10^{26} $&
%  $\underline{-0.164}~372~424$&
  $\underline{-0.163}~675~239$&
  $\underline{-0.164~59}9~045$&
  $\underline{-0.164~598~93}8$\\
$ 20 $&
  $\underline{}-8.876~881 \times 10^{28} $&
%  $\underline{-0.164}~393~858$&
  $\underline{-0.165}~109~383$&
  $\underline{-0.164~598}~656$&
  $\underline{-0.164~598~939}$\\
$ 21 $&
  $\underline{} 1.625~237 \times 10^{31} $&
%  $\underline{-0.164}~412~410$&
  $\underline{-0.163}~836~808$&
  $\underline{-0.164~598}~854$&
  $\underline{-0.164~598~939}$\\
$ 22 $&
  $-3.262~042 \times 10^{33} $&
%  $\underline{-0.164}~428~572$&
  $\underline{-0.165}~027~568$&
  $\underline{-0.164~59}9~254$&
  $\underline{-0.164~598~939}$\\
$ 23 $&
  $7.148~693 \times 10^{35} $&
%  $\underline{-0.164}~442~737$&
  $\underline{-0.163}~959~721$&
  $\underline{-0.164~59}9~016$&
  $\underline{-0.164~598~939}$\\
$ 24 $&
  $-1.704~211 \times 10^{38} $&
%  $\underline{-0.164}~455~218$&
  $\underline{-0.164}~963~933$&
  $\underline{-0.164~598}~550$&
  $\underline{-0.164~598~939}$\\
$ 25 $&
  $4.404~583 \times 10^{40} $&
%  $\underline{-0.164}~466~272$&
  $\underline{-0.164}~055~347$&
  $\underline{-0.164~598}~862$&
  $\underline{-0.164~598~939}$\\
\hline
exact&
  ${-0.164~598~939}$&
  ${-0.164~598~939}$&
  ${-0.164~598~939}$&
  ${-0.164~598~939}$\\
\hline
\hline
\end{tabular}
\caption{\label{TableB2}
Evaluation of the perturbation series for $S_{\rm B}$
given in Eq.~(\ref{SRB})
for $g_{\rm B}=1$. Results are given
in terms of the scaled function
${\bar S}_{\rm B}$ given by
$S_{\rm B} = 10^{-1} \times (-e^2 \, B^2)/(8 \, \pi^2)
\times {\bar S}_{\rm B}$.
The apparent convergence is indicated by
underlining the decimal figures that have stabilized in increasing
transformation order.}
\end{center}
\end{minipage}
\end{center}
\end{table}

The data in the tables is presented to a numerical accuracy of $10^{-9}$.
The convergence of the transforms is indicated by underlining those
decimal places which appear to have converged to the nonperturbative
result.  The numerical data in the Tables~\ref{TableB1} and~\ref{TableB2}
indicate that the most favourable numerical 
results are obtained with the Weniger delta-transformation.
For the numerical calculations, exact rational arithmetics
was used~\cite{Wo1988}. Numerical 
experiments for $g_{\rm B}$ larger than $1$ (not shown) and
data presented in~\cite{JeBeWeSo2000} indicate that
the Weniger delta transformation appears to resum the divergent
series~(\ref{Bperser}) at least up to $g_{\rm B} = 200$, and to
arbitrary precision. The other resummation prescriptions -- Pad\'{e},
and the Levin d-transformation -- {\em fail} in this domain of strong
coupling.

In the case of a background electric field, Eq.~(\ref{EffAct}) yields 
the following expression
\begin{equation}
\label{SRE}
S_{\rm E} \equiv
\Delta {\mathcal L} (E, B = 0) = 
\frac{e^2 E^2}{8 \pi^2} \int_0^\infty \frac{ds}{s^2}
\left\{\coth s - \frac{1}{s} - \frac{s}{3} \right\}
\exp\left(- {\mathrm i}\,
\frac{(m_{\rm e} - {\mathrm i}\epsilon)^2}{e\,E}\,s\right)\,.
\end{equation}
In analogy to the expansion parameter for the magnetic case from Eq.~(\ref{gbdef}),
we can define the effective coupling as
\begin{equation}
\label{gedef}
g_{\rm E} = \frac{e^2\,E^2}{m_{\rm e}^4}\,.
\end{equation}
The expansion of the effective action in powers of $g_{\rm E}$ leads
to the formal power series
\begin{eqnarray}
\label{Eperser}
S_{\rm E} & \sim &
- \frac{2 e^2 E^2 }{ \pi^2} \, g_{\rm E} \,
\sum_{n=0}^\infty \;
\frac{ 4^n \; \left| {\cal B}_{2n+4}\right| }{(2n+4)(2n+3)(2n+2)}
\; g_{\rm E}^n \nonumber\\
& = & - \frac{2 e^2 E^2 }{\pi^2} \, g_{\rm E} \,
\left[
\frac{1}{720}
+ \frac{1}{1260} g_{\rm E}
+ \frac{1}{630} \, g_{\rm E}^2 \right. \nonumber\\[2ex]
& & \;\;\;\;\;
\left. + \ldots + 2.33 \times 10^{107} g_{\rm E}^{50} + \ldots \right]\,.
\end{eqnarray}
The perturbation series is nonalternating.
Because the remainder estimate $\omega_n = a_{n+1}$
[see Eq.~(\ref{d_Est})] is in general not valid for nonalternating series,
we cannot assume that any of the previously discussed
resummation methods (Pad\'{e}, or the delta or d-transformation) 
are able to resum the 
perturbation series Eq.~(\ref{Eperser}). This is confirmed
by the explicit numerical data presented in Table~\ref{TableE4}.

%
% Table 4
%
\begin{table}[htb!]
\begin{center}
\begin{minipage}{16cm}
\begin{center}
\label{TableE4}
\begin{tabular}{lrrrr}
\hline
\hline
\multicolumn{1}{c}{$n$}%
& \multicolumn{1}{c}{$s_n$}%
& \multicolumn{1}{c}{$\lbrack \nu - \Ent {\nu/2} / \Ent {\nu/2} \rbrack$ }%
& \multicolumn{1}{c}{$d_{n}^{(0)} \bigl(1, s_0 \bigr)$ }%
& \multicolumn{1}{c}{$\delta_{n}^{(0)} \bigl(1, s_0 \bigr)$ } \\
\hline
$1$&
  $-0.349~206~349$&
%  $-0.270~879~691$&
  $-0.349~206~349$&
  $-0.095~238~095$&
  $-0.095~238~095$\\
$2$&
  $-0.603~174~603$&
%  $-0.227~975~543$&
  $-0.095~238~095$&
  $-0.406~735~069$&
  $-0.406~735~069$\\
$3$&
  $-1.680~615~680$&
%  $-0.194~054~394$&
  $-0.270~879~691$&
  $-0.187~029~818$&
  $-0.221~607~947$\\
$4$&
  $-9.534~805~934$&
%  $-0.162~612~692$&
  $ 0.130~519~638$&
  $-6.833~824~219$&
  $-0.003~699~447$\\
$5$&
  $-97.056~173~456$&
%  $-0.130~635~183$&
  $-0.227~975~543$&
  $-0.305~826~743$&
  $-0.718~038~750$\\
$\dots$ &
  $\dots$ &
  $\dots$ &
  $\dots$ &
  $\dots$ \\
$ 20 $&
  $-8.984~372~007 \times 10^{28} $&
%  $-0.805~807~028$&
  $-0.345~709~493$&
  $-0.214~873~519$&
  $-0.213~414~911$\\
$ 21 $&
  $-1.643~099~018 \times 10^{31} $&
%  $-0.716~045~253$&
  $ 0.132~431~882$&
  $-0.045~964~685$&
  $-0.148~354~420$\\
$ 22 $&
  $-3.294~725~508 \times 10^{33} $&
%  $-0.652~011~235$&
  $-0.328~142~128$&
  $-1.274~287~822$&
  $-0.037~304~526$\\
$ 23 $&
  $-7.214~261~175 \times 10^{35} $&
%  $-0.603~844~031$&
  $ 0.248~593~926$&
  $-0.358~736~820$&
  $-0.367~757~660 $\\
$ 24 $&
  $-1.718~574~682 \times 10^{38} $&
%  $-0.566~160~116$&
  $-0.313~562~944$&
  $-0.237~543~483$&
  $-1.233~623~635$\\
$ 25 $&
  $-4.438~810~865 \times 10^{40} $&
%  $-0.535~770~215$&
  $ 0.434~092~399$&
  $-0.109~659~997$&
  $-0.499~971~155$\\
\hline
\hline
\end{tabular}
\caption{{\em Failure} of the evaluation
of the perturbation series for $S_{\rm E}$
(electric background field) given in Eq.~(\ref{SRE})
for $g_{\rm E}=1$. Results are given
in terms of the scaled function
${\bar S}_{\rm E}$ given by
$S_{\rm E} = 10^{-1} \times (-e^2 \, E^2)/(8 \, \pi^2)
\times {\bar S}_{\rm E}$. Observe the apparent lack of
convergence.}
\end{center}
\end{minipage}
\end{center}
\end{table}

There exists a
``nonperturbative'' imaginary part for the uniform background electric
field which is {\em a priori} not 
contained in the perturbation series whose coefficients are all real
and positive [see Eq.~(\ref{Eperser})].  The imaginary part can be
directly inferred from Eq.~(\ref{SRE}) 
by residue calculus, the result is ~\cite{ItZu1980,DuHa1999}
\begin{equation}
\label{impartE}
{\rm Im} \, S_{\rm E} = \frac{e^2 E^2}{8\,\pi^3} \,
\sum_{n=1}^{\infty} \frac{1}{n^2} \,
\exp\left(- \frac{n\,\pi\,m^2_{\rm e}}{e\,E}\right)\,.
\end{equation}
The characteristic factor
\begin{equation}
{\rm exp}\left(-\frac{1}{g_{\rm E}}\right)
\end{equation}
is nonperturbative since Eq.~(\ref{impartE}) has an essential singularity
in the limit $g_{\rm E} \to 0$, and it is reminiscent of characteristic
expressions occurring in the description of quantum mechanical
tunneling processes~(see p.~195 in~\cite{ItZu1980}).  
In our context, the imaginary part (\ref{impartE}) describes the 
pair production in an electric background field. 
As is well known, the vacuum in very 
strong electric fields becomes unstable with regard to particle-antiparticle
pair production (see also the elucidating discussion in~\cite{GrRe1992}).

Spontaneous pair production becomes considerable only in 
strong external electric fields. Ordinary field strengths are much
smaller than the critical field $E_{\rm crit}$ for which
\begin{equation}
\frac{e\,E_{\rm crit}}{m^2_{\rm e}} = 1\,.
\end{equation}
The critical field strength $E_{\rm crit}$ can be written
(including $\hbar$ and $c$) as the ratio of the rest mass of the 
electron to the characteristic length scale which is set by the
Compton wavelength of the electron:
\begin{equation}
\label{ecritdef}
E_{\rm crit} \; = \; \frac{m_{\rm e} c^2}{e \hbar /m_{\rm e} c}
\; = \; \frac{m_{\rm e}^2 c^3}{e \hbar} \; \sim \; 10^{16} 
\frac{\rm V}{\rm cm} \;.
\end{equation}
This  critical field strengths are comparable to those felt by an
electron in a hydrogenlike atom whose nucleus has the charge $Z
\approx 137$. 

Nonalternating divergent perturbation series require considerable
effort in their resummation. It has been shown in~\cite{Je2000prd} that
a generalized Borel-Pad\'{e} method can be used to reconstruct the imaginary
part of the full nonperturbative result.
This method depends on the special integration contours in the complex plane,
introduced in Secs.~\ref{sec_bp} 
and~\ref{sec_cm}. In Table~\ref{tableSE}, 
we present results for the generalized Borel transforms 
constructed according to Eqs.~(1) -- (6) of~\cite{Je2000prd}).

As for the complex resonances of the Stark effect 
(see the numerical results in Sec.~\ref{sec_nc}), 
it is possible to reconstruct the nonperturbative imaginary 
part (pair production rate) from the purely real perturbative 
coefficients in Eq.~(\ref{Eperser}).

%
% Table SE
%
\begin{table}[htb!]
\begin{center}
\begin{minipage}{16.0cm}
\begin{center}
\begin{tabular}{ccl}
\hline
\hline
\multicolumn{1}{c}{\rule[-3mm]{0mm}{8mm}{$n$}} &
\multicolumn{1}{c}{\rule[-3mm]{0mm}{8mm}{partial sum}} &
\multicolumn{1}{c}{\rule[-3mm]{0mm}{8mm}{${\cal T}S_n(g_{\rm E})$}} \\
\hline
2  & $0.001~146~032$ &  $0.001~144~848 + {\rm i}\,7.70 \times 10^{-17}$ \\
3  & $0.001~146~705$ &  $0.001~146~639 + {\rm i}\,8.22 \times 10^{-11}$ \\
4  & $0.001~146~951$ &  $0.001~147~113 + {\rm i}\,3.54 \times 10^{-8}$ \\
5  & $0.001~147~087$ &  $0.001~147~264 + {\rm i}\,1.93 \times 10^{-8}$ \\
6  & $0.001~147~195$ &  $0.001~147~173 + {\rm i}\,3.15 \times 10^{-7}$ \\
7  & $0.001~147~310$ &  $0.001~147~113 + {\rm i}\,2.58 \times 10^{-7}$ \\
8  & $0.001~147~469$ &  $0.001~147~162 + {\rm i}\,2.30 \times 10^{-7}$ \\
9  & $0.001~147~743$ &  $0.001~147~165 + {\rm i}\,2.63 \times 10^{-7}$ \\
10 & $0.001~148~327$ &  $0.001~147~144 + {\rm i}\,2.53 \times 10^{-7}$ \\
11 & $0.001~149~825$ &  $0.001~147~157 + {\rm i}\,2.46 \times 10^{-7}$ \\
12 & $0.001~154~375$ &  $0.001~147~155 + {\rm i}\,2.56 \times 10^{-7}$ \\
13 & $0.001~170~560$ &  $0.001~147~151 + {\rm i}\,2.51 \times 10^{-7}$ \\
14 & $0.001~237~137$ &  $0.001~147~156 + {\rm i}\,2.51 \times 10^{-7}$ \\
15 & $0.001~550~809$ &  $0.001~147~153 + {\rm i}\,2.53 \times 10^{-7}$ \\
16 & $0.003~228~880$ &  $0.001~147~154 + {\rm i}\,2.51 \times 10^{-7}$ \\
17 & $0.013~345~316$ &  $0.001~147~154 + {\rm i}\,2.52 \times 10^{-7}$ \\
18 & $0.081~610~937$ &  $0.001~147~153 + {\rm i}\,2.52 \times 10^{-7}$ \\
19 & $0.594~142~371$ &  $0.001~147~154 + {\rm i}\,2.52 \times 10^{-7}$ \\
20 & $4.852~426~276$ &  $0.001~147~154 + {\rm i}\,2.52 \times 10^{-7}$ \\
\hline
\multicolumn{1}{l}{exact} &
  $0.001~147~154$ & $0.001~147~154 + {\rm i}\,2.52 \times 10^{-7}$ \\
\hline
\hline
\end{tabular}
\caption{\label{tableSE} Resummation of the asymptotic series for the
QED effective action~(\protect{\ref{Eperser}}) in a constant
background electric field for $g_{\rm E} = 0.05$.  Results in the
third column are obtained by the method discussed
in Secs.~\ref{sec_bp} and~\ref{sec_cm}
along the integration contour $C_{+1}$ (see Fig.~\ref{figg1}).  The
partial sums in the second column are obtained from the asymptotic
series~(\ref{Eperser}).}
\end{center}
\end{minipage}
\end{center}
\end{table}

%
% The Double--Well Problem
%
\subsection{The Double--Well Problem}
\label{DivMATIns}

We discuss in the current section specific results,
recently obtained in~\cite{JeZJ2001}, for the 
quantum-mechanical double-well problem and the related multi-instanton
expansion.
The energy levels of the double-well potential receive, beyond
perturbation theory, contributions which are non-analytic in the coupling
strength; these are related to instanton effects. For example, the
separation between the energies of odd- and even-parity states is given
at leading order by the one-instanton contribution, which, in the
path integral formalism, corresponds to a tunneling of the particle
from one minimum of the potential to the other minimum. In
order to determine
the energies (of individual levels as well as the separations)
more accurately, multi-instanton configurations have also to
be taken into account. The multi-instanton configurations 
describe a particle that tunnels between the minima more than
once. For even instantons, the particle returns to the minimum
from which it started the motion.

For the double-well problem, the energy eigenvalues
of the states at nonvanishing coupling $g \neq 0$ cannot 
{\em in principle} be obtained by analytic continuation from the 
unperturbed situation at vanishing coupling $g=0$ because a potential with 
degenerate minima introduces a degeneracy in the spectrum: for any
{\em one} unperturbed state, {\em two} states emerge when the perturbation
is switched on. These two states are separated 
by an energy shift which is nonperturbative and nonanalytic in the coupling,
i.e.~vanishing to any order in perturbation theory.
Therefore, the two states are described by the same 
perturbation series and yet differ in their energy by instanton
contributions. Specifically, we consider
the case of the double-well potential with the hamiltonian
\begin{equation}
\label{Hdw}
H = - \frac{g}{2} \, 
\frac{\partial^2}{\partial q^2} + \frac{1}{g} \, V(q)\,,
\quad V(q) = \frac{1}{2}\,q^2\,(1 - q)^2\,.
\end{equation}
Of course, the alternative formulation 
$H = - \partial^2/\partial q^2 + 
\frac{1}{2}\,q^2\,(1 - \sqrt{g} q)^2$ gives rise to the same 
energy levels, but the form (\ref{Hdw}) illustrates that the coupling
plays the formal r\^{o}le of $\hbar$.
The hamiltonian (\ref{Hdw}) can be obtained from 
$H = - \partial^2/\partial q^2 +
\frac{1}{2}\,q^2\,(1 - \sqrt{g} q)^2$ by the scaling 
$q \to q/\sqrt{g}$.
It has been conjectured~\cite{ZJ1981jmp,ZJ1981npb,ZJ1983npb,ZJ1984jmp} 
that an asymptotic expansion
for the energy eigenvalue can be obtained by finding a solution
to the equation
\begin{equation}
\label{quantization}
\frac{1}{\sqrt{2\pi}} \, \Gamma\left( \frac{1}{2} - D(E,g) \right) \,
\left(- \frac{2}{g} \right)^{D(E,g)} \, \exp[ -A(E,g)/2 ] = \pm {\rm i}\,,
\end{equation}
which can be understood as a modified Bohr-Sommerfeld quantization
condition. 
The plus and minus signs apply to even- and odd-parity
states, respectively. 
The conjecture (\ref{quantization}), whose validity has been
proven in~\cite{DeDi1991}, 
has found a natural explanation in
the framework of Ecalle's theory of resurgent 
functions~\cite{Ph1988,CaNoPh1993,Bo1994}.          
The functions $D(E,g)$, $A(E,g)$ constitute power series
in both variables. The function $D(E,g)$ describes the perturbative 
expansion; its evaluation is discussed in~\cite{ZJ1981jmp,ZJ1984jmp}. 
The first terms read
\begin{equation}
D(E,g) = E + g\,\left(3 \, E^2 + \frac{1}{4} \right) +
  g^2 \, \left(35 \, E^3 + \frac{25}{4} \, E \right) +
  \mathcal{O}(g^2) \,.
\end{equation}
The ground and the first excited state are both described
by the same perturbation series which can be found by
inverting the equation $D(E,g) = 1/2$. When the energy is expressed
in terms of the naive perturbation series in $g$,
the function $[D(E,g) - N - 1/2]$ then vanishes in any order of perturbation
theory, i.e.~in all orders in $g$. Here,
$N$ is the quantum number of the unperturbed 
state which is a harmonic oscillator eigenstate.
The function $A(E,g)$ essentially describes instanton 
contributions~\cite{ZJ1984jmp}; its first terms read
\begin{equation}
A(E,g) = \frac{1}{3\,g} + g\,\left(17 \, E^2 + \frac{19}{12} \right) +
  g^2 \, \left(227 \, E^3 + \frac{187}{4} \, E \right) +
  \mathcal{O}(g^2) \,.
\end{equation}
A solution to the equation (\ref{quantization}) can be found by 
systematically expanding the energy eigenvalue $E(g)$ in powers of
$g$ and in the two quantities
\begin{equation}
\label{xidef}
\lambda(g) = \ln\left( - \frac{2}{g} \right) \quad
\mbox{and} \quad \xi(g) = \frac{\exp[-1/(6 g)]}{\sqrt{\pi g}}\,.
\end{equation}
Terms of order $\xi(g)^n$ belong to the $n$-instanton contribution.
The energy eigenvalue for nonvanishing perturbation $g\neq 0$ can be 
described by two quantum numbers: the unperturbed quantum number $N$
and the positive or negative parity of the state. We have (the upper
index denotes the instanton order)
\begin{equation}
\label{ExpansionIntoInstantons}
E_{N,\pm}(g) = \sum_{n=0}^{\infty} E^{(n)}_{N,\pm}(g)
\end{equation}
where the perturbation series (zero-instanton contribution) is given as
\begin{equation}
\label{PerturbationSeries}
E^{(0)}_{N,\pm}(g) = \sum_{K=0}^{\infty} E^{(0)}_{N,K} \, g^K \,,
\end{equation}
where the right-hand side is parity independent. For 
$n > 0$, the instanton contribution reads
\begin{equation}
\label{InstantonContribution}
E^{(n)}_{N,\pm}(g) = \left( \frac{2}{g} \right)^{N n} \,
\xi(g)^n \, \sum_{k = 0}^{n-1} \lambda(g)^k \sum_{l=0}
\epsilon^{(N,\pm)}_{nkl} \, g^l \,.
\end{equation}
The lower indices $n$, $k$ and $l$ of
the $\epsilon$ coefficients denote the instanton
order, the power of the logarithm and the power of $g$, respectively.
Some of the results that will be used in the sequel read,
\begin{eqnarray}
\label{Results}
\epsilon^{(0,+)}_{100} &=& - \epsilon^{(0,-)}_{100} = -1 \,, \quad
\epsilon^{(0,+)}_{101} = - \epsilon^{(0,-)}_{101} = \frac{71}{12} \,,
\nonumber\\[2ex]
\epsilon^{(0,+)}_{101} &=& - \epsilon^{(0,-)}_{101} = \frac{6299}{288} 
\,, \quad
\epsilon^{(0,+)}_{210} = \epsilon^{(0,-)}_{210} = 1 \,,
\nonumber\\[2ex]
\epsilon^{(0,+)}_{211} &=& \epsilon^{(0,-)}_{211} = -\frac{53}{6} 
\,, \quad
\epsilon^{(0,+)}_{212} = \epsilon^{(0,-)}_{212} = -\frac{1277}{72} \,,
\nonumber\\[2ex]
\epsilon^{(0,+)}_{200} &=& \epsilon^{(0,-)}_{200} = \gamma 
\,, \quad
\epsilon^{(0,+)}_{201} = 
\epsilon^{(0,-)}_{201} = -\frac{23}{2} - \frac{53}{6} \, \gamma \,,
\nonumber\\[2ex]
\epsilon^{(0,+)}_{202} &=& 
\epsilon^{(0,-)}_{202} = \frac{13}{12} - \frac{1277}{72} \, \gamma \,,
\end{eqnarray}
where $\gamma = 0.57221\dots$ is Euler's constant.
Odd-instanton contributions have opposite sign for opposite-parity
states and are responsible, in particular, for the energy difference 
of the ground state with quantum numbers
$(0,+)$ and the first excited state 
with quantum numbers $(0,-)$.
The dominant contribution to the separation of the two
lowest energy levels is given by the one-instanton contribution:
\begin{equation}
\label{Separation}
E_{0,-}(g) - E_{0,+}(g) \sim 2 \, \xi(g) \, 
\left(1 - \frac{71}{12} \, g - \frac{6299}{288} \, g^2 +
\mathcal{O}(g^3)\right) + \mathcal{O}(\xi(g)^3)\,.
\end{equation}
By contrast, even-instanton contributions have like sign for 
opposite-parity 
states and are responsible, in particular, for the displacement
of the mean value $(1/2) \, [E_{0,-}(g) + E_{0,+}(g)]$ from the 
value of the generalized Borel sum of the perturbation series
$\mathcal{B}\left(\sum_{K=0}^{\infty} E^{(0)}_{0,K} \, g^K\right)$
(for the evaluation of the generalized Borel sum of a nonalternating
divergent series, see Sec.~\ref{degen}).
The dominant contribution to the displacement comes from 
the two-instanton effect, and we have
\begin{eqnarray}
\label{Displacement}
& & \frac{1}{2} \, [E_{0,-}(g) + E_{0,+}(g)] -
\mathrm{Re} \left\{
\mathcal{B}\left(\sum_{K=0}^{\infty} E^{(0)}_{0,K} \, g^K\right)
\right\} \sim \nonumber\\[2ex]
& & \xi(g)^2 \, \left\{ \ln\left(\frac{2 \mathrm{e}^\gamma}{g}\right)
+ \! g \left[ - \frac{53}{6}  
\ln\left(\frac{2 \mathrm{e}^\gamma}{g}\right) \! - \! \frac{23}{2} \right]
+ \! g^2 \left[ - \frac{1277}{72} 
\ln\left(\frac{2 \mathrm{e}^\gamma}{g}\right) \! + \!
\frac{13}{12} \right] \! + \! \mathcal{O}(g^3 \ln(g)) \right\} 
\nonumber\\[2ex]
&& + \mathcal{O}(\xi(g)^4)\,. 
\end{eqnarray}
The function~\cite{ZJ1981npb}
\begin{equation}
\label{DefinitionOfDelta}
\Delta(g) = 4 \, \frac{
\frac{\displaystyle 1}{\displaystyle 2} \, [E_{0,-}(g) + E_{0,+}(g)] - 
\mathrm{Re} \left\{
\mathcal{B}\left(\sum_{K=0}^{\infty} E^{(0)}_{0,K} \, g^K\right)\right\}}
{\left[E_{0,-}(g) - E_{0,+}(g)\right]^2 \, \ln\left(
\frac{\displaystyle 2 \mathrm{e}^\gamma}{\displaystyle g}\right)}
\end{equation}
relates the multi-instanton contributions to the energy eigenvalues,
which can be evaluated numerically,
and to the (generalized) Borel sum of the perturbation series
which is evaluated by analytic continuation of the 
integration path into the complex plane (see~\cite{ZJ1996}). 
The calculation of $\Delta(g)$ at small coupling is 
problematic because of severe numerical cancellations.
From the equations (\ref{Separation}), (\ref{Displacement}) and
(\ref{DefinitionOfDelta}), we obtain the following asymptotics for 
$\Delta(g)$,
\begin{eqnarray}
\label{DeltaAsymptotics}
\Delta(g) & \sim & 1 + g\,\left[ \frac{71}{6} +
\left( -\frac{53}{6} \, 
\ln\left(\frac{2\mathrm{e}^\gamma}{g}\right) - \frac{23}{2} \right)
\bigg/
\ln\left(\frac{2\mathrm{e}^\gamma}{g}\right) \right] \nonumber\\[2ex]
& & + g^2 \,\left[\frac{10711}{72} +
\left( \frac{1277}{72} \,
\ln\left(\frac{2\mathrm{e}^\gamma}{g}\right) - \frac{13}{12} \right)
\bigg/ \ln\left(\frac{2\mathrm{e}^\gamma}{g}\right)    
\right] + \mathcal{O}(g^3) \,.
\end{eqnarray}
If we additionally perform an expansion in inverse powers
of $\ln(2/g)$ and keep only the first few terms in 
$\{1/\ln(2/g)\}$ in each term in the $g$-expansion, the result reads
\begin{eqnarray}
\label{DeltaAsymptotics2}
\Delta(g) & \sim & 1 + 3 g - \frac{23}{2} \frac{g}{\ln(2/g)} \,
\left[1 - \frac{\gamma}{\ln(2/g)} + \frac{\gamma^2}{\ln^2(2/g)} +
\mathcal{O}\left(\frac{1}{\ln^3(2/g)}\right) \right] 
+ \frac{53}{2} g^2 \nonumber\\[2ex] 
&& - 135 \, \frac{g^2}{\ln(2/g)} \,
\left[1 - \frac{\gamma}{\ln(2/g)} + \frac{\gamma^2}{\ln^2(2/g)}  + 
\mathcal{O}\left(\frac{1}{\ln^3(2/g)}\right) \right]
+ \mathcal{O}\left(g^3\right) \,.
\end{eqnarray}
The higher-order corrections, which are only
logarithmically suppressed with respect to
the leading terms $1 + 3g$, change the numerical values 
quite significantly,
even at small coupling. In Table~\ref{table1} we present numerical
results for the function $\Delta(g)$ at small coupling; these are
in agreement with the first few asymptotic terms listed in equation
(\ref{DeltaAsymptotics}) up to numerical accuracy. Of course, 
for strong coupling, significant deviations
from the leading asymptotics must be expected due to higher-order
effects; these are indeed
observed. For example, at $g=0.1$ the numerically
determined value reads $\Delta(0.1) = 0.87684(1)$ whereas the first
asymptotic terms given in equation (\ref{DeltaAsymptotics}) 
sum up to a numerical value of $0.86029$.  

%
% table1
%
\begin{table}[tbh]
\begin{center}
\begin{minipage}{15cm}
\begin{center}
\caption{\label{table1} Comparison of numerical values for
the function $\Delta(g)$ defined in equation (\ref{DefinitionOfDelta})
in the region of small coupling to values obtained by
calculating the first few terms in its asymptotic expansion 
given in (\ref{DeltaAsymptotics}).}
\vspace*{0.3cm}
\begin{tabular}{cr@{.}lr@{.}lr@{.}lr@{.}lr@{.}lr@{.}l%
r@{.}lr@{.}lr@{.}lr@{.}lr@{.}l}
\hline
\hline
\rule[-3mm]{0mm}{8mm} coupling $g$ &
 $0$ & $005$ &
 $0$ & $006$ &
 $0$ & $007$ &
 $0$ & $008$ &
 $0$ & $009$ \\
\hline
$\Delta(g)$ num. &
\rule[-3mm]{0mm}{8mm}
 $1$ & $0063(5)$ &
 $1$ & $0075(5)$ &
 $1$ & $00832(5)$ &
 $1$ & $00919(5)$ &
 $1$ & $00998(5)$ \\
$\Delta(g)$ asymp. &
\rule[-3mm]{0mm}{8mm}
 $1$ & $00640$ &
 $1$ & $00739$ &
 $1$ & $00832$ &
 $1$ & $00919$ &
 $1$ & $01001$ \\
\hline
\hline
\end{tabular}
\end{center}
\end{minipage}
\end{center}
\end{table}

The higher-order corrections to the two-instanton effect are related 
to the corrections to the leading factorial growth of the perturbative
coefficients. This can be seen by expressing that the imaginary part 
of the perturbation series, when continued analytically from 
negative to positive coupling, has to cancel with the imaginary
part of the two-instanton contribution which is generated by the 
logarithms $\ln(-2/g)$. The corrections of order $g\,\ln(-2/g)$
and $g^2\,\ln(-2/g)$ yield the $1/K$-- and $1/K^2$--corrections
to the leading factorial growth of the perturbative coefficients.
From the results for $\epsilon^{(0,\pm)}_{21j}$ ($j = 0,1,2$) given in 
equation (\ref{Results}), we obtain
\begin{equation}
\label{Corrections}
E^{0}_{0,K} \sim - \frac{3^{K+1} \, K!}{\pi} \left[ 1 -
\frac{53}{18} \, \frac{1}{K} -
\frac{1277}{648} \, \frac{1}{K^2} + \mathcal{O}\left(\frac{1}{K^3}\right)
\right]\,.
\end{equation}
The analytic results should be checked against explicit values
of the perturbative coefficients.
We have determined the first 200 perturbative coefficients 
$E^{(0)}_{0,K}$ ($K = 0,\dots, 200$) of the perturbation in 
the form of rational numbers, i.e.~to formally infinite
numerical accuracy. This allows to verify the 
$1/K$-- and $1/K^2$--corrections to the leading factorial growth in
equation (\ref{Corrections}) to high accuracy, for example by 
employing Richardson extrapolation~\cite{Ri1927}.
Using the 160th through the 200th perturbation coefficient
as input data for the Richardson algorithm,
the coefficients of the leading, of the $1/K$-subleading and of the 
$1/K^2$ suppressed corrections are found to be consistent with the 
analytic results given in equation (\ref{Corrections}) up to a relative
numerical accuracy of $10^{-26}$, $10^{-23}$ and $10^{-20}$, 
respectively. 
For completeness, we give here the numerical
values of the 198th through the 200th perturbative coefficients,
to 30 decimals. These read:
\begin{eqnarray}
E^{(0)}_{0,198} = -5.50117\,76962\,88587\,93527\,75694\,38632 \times
  10^{464}\,, \nonumber\\
E^{(0)}_{0,199} = -3.28445\,39841\,65780\,00616\,21912\,32835 \times
  10^{467}\,, \nonumber\\
E^{(0)}_{0,200} = -1.97082\,14193\,09543\,76979\,53006\,07410 \times
  10^{470}\,.
\end{eqnarray}
Values for all 200 coefficients are available~\cite{JeHome}.

It is an interesting consequence of the expansion 
(\ref{ExpansionIntoInstantons}) that the energy difference 
$(E_{0,-} - E_{0,+})$, at small coupling, is described 
to high accuracy by the one-instanton contribution ($n=1$ in 
equation (\ref{InstantonContribution})).
For $g=0.001$, we obtain to 180 decimals, 
\begin{eqnarray}
\label{Result0plus}
E_{0,+}(0.001) &= 
0.&49899\,54548\,62109\,17168\,91308\,39481\,92163\,68209\,47240\,
\nonumber\\
& & 20809\,66532\,93278\,69722\,01391\,
                            \underline{15135\,28505\,38294\,45798}\,
\nonumber\\
& & 
\underline{45759\,95999\,06739\,55175\,84722\,67802\,81306\,96906\,01325}\,
\nonumber\\
& &
\underline{25943\,77289\,94365\,88255\,24440\,17437\,12789\,27978\,99793}\,,
\end{eqnarray}
whereas
\begin{eqnarray}   
\label{Result0minus}
E_{0,-}(0.001) &=
0.&49899\,54548\,62109\,17168\,91308\,39481\,92163\,68209\,47240\,
\nonumber\\
& & 20809\,66532\,93278\,69722\,01391\,
                     \underline{29839\,92959\,55803\,70812}\,
\nonumber\\
& & 
\underline{27749\,92448\,48259\,36743\,64757\,68328\,84835\,35511\,34663}\,
\nonumber\\ 
& &
\underline{06309\,82331\,51885\,23308\,08622\,84780\,52722\,10103\,67282}\,.
\end{eqnarray}
Decimals which differ in the two energy levels are underlined.
The results have been obtained by lattice extrapolation using a
modified Richardson algorithm which is constructed according to 
ideas outlined in~\cite{We1989}. Calculations were performed on 
IBM RISC/6000 workstations while making extensive use of 
multiprecision libraries~\cite{Ba1990tech,Ba1993,Ba1994tech}.
We define $\mathcal{P}_M(g)$
as the $M$th partial sum of the one-instanton contribution
$E^{(1)}_{0,-}(g) - E^{(1)}_{0,+}(g)$,
\begin{equation}
\mathcal{P}_M(g) =
2 \, \xi(g) \, \sum_{j = 0}^{M} \epsilon^{(0,-)}_{10j} \, g^j \,.
\end{equation}               
Using exact rational expressions for the 
coefficients $\epsilon^{(0,-)}_{10j}$ ($j \leq 141$),
we obtain
\begin{eqnarray}
& & \mathcal{P}_{140}(0.001) \times 10^{71} = \nonumber\\[1ex]
& & \;    1.47046\,44541\,75092\,50138\,19899\,%
64494\,15198\,15678\,00350\,05260\,35283\nonumber\\
& & \;\;\;\; 86053\,33378\,03660\,50415\,75193\,%
50528\,41826\,73433\,99328\,21246\,74888\,, \\[3ex]
& & \mathcal{P}_{141}(0.001) \times 10^{71} = \nonumber\\[1ex]
& & \; 1.47046\,44541\,75092\,50138\,19899\,%
64494\,15198\,15678\,00350\,05260\,35283\nonumber\\
& & \;\;\;\; 86053\,33378\,03660\,50415\,75193\,%
50528\,41826\,73433\,99328\,21246\,74887\,.    
\end{eqnarray}
These values are in excellent agreement with the 
numerically determined energy difference (see the 
results presented above in equations (\ref{Result0plus}) and
(\ref{Result0minus}))
\begin{eqnarray}
& & [E_{0,-}(0.001) - E_{0,+}(0.001)] \times 10^{71} = \nonumber\\[1ex]
& & \; 1.47046\,44541\,75092\,50138\,19899\,%
64494\,15198\,15678\,00350\,05260\,35283 \nonumber\\
& & \;\;\;\; 86053\,33378\,03660\,50415\,75193\,%
50528\,41826\,73433\,99328\,21246\,74887\,.       
\end{eqnarray}
The first 70 decimals in equations (\ref{Result0plus}) and
(\ref{Result0minus}) are the same because the one-instanton 
contribution is of the order of $1.4\times 10^{-71}$.
The accuracy to which the one-instanton contribution describes
the energy difference $E_{0,-}(0.001) - E_{0,+}(0.001)$ is limited
by the three-instanton effect which for $g=0.001$ is of the order of
$8\times 10^{-212}$. Note that the two-instanton effect 
(which for $g=0.001$ is of the
order of $4\times 10^{-142}$) does not limit the accuracy to which 
the one-instanton contribution describes the energy {\em difference}
because it has the same sign for opposite-parity states.

We have demonstrated that the behavior of the 
characteristic function $\Delta(g)$
defined in equation (\ref{DefinitionOfDelta}) 
at small coupling is consistent with higher-order corrections to the 
one- and two-instanton contributions, specifically with the 
instanton expansion of the energy levels
governed by the equations
(\ref{ExpansionIntoInstantons}) and (\ref{InstantonContribution}), 
with the assumption that the instanton contributions
given by equation (\ref{InstantonContribution}) should
be Borel summed, with the 
explicit results for the higher-order coefficients
listed in (\ref{Results}) and the analytically derived 
asymptotics for the function $\Delta(g)$ 
given in equation (\ref{DeltaAsymptotics}).
The corrections of relative order $1/K^m$ 
to the leading factorial growth
of the perturbative coefficients -- see equation (\ref{Corrections}) --
are consistent with the analytically evaluated $g^m\,\ln(-2/g)$--corrections
to the two-instanton effect and with the explicit values for 
the first 200 terms in the perturbation series (\ref{PerturbationSeries}).
The nonperturbative energy difference $E_{0,-}(g) - E_{0,+}(g)$
at small coupling $g$ is described, to high accuracy, by the
one-instanton contribution only.

%
% Results
%
\section{Divergent Series: Some Conclusions}
\label{sec_co}

{\em A priori},
it may seem rather unattractive to assume that the
quantum electrodynamic perturbation series may be divergent
even after the regularization and the renormalization.
However, as shown by explicit nontrivial 30-loop calculations
of renormalization group $\gamma$ functions in a six-dimensional
$\phi^3$ theory, and in a Yukawa
theory (presented recently in~\cite{BrKr2000,BrKr2001}),
we believe that the ultimate divergence of the perturbative
expansion can be regarded as a matter-of-fact, clearly demonstrated
by explicit high-order calculations. Therefore,
it appears meaningful to explore the physical implications of
this divergence.

We have discussed four physical applications of 
resummation methods for divergent series:
(i) the energy displacement of a hydrogen
atom in a background electric field (Sec.~\ref{sec_nc}),
(ii) zero-dimensional (model) field theories (Sec.~\ref{degen}),
(iii) the QED effective Lagrangian (Sec.~\ref{DivQEDEff}),
and (iv) the energy levels of the double-well
potential (Sec.~\ref{DivMATIns}).
The mathematical structure of these problems can be characterized as follows,
in the order of increasing complexity:
\begin{itemize}
\item The QED effective Lagrangian for a background
magnetic field, expressed as a perturbation
series in $\alpha$, is manifestly Borel 
summable~\cite{Og1956,GrGrSi1970,DuHa1999},
and the nonperturbative result can be inferred by a number
of different resummation methods, as discussed in Sec.~\ref{DivQEDEff}.
\item The QED effective Lagrangian for a background
electric field, as well as the perturbation
series for a hydrogen atom in an electric field, is not Borel summable.
The same applies to the perturbation series for a zero-dimensional
field theory with degenerate minima.
However, these three problems admit a treatment
according to the concept of {\em generalized} (distributional) Borel
summability~\cite{CaGrMa1986}.
Numerical results (see~\cite{Je2000prd,Je2001pra} and Secs.~\ref{sec_nc}
and~\ref{DivQEDEff}) confirm the distributional Borel summability,
and the use of additional asymptotic information about the
perturbative coefficients [``leading renormalon poles'',
see Eqs.~(\ref{replacement000}) and~(\ref{replacement301}) in
Sec.~\ref{sec_bp}] accelerates the convergence of the
Borel-transformed perturbation series.
\item The double-well potential represent a very problematic case:
the full nonperturbative solution to the eigenvalue problem cannot
be obtained {\em in principle} from perturbation theory. The reason
is the following: the perturbation introduces an additional
degeneracy in the spectrum: each unperturbed level splits into
{\em two} energy levels when the perturbation is nonvanishing.
Both of these levels are described by one and the same naive
perturbation series (see Sec.~\ref{DivMATIns}).
The energy difference is nonperturbative and nonanalytic
in the coupling strength, and finds an explanation in
the theory of instantons.
\end{itemize}

For the Stark effect and the QED effective Lagrangian 
(electric field), the existence of nonperturbative
contributions is intimately linked
with the failure of the Carleman criterion 
(see for example~\cite{GrGrSi1970}, Theorems XII.17 and XII.18 and the
definition on p.~43 in~\cite{ReSi1978},
p.~410 in~\cite{BeOr1978}, Ch.~\ref{ConDivInt} of this
Thesis or the elucidating discussion
in Ref.~\cite{Fi1997}).
The Carleman criterion determines,
roughly speaking, if nonanalytic contributions exist for a given
effect which is described by a specified perturbation series.
In this sense, the divergence of the perturbative expansion
is physically important: for the QED effective action,
it allows for the existence of nonanalytic,
nonperturbative contributions like the pair-production
amplitude which is not contained in the real perturbation series
(\ref{Eperser}) and cannot be obtained on the basis of perturbation
theory alone. 

Also, we would like to illustrate here the
utility of resummation methods in those cases where
perturbation theory breaks down at large coupling.
As explained in Secs.~\ref{sec_nc},~\ref{degen} and~\ref{DivQEDEff},
even in situations where the
perturbation series diverges strongly, it can still be used
to obtain meaningful physical results if it is
combined with a suitable resummation method.
In a relatively weak field, it is possible to obtain more
accurate numerical results by resummation
than by optimal truncation of the
perturbation series~(see also~\cite{Je2000prd}). In a strong field,
it is possible to obtain physically correct
results by resummation even though the perturbation series
diverges strongly (see the discussion in Sec.~\ref{sec_nc}
and the data in Tables~\ref{tablestark1},~\ref{tablestark2} 
and~\ref{sailer}).
By resummation, the perturbation series which is
inherently a weak-coupling expansion can be given a
physical interpretation even
in situations where the coupling is large.
Returning to the analogy to quantum field theory,
one might be tempted to suggest that physically complete
results can in many cases be
obtained after regularization, renormalization
{\em and} resummation.

Can the full nonperturbative result (for energy levels etc.)
be inferred in {\em all} cases by a resummation of the divergent
perturbation series? The answer is, unfortunately, no.
This is demonstrated in Sec.~\ref{DivMATIns} by way of example.
In theories with degenerate minima (such as
the double-well problem), corrections to
energy levels are caused by so-called instanton
contributions that follow naturally by an expansion of
the path integral around nontrivial saddle points.
The $n$-instanton contribution is characterized by
a nonperturbative factor $\xi(g)^n$ where [see Eq.~(\ref{xidef})]
$\xi(g) = \exp[-1/(6 g)]/\sqrt{\pi g}$, and $g$ is the coupling.
The expansion of $\xi(g)$ in powers of $g$ vanishes
in all orders of $g$.
The nonperturbative energy splitting between even- and
odd-parity states is investigated by considering
the function $\Delta(g)$ defined in Eq.~(\ref{DefinitionOfDelta}).
This function would vanish if the generalized Borel sum of
the perturbation series could reproduce energy levels exactly.
However, it does not vanish, as demonstrated by the data in
Tab.~\ref{table1} (the numerical data can be interpreted naturally
by considering so-called two-instanton effects).
At the same time, in a zero-dimensional theory
with degenerate minima (considered in Sec.~\ref{degen}),
resummation is successful {\em despite} the existence of a
nontrivial saddle point. The reason for the success of
the resummation in this model problem appears
to lie in the fact that
no degeneraces are introduced
by the additional saddle point. The perturbation series
determines the generating functional uniquely,
which is not the case in the double-well potential where
two distinct energy levels (of opposite parity)
share the same perturbation series.
This intriguing situation will warrant further investigation
in the near future.

%
% Conclusions
%
\chapter{Conclusions}
\label{Conclu}

\typeout{===========}
\typeout{Conclusions}
\typeout{===========}
\typeout{}

We proceed to the interpretation
of the results obtained in this Thesis, which has a dual
subject: ``quantum electrodynamic bound-state calculations
and large-order perturbation theory''.

The significance of quantum electrodynamics as one of
the most appealing physical 
theories (a ``jewel''  of theoretical physics according to Richard Feynman) 
does not require any further explanation. The esthetic appeal
of bound-state quantum electrodynamics stems from the accuracy
of the experimental verifications, the significance of the 
theory for the determination of the fundamental constants, and the 
conceptual complexity of the calculations
which derives from the apparent simplicity of the 
physical systems under study, when an accurate understanding
is required in higher orders of perturbation theory.
However, the divergence of the perturbative expansion in
large orders, which persists even after the regularization
and renormalization, raises fundamental 
questions regarding the internal consistency of the theoretical
predictions. The problems associated with the divergence of the 
perturbative expansion do not only plague quantum electrodynamics,
but even occur in ordinary quantum mechanical perturbation theory.
This Thesis is divided into two Parts, which represent
two attempts to advance our understanding of the higher-order corrections
as well as of the conceptual questions raised by the divergence of the
perturbative expansion. 

In the first Part, we discussed aspects of 
certain quantum electrodynamic 
bound--state calculations, which have been carried out
over the past years. 
Not all of these calculations~\cite{JePa1996,JeSoMo1997,%,
JeSoIvKa1997,KaJeIvSo1998,JeSoIvKa1998,%,
KaIvJeSo1998,KaJeIvSo1998EPJ,GiEtAl1998,%,
JeMoSo1999,JeMoSo1999asil,JeMoSo1999ieee,JeMoSo2001pra,%
GoEtAl2001,SoEtAl2001,JeMoSo2001hyp,%,
JePa2002,JeSoIn2002,JeNa2002} 
could be described in full detail.
Both the experimental accuracy
as well as the accuracy of the theoretical
calculations have improved
dramatically in recent years~\cite{NiEtAl2000}.
On the theoretical side, 
the calculations at low nuclear charge number may be
based on either of two methods: 
the analytic approach that is based on the $Z\alpha$-expansion
and the numerical approach in which all electron
propagators are kept in exact relativistic form. 
Progress has been achieved due to advances in the analytic approach 
that profit from an adequate formulation of the 
problem,
and due to the development of numerical techniques
which have led to the highly accurate numerical evaluation of
QED corrections in the realm of low nuclear charge number.
This has led to the favourable
situation where analytic and numerical calculations can be
checked against each other. In view of the complexity of 
the calculations and their importance for the 
determination of the fundamental constants~\cite{MoTa2000}, 
the existence of independent cross-checks for the 
calculations is highly desirable.

Numerical calculations at low $Z$ have been
made possible by convergence acceleration methods described
in the second Part of this Thesis (notably by the combined
nonlinear-condensation transformation).
These techniques have led to the solution of the severe numerical
difficulties associated with the singularity of the 
propagators for equal radial arguments, and to a reduction 
in computing time by three orders of magnitude.   
Results obtained for the one-photon self energy are 
several orders of magnitude more accurate than previous
calculations of the effect. They represent the first {\em direct}
evaluations of the one-loop self energy at the nuclear charge
numbers $Z=1$ and $Z=2$, which are of crucial importance 
for precision spectroscopy (atomic hydrogen and hydrogenlike helium).

The analytic calculations of the one-loop self energy
for higher excited states  
will be complemented in the near future 
by numerical calculations based on the new 
techniques~\cite{LBInPriv2002}, leading to 
a cross-check of the type mentioned above.
Severe difficulties associated with the multitude of analytic terms 
due to the more complicated wavefunctions
of the highly excited states
and the associated angular momentum algebra 
have been a problem for self energy calculations in this area.

The calculations for the 
bound-state two-loop self energy report on the 
first evaluation of the highly problematic nonlogarithmic higher-order
two-loop binding correction $B_{60}$. Let us recall that analytic 
work on the corresponding $A_{60}$ correction for the 
less involved one-loop problem has extended over
three decades~\cite{ErYe1965a,ErYe1965b,Er1971,Sa1981,Pa1993}.
The development of analytic methods for the evaluation of 
$B_{60}$ leads to an improved understanding of the scaling 
of the two-loop effect at low $Z$ and will enable a 
detailed comparison of analytic and possibly available
numerical results for the two-loop effect in the future.
At the same time, these results
lead to improved predictions for the hydrogen and helium
fine structure.

The spin-dependence of quantum electrodynamic corrections 
is a conceptually important issue.
We focus on spinless particles and 
present a simplified derivation of the
generalized Breit hamiltonian for a system of two spinless particles.
Although scalar QED is a renormalizable theory, we found the 
issue of bound-state calculations within this theory insufficiently
addressed in the literature. Our calculations for the leading-order 
self energy and the relativistic recoil correction 
are the first corrections of the ``self energy type''
to be evaluated within bound-state QED
involving only scalar particles. 

As a further application, we would 
like to mention the recent investigation~\cite{JeNa2002}.

In the second Part of the Thesis, we investigate
convergence acceleration and resummation methods for 
divergent series.
The convergence acceleration techniques 
address the fundamental problem of the
acceleration of a slowly convergent, nonalternating series. 
Severe numerical instabilities associated with the formation 
of higher-order weighted differences for
nonalternating series have been a major 
obstacle for algorithms that try to accomplish this task.
The methods find applications in areas as diverse
as DNA sequence analysis and experimental mathematics.

Recent highly nontrivial 30-loop calculations~\cite{BrKr2000,BrKr2001}
have convincingly demonstrated the divergence of the perturbative
expansions in quantum field theory originally 
conjectured by Dyson~\cite{Dy1952}.
We discuss nonlinear sequence transformations
as an alternative to the ``usual'' Borel method for the resummation of 
divergent series that result from perturbative expansions in quantum 
field theory. These transformations have favourable asymptotic properties
and lead in many cases to better numerical results than Pad\'{e}
approximants. We also discuss generalizations of the Borel
method which lead to consistent results even if the 
full, nonperturbative physical energy level or 
vacuum-to-vacuum amplitude acquires an imaginary part
due to quantum mechanical tunneling. This imaginary part,
which represents the autoionization decay width (Stark 
effect) and the electron-positron pair-production amplitude
(in the case of the QED effective action), can be derived
starting only from the real (not complex!) perturbative 
coefficients. In the double-well problem, the situation is 
different. Even- and odd-parity states acquire an energy  
separation which is nonperturbative and nonanalytic in the 
coupling strength and is an effect which {\em in 
principle} cannot be derived from perturbation theory 
alone. The energy separation is given by so-called 
multi-instantons 
which follow naturally from a path integral representation 
of the partition function and correspond to configurations where the
particle tunnels repeatedly between the two minima of the 
double-well potential. The results
lead to a better understanding of the 
energy shifts due to multi-instanton effects, by evaluating higher-order
corrections to the two-instanton effect, and they also 
lead to an accurate verification of the instanton expansion
via a comparison to numerically determined energy levels.

The common theme of all investigations discussed in
is to {\em to explore the
predictive limits of quantum theory}.
It has been the aim of this Thesis to present results for
some of the essential QED corrections which influence the spectrum
of bound systems and are of current experimental interest, as well
as to work towards a solution of the questions regarding the
predictive limits of field theories set by the ultimate divergence of the
perturbative expansion.
Certain methods developed along this endeavour 
have meanwhile found applications in other areas
(applied biophysics and mathematics).

%
% Danksagung
%

\newpage

\pagestyle{plain}

\parskip1.5ex plus0.5ex minus0.5ex
\parindent0ex

\begin{center}
{\large\sc Danksagung}
\end{center}

\noindent
Herrn Prof.~Dr.~Gerhard Soff m\"{o}chte ich sehr
herzlich f\"{u}r die immerw\"{a}hrende Unterst\"{u}tzung
meiner Arbeit danken und dabei die vielen hilfreichen Diskussionen
nicht unerw\"{a}hnt lassen. Sein nie nachlassendes Interesse an 
den zu untersuchenden Fragestellungen, seine Hilfsbereitschaft sowie 
die vielfache Unterst\"{u}tzung meiner Forschungsaktivit\"{a}ten in
Rat und Tat haben wesentlich zum Gelingen dieser
Arbeit beigetragen.  

Weiterhin gilt mein Dank allen Kollegen aus der Arbeitsgruppe
``Theorie der Hadronen und Kerne'' am Institut f\"{u}r Theoretische
Physik der Technischen Universit\"{a}t Dresden. 
Die freundliche Atmosph\"{a}re am Institut hat sich sehr 
positiv auf das Gelingen der vielf\"{a}ltigen Forschungsprojekte ausgewirkt.
In diesem Zusammenhang m\"ochte ich besonders den Institutssekret\"arinnen
G.~Sch\"adlich, G.~Latus und U.~W\"achtler an dieser Stelle
recht herzlich danken.

Es ist mir eine besondere Freude, die  
wissenschaftlichen Kooperations--Partner an dieser Stelle zu erw\"{a}hnen, 
mit denen ich das Vergn\"{u}gen und das Gl\"{u}ck hatte, im Laufe 
der Arbeiten zusammenzukommen. Die Diskussionen waren Anla\ss{} zu
h\"{o}chster intellektueller Freude und trugen wesentlich zum Genu\ss{}
der wissenschaftlichen Arbeit bei. Man k\"{o}nnte behaupten,
ein wichtiger Aspekt Wissenschaft werde in guter N\"{a}herung gerade
durch die Diskussion und den sozialen Proze\ss{} des
Austauschs der Gedanken {\em definiert}, und/oder
der Austausch der Gedanken {\em sei} die Wissenschaft.
Erw\"{a}hnt seien an dieser Stelle 
Frau Prof.~Dr. Emanuela Caliceti,
Herr Prof.~Dr. Victor Elias,
Herr Dr.~Holger Gies,
Herr Prof.~Dr. Vincenzo Grecchi,
Herr Prof.~Dr. Barry R. Holstein,
Herr Prof.~Dr. Paul J. Indelicato,
Herr Dr.~F.~Krauss, 
Herr Prof.~Dr. Darrell R. Lamm,
Herr Dr.~Michael Meyer--Hermann,
Herr Dr.~Peter J. Mohr,
Herr Priv.--Doz. Dr.~Krzysztof Pachucki,
Herr Priv.--Doz. Dr.~G\"{u}nter Plunien,
Herr Prof.~Dr.~Vladimir M. Shabaev,
Herr Prof.~Dr.~Sreeram Valluri,
Herr Priv.--Doz.~Dr.~Ernst Joachim Weniger,
Herr Dr.~Vladimir A. Yerokhin,
sowie in besonderem Ma\ss{}e
Herr Prof.~Dr. Jean Zinn--Justin.

Herrn Dr.~J.~Sims vom National Institute of
Standards and Technology sei herzlich gedankt f\"{u}r seine Hilfe bei
der Entwicklung von parallelen Rechenprogrammen mit Hilfe des MPI Message
Passing Interface.

Besonderer Dank gilt auch
meinen Eltern f\"{u}r wichtige Unterst\"{u}tzung und Hilfe.
Schlie\ss{}lich gilt Sabine mein ganz besonderer Dank f\"{u}r Ihre
immerw\"{a}hrende Unterst\"{u}tzung, ihre Geduld und ihre Hingabe.\\[2ex]

\hrule
\vspace*{0.2cm}
\hrule
\vspace*{0.2cm}
\hrule

This work is based on a Habilitation Thesis which 
was submitted to the
University of Technology, Dresden, on 26 June 2002.
It was defended on 14 April 2003.
The current version is not identical to the habilitation thesis
submitted to Dresden University of Technology;
it contains hyperreferences and updates, and 
in addition a number of typographical errors present in the 
original Thesis have been eliminated. Numerical results in 
Eqs.~(\ref{FH2P12})---(\ref{FL5G92}) are in agreement
with recently available data as reported in
[U.~D.~Jentschura, E.-O.~Le~Bigot, P. Indelicato, P. J. Mohr and G. Soff,
Asymptotic Properties of Self--Energy Coefficients, 
e-print physics/0304042, Phys. Rev. Lett. {\bf 90}, 163001 (2003), and 
E.-O.~Le~Bigot, U.~D.~Jentschura, P. Indelicato, P. J. Mohr and G. Soff,
e-print physics/0304068, Phys. Rev. A {\bf 68}, 042101 (2003)].
A number of biographic references 
(e.g.~\cite{JeAkMoSaSo2003,AkSaJeBeSoMo2003,JeEtAl2003,BoMaRe2003,%
MaRe2003,BoMaRe2004}) 
have been updated, and some section headings have been 
edited in order to be conformal with the hypertext standard 
for electronic documents (mathematical expressions cannot be 
used within pdf-bookmarks).

\hrule
\vspace*{0.2cm}
\hrule
\vspace*{0.2cm}
\hrule

This thesis is the third in the traditional threefold sequence
``diploma(master)--dissertation(PhD)--habilitation''
that is being followed in Continental Europe.
Copies of the author's PhD thesis are available as U.~D. Jentschura,
``Quantum Electrodynamic Radiative Corrections in Bound Systems
(Dresdner Forschungen: Theoretische Physik, Band 2)'',
w.e.b. Universit\"{a}tsverlag, Dresden, 1999 (ISBN: 3-933592-65-8), 225 pages
(address of w.e.b. Publishers: Bergstra\ss{}e 78, 01069 Dresden, Germany,
world-wide web address: {\tt http://www.web-univerlag.de}, 
electronic-mail address: {\tt mail@web-univerlag.de}).
This habilitation thesis is also available -- in the form of a book --
from Shaker Verlag GmbH, Postfach 101818, 52018 Aachen,
Germany (world-wide web address: {\tt http://www.shaker.de},
electronic-mail address: {\tt info@shaker.de}). 
Further inquiries and questions are always wellcome 
(email: jentschura@physik.uni-freiburg.de or
U. D. Jentschura, Physikalisches Institut, Universit\"at Freiburg,
Hermann--Herder--Stra\ss{}e 3, 79104 Freiburg im Breisgau, Germany).

\vfill


\newpage

\begin{thebibliography}{100}

\bibitem{JePa1996}
U.~D. Jentschura and K. Pachucki, Phys. Rev. A {\bf 54},  1853  (1996).

\bibitem{JeSoMo1997}
U.~D. Jentschura, G. Soff, and P.~J. Mohr, Phys. Rev. A {\bf 56},  1739
  (1997).

\bibitem{JeSoIvKa1997}
U.~D. Jentschura, G. Soff, V.~G. Ivanov, and S.~G. Karshenbo\u{\i}m, Phys. Rev.
  A {\bf 56},  4483  (1997).

\bibitem{KaJeIvSo1998}
S.~G. Karshenbo\u{\i}m, U.~D. Jentschura, V.~G. Ivanov, and G. Soff, Phys.
  Lett. B {\bf 424},  397  (1998).

\bibitem{JeSoIvKa1998}
U.~D. Jentschura, G. Soff, V. Ivanov, and S.~G. Karshenbo\u{\i}m, Phys. Lett. A
  {\bf 241},  351  (1998).

\bibitem{KaIvJeSo1998}
S.~G. Karshenbo\u{\i}m, V.~G. Ivanov, U.~D. Jentschura, and G. Soff, Zh.
  \'{E}ksp. Teor. Fiz. {\bf 113},  409  (1998), [JETP {\bf 86}, 226 (1998)].

\bibitem{KaJeIvSo1998EPJ}
S.~G. Karshenbo\u{\i}m, U.~D. Jentschura, V.~G. Ivanov, and G. Soff, Eur. Phys.
  J. D {\bf 2},  209  (1998).

\bibitem{GiEtAl1998}
I.~F. Ginzburg, U.~D. Jentschura, S.~G. Karshenbo\u{\i}m, F. Krauss, V.~G.
  Serbo, and G. Soff, Phys. Rev. C {\bf 58},  3565  (1998).

\bibitem{JeMoSo1999}
U.~D. Jentschura, P.~J. Mohr, and G. Soff, Phys. Rev. Lett. {\bf 82},  53
  (1999).

\bibitem{JeMoSo1999asil}
U.~D. Jentschura, P.~J. Mohr, and G. Soff,  in {\em AIP Conf. Proc. 457},
  edited by D.~H.~E. Dubin and D. Schneider (A. I. P., Woodbury, NY, 1999), p.\
  40.

\bibitem{JeMoSo1999ieee}
U.~D. Jentschura, P.~J. Mohr, and G. Soff, IEEE Trans. Instrum. Meas. {\bf 48},
   186  (1999).

\bibitem{JeMoSo2001pra}
U.~D. Jentschura, P.~J. Mohr, and G. Soff, Phys. Rev. A {\bf 64},  042512
  (2001).

\bibitem{GoEtAl2001}
I.~A. Goidenko, L.~N. Labzowsky, A.~V. Nefiodov, U.~D. Jentschura, G. Plunien,
  S. Zschocke, and G. Soff, Phys. Scr. T {\bf 92},  426  (2001).

\bibitem{SoEtAl2001}
G. Soff, I. Bednyakov, T. Beier, F. Erler, I.~A. Goidenko, U.~D. Jentschura,
  L.~N. Labzowsky, A.~V. Nefiodov, G. Plunien, R. Sch\"{u}tzhold, and S.
  Zschocke, Hyp. Int. {\bf 132},  75  (2001).

\bibitem{JeMoSo2001hyp}
U.~D. Jentschura, P.~J. Mohr, and G. Soff, Hyp. Int. {\bf 132},  373  (2001).

\bibitem{JePa2002}
U.~D. Jentschura and K. Pachucki, J. Phys. A {\bf 35},  1927  (2002).

\bibitem{JeSoIn2002}
U.~D. Jentschura, G. Soff, and P.~J. Indelicato, J. Phys. B {\bf 35},  2459
  (2002).

\bibitem{JeNa2002}
U.~D. Jentschura and I. Nandori, Phys. Rev. A {\bf 66},  022114  (2002).

\bibitem{UrKrJeSo1998}
J. Urban, F. Krauss, U. Jentschura, and G. Soff, Nucl. Phys. B {\bf 523},  40
  (1998).

\bibitem{JeGiVaLaWe2002}
U.~D. Jentschura, H. Gies, S.~R. Valluri, D.~R. Lamm, and E.~J. Weniger, Can.
  J. Phys. {\bf 80},  267  (2002).

\bibitem{NaSaJeSo2002}
I. Nandori, K. Sailer, U.~D. Jentschura, and G. Soff, J. Phys. G {\bf 28},  607
   (2002).

\bibitem{LaVaJeWe2002}
D.~R. Lamm, S.~R. Valluri, U.~D. Jentschura, and E.~J. Weniger, Phys. Rev.
  Lett. {\bf 88},  089101  (2002).

\bibitem{JeMoSoWe1999}
U.~D. Jentschura, P.~J. Mohr, G. Soff, and E.~J. Weniger, Comput. Phys. Commun.
  {\bf 116},  28  (1999).

\bibitem{JeBeWeSo2000}
U.~D. Jentschura, J. Becher, E.~J. Weniger, and G. Soff, Phys. Rev. Lett. {\bf
  85},  2446  (2000).

\bibitem{Je2000prd}
U.~D. Jentschura, Phys. Rev. D {\bf 62},  076001  (2000).

\bibitem{JeWeSo2000}
U.~D. Jentschura, E.~J. Weniger, and G. Soff, J. Phys. G {\bf 26},  1545
  (2000).

\bibitem{JeSo2001}
U.~D. Jentschura and G. Soff, J. Phys. A {\bf 34},  1451  (2001).

\bibitem{Je2001pra}
U.~D. Jentschura, Phys. Rev. A {\bf 64},  013403  (2001).

\bibitem{JeZJ2001}
U.~D. Jentschura and J. Zinn-Justin, J. Phys. A {\bf 34},  L253  (2001).

\bibitem{JeAkMoSaSo2003}
U.~D. Jentschura, S.~V. Aksenov, P.~J. Mohr, M.~A. Savageau, and G. Soff,  in
  {\em Proceedings of the Nano\-tech 2003 Conference (Vol. 2), see also e-print
  math.NA/0202009} (Computational Publications, Boston, 2003), pp.\ 535--537.

\bibitem{AkSaJeBeSoMo2003}
S.~V. Aksenov, M.~A. Savageau, U.~D. Jentschura, J. Becher, G. Soff, and P.~J.
  Mohr, Comput. Phys. Commun. {\bf 150},  1  (2003).

\bibitem{Pa1993}
K. Pachucki, Ann. Phys. (N.Y.) {\bf 226},  1  (1993).

\bibitem{Be1947}
H.~A. Bethe, Phys. Rev. {\bf 72},  339  (1947).

\bibitem{ItZu1980}
C. Itzykson and J.~B. Zuber, {\em Quantum Field Theory} (McGraw-Hill, New York,
  NY, 1980).

\bibitem{BeEtAl1997}
B. de~Beauvoir, F. Nez, L. Julien, B. Cagnac, F. Biraben, D. Touahri, L.
  Hilico, O. Acef, A. Clairon, and J.~J. Zondy, Phys. Rev. Lett. {\bf 78},  440
   (1997).

\bibitem{MyTa1999}
E.~G. Myers and M.~R. Tabutt, Phys. Rev. A {\bf 61},  010501  (1999).

\bibitem{MyEtAl1999}
E.~G. Myers, H.~S. Margolis, J.~K. Thompson, M.~A. Farmer, J.~D. Silver, and
  M.~R. Tabutt, Phys. Rev. Lett. {\bf 82},  4200  (1999).

\bibitem{StGeHe2000}
C.~H. Storry, M.~C. George, and E.~A. Hessels, Phys. Rev. Lett. {\bf 84},  3274
   (2000).

\bibitem{GeLoHe2001}
M.~C. George, L.~D. Lombardi, and E.~A. Hessels, Phys. Rev. Lett. {\bf 87},
  173002  (2001).

\bibitem{PaKa1995}
K. Pachucki and S.~G. Karshenbo\u{\i}m, J. Phys. B {\bf 28},  L221  (1995).

\bibitem{UdEtAl1997}
T. Udem, A. Huber, B. Gross, J. Reichert, M. Prevedelli, M. Weitz, and T.~W.
  H\"{a}nsch, Phys. Rev. Lett. {\bf 79},  2646  (1997).

\bibitem{NiEtAl2000}
M. Niering, R. Holzwarth, J. Reichert, P. Pokasov, T. Udem, M. Weitz, T.~W.
  H\"{a}nsch, P. Lemonde, G. Santarelli, M. Abgrall, P. Laurent, C. Salomon,
  and A. Clairon, Phys. Rev. Lett. {\bf 84},  5496  (2000).

\bibitem{Je1999}
U.~D. Jentschura, {\em {\rm Ph. D. thesis, Dresden Unversity of Technology;}
  published as ``Quantum Electrodynamic Radiative Corrections in Bound
  Systems'', Dresdner Forschungen: Theoretische Physik, Band 2} (w.e.b.
  Universit\"{a}tsverlag, Dresden, 1999).

\bibitem{Ka1997}
S.~G. Karshenboim, Z. Phys. D {\bf 39},  109  (1997).

\bibitem{HaPr2000}
T.~W. H\"{a}nsch, private communication (2000).

\bibitem{CeEtAl1996}
C.~L. Cesar, D.~G. Fried, T.~C. Killian, A.~D. Polcyn, J.~C. Sandberg, I.~A.
  Yu, T.~J. Greytak, D. Kleppner, and J.~M. Doyle, Phys. Rev. Lett. {\bf 77},
  255  (1996).

\bibitem{KiEtAl1998}
T.~C. Killian, D.~G. Fried, L. Willmann, D. Landhuis, S.~C. Moss, T.~J.
  Greytak, and D. Kleppner, Phys. Rev. Lett. {\bf 81},  3807  (1998).

\bibitem{JoSo1985}
W.~R. Johnson and G. Soff, At. Data Nucl. Data Tables {\bf 33},  405  (1985).

\bibitem{MoPlSo1998}
P.~J. Mohr, G. Plunien, and G. Soff, Phys. Rep. {\bf 293},  227  (1998).

\bibitem{Mo1996}
P.~J. Mohr,  in {\em Atomic, Molecular, and Optical Physics Handbook}, edited
  by G.~W.~F. Drake (A. I. P., Woodbury, NY, 1996), pp.\ 341--351.

\bibitem{EiGrSh2001}
M.~I. Eides, H. Grotch, and V.~A. Shelyuto, Phys. Rep. {\bf 342},  63  (2001).

\bibitem{WhWa1944}
E.~T. Whittaker and G.~N. Watson, {\em A course of modern analysis} (Cambridge
  University Press, Cambridge, UK, 1944).

\bibitem{Er1987}
H. Erdelyi, {\em Asymptotic Expansions} (Dover, New York, NY, 1987).

\bibitem{Fe1948}
R.~P. Feynman, Phys. Rev. {\bf 74},  1430  (1948).

\bibitem{Fe1949}
R.~P. Feynman, Phys. Rev. {\bf 76},  769  (1949).

\bibitem{FrWe1949}
J.~B. French and V.~F. Weisskopf, Phys. Rev. {\bf 75},  1240  (1949).

\bibitem{KrLa1949}
N.~M. Kroll and W.~E. Lamb, Phys. Rev. {\bf 75},  388  (1949).

\bibitem{Sc1949}
J. Schwinger, Phys. Rev. {\bf 75},  898  (1949).

\bibitem{FuMiTo1949}
H. Fukuda, Y. Miyamoto, and S. Tomonaga, Prog. Theor. Phys. (Kyoto) {\bf 4},
  47  (1949).

\bibitem{Ba1951}
M. Baranger, Phys. Rev. {\bf 84},  866  (1951).

\bibitem{KaKlSc1952}
R. Karplus, A. Klein, and J. Schwinger, Phys. Rev. {\bf 86},  288  (1952).

\bibitem{BaBeFe1953}
M. Baranger, H.~A. Bethe, and R.~P. Feynman, Phys. Rev. {\bf 92},  482  (1953).

\bibitem{FrYe1958}
H.~M. Fried and D.~R. Yennie, Phys. Rev. {\bf 112},  1391  (1958).

\bibitem{FrYe1960}
H.~M. Fried and D.~R. Yennie, Phys. Rev. Lett. {\bf 4},  583  (1960).

\bibitem{La1960}
A.~J. Layzer, Phys. Rev. Lett. {\bf 4},  580  (1960).

\bibitem{La1961a}
A.~J. Layzer, J. Math. Phys. {\bf 2},  292  (1961).

\bibitem{La1961b}
A.~J. Layzer, J. Math. Phys. {\bf 2},  308  (1961).

\bibitem{ErYe1965a}
G.~W. Erickson and D.~R. Yennie, Ann. Phys. (N.Y.) {\bf 35},  271  (1965).

\bibitem{ErYe1965b}
G.~W. Erickson and D.~R. Yennie, Ann. Phys. (N.Y.) {\bf 35},  447  (1965).

\bibitem{Er1971}
G.~W. Erickson, Phys. Rev. Lett. {\bf 27},  780  (1971).

\bibitem{Sa1981}
J. Sapirstein, Phys. Rev. Lett. {\bf 47},  1723  (1981).

\bibitem{KlMa1973}
S. Klarsfeld and A. Maquet, Phys. Lett. B {\bf 43},  201  (1973).

\bibitem{BeBrSt1950}
H.~A. Bethe, L.~M. Brown, and J.~R. Stehn, Phys. Rev. {\bf 77},  370  (1950).

\bibitem{Ha1956}
J.~M. Harriman, Phys. Rev. {\bf 101},  594  (1956).

\bibitem{ScTi1959}
C. Schwartz and J.~J. Tieman, Ann. Phys. (N.Y.) {\bf 6},  178  (1959).

\bibitem{Li1968}
M. Lieber, Phys. Rev. {\bf 174},  2037  (1968).

\bibitem{Hu1969}
R.~W. Huff, Phys. Rev. {\bf 186},  1367  (1969).

\bibitem{Er1977}
G.~W. Erickson, J. Phys. Chem. Ref. Data {\bf 6},  831  (1977).

\bibitem{DrSw1990}
G.~W.~F. Drake and R.~A. Swainson, Phys. Rev. A {\bf 41},  1243  (1990).

\bibitem{Mo1992}
P.~J. Mohr, Phys. Rev. A {\bf 46},  4421  (1992).

\bibitem{Mo1974a}
P.~J. Mohr, Ann. Phys. (N.Y.) {\bf 88},  26  (1974).

\bibitem{Mo1974b}
P.~J. Mohr, Ann. Phys. (N.Y.) {\bf 88},  52  (1974).

\bibitem{Mo1982}
P.~J. Mohr, Phys. Rev. A {\bf 26},  2338  (1982).

\bibitem{Mo1973phd}
P.~J. Mohr, Radiative Corrections in Hydrogen-like Systems, PhD thesis,
  University of California, Berkeley (1973).

\bibitem{MoKi1992}
P.~J. Mohr and Y.~K. Kim, Phys. Rev. A {\bf 45},  2727  (1992).

\bibitem{MoTa2000}
P.~J. Mohr and B.~N. Taylor, Rev. Mod. Phys. {\bf 72},  351  (2000).

\bibitem{MaSt2000}
A.~V. Manohar and I.~W. Stewart, Phys. Rev. Lett. {\bf 85},  2248  (2000).

\bibitem{Pa1998priv}
The numerical value of $A_{60} = -30.924\,15(1)$ has been obtained by K.
  Pachucki after reevaluation of certain poorly convergent one-dimensional
  numerical integrals in his calculation \cite{Pa1993}.

\bibitem{Pa1999}
K. Pachucki, J. Phys. B {\bf 32},  137  (1999).

\bibitem{Wo1988}
S. Wolfram, {\em Mathematica-A System for Doing Mathematics by Computer}
  (Addison-Wesley, Reading, MA, 1988).

\bibitem{Ol1974}
F.~W.~J. Olver, {\em Asymptotics and Special Functions} (Academic Press, New
  York, NY, 1974).

\bibitem{Ba1953vol1}
H. Bateman, {\em Higher Transcendental Functions} (McGraw-Hill, New York, NY,
  1953), Vol.~1.

\bibitem{LBInMo2001}
E.-O. Le~Bigot, P. Indelicato, and P.~J. Mohr, Phys. Rev. A {\bf 64},  052508
  (2001).

\bibitem{InMo1998hyp}
P. Indelicato and P.~J. Mohr, Hyp. Int. {\bf 114},  147  (1998).

\bibitem{LBInPriv2002}
E.-O. Le~Bigot and P. Indelicato, private communication (2002).

\bibitem{ErYe1965ab}
G.~W. Erickson and D.~R. Yennie, Ann. Phys. (N.Y.) {\bf 35},  271, 447  (1965).

\bibitem{SaYe1990}
J. Sapirstein and D.~R. Yennie,  in {\em Quantum Electrodynamics}, Vol.~7 of
  {\em Advanced Series on Directions in High Energy Physics}, edited by T.
  Kinoshita (World Scientific, Singapore, 1990), pp.\ 560--672.

\bibitem{Pa2001}
K. Pachucki, Phys. Rev. A {\bf 63},  042503  (2001).

\bibitem{ApBr1970}
T. Appelquist and S.~J. Brodsky, Phys. Rev. A {\bf 2},  2293  (1970).

\bibitem{Pa1994prl}
K. Pachucki, Phys. Rev. Lett. {\bf 72},  3154  (1994).

\bibitem{EiKaSh1995}
M.~I. Eides and V.~A. Shelyuto, Phys. Rev. A {\bf 52},  954  (1995).

\bibitem{Pa1998}
K. Pachucki, J. Phys. B {\bf 31},  5123  (1998).

\bibitem{BB1984}
I. Bia\l{}ynicki-Birula,  in {\em Quantum Electrodynamics and Quantum Optics},
  edited by A.~O. Barut (Plenum, New York, 1984), pp.\ 41--61.

\bibitem{CaLe1986}
W.~E. Caswell and G.~P. Lepage, Phys. Lett. B {\bf 167},  437  (1986).

\bibitem{Re1972a}
E. Remiddi, Nuovo Cim. A {\bf 11},  825  (1972).

\bibitem{Re1972b}
E. Remiddi, Nuovo Cim. A {\bf 11},  865  (1972).

\bibitem{YeFrSu1961}
D.~R. Yennie, S.~C. Frautschi, and H. Suura, Ann. Phys. (N.Y.) {\bf 13},  379
  (1961).

\bibitem{MaRe2001}
P. Mastrolia and E. Remiddi, private communication (2001).

\bibitem{BoMaRe2003}
R. Bonciani, P. Mastrolia, and E. Remiddi, Nucl. Phys. B {\bf 661},  289
  (2003).

\bibitem{MaRe2003}
P. Mastrolia and E. Remiddi, Nucl. Phys. B {\bf 664},  341  (2003).

\bibitem{BoMaRe2004}
R. Bonciani, P. Mastrolia, and E. Remiddi, Nucl. Phys. B {\bf 676},  399
  (2004).

\bibitem{ReVa2000}
E. Remiddi and J.~A.~M. Vermaseren, Int. J. Mod. Phys. A {\bf 15},  725
  (2000).

\bibitem{GeRe2001}
T. Gehrmann and E. Remiddi, Comput. Phys. Commun. {\bf 141},  296  (2001).

\bibitem{SwDr1991a}
R.~A. Swainson and G.~W.~F. Drake, J. Phys. A {\bf 24},  79  (1991).

\bibitem{SwDr1991b}
R.~A. Swainson and G.~W.~F. Drake, J. Phys. A {\bf 24},  95  (1991).

\bibitem{BeSa1957}
H.~A. Bethe and E.~E. Salpeter, {\em Quantum Mechanics of One- and Two-Electron
  Atoms} (Springer, Berlin, 1957).

\bibitem{JeEtAl2003}
U.~D. Jentschura, E.-O. Le~Bigot, P. Indelicato, P.~J. Mohr, and G. Soff, Phys.
  Rev. Lett. {\bf 90},  163001  (2003).

\bibitem{AfEtAl1993}
L.~G. Afanasyev, A.~S. Chvyrov, O.~E. Gorchakov, M.~A. Ivanov, V.~V. Karpukhin,
  A.~V. Kolomyichenko, V.~I. Komarov, V.~V. Kruglov, A.~V. Kuptsov, L.~L.
  Nemenov, M.~V. Nikitin, Z.~P. Pustylnik, A.~V. Kulikov, S.~V. Trusov, V.~V.
  Yazkov, G.~G. Mkrtchyan, and A.~P. Kurov, Phys. Lett. B {\bf 308},  200
  (1993).

\bibitem{AfEtAl1994}
L.~G. Afanasyev, A.~S. Chvyrov, O.~E. Gorchakov, V.~V. Karpukhin, A.~V.
  Kolomyichenko, V.~I. Komarov, V.~V. Kruglov, A.~V. Kuptsov, L.~L. Nemenov,
  M.~V. Nikitin, Z.~P. Pustylnik, A.~V. Kulikov, S.~V. Trusov, and V.~V.
  Yazkov, Phys. Lett. B {\bf 308},  478  (1994).

\bibitem{HeHeTrBa2001}
A. Heim, K. Hencken, D. Trautmann, and G. Baur, J. Phys. B {\bf 34},  3763
  (2001).

\bibitem{KlRa1975}
A. Klein and J. Rafelski, Phys. Rev. D {\bf 11},  300  (1975).

\bibitem{BaLa1975}
M. Bawin and J.~P. Lavine, Phys. Rev. D {\bf 12},  1192  (1975).

\bibitem{KlRa1975reply}
A. Klein and J. Rafelski, Phys. Rev. D {\bf 12},  1194  (1975).

\bibitem{FlSo1984}
W. Fleischer and G. Soff, Z. Naturforschung {\bf 101},  703  (1984).

\bibitem{LaPa2001}
G. Lach and K. Pachucki, Phys. Rev. A {\bf 64},  042510  (2001).

\bibitem{BeLiPi1991}
V.~B. Berestetskii, E.~M. Lifshitz, and L.~P. Pitaevskii, {\em
  Quantenelektrodynamik (Band IV der Lehrbuchreihe \"uber Theoretische Physik
  von L.~D.~Landau und E.~M.~Lifshitz)}, 7 ed. (Akademie-Verlag, Berlin, 1991).

\bibitem{BaGl1955}
W.~A. Barker and F.~N. Glover, Phys. Rev. {\bf 99},  317  (1955).

\bibitem{BrItZJ1969}
E. Brezin, C. Itzykson, and J. Zinn-Justin, Phys. Rev. D {\bf 1},  2349
  (1969).

\bibitem{Na1972}
A. Nandy, Phys. Rev. D {\bf 5},  1531  (1972).

\bibitem{Ow1994}
D.~A. Owen, Found. Phys. {\bf 24},  273  (1994).

\bibitem{HaOw1994}
M. Halpert and D.~A. Owen, J. Phys. G {\bf 20},  51  (1994).

\bibitem{He1950}
W. Heitler, {\em Quantum Theory of Radiation} (Oxford University Press, New
  York, 1950).

\bibitem{Sa1967Adv}
J.~J. Sakurai, {\em Advanced Quantum Mechanics} (Addison-Wesley, Reading, MA,
  1967).

\bibitem{Pu1957}
G. Pustovalov, Zh. \'{E}ksp. Teor. Fiz. {\bf 32},  1519  (1957), [JETP {\bf 5},
  1234 (1957)].

\bibitem{BoRi1982}
E. Borie and G.~A. Rinker, Rev. Mod. Phys. {\bf 54},  67  (1982).

\bibitem{EiSo2000plb}
D. Eiras and J. Soto, Phys. Rev. Lett. {\bf 491},  101  (2000).

\bibitem{EiSo2000prd}
D. Eiras and J. Soto, Phys. Rev. D {\bf 61},  114027  (2000).

\bibitem{LaBu1998}
P. Labelle and K. Buckley, e-print hep-ph/9804201.

\bibitem{GaLyRuGa2001}
J. Gasser, V.~E. Lyubovitskij, A. Rusetzky, and A. Gall, Phys. Rev. D {\bf 64},
   016008  (2001).

\bibitem{ArShYe1995pra}
A.~N. Artemyev, V.~M. Shabaev, and V.~A. Yerokhin, Phys. Rev. A {\bf 52},  1884
   (1995).

\bibitem{ArShYe1995}
A.~N. Artemyev, V.~M. Shabaev, and V.~A. Yerokhin, J. Phys. B {\bf 28},  5201
  (1995).

\bibitem{ShEtAl1998jpb}
V.~M. Shabaev, A.~N. Artemyev, T. Beier, and G. Soff, J. Phys. B {\bf 31},
  L337  (1998).

\bibitem{ShEtAl1998}
V.~M. Shabaev, A.~N. Artemyev, T. Beier, G. Plunien, V.~A. Yerokhin, and G.
  Soff, Phys. Rev. A {\bf 57},  4235  (1998).

\bibitem{PaGr1995}
K. Pachucki and H. Grotch, Phys. Rev. A {\bf 51},  1854  (1995).

\bibitem{Ye1998}
A.~S. Yelkhovsky, JETP {\bf 86},  472  (1998).

\bibitem{BeSo1988}
T. Beier and G. Soff, Z. Phys. D {\bf 8},  129  (1988).

\bibitem{ScGrSo1993}
S.~M. Schneider, W. Greiner, and G. Soff, J. Phys. B {\bf 26},  L529  (1993).

\bibitem{Ka1993log}
S.~G. Karshenboim, JETP {\bf 76},  541  (1993), [ZhETF {\bf 103}, 1105 (1993)].

\bibitem{GoLaNePlSo1999}
I. Goidenko, L. Labzowsky, A. Nefiodov, G. Plunien, and G. Soff, Phys. Rev.
  Lett. {\bf 83},  2312  (1999).

\bibitem{MaSa1998b}
S. Mallampalli and J. Sapirstein, Phys. Rev. Lett. {\bf 80},  5297  (1998).

\bibitem{Ye2000}
V.~A. Yerokhin, Phys. Rev. A {\bf 62},  012508  (2000).

\bibitem{YeSh2001}
V.~A. Yerokhin and V.~M. Shabaev, Phys. Rev. A {\bf 64},  062507  (2001).

\bibitem{StElAl2000}
T. St\"{o}hlker, P.~H. Mokler, F. Bosch, R.~W. Dunford, F. Franzke, O. Klepper,
  C. Kozhuharov, T. Ludziejewski, F. Nolden, H. Reich, P. Rymuza, Z. Stachura,
  M. Steck, P. Swiat, and A. Warczak, Phys. Rev. Lett. {\bf 85},  3109  (2000).

\bibitem{MoSo1993}
P.~J. Mohr and G. Soff, Phys. Rev. Lett. {\bf 70},  158  (1993).

\bibitem{Ha1949}
G.~H. Hardy, {\em Divergent Series} (Clarendon Press, Oxford, UK, 1949).

\bibitem{We1989}
E.~J. Weniger, Comput. Phys. Rep. {\bf 10},  189  (1989).

\bibitem{FrGrSi1985}
V. Franceschini, V. Grecchi, and H.~J. Silverstone, Phys. Rev. A {\bf 32},
  1338  (1985).

\bibitem{CaGrMa1986}
E. Caliceti, V. Grecchi, and M. Maioli, Commun. Math. Phys. {\bf 104},  163
  (1986).

\bibitem{CaGrMa1993}
E. Caliceti, V. Grecchi, and M. Maioli, Commun. Math. Phys. {\bf 157},  347
  (1993).

\bibitem{Ca2000}
E. Caliceti, J. Phys. A {\bf 33},  3753  (2000).

\bibitem{Ca1926}
T. Carleman, {\em Les Fonctions Quasi-Analytiques} (Gauthiers-Villars, Paris,
  1926).

\bibitem{KnRe2001}
J.~L. Kneur and D. Reynaud, e-print hep-th/0111120.

\bibitem{GrGrSi1970}
S. Graffi, V. Grecchi, and B. Simon, Phys. Lett. B {\bf 32},  631  (1970).

\bibitem{ReSi1978}
M. Reed and B. Simon, {\em Methods of Modern Mathematical Physics IV: Analysis
  of Operators} (Academic Press, New York, 1978).

\bibitem{BeOr1978}
C.~M. Bender and S.~A. Orszag, {\em Advanced Mathematical Methods for
  Scientists and Engineers} (McGraw-Hill, New York, NY, 1978).

\bibitem{Fi1997}
J. Fischer, Int. J. Mod. Phys. A {\bf 12},  3625  (1997).

\bibitem{LaLi1958}
L.~D. Landau and E.~M. Lifshitz, {\em Quantum Mechanics {\em (Volume 3 of the
  Course of Theoretical Physics)}} (Pergamon Press, London, 1958).

\bibitem{LaLi1979}
L.~D. Landau and E.~M. Lifshitz, {\em Quantenmechanik (Band III der
  Lehrbuchreihe \"uber Theoretische Physik von L.~D.~Landau und
  E.~M.~Lifshitz)} (Akademie-Verlag, Berlin, 1979).

\bibitem{CaGrMa1988}
E. Caliceti, V. Grecchi, and M. Maioli, Commun. Math. Phys. {\bf 113},  625
  (1988).

\bibitem{CaGrMa1996}
E. Caliceti, V. Grecchi, and M. Maioli, Commun. Math. Phys. {\bf 176},  1
  (1996).

\bibitem{ZJ1981jmp}
J. Zinn-Justin, J. Math. Phys. {\bf 22},  511  (1981).

\bibitem{ZJ1981npb}
J. Zinn-Justin, Nucl. Phys. B {\bf 192},  125  (1981).

\bibitem{ZJ1983npb}
J. Zinn-Justin, Nucl. Phys. B {\bf 218},  333  (1983).

\bibitem{ZJ1984jmp}
J. Zinn-Justin, J. Math. Phys. {\bf 25},  549  (1984).

\bibitem{Ph1988}
F. Pham, {\em Resurgence, Quantized Canonical Transformation and
  Multi-Instanton}, Algebraic Analysis, vol. II (1988); C. R. Acad. Sci. Paris
  {\bf 309}, 999 (1989).

\bibitem{DeDi1991}
E. Delabaere and H. Dillinger, Ph.D. thesis, University of Nice, Nice, 1991
  (unpublished).

\bibitem{CaNoPh1993}
B. Candelpergher, J.~C. Nosmas, and F. Pham, {\em Approche de la
  R\'{e}surgence} (Hermann, Paris, 1993).

\bibitem{Ba1975}
G.~A. Baker, {\em Essentials of Pad\'{e} approximants} (Academic Press, New
  York, 1975).

\bibitem{BaGr1996}
G.~A. Baker and P. Graves-Morris, {\em Pad\'{e} Approximants}, 2nd ed.
  (Cambridge University Press, Cambridge, 1996).

\bibitem{BrRZ1991}
C. Brezinski and M. Redivo-Zaglia, {\em Extrapolation Methods} (North-Holland,
  Amsterdam, 1991).

\bibitem{Ba1965}
G.~A. Baker, Adv. Theor. Phys. {\bf 1},  1  (1965).

\bibitem{BaGa1970}
{\em The Pad\'e Approximant in Theoretical Physics}, edited by G.~A. Baker and
  J.~L. Gammel (Academic Press, New York, 1970).

\bibitem{Ba1972}
J.~L. Basdevant, Fortschr. Physik {\bf 20},  283  (1972).

\bibitem{Ba1990}
G.~A. Baker, {\em Quantitative theory of critical phenomena} (Academic Press,
  San Diego, 1990).

\bibitem{AbSt1972}
M. Abramowitz and I.~A. Stegun, {\em Handbook of Mathematical Functions}, 10
  ed. (National Bureau of Standards, Washington, D. C., 1972).

\bibitem{WeCiVi1993}
E.~J. Weniger, J. {\v C}{\'\i}{\v z}ek, and F. Vinette, J. Math. Phys. {\bf
  34},  571  (1993).

\bibitem{Br1980}
C. Brezinski, {\em Pad\'{e}-type Approximation and General Orthogonal
  Polynomials} (Birkh\protect{\"{a}}user, Basel, 1980).

\bibitem{Br1977}
C. Brezinski, {\em Acc\'el\'eration de la Convergence en Analyse Num\'erique}
  (Springer, Berlin, 1977).

\bibitem{Br1978}
C. Brezinski, {\em Algorithmes d'Acc\'el\'eration de la Convergence -- \'Etude
  Num\'erique} (Editions Technip, Paris, 1978).

\bibitem{Br1997}
C. Brezinski, {\em Projection Methods for Systems of Equations} (Elsevier,
  Amsterdam, 1997).

\bibitem{De1988}
J.-P. Delahaye, {\em Sequence Transformations} (Springer, Berlin, 1988).

\bibitem{LiLuSh1995}
C.~B. Liem, T. L\"{u}, and T.~M. Shih, {\em The Splitting Extrapolation Method}
  (World Scientific, Singapore, 1995).

\bibitem{MaSh1983}
G.~I. Marchuk and V.~V. Shaidurov, {\em Difference Methods and Their
  Extrapolations} (Springer, New York, 1983).

\bibitem{Wa1996}
G. Walz, {\em Asymptotics and Extrapolation} (Akademie-Verlag, Berlin, 1996).

\bibitem{Wi1981}
J. Wimp, {\em Sequence Transformations and Their Applications} (Academic Press,
  New York, NY, 1981).

\bibitem{Br1985jcam}
C. Brezinski, J. Comput. Appl. Math. {\bf 12--13},  19  (1985).

\bibitem{GaGu1974}
D.~S. Gaunt and A.~J. Guttmann,  in {\em Phase Transitions and Critical
  Phenomena 3}, edited by C. Domb and M.~S. Green (Academic Press, London,
  1974), p.\ 181.

\bibitem{Gu1989}
A.~J. Guttmann,  in {\em Phase Transitions and Critical Phenomena}, edited by
  C. Domb and J.~L. Lebowitz (Academic Press, London, 1989), pp.\ 3--234.

\bibitem{Wy1956a}
P. Wynn, Math. Tables Aids Comput. {\bf 10},  91  (1956).

\bibitem{Ca1977}
B.~C. Carlson, {\em Special Functions of Applied Mathematics} (Academic Press,
  New York, 1977).

\bibitem{Le1973}
D. Levin, Int. J. Comput. Math. B {\bf 3},  371  (1973).

\bibitem{SmFo1979}
D.~A. Smith and W.~F. Ford, SIAM J. Numer. Anal. {\bf 16},  223  (1979).

\bibitem{SmFo1982}
D.~A. Smith and W.~F. Ford, Math. Comput. {\bf 38},  481  (1982).

\bibitem{We1994a}
E.~J. Weniger,  in {\em Nonlinear Numerical Methods and Rational
  Approximation}, edited by A. Cuyt (Kluwer, Doordrecht, 1994), Vol.~{II}, pp.\
  269--282.

\bibitem{HoWe1995}
H.~H.~H. Homeier and E.~J. Weniger, Comput. Phys. Commun. {\bf 92},  1  (1995).

\bibitem{We1996c}
E.~J. Weniger, Comput. Phys. {\bf 10},  496  (1996).

\bibitem{We1990}
E.~J. Weniger, J. Comput. Appl. Math. {\bf 32},  291  (1990).

\bibitem{WeCi1990}
E.~J. Weniger and J. {\v C}{\' \i}{\v z}ek, Comput. Phys. Commun. {\bf 59},
  471  (1990).

\bibitem{RoBhBh1996}
D. Roy, R. Bhattacharya, and S. Bhowmick, Comput. Phys. Commun. {\bf 93},  159
  (1996).

\bibitem{BhRoBh1997}
R. Bhattacharya, D. Roy, and S. Bhowmick, Comput. Phys. Commun. {\bf 101},  213
   (1997).

\bibitem{We1997}
E.~J. Weniger, Phys. Rev. A {\bf 56},  5165  (1997).

\bibitem{We1992}
E.~J. Weniger, Numer. Algor. {\bf 3},  477  (1992).

\bibitem{WeCiVi1991}
E.~J. Weniger, J. {\v C}{\'\i}{\v z}ek, and F. Vinette, Phys. Lett. A {\bf
  156},  169  (1991).

\bibitem{We1996a}
E.~J. Weniger, Int. J. Quantum Chem. {\bf 57},  265  (1996).

\bibitem{We1996b}
E.~J. Weniger, Ann. Phys. (N.Y.) {\bf 246},  133  (1996).

\bibitem{We1996d}
E.~J. Weniger, Phys. Rev. Lett. {\bf 77},  2859  (1996).

\bibitem{RoBhBh1998}
D. Roy, R. Bhattacharya, and S. Bhowmick, Comput. Phys. Commun. {\bf 113},  131
   (1998).

\bibitem{Si1981}
A. Sidi, J. Comput. Appl. Math. {\bf 7},  37  (1981).

\bibitem{vW1965}
A. van Wijngaarden,  in {\em Cursus: Wetenschappelijk Rekenen B, Process
  Analyse} (Stichting Mathematisch Centrum, Amsterdam, 1965), pp.\ 51--60.

\bibitem{Bw1991}
T.~J. Bromwich, {\em An Introduction to the Theory of Infinite Series}, 3 ed.
  (Chelsea, New York, NY, 1991).

\bibitem{Kn1964}
K. Knopp, {\em Theorie und Anwendung der unendlichen Reihen} (Springer, Berlin,
  1964).

\bibitem{Da1969}
J.~W. Daniel, Math. Comput. {\bf 23},  91  (1969).

\bibitem{JoKoKe1992}
N.~L. Johnson, S. Kotz, and A.~W. Kemp, {\em Univariate discrete
  distributions}, {\em Wiley series in probability and mathematical statistics}
  (J. Wiley \& Sons, New York, 1992).

\bibitem{Ma1983}
B.~B. Mandelbrot, {\em The fractal geometry of nature} (W. H. Freeman, San
  Francisco, 1983).

\bibitem{Go1953}
I.~J. Good, Biometrika {\bf 40},  237  (1953).

\bibitem{Tr1945}
C. Truesdell, Ann. Math. {\bf 46},  144  (1945).

\bibitem{KuTo1992}
K.~B. Kulasekera and D.~W. Tonkyn, Commun. Stat. Sim. {\bf 21},  499  (1992).

\bibitem{DoLu1997}
L.~G. Doray and A. Luong, Commun. Stat. Sim. {\bf 26},  1075  (1997).

\bibitem{GaYc1955}
G. Gamow and M. Y\v{c}as, Proc. Natl. Acad. Sci. USA {\bf 41},  1011  (1955).

\bibitem{Maetal1994}
R.~N. Mantegna, S.~V. Buldyrev, A.~L. Goldberger, S. Havlin, C.~K. Peng, M.
  Simons, and H.~E. Stanley, Phys. Rev. Lett. {\bf 73},  3169  (1994).

\bibitem{KoMa1995}
A.~K. Konopka and C. Martindale, Science {\bf 268},  789  (1995).

\bibitem{MaKo1996}
C. Martindale and A.~K. Konopka, Computers Chem. {\bf 20},  35  (1996).

\bibitem{AkHome}
S.~V. Aksenov, internet homepage:\\ {\tt http://aksenov.freeshell.org/}.

\bibitem{JeHome}
See the URL {\tt http://tqd1.physik.uni-freiburg.de/\~{}ulj}.

\bibitem{Ba1994tech}
D. H. Bailey, {\it A Fortran-90 based multiprecision system}, NASA Ames Tech.
  Rep. RNR-94-013.

\bibitem{BaBoGi1994}
D.~H. Bailey, J.~M. Borwein, and R. Girgensohn, Exp. Math. {\bf 3},  17
  (1994).

\bibitem{Ba1990tech}
D. H. Bailey, {\it A portable high performance multiprecision package}, NASA
  Ames Tech. Rep. RNR-90-022.

\bibitem{Ba1993}
D.~H. Bailey, ACM Trans. Math. Soft. {\bf 19},  288  (1993).

\bibitem{Be1999master}
J. Becher, {\em Master Thesis: Numerical Convergence Acceleration Techniques
  and Their Application in Theoretical Physics (in German).\\ {\rm Available at
  {\tt http://www.physik.tu-dresden.de/publik/diplom.htm\#Dipl\_{}ITP}}\\}
  (Technische Universit\"{a}t Dresden, 1999, unpublished).

\bibitem{DeKePaGl1984}
J.~P. Dempsey, L.~M. Keer, N.~B. Patel, and M.~L. Glasser, ASME J. Appl. Mech.
  {\bf 51},  324  (1984).

\bibitem{Ma1997}
D.~A. Macdonald, Math. Comput. {\bf 66},  1619  (1997).

\bibitem{BaGa1995}
P. Baratella and R. Gabutti, J. Comput. Appl. Math. {\bf 62},  181  (1995).

\bibitem{BoDe1992}
J. Boersma and J.~P. Dempsey, Math. Comput. {\bf 59},  157  (1992).

\bibitem{Ga1991}
W. Gautschi, Math. Comput. {\bf 57},  325  (1991).

\bibitem{DeLiDe1990}
K.~M. Dempsey, D. Liu, and J.~P. Dempsey, Math. Comput. {\bf 55},  693  (1990).

\bibitem{ApKa1926}
P. Appell and J.~K. de~Feriet, {\em Fonctions Hyperg\'{e}om\'{e}triques et
  Hypersph\'{e}riques Polyn\^{o}mes d'Hermite} (Gauthiers-Villars, Paris,
  1926).

\bibitem{CoGaMi2001}
F.~D. Colavecchia, G. Gasaneo, and J.~E. Miraglia, Comput. Phys. Commun. {\bf
  138},  29  (2001).

\bibitem{BuCh1940}
J.~L. Burchnall and T.~W. Chaundy, Quart. J. Math. (Oxford) {\bf 11},  248
  (1940).

\bibitem{BuCh1941}
J.~L. Burchnall and T.~W. Chaundy, Quart. J. Math. (Oxford) {\bf 12},  112
  (1941).

\bibitem{LGZJ1990}
J.~C. LeGuillou and J. Zinn-Justin, {\em Large-Order Behaviour of Perturbation
  Theory} (North-Holland, Amsterdam, 1990).

\bibitem{St1913}
J. Stark, Berl. Ber. {\bf 1913},  932  (1913).

\bibitem{Su1913}
A. LoSurdo, Atti R. Accad. Lincei {\bf 22},  664  (1913).

\bibitem{Si1978}
H. Silverstone, Phys. Rev. A {\bf 18},  1853  (1978).

\bibitem{BeGrHaSi1979}
L. Benassi, V. Grecchi, E. Harrell, and B. Simon, Phys. Rev. Lett. {\bf 42},
  704  (1979).

\bibitem{DuHa1999}
G.~V. Dunne and T.~M. Hall, Phys. Rev. D {\bf 60},  065002  (1999).

\bibitem{SiKo1979}
H.~S. Silverstone and P.~M. Koch, J. Phys. B {\bf 12},  L537  (1979).

\bibitem{Op1928}
J.~R. Oppenheimer, Phys. Rev. {\bf 31},  66  (1928).

\bibitem{Al1969}
M.~H. Alexander, Phys. Rev. {\bf 178},  34  (1969).

\bibitem{HeInBr1974}
M. Hehenberger, H.~V. McIntosh, and E. Br\"{a}ndas, Phys. Rev. A {\bf 10},
  1494  (1974).

\bibitem{GuNi1975}
N.~A. Gushina and V.~K. Nikulin, Chem. Phys. {\bf 10},  23  (1975).

\bibitem{FrBr1975}
P. Froelich and E. Br\"{a}ndas, Phys. Rev. A {\bf 12},  1  (1975).

\bibitem{DaKo1976}
R.~J. Damburg and V.~V. Kolosov, J. Phys. B {\bf 9},  3149  (1976).

\bibitem{YaTaSi1977}
T. Yamabe, A. Tachibana, and H.~J. Silverstone, Phys. Rev. A {\bf 16},  877
  (1977).

\bibitem{DaKo1978}
R.~J. Damburg and V.~V. Kolosov, J. Phys. B {\bf 11},  1921  (1978).

\bibitem{GrGr1978b}
S. Graffi and V. Grecchi, Commun. Math. Phys. {\bf 62},  83  (1978).

\bibitem{HeSi1978}
I.~W. Herbst and B. Simon, Phys. Rev. Lett. {\bf 41},  67  (1978).

\bibitem{BeGrHaSi1979err}
L. Benassi, V. Grecchi, E. Harrell, and B. Simon, Phys. Rev. Lett. {\bf 42},
  1430 (Erratum)  (1979).

\bibitem{DaKo1979}
R.~J. Damburg and V.~V. Kolosov, J. Phys. B {\bf 12},  2637  (1979).

\bibitem{SiAdCiOt1979}
H. Silverstone, B.~G. Adams, J. $\check{\rm C}{\rm i}\check{\rm z}$ek, and P.
  Otto, Phys. Rev. Lett. {\bf 43},  1498  (1979).

\bibitem{AvEtAl1979}
J.~E. Avron, B.~G. Adams, J. $\check{\rm C}{\rm i}\check{\rm z}$ek, M. Clay,
  M.~L. Glasser, P. Otto, J. Paldus, and E. Vrscay, Phys. Rev. Lett. {\bf 43},
  691  (1979).

\bibitem{AdEtAl1980}
B.~G. Adams, J.~E. Avron, J. $\check{\rm C}{\rm i}\check{\rm z}$ek, P. Otto, J.
  Paldus, R.~K. Moats, and H.~J. Silverstone, Phys. Rev. A {\bf 21},  1914
  (1980).

\bibitem{BeGr1980}
L. Benassi and V. Grecchi, J. Phys. B {\bf 13},  911  (1980).

\bibitem{LKBa1980a}
E. Luc-Koenig and A. Bachelier, J. Phys. B {\bf 13},  1743  (1980).

\bibitem{LKBa1980b}
E. Luc-Koenig and A. Bachelier, J. Phys. B {\bf 13},  1769  (1980).

\bibitem{SiHaGr1981}
H.~J. Silverstone, E. Harrell, and C. Grot, Phys. Rev. A {\bf 24},  1925
  (1981).

\bibitem{MaChRe1983}
A. Maquet, S.~I. Chu, and W.~P. Reinhardt, Phys. Rev. A {\bf 27},  2946
  (1983).

\bibitem{FaRe1983}
D. Farrelly and W.~P. Reinhardt, J. Phys. B {\bf 16},  2103  (1983).

\bibitem{GlNgYaNa1985}
W.~L. Glab, K. Ng, D. Yao, and M.~H. Nayfeh, Phys. Rev. A {\bf 31},  3677
  (1985).

\bibitem{Ko1987}
V.~V. Kolosov, J. Phys. B {\bf 20},  2359  (1987).

\bibitem{Ko1989}
V.~V. Kolosov, J. Phys. B {\bf 22},  833  (1989).

\bibitem{AlSi1989}
G. Alvarez and H.~J. Silverstone, Phys. Rev. Lett. {\bf 63},  1364  (1989).

\bibitem{AlDaSi1991}
G. Alvarez, R.~J. Damburg, and H.~J. Silverstone, Phys. Rev. A {\bf 44},  3060
  (1991).

\bibitem{Fe1992}
F.~M. Fern\'{a}ndez, J. Phys. A {\bf 25},  495  (1992).

\bibitem{NiTh1992}
C.~A. Nicolaides and S.~I. Themelis, Phys. Rev. A {\bf 45},  349  (1992).

\bibitem{AlSi1994}
G. Alvarez and H.~J. Silverstone, Phys. Rev. A {\bf 50},  4679  (1994).

\bibitem{ZaCiSk2000}
J. Zamastil, J. $\check{\rm C}{\rm i}\check{\rm z}$ek, and L. Sk\'{a}la, Phys.
  Rev. Lett. {\bf 84},  5683  (2000).

\bibitem{Ko1978}
P.~M. Koch, Phys. Rev. Lett. {\bf 41},  99  (1978).

\bibitem{BeEtAl1984}
T. Bergeman, C. Harvey, K.~B. Butterfield, H.~C. Bryant, D.~A. Clark, P.~A.~M.
  Gram, D. MacArthur, M. Davis, J.~B. Donahue, J. Dayton, and W.~W. Smith,
  Phys. Rev. Lett. {\bf 53},  775  (1984).

\bibitem{GlNa1985}
W.~L. Glab and M.~N. Nayfeh, Phys. Rev. A {\bf 31},  530  (1985).

\bibitem{RoWe1986}
H. Rottke and K.~H. Welge, Phys. Rev. A {\bf 33},  301  (1986).

\bibitem{SeZJ1979}
R. Seznec and J. Zinn-Justin, J. Math. Phys. {\bf 20},  1398  (1979).

\bibitem{LGZJ1983}
J.~C. LeGuillou and J. Zinn-Justin, Ann. Phys. (N.Y.) {\bf 147},  57  (1983).

\bibitem{GuKoSu1995}
R. Guida, K. Konishi, and H. Suzuki, Ann. Phys. (N.Y.) {\bf 241},  152  (1995).

\bibitem{GuKoSu1996}
R. Guida, K. Konishi, and H. Suzuki, Ann. Phys. (N.Y.) {\bf 249},  109  (1996).

\bibitem{Pi1999}
M. Pindor, e-print hep-th/9903151.

\bibitem{Raczka1991}
P.~A. R\c{a}czka, Phys. Rev. D {\bf 43},  R9  (1991).

\bibitem{ZJ1996}
J. Zinn-Justin, {\em Quantum Field Theory and Critical Phenomena}, 3rd ed.
  (Clarendon Press, Oxford, 1996).

\bibitem{Wi1966}
A. van Wijngaarden, B.I.T. {\bf 6},  66  (1966).

\bibitem{CaFi1999}
I. Caprini and J. Fischer, Phys. Rev. D {\bf 60},  054014  (1999).

\bibitem{CaFi2000}
I. Caprini and J. Fischer, Phys. Rev. D {\bf 62},  054007  (2000).

\bibitem{LGZJ1980}
J.~C. LeGuillou and J. Zinn-Justin, Phys. Rev. B {\bf 21},  3976  (1980).

\bibitem{BeLiPi1973}
V.~B. Berestetskii, E.~M. Lifshitz, and L.~P. Pitaevskii, {\em
  Quantenelektrodynamik (Band IVb der Lehrbuchreihe \"uber Theoretische Physik
  von L.~D.~Landau und E.~M.~Lifshitz)}, 1 ed. (Akademie-Verlag, Berlin, 1973).

\bibitem{GrRe1992}
W. Greiner and J. Reinhardt, {\em Quantum Electrodynamics} (Springer, New York,
  1992).

\bibitem{DiGi2000}
W. Dittrich and H. Gies, {\em Probing the Quantum Vacuum -- Tracts in Physics
  Vol. 166} (Springer, Berlin, Heidelberg, New York, 2000).

\bibitem{HeEu1936}
W. Heisenberg and H. Euler, Z. Phys. {\bf 98},  714  (1936).

\bibitem{We1936}
V. Weisskopf, Dan. Mat. Fys. Medd. {\bf 14},  1  (1936).

\bibitem{Sc1951}
J. Schwinger, Phys. Rev. {\bf 82},  664  (1951).

\bibitem{BBBB1970}
Z. Bia\l{}ynicka-Birula and I. Bia\l{}ynicki-Birula, Phys. Rev. D {\bf 2},
  2341  (1970).

\bibitem{DiRe1985}
W. Dittrich and M. Reuter, {\em Effective Lagrangians in Quantum
  Electrodynamics -- Lecture Notes in Physics Vol. 220} (Springer, Berlin,
  Heidelberg, New York, 1985).

\bibitem{Og1956}
V.~I. Ogievetsky, Proc. Acad. Sci. USSR {\bf 109},  919  (1956), [in Russian].

\bibitem{Li1976lett}
L.~N. Lipatov, Pis'ma Zh. \'{E}ksp. Teor. Fiz. {\bf 24},  179  (1976), [JETP
  Lett. {\bf 24}, 157 (1976)].

\bibitem{Li1976}
L.~N. Lipatov, Zh. \'{E}ksp. Teor. Fiz. {\bf 71},  2010  (1976), [JETP {\bf
  44}, No.~6 (1977)].

\bibitem{Li1977}
L.~N. Lipatov, Zh. \'{E}ksp. Teor. Fiz. {\bf 72},  411  (1977), [JETP {\bf 45},
  216 (1977)].

\bibitem{ItPaZu1977}
C. Itzykson, G. Parisi, and J.~B. Zuber, Phys. Rev. D {\bf 16},  996  (1977).

\bibitem{BaItZuPa1978}
R. Balian, C. Itzykson, J.~B. Zuber, and G. Parisi, Phys. Rev. D {\bf 17},
  1041  (1978).

\bibitem{BrKr2000}
D. Broadhurst and D. Kreimer, Phys. Lett. B {\bf 475},  63  (2000).

\bibitem{BrKr2001}
D. Broadhurst and D. Kreimer, Nucl. Phys. B {\bf 600},  403  (2001).

\bibitem{Bo1994}
L. $\mathrm{Boutet~de~Monvel}$ (Ed.), {\em M\'{e}thodes R\'{e}surgentes}
  (Hermann, Paris, 1994).

\bibitem{Ri1927}
L.~F. Richardson, Phil. Trans. Roy. Soc. London A {\bf 226},  229  (1927).

\bibitem{Dy1952}
F.~J. Dyson, Phys. Rev. {\bf 85},  32  (1952).

\end{thebibliography}
\end{document}